\documentclass[12pt]{article}
\usepackage{graphicx,psfrag,color}
\usepackage{rotating}
\usepackage{xr}
\renewcommand{\arraystretch}{1.25}

\def\bm#1{\mbox{\boldmath$#1$\unboldmath}}

\newcommand{\bff}[1]{{\bf {#1}}}

\newcommand{\PG}{\hat \Psi}

\usepackage{amsmath,amssymb}
\numberwithin{equation}{section}
\allowdisplaybreaks

\newcommand{\refI}[1]{I:\ref{#1}}

\newcounter{MBQ}

\newcounter{MBP}

\catcode`@=11
\newcount\@tempcntc
\def\@citex[#1]#2{\if@filesw\immediate\write\@auxout{\string\citation{#2}}\fi
  \@tempcnta\z@\@tempcntb\m@ne\def\@citea{}\@cite{\@for\@citeb:=#2\do
    {\@ifundefined
       {b@\@citeb}{\@citeo\@tempcntb\m@ne\@citea\def\@citea{,}{\bf ?}\@warning
       {Citation `\@citeb' on page \thepage \space undefined}}%
    {\setbox\z@\hbox{\global\@tempcntc0\csname b@\@citeb\endcsname\relax}%
     \ifnum\@tempcntc=\z@ \@citeo\@tempcntb\m@ne
       \@citea\def\@citea{,}\hbox{\csname b@\@citeb\endcsname}%
     \else
      \advance\@tempcntb\@ne
      \ifnum\@tempcntb=\@tempcntc
      \else\advance\@tempcntb\m@ne\@citeo
      \@tempcnta\@tempcntc\@tempcntb\@tempcntc\fi\fi}}\@citeo}{#1}}
\def\@citeo{\ifnum\@tempcnta>\@tempcntb\else\@citea\def\@citea{,}%
  \ifnum\@tempcnta=\@tempcntb\the\@tempcnta\else
   {\advance\@tempcnta\@ne\ifnum\@tempcnta=\@tempcntb \else \def\@citea{--}\fi
    \advance\@tempcnta\m@ne\the\@tempcnta\@citea\the\@tempcntb}\fi\fi}
\catcode`@=12

\begin{document}

\begin{titlepage}

\begin{flushright}
TUM-HEP-1525/24\\
September 08, 2024
\end{flushright}

\vskip1cm
\begin{center}
\Large\bf\boldmath Third-order correction to top-quark
pair production near threshold II. Potential contributions
\end{center}

\vspace{1cm}
\begin{center}
{\sc M.~Beneke}$^a$ and {\sc Y. Kiyo}$^b$\\[5mm]
  {\it $^a$ Physik Department T31, James-Franck-Stra\ss e~1,}\\
  {\it Technische Universit\"at M\"unchen, D--85748 Garching, Germany}\\[0.1cm]
  {\it $^b$ Department of Physics, Juntendo University,}\\[0.0cm]
  {\it Inzai, Chiba 270-1695, Japan}\\[0.3cm]
\end{center}

\vspace{1.6cm}
\begin{abstract}
\noindent 
We provide a detailed account of the methods and calculations for the 
third-order corrections to the S-wave Green function from heavy-quark 
potentials other than the Coulomb potential. The results of this 
paper are relevant to the top-antitop
threshold production process in next-to-next-to-next-to-leading (NNNLO)
order and to the determination of the bottom-quark mass from 
high-moment sum rules, and have been employed in corresponding previous
publications. Further to the third-order calculation, we discuss in 
detail three refinements necessary to obtain reliable third-order 
results for the top threshold: finite-width effects, pole resummation, 
and the implementation of the potential-subtracted mass scheme. A 
detailed numerical analysis of residual scale dependence and the 
size of various contributions to the top production cross section 
is provided. The S-wave energy levels and wave functions at the 
origin for heavy quarkonium states of arbitrary principal quantum number 
$n$ are collected in an appendix.
\end{abstract}
\end{titlepage}

\pagenumbering{roman}
\tableofcontents
\newpage
\pagenumbering{arabic}

\setcounter{section}{0}
\setcounter{figure}{0}
\setcounter{footnote}{0}

\section{Introduction}

The present paper constitutes the sequel to \cite{Beneke:2013jia}, 
henceforth referred to as (paper) I, in which we presented the 
framework and matching coefficients for the computation of the 
top-quark pair production cross section in the threshold 
region to third order in a systematic expansion in the 
strong coupling $\alpha_s$ and the small relative velocity $v$ of the top 
quarks. The presence of the strong Coulomb force requires a summation 
of certain parts of ordinary Feynman diagrams to all orders in 
perturbation theory, treating $\alpha_s/v$ as $O(1)$. This is 
best done in the framework of non-relativistic effective theories 
as discussed in paper~I. Once the hard and soft matching coefficients 
have been extracted, what remains to be done for the production 
of a top-quark pair through the vector coupling, which dominates 
the total cross section, is to compute 
the correlation function
\begin{equation}
\label{eq:GPNRQCDII}
G(E) =\frac{i}{2 N_c (d-1)} \int d^{d} x\, e^{iEx^0}\,
\langle 0| \,T(\,
[\chi^{\dag}\sigma^i\psi](x)\,
[\psi^{\dag}\sigma^i\chi](0))
|0\rangle_{| \rm PNRQCD}
\end{equation}
in potential non-relativistic QCD (PNRQCD), see (\refI{eq:GPNRQCD}). 
($N_c=3$ denotes the number 
of colours, and $d=4-2\epsilon$ the number of space-time dimensions in 
dimensional regularization.) The perturbation expansion of PNRQCD 
is rather different from the familiar free-field expansion, since the 
unperturbed Lagrangian includes the Coulomb interaction. As long as only 
instantaneous potential interactions are involved, PNRQCD perturbation
theory is similar to perturbation theory in quantum mechanics with 
a non-trivial unperturbed Hamiltonian $H_0$.

The expansion to third order is expressed in the form
\begin{equation}
\label{greenexpandII}
G(E) = G_0(E) + \delta_1 G(E) + \delta_2 G(E) + \delta_3 G(E) 
+ \ldots
\end{equation}
with $G_0(E) = \langle \bff{0}| \hat G_{0}(E) |\bff{0}\rangle$. 
Here we use quantum-mechanical operator notation whereby 
$\hat G_{0}(E)$ denotes the Green function operator of the unperturbed 
Hamiltonian. If $\delta V_n$ represents the sum of all $n$th order 
perturbation potentials, the third-order correction to the 
correlation function is given by
\begin{eqnarray}
\delta_3 G(E) &=& \langle \bff{0}| \hat G_0(E)i \delta V_1 
i\hat G_0(E) i \delta V_1 
i\hat G_0(E) i\delta V_1 i\hat G_0(E)|\bff{0}\rangle 
\nonumber\\ && +\, 
2 \langle \bff{0}| \hat G_0(E)i \delta V_1 
i\hat G_0(E) i\delta V_2 i\hat G_0(E)|\bff{0}\rangle + 
 \langle \bff{0}| \hat G_0(E)i\delta V_3 
i\hat G_0(E)|\bff{0}\rangle \nonumber\\ 
&&+ \, \delta^{us}G(E)\,,
\label{masterthirdorderII}
\end{eqnarray} 
see (\refI{masterthirdorder}). The main result of the present paper 
is the computation of the  the single and double potential insertions 
terms in the second line of this equation, which completes the third-order 
calculation since the other terms are already 
known~\cite{Beneke:2005hg,Beneke:2008cr}.

Although the single and double insertions appear less 
involved than the triple insertion of the first-order potential in the 
first line of (\ref{masterthirdorderII}), this is in
fact not so. The point is that $\delta V_1$ is the next-to-leading order 
(NLO) correction to the Coulomb potential and therefore 
the triple insertion contains only finite integrals, which can be converted 
to sums or done numerically. On the contrary, 
the higher-order potentials $\delta V_n$ are more singular, 
which leads to ultraviolet divergent integrals in the single 
and double insertion. Moreover, the 
potentials are themselves singular and correspond to regulated 
expressions. All calculations will be performed in dimensional 
regularization with $d=4-2\epsilon$, which is the only practical scheme for 
multi-loop computations of the hard and soft matching coefficients. 
But since $i\hat G_0(E)$ is available only in $d=4$ dimensions this requires 
a series of subtractions to isolate the divergent subgraphs from 
a finite remainder. These subtractions must be consistent with 
the computation of the matching coefficients and the ultrasoft 
contribution $\delta^{us}G(E)$ in dimensional regularization. 
Overcoming these difficulties is our main technical result and 
represents a highly non-trivial application of dimensional regularization.

Section~\ref{sec:insertions} constitutes the technical
core of the paper, where we calculate the cross section up to the third order 
in perturbation theory, the new result being the non-Coulomb potential 
contributions up to the third order. We present a fair amount of details, 
since the methods may be non-standard, of the calculation of the single, 
double and triple insertions of the various potentials. 
Further details and the corresponding third-order corrections 
to the bound state poles of the correlation function are delegated 
to appendices \ref{app:details} and \ref{sec:enpsicorr}. The literal 
expansion (\ref{greenexpandII}) does not provide an accurate 
representation of the cross section in the threshold region and 
several refinements are required. First, the top quark width $\Gamma_t$ 
must be accounted for, which in the present approximation simply amounts 
to the replacement $E\to E +i\,\Gamma_t$ at the very end of the 
calculation.\footnote{See the introduction 
of paper I for a discussion of the limits of this approximation and 
a systematic treatment.} Second, the top quark pole mass must be 
replaced by a mass parameter defined in a renormalization scheme that 
is less sensitive to infrared contributions than the pole scheme, 
since otherwise there would be large but spurious 
shifts of the peak location of the threshold cross section. Finally, 
the expansion (\ref{greenexpandII}) breaks down for small width when 
$E$ is near the location of the bound-state poles for zero-width. This 
problem can be cured by a procedure that we call ``pole resummation'' 
\cite{Beneke:1999qg}. The implementation of these necessary refinements 
are discussed in section~\ref{sec:refine}.

A preliminary result without computational details for the non-Coulomb potential 
contributions has been presented 
already in~\cite{Beneke:2008ec,Schuller:2008}. 
Meanwhile, the then missing three-loop correction $c_3$ to the 
hard matching coefficient of the non-relativistic vector 
current has become available \cite{Marquard:2014pea}, and this 
and the result of the present paper have been used to obtain 
a precise determination of the bottom-quark mass from high-moment 
sum rules \cite{Beneke:2014pta}, as well as a precise prediction 
of the top-antitop production cross section in the threshold region  
\cite{Beneke:2015kwa}. The results have also been made available 
in the code \texttt{QQbar\_threshold} based on 
the {\sc mathematica/C++} software~\cite{Beneke:2016kkb}.  
In section~\ref{sec:results} of the present paper we put all 
third-order results together and provide a numerical analysis of the 
top threshold cross section, which explains the findings of the 
short communication~\cite{Beneke:2015kwa} and the implementation and 
parameter choices made there. We perform a study of 
the residual scale dependence and of the 
size of various contributions to the top production cross section. 
We also compare different mass schemes (pole, PS and $\overline{\rm MS}$) 
and analyze the effect of pole resummation. 

We conclude in section~\ref{sec:conclusion}. In appendix~\ref{app:glossary}, 
for convenience, we provide a glossary of definitions, symbols and 
special functions that appear throughout the main text. Further appendices 
collect expressions for the $S$-wave quarkonium 
energy levels and wave functions at the origin, as well as further 
technical details of the calculation of potential insertions 
as already mentioned. 


\section{Calculation of the potential insertions}
\label{sec:insertions}

\subsection{Definitions}

Before proceeding to the calculation of potential insertions 
in the Coulomb background, we set up a notation for dealing with 
insertions of dimensionally regulated potentials with divergent 
coefficients. 

We recall from paper~I that the propagator is given by the 
Green function of the Hamiltonian including the lowest-order 
Coulomb potential:
\begin{eqnarray}
G_0(\bff{r},\bff{r}^\prime;E) &=& 
\langle \bff{r} |\hat{G}_0(E)|\bff{r}^\prime\rangle = 
\langle \bff{r}|\frac{1}{H_0 - E-i \epsilon} |\bff{r}^\prime\rangle.
\end{eqnarray}
The corresponding momentum-space Green function will be denoted 
by $\tilde G_0(\bff{p},\bff{p}^\prime;E)$.\footnote{Since we
 will switch back-and-forth between position and momentum space in 
this paper, we now use a tilde to indicate momentum space.}  
Some explicit expressions can be found in I, section~\ref{sec:CoulGreen}. 
A potential insertion then takes the form of an integral
\begin{equation}
\langle \bff{0}| \hat G_0(E) \delta V \hat G_0(E)|\bff{0}\rangle 
=  \int\prod_{i=1}^4 
\Bigg[\frac{d^{d-1}{\bf p}_i}{(2\pi)^{d-1}}\Bigg]
\tilde{G}_0({\bf p}_1,{\bf p}_2;E) \delta V(\bff{p}_2,\bff{p}_3)
\tilde{G}_0({\bf p}_3,{\bf p}_4;E)\,,
\label{examplesingleinsertion}
\end{equation}
with obvious generalization to multiple insertions of potentials. 
Two remarks should be made. First, we always work with spin-triplet and 
colour-singlet projected potentials. Hence, the spin-algebra and colour 
algebra on (\ref{eq:GPNRQCDII}) is trivial and yields a factor 
$2 N_c (d-1)$ that cancels the corresponding 
normalization factor. Second, the local, non-derivative production 
current in (\ref{eq:GPNRQCDII}) implies that the ``outer'' momentum 
arguments of an insertion ($\bff{p}_1$ and $\bff{p}_4$ in 
(\ref{examplesingleinsertion})) are simply integrated over 
all of momentum space, 
which sets the conjugate position argument to zero as required.

In an expression such as (\ref{examplesingleinsertion}) $\delta V$  
often has divergent coefficients, and also some of the momentum 
integrals may be divergent. The whole expression then has to be evaluated 
correctly in dimensional regularization in an expansion in 
$\epsilon$ including the finite part. In 
the following we introduce a notation that simplifies the organization 
of the calculation.

\subsubsection{Single insertions}

The typical potential is of the form 
\begin{equation}
\delta V(\bff{p},\bff{p}-\bff{q}) =\frac{1}{({\bf
q}^2)^x}\bigg(\frac{\mu^2}{{\bf q}^2}\bigg)^{\!a\epsilon} \!\!
\times w(\epsilon)\,,
\end{equation}
where $x,a$ are integers and $\epsilon=(4-d)/2$.
Stripping off the coefficient and keeping only the momentum dependence, 
we define the single-insertion function 
\begin{eqnarray}
I[x+a\epsilon]&=& \int\prod_{i=1}^4\Bigg[\frac{d^{d-1}{\bf
p}_i}{(2\pi)^{d-1}}\Bigg]\tilde{G}_C({\bf p}_1,{\bf
p}_2;E)\frac{1}{({\bf q}_{23}^2)^x}\bigg(\frac{\mu^2}{{\bf
q}_{23}^2}\bigg)^{\!a\epsilon}
 \tilde{G}_C({\bf p}_3,{\bf p}_4;E),
\label{singleinsertionI}
\end{eqnarray}
where ${\bf q}_{ij}={\bf p}_i-{\bf p}_j$ and $a$ is an integer.
Using this notation and the results of paper I,  
sections~\ref{sec:potentials} and \ref{sec:eqofmotion}, 
four types of single insertions are needed after applying the equation 
of motion relations:
\begin{equation}
I[1+a\epsilon], I[1/2+a\epsilon], I[a\epsilon], I[\delta]\,.
\end{equation}
Here $I[\delta]$ denotes the special case, where the potential  
$\mu^{2 a\epsilon}/
({\bf q}_{23}^2)^{x+a\epsilon}$ in (\ref{singleinsertionI}) is replaced 
by the ``contact potential'' $(2\pi)^{d-1}\delta^{(d-1)}(\bff{q}_{23})$.

Some of the potential insertions are multiplied by a divergent
coefficient function $w(\epsilon)$. This means that one should calculate
$I[x+a\epsilon]$ to order $\epsilon$. However, these
divergent coefficient functions always appear in conjunction with a 
counterterm with a slightly different momentum dependence, such that 
the potential expanded in $\epsilon$ is finite. We therefore consider 
the expression 
\begin{equation}
\frac{1}{({\bf q}^2)^x} 
\left[\bigg(\frac{{\mu }^2}{\bf{q}^2} \bigg)^{\!a \epsilon }\,
w(\epsilon)-\frac{w^{(1/\epsilon)}}{\epsilon}\right],
\label{wdiff}
\end{equation}
with 
\begin{equation}
w(\epsilon)=\frac{w^{(1/\epsilon)}}{\epsilon}+w+
w^{(\epsilon)}\epsilon+O(\epsilon^2)
\end{equation}
and define the corresponding counterterm-including single-insertion 
function as 
\begin{eqnarray}\label{eq:J-definition1}
J[x+a\epsilon;w(\epsilon)]&=&
\frac{1}{\epsilon}w^{(1/\epsilon)}\bigg(I[x+a\epsilon]-I[x]\bigg)
+\bigg(w+ w^{(\epsilon)}\epsilon\bigg)I[x+a\epsilon].
\end{eqnarray}
The advantage of this expression is that it avoids the need to 
calculate in $I[x+a\epsilon]$ those order $\epsilon$ terms, 
which are independent of $a$, since they drop out 
in the difference in brackets in the first term. These would indeed 
be difficult to obtain, since they depend on the unknown 
$O(\epsilon)$ term in the Coulomb Green function. 
The second term is multiplied by a finite 
series. Hence the $O(\epsilon)$ term of $I[x+a\epsilon]$ alone is 
indeed never required.

Note that in general the factor $({\mu}^2/\bff{q}^2)^{a \epsilon }$
in (\ref{wdiff}) 
cannot be expanded in $\epsilon$ before the integration over the 
momenta $\bff{p}_i$, since these integrations can be divergent.
In the case of an insertion of the Coulomb potential ($x=1$), however, the 
calculation can be simplified, because the imaginary part of this 
insertion is always finite. In this case the square bracket in 
(\ref{wdiff}) {\em can} be expanded and yields an expression of 
the form $w^{(c)}+w^{(L)} L_q+\cdots$ with
$L_q=\ln(\mu^2/\bff{q}^2)$. We then define 
\begin{eqnarray}
&&J^{(C)}[1;w^{(c)}+w^{(L)}
L_q+w^{(L^2)}L_q^2+w^{(L^3)}L_q^3+\ldots]=\int\prod_{i=1}^4
\Bigg[\frac{d^{3}{\bf
p}_i}{(2\pi)^{3}}\Bigg]\tilde{G}_0({\bf p}_1,{\bf p}_2;E)
\nonumber
\\  &&\hspace{0.6cm}\times\,\frac{1}{{\bf q}_{23}^2}\bigg(w^{(c)}+w^{(L)} \ln\frac{\mu^2}{{\bf q}_{23}^2}+w^{(L^2)} \ln^2\frac{\mu^2}{{\bf q}_{23}^2}+w^{(L^3)}
\ln^3\frac{\mu^2}{{\bf q}_{23}^2}+\ldots\bigg)
 \,\tilde{G}_0({\bf p}_3,{\bf p}_4;E)\,, \;\;\qquad
\end{eqnarray}
where the superscript ``$(C)$'' is added to mark that the
expanded Coulomb potential is used. 
In practice it will be convenient to generate the insertion of $n$ powers 
of logarithms from the $n$th derivative of $1/({\bff q}^2)^u$ with respect 
to $u$.

\subsubsection{Double and triple insertions}

The basic functions for multiple insertions are defined in an analogous 
way by 
\begin{eqnarray}
I[x+a\epsilon,y+b\epsilon,\cdots]&=&
\int\prod_{i=1}^6\Bigg[\frac{d^{d-1}{\bf
p}_i}{(2\pi)^{d-1}}\Bigg]\tilde{G}_0({\bf p}_1,{\bf
p}_2;E)\frac{1}{({\bf q}_{23}^2)^x}\bigg(\frac{\mu^2}{{\bf
q}_{23}^2}\bigg)^{\!a\epsilon}
 \tilde{G}_0({\bf p}_3,{\bf p}_4;E)
 \nonumber \\ 
&& \times\frac{1}{({\bf q}_{45}^2)^y}\bigg(\frac{\mu^2}{{\bf
q}_{45}^2}\bigg)^{\!b\epsilon}
 \tilde{G}_0({\bf p}_5,{\bf
p}_6;E) \cdots \, .
\end{eqnarray}
Since the non-Coulomb potentials arise first at the second order, 
any double insertion must contain at least once the NLO Coulomb 
potential. Hence, we encounter four types of double insertions, 
\begin{equation}
I[1+a\epsilon,1+\epsilon],I[1/2+\epsilon,1+\epsilon],I[0,1+\epsilon],I[\delta,1+\epsilon]\,,
\end{equation}
and the triple insertion
\begin{equation}
I[1+\epsilon,1+\epsilon,1+\epsilon]\,,
\end{equation}
of three NLO Coulomb potentials.

When the double insertion consists of a non-Coulomb and 
the NLO Coulomb potential, the coefficient $w(\epsilon)$ of the 
non-Coulomb potential has no $1/\epsilon$ pole, since it is tree-level. The 
counterterm therefore comes from charge renormalization in 
the Coulomb potential. Thus, the counterterm-including insertion 
function appropriate to the double insertion of a tree-level 
non-Coulomb and the NLO Coulomb potential (\refI{eq:vcoulombNLO}) 
is defined as 
\begin{eqnarray}
J[x+a\epsilon,1+\epsilon;w(\epsilon)]&=&
\frac{w(\epsilon)\beta_0}{\epsilon}\, \bigg(I[x+a\epsilon,1+\epsilon] - 
I[x+a\epsilon,1]\bigg) 
\nonumber\\ 
&& + \,w(\epsilon) a_1(\epsilon)I[x+a\epsilon,1+\epsilon]\,.
\label{eq:defJdouble}
\end{eqnarray}
Here $w(\epsilon)$ is the coefficient of the potential with power 
$x+a\epsilon$ of momentum transfer squared. 

In the case of only Coulomb potential insertions, 
the calculation can again be simplified, because the
imaginary part of these insertions is always finite. Hence, one does
not need the $O(\epsilon)$ dependence of the potential, and can use
instead the finite part of the $\epsilon$ expanded coefficient
function. We then define 
\begin{eqnarray}
&& \hspace*{-0.2cm} 
J^{(C)}[1,1;w^{(c)}+w^{(L)}
L_q+w^{(L^2)}L_q^2]=\int\prod_{i=1}^6\Bigg[\frac{d^{3}{\bf
p}_i}{(2\pi)^3}\Bigg]\tilde{G}_0({\bf p}_1,{\bf
p}_2;E)\frac{1}{{\bf q}_{23}^2}\bigg(a_1+\beta_0
\ln\frac{\mu^2}{{\bf q}_{23}^2}\bigg)\nonumber
\\  && \hspace{1cm}\times\,\tilde{G}_0({\bf p}_3,{\bf
p}_4;E)\frac{1}{{\bf q}_{45}^2}\bigg(w^{(c)}+w^{(L)}
\ln\frac{\mu^2}{{\bf q}_{45}^2}+w^{(L^2)} \ln^2\frac{\mu^2}{{\bf
q}_{45}^2}\bigg)
 \tilde{G}_0({\bf p}_5,{\bf
p}_6;E)\,,
\label{eq:doubleJc}
\end{eqnarray}
where again ${\bf q}_{ij}={\bf p}_i-{\bf p}_j$, to account for the double 
insertion of the NLO with the NNLO Coulomb potential.

At the third order, the triple insertion necessarily involves three 
NLO Coulomb potentials, and is finite. This leads to the definition
\begin{eqnarray}
&&J^{(C)}[1,1,1]=\int\prod_{i=1}^8\Bigg[\frac{d^{3}{\bf
p}_i}{(2\pi)^{3}}\Bigg]\tilde{G}_0({\bf p}_1,{\bf
p}_2;E)\frac{1}{{\bf q}_{23}^2}\bigg(a_1+\beta_0
\ln\frac{\mu^2}{{\bf q}_{23}^2}\bigg)
 \tilde{G}_0({\bf p}_3,{\bf
p}_4;E) \nonumber
\\  && \hspace{0.6cm} \times\,\frac{1}{{\bf q}_{45}^2}\bigg(a_1+\beta_0 \ln\frac{\mu^2}{{\bf
q}_{45}^2}\bigg)
 \tilde{G}_0({\bf p}_5,{\bf
p}_6;E)\frac{1}{{\bf q}_{67}^2}\bigg(a_1+\beta_0
\ln\frac{\mu^2}{{\bf q}_{67}^2}\bigg)\tilde{G}_0({\bf p}_7,{\bf
p}_8;E)\,.\qquad
\label{eq:tripledef}
\end{eqnarray}

\subsubsection{General remarks}

With these definitions of the $J$-functions we can 
express the final result only in terms of $J$-functions, even if the
coefficient of the potential is not divergent (by setting the
corresponding coefficient $w^{(1/\epsilon)}$ to zero in the argument of the
$J$-function), so the $I$-functions are only needed in intermediate
steps. Since the heavy-quark production cross section is given 
by the imaginary part of the correlation function, 
for simplicity, if not stated otherwise, {\em we omit terms in the 
results for the $J$-functions, which do not contribute to the 
imaginary part. 
An exception are terms of the form $E\times J[\ldots]$}, 
for which, since $E$ is complex, the real part of the $J$-function 
must also be determined. The $1/\epsilon$ poles in the computation of the 
potential insertion must cancel with poles in the hard matching 
coefficients which multiply the LO and NLO correlation function. 
Therefore, the divergent parts of the $J$-functions must be 
factorized such that they multiply the LO and NLO Green function 
as well. After the pole cancellation the  LO and NLO Green functions 
can be evaluated in four dimensions.

The insertion $J$-functions are analytic functions of $E$ or, 
equivalently the variable $\lambda=\alpha_s C_F/(2\sqrt{-E/m})$ 
with poles as $\lambda\rightarrow n$, which correspond to the 
S-wave Coulomb bound states in the correlation function.  To extract the 
third-order corrections to the energy levels and wave functions 
at the origin, we also need the singular terms of the 
$J$-functions as $\lambda\rightarrow n$. The notation for the 
singular part of $J[\ldots]$ and $J^{(C)}[\ldots]$ will be 
$\hat J[\ldots]$ and $\hat J^{(C)}[\ldots]$, respectively.

In the results several shorthand notations will be used. The logarithms of
$\lambda$ are written as $L_{\lambda}= -\frac{1}{2}\,
\ln\left(-4 m E/\mu^2\right) = \ln(\lambda\mu/(m
\alpha_s C_F))$, in the limit $\lambda\rightarrow n$ the corresponding
logs are written as $L_{n}=\ln(n\mu/(m \alpha_s C_F))$.  $\Psi(x)$
is the Euler Psi-function and $\Psi_i$ denotes its $i$th derivative. 
$\gamma_E=0.577216...$ is the 
Euler-Mascheroni number. This always appears together with the 
Psi-function and is therefore combined in the following way:
$\PG(x)=\gamma_E+\Psi(x)$. These and similar definitions are 
summarized in Appendix~\ref{app:glossary} for quick reference.

\subsection{Single insertions}

\subsubsection{Coulomb potential}
\label{sec:singleC}

We begin with the single insertion of the Coulomb potential, which 
appears at first, second and third order. However, in every order 
one more power of $L_q=\ln(\mu^2/\bff{q}^2)$ appears, which leads to 
increasingly complicated analytic expressions, as will be seen 
below. We will discuss the case of the Coulomb potential in some 
detail to exemplify the general methods of calculation.

\begin{figure}[t]
\begin{center}
\makebox[0cm]{ \scalebox{0.4}{\rotatebox{0}{
     \includegraphics{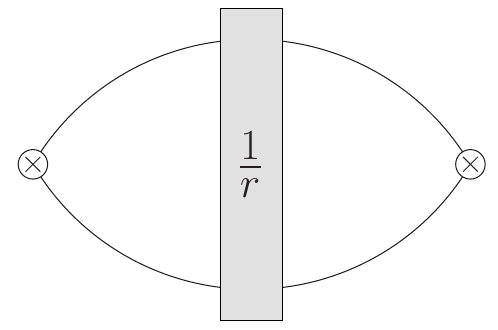}}}\hspace{1cm}
\scalebox{0.4}{\rotatebox{0}{
     \includegraphics{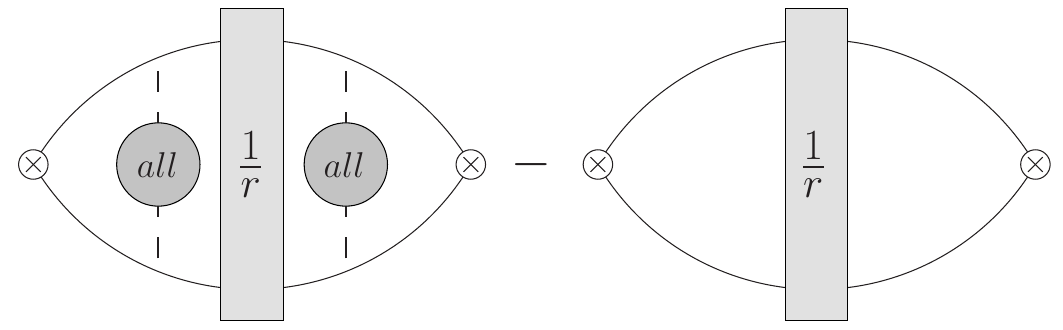}}}
     }
\end{center}
\hspace{3.5cm}(a)\hspace{5.9cm}(b)\hspace{3.05cm}
 \caption{\label{fig:Csubdiagrams} The two parts of the
Coulomb potential insertion which have to be separated.}
\end{figure}

The imaginary part of the Coulomb single insertion is finite, but the 
real part is ultraviolet divergent. Power counting shows that the divergence
arises only from an overall divergence in the two-loop diagram, which 
corresponds to the insertion with no further Coulomb exchanges in 
the Green functions $\tilde G_0$ to the left and right of the insertion. 
We therefore split the calculation into two parts, as shown in 
figure \ref{fig:Csubdiagrams}. Diagram a is divergent and must 
be done in $d$ dimensions. Diagram b is finite and we can set 
$d=4$ from the start. Note that it would in fact not be possible 
to compute diagram b in $d$ dimensions, since we do not know the 
Coulomb Green function $\tilde G_0$ in $d$ dimensions. On the other 
hand, if we split 
\begin{equation}
\tilde G_0 = \tilde G_0^{(\leq k ex)} + \tilde G_0^{(>k ex)},
\end{equation}
where the first term includes all ladder diagrams with up to $k$ 
rungs, then this first term {\em can} be computed in $d$ dimensions, 
since it corresponds to a {\em finite$\,$} sum of ordinary, 
dimensionally regulated Feynman diagrams. On the other hand, 
$\tilde G_0^{(>k ex)}$ is obtained through the solution of the 
Lippmann-Schwinger equation (\refI{eq:lippmann-schwinger}), 
which is only known in $d=4$ dimensions. For the insertion function, 
the separation into the two terms a and b implies writing  
\begin{eqnarray}
I[1+u]&=&(\mu^2)^u\!\int\prod_{i=1}^4\Bigg[\frac{d^{d-1}{\bf
p}_i}{(2\pi)^{d-1}}\Bigg]\frac{\tilde{G}_0^{(0ex)}({\bf p}_1,{\bf
p}_2;E)\tilde{G}_0^{(0ex)}({\bf p}_3,{\bf p}_4;E)}{({\bf
q}_{23}^2)^{1+u}}
\label{eq:i1u}
\\ &&
\nonumber \hspace{-1.7cm}
+\,(\mu^2)^u\!\int\prod_{i=1}^4\Bigg[\frac{d^{3}{\bf
p}_i}{(2\pi)^{3}}\Bigg]\frac{\tilde{G}_0({\bf p}_1,{\bf
p}_2;E)\tilde{G}_0({\bf p}_3,{\bf p}_4;E)-\tilde{G}_0^{(0ex)}({\bf
p}_1,{\bf p}_2;E)\tilde{G}_0^{(0ex)}({\bf p}_3,{\bf p}_4;E)}{({\bf
q}_{23}^2)^{1+u}}\,,
\end{eqnarray}
where we generalized the argument $1+a\epsilon$ of $I$ to $1+u$ for reasons 
that will become clear shortly.

Turning to the computation of diagram a, we first recall that 
the Coulomb Green function without gluon exchange is given by
\begin{eqnarray}
\tilde{G}_0^{(0ex)}({\bf p}_1,{\bf p}_2;E)=
\frac{m (2\pi)^{d-1}\delta^{(d-1)}({\bf p}_1-{\bf p}_2)}{{\bf p}_1^2-mE}\,,
\end{eqnarray}
see (\refI{eq:schwingerrep}). The contribution from part a is 
therefore 
\begin{equation}
I_a[1+u] = (\mu^2)^u\!\int\frac{d^{d-1}{\bf p}_1}{(2\pi)^{d-1}}
\frac{d^{d-1}{\bf p}_2}{(2\pi)^{d-1}}
\frac{m^2}{({\bf p}_1^2-mE)[(\bff{p}_1^2-\bff{p}_2^2)]^{1+u} ({\bf
p}_2^2-mE)}\,.
\label{eq:iscparta}
\end{equation}
Similar diagrams but with additional gluon exchanges will appear
in the following sections. The calculation of these multiple diagrams
in $d$ dimensions is usually straightforward with standard methods (Feynman
parameters, integration by parts and Mellin Barnes representations).
In the present case we find (using, for instance Feynman parameters) 
\begin{equation}
I_a[1+u] = \frac{m^2}{64\pi^3} \left(-\frac{m E}{\mu^2}\right)^{\!-u} 
\!\left(-\frac{m E}{4\pi}\right)^{\!-2\epsilon}\,
\frac{\Gamma(\frac{1}{2}-\epsilon-u)\Gamma(\frac{1}{2}+\epsilon+u)^2
\Gamma(u+2\epsilon)}{\Gamma(\frac{3}{2}-\epsilon)
\Gamma(1+2\epsilon+2 u)}\,.
\label{eq:ia1}
\end{equation}
This expression can be used to compute both, $J_a^{(C)}[1;\ldots]$ and 
$J_a[1+a\epsilon;w(\epsilon)]$. 
For $J_a^{(C)}[1;\ldots]$, we set $\epsilon$ to zero and use 
\begin{equation}
J^{(C)}[1; L_q^n] = \frac{d^n}{du^n} I[1+u]_{\big| u=0}.
\label{eq:JcIrel}
\end{equation}
This results in 
\begin{eqnarray}
&&J_a^{(C)}[1;w^{(c)}+w^{(L)} L_q+w^{(L^2)}L_q^2+w^{(L^3)}L_q^3]
=\frac{m^2}{16\pi^2}\Bigg[w^{(c)}L_\lambda+w^{(L)}L_\lambda^2
\nonumber \\
&&\hspace{1cm}+\,w^{(L^2)}\bigg(\pi^2L_\lambda+
\frac{4}{3}L_\lambda^3\bigg)+w^{(L^3)}
\bigg(3\pi^2L_\lambda^2+2L_\lambda^4\bigg)\Bigg],
\end{eqnarray}
where we have left out non-logarithmic terms, which do not contribute to the 
imaginary part of $J_a^{(C)}[1;\ldots]$. However, we also need the 
divergent real part since the application of the equation-of-motion 
relation (\refI{eompoverq}) to the $\bff{p}^2/(m^2 \bff{q}^2)$ potential 
leads to the product $E\times J_a[1+a\epsilon;w(\epsilon)]$ of 
complex energy (since $E$ contains the top-quark width) and the 
Coulomb insertion. To compute $J_a[1+a\epsilon;w(\epsilon)]$ 
we set $u=a\epsilon$ in (\ref{eq:ia1}) and use the $J$-function definition 
(\ref{eq:J-definition1}). Expanding in $\epsilon$, we obtain
\begin{eqnarray}\label{eq:Ja(1+aeps)}
&& J_a[1+a\epsilon;w(\epsilon)]
= \frac{m^2}{16\pi^2} \Bigg[
-\frac{a w^{(1/\epsilon)}}{4(2+a)\epsilon^2}+
\frac{w-aw^{(1/\epsilon)}}{2(2+a)\epsilon}
\\
&&\hspace{1cm}+\,\frac{1}{2+a}\bigg[
\frac{w^{(\epsilon)}}{2}+w\bigg(1+(2+a)L_\lambda\bigg)
+aw^{(1/\epsilon)}\bigg(\frac{\pi^2}{24}\,(11+6a)-1+
(2+a)L_\lambda^2\bigg)\bigg]\Bigg].
\nonumber
\end{eqnarray}
Note that the real constant terms are included here, since the expression 
is eventually multiplied by the complex quantity $E$. Furthermore, since the 
$\bff{p}_1$, $\bff{p}_2$ integrations in  (\ref{eq:iscparta}) are 
now $d$-dimensional, (\ref{eq:ia1}) 
must be multiplied by a factor $(\tilde \mu^2)^{2\epsilon}$ to 
obtain  $J_a[1+a\epsilon;w(\epsilon)]$, 
where $\tilde\mu^2 = \mu^2 e^{\gamma_E}/(4\pi)$ ensures 
that $\mu$ corresponds to the $\overline{\rm MS}$ subtraction scale. 
The $\tilde \mu^{2\epsilon}$ factors from the loop momentum integration 
measure can always be inferred from 
dimensional analysis and hence, in general, we do not write them   
explicitly, until an expression is expanded in $\epsilon$. 
Due to the inclusion of the counterterm, the pole part of 
(\ref{eq:Ja(1+aeps)}) is local as it should be. After multiplication 
with $E$, these poles give rise to finite-width divergences 
proportional to $\Gamma_t$ in the imaginary part of the correlation 
function.

By construction part b of figure \ref{fig:Csubdiagrams} is 
finite and can be computed in four dimensions. We always find it 
simpler to compute such finite expressions involving all-order 
summed Coulomb exchanges with coordinate- rather than momentum-space 
Coulomb Green functions. After introducing the coordinate 
representation of the Green functions in the second line of 
(\ref{eq:i1u}) and performing the momentum integrations, 
we arrive at
\begin{equation}
I_b[1+u] = \frac{(\mu^2)^u}{4\pi\Gamma(1+2u)\cos(\pi u)}
\int d^{3}{\bf r} \left(\bff{r}^2\right)^{u-\frac{1}{2}}
\left(G_0(0,\bff{r};E)^2-
G_0^{(0ex)}(0,\bff{r};E)^2\right).
\label{eq:pb1}
\end{equation} 
We now use the representation (\refI{eq:greenint}) 
for $G_0(0,\bff{r};E)$. The zero-exchange diagram that is subtracted 
in part b is simply the limit $\alpha_s\to 0$ of the full expression. 
Hence the combination of coordinate-space Green functions 
required for the evaluation of part b is 
\begin{eqnarray}
&& G_0(0,\bff{r};E)^2-G_0^{(0ex)}(0,\bff{r};E)^2=
\nonumber \\
&& \hspace*{1cm}
- \frac{m^3E}{4\pi^2}\int_{0}^{\infty}dt_{1}\int_{0}^{\infty}dt_{2}\,
e^{-2\sqrt{-mE}r(1+t_{1}+t_{2})}
\Bigg[\left(\frac{1+t_{1}}{t_{1}}\right)^{\!\lambda}
\left(\frac{1+t_{2}}{t_{2}}\right)^{\!\lambda}-1\Bigg]\quad
\end{eqnarray}
with $r=|\bff{r}|$. Inserting this into (\ref{eq:pb1}) and performing 
the trivial angular integration, we obtain 
\begin{eqnarray}
I_b[1+u] &=& - \frac{m^3E}{4\pi^2}
\frac{1}{\Gamma(1+2u)\cos\pi u} \,(\mu^2)^u
\int_0^\infty  dr \int_{0}^{\infty}dt_{1} \int_{0}^{\infty}dt_{2}
 \nonumber 
\\&&
e^{-2\sqrt{-mE}r(1+t_{1}+t_{2})} \, r^{2u+1}
\left(\left(\frac{1+t_{1}}{t_{1}}\right)^{\lambda}
\left(\frac{1+t_{2}}{t_{2}}\right)^{\lambda}-1\right).
\end{eqnarray}
The $r$ integral is elementary and we remain with a two-fold integral
over the parameters from the Green function representation. After
the substitution $z=(1+t_1+t_2)/(t_1t_2)$ and
$y=t_1/(1+t_1+t_2)$ the result is
\begin{eqnarray}
\label{eq:Coul-J}
I_b[1+u] &=& 
\frac{m^{2}}{16\pi^2}
\left(\frac{-4mE}{\mu^2}\right)^{\!-u} \frac{1+2 u}{\cos\pi u}\,j(u)
\end{eqnarray}
with 
\begin{equation}
j(u) = 
\int_{0}^{\infty}dz \left((1+z)^{\lambda} -1\right)z^{2 u-1}
\int_{0}^{1}dy \,\frac{y^{2u}(1-y)^{2u}}{(1+yz)^{2u+1}}
\label{eq:defj}
\end{equation}
The third-order Coulomb potential involves up to three powers of 
$L_q$. Recalling (\ref{eq:JcIrel}), we therefore need to compute 
up to three derivatives of $j(u)$ at $u=0$. The $n$th derivative 
is 
\begin{equation}
j^{(n)} = 
2^n \int_{0}^{\infty}dz \,\frac{(1+z)^{\lambda} -1}{z}
\int_{0}^{1}dy \,\frac{1}{1+yz} \ln^n\!\left(\frac{z y (1-y)}{1+y z}
\right).
\label{jder}
\end{equation}
These integrals can be done analytically, though with increasing 
effort for the higher derivatives. For some integrals we used the 
function HypExpInt from the HypExp program 
package \cite{Huber:2005yg,Huber:2007dx}. The results up to
the third derivative are implicit in the expressions for $j_{0,1,2}$ 
given below.

We describe here explicitly the computation of the first derivative. 
The $y$-integral in (\ref{jder}) can be expressed in terms of 
dilogarithms. Then, introducing the variable $t=z/(1+z)$, we find 
\begin{equation}
j^{(1)} = 
\int_{0}^1 dt \,\frac{1-(1-t)^{-\lambda}}{t^2} 
\left(4\,\mbox{Li}_2(t)+2 \ln t\ln (1-t)\right).
\label{jder11}
\end{equation}
Now define 
\begin{equation}
f(a,b) = \int_0^1dt\,t^a\left((1-t)^b-(1-t)^{b-\lambda}\right),
\end{equation}
which can be expressed in terms of $\Gamma$-functions, and use 
the series representation of $\mbox{Li}_2(t)$, to 
convert (\ref{jder11}) into 
\begin{eqnarray}
j^{(1)} &=& 2\frac{\partial^2}{\partial\delta\partial\varepsilon}
f(-2+\delta,\varepsilon)_{\big|\delta=\varepsilon=0} 
+4\sum_{n=1}^\infty\frac{f(-2+n,0)}{n^2}
\nonumber\\
&=& 
4\,{}_4F_3(1,1,1,1;2,2,1-\lambda;1)-\frac{\pi^2}{6}+\PG(1-\lambda)^2
-2\lambda\PG(1-\lambda)\Psi_1(1-\lambda)
\nonumber\\
&& +\,2\PG(1-\lambda)-(3+2\lambda)\Psi_1(1-\lambda) +\lambda\Psi_2(1-\lambda)
\,.
\label{jder12}
\end{eqnarray}

With these methods we can finally express part b of the single insertion 
of the third-order Coulomb potential as
\begin{eqnarray}
&& J_b^{(C)}[1;w^{(c)}+w^{(L)} L_q+w^{(L^2)}L_q^2+w^{(L^3)}L_q^3]
=\frac{m^2}{16\pi^2}\Bigg[w^{(c)}j_0+w^{(L)}\bigg(j_1+2j_0L_{\lambda}\bigg)
 \nonumber\\ 
&&\hspace{1cm}+w^{(L^2)}\bigg(4j_0L_\lambda^2+4j_1L_\lambda+j_2\bigg)
+w^{(L^3)}\bigg(8j_0L_{\lambda}^3+12j_1L_{\lambda}^2+
6j_2L_{\lambda}+j_3\bigg)\Bigg],\qquad
\label{eq:JbC}
\end{eqnarray}
with
\begin{eqnarray}
j_0 &=&j^{(0)} = -\PG(1-\lambda)+\lambda\Psi_1(1-\lambda)\,,
\\
j_1 &=& j^{(1)} +2 j^{(0)} = 4\,{}_4F_3(1,1,1,1;2,2,1-\lambda;1)
-\frac{\pi^2}{6}+\PG(1-\lambda)^2
 \nonumber\\  
&& -\,2\lambda\PG(1-\lambda)\Psi_1(1-\lambda)-3\Psi_1(1-\lambda)+
\lambda\Psi_2(1-\lambda)\,,
\\  \nonumber
j_2&=& j^{(2)}+4 j^{(1)} + \pi^2 j^{(0)} = 48+\frac{88}{3}\zeta_3
-8\Big[2+\PG(1-\lambda)\Big]\,_4F_3(1,1,1,1;2,2,1-\lambda;1)
\\ \nonumber &&-32\,{}_5F_4(1,1,1,1,1;2,2,2,1-\lambda;1)+
\PG(1-\lambda)\bigg[16(\lambda\zeta_3-1)-3\pi^2
\\ \nonumber
&&-\,4(1+2\lambda)\Psi_1(1-\lambda)+4\lambda\Psi_2(1-\lambda)+
4\lambda\Psi_1(1-\lambda)\PG(1-\lambda)-\frac{4}{3}\PG(1-\lambda)^2\bigg]
\\ \nonumber &&+\,\Psi_1(1-\lambda)\bigg[3\pi^2\lambda+32\lambda-8+
4\lambda\Psi_1(1-\lambda)\bigg]+\frac{32}{3}\Psi_2(1-\lambda)
-\frac{8}{3}\lambda\Psi_3(1-\lambda)
\\
&&+\,16\sum_{k=1}^{\infty}\frac{(k-\lambda)\PG(k-\lambda)}{k^3}
+8\sum_{k=1}^{\infty}\frac{\Gamma(k)\Gamma(1-\lambda)
\left[\PG(k-\lambda+1)-2\PG(k)\right]}{(k+1)^2\Gamma(k-\lambda+1)}\,.
\label{eq:j2}
\end{eqnarray}
The expressions $j_{0,1,2}$ have already been computed for the 
NNLO results of \cite{Beneke:1999qg}. 
The hypergeometric functions ${}_{q+1}F_q(1,1,1,\ldots;2,2,\ldots,
1-\lambda;1)$ 
can be expressed in terms of nested harmonic sums by generalizing 
from the case $q=3$  explained in the appendix of~\cite{Beneke:2011mq}, 
which can be useful to construct the analytic continuation in 
$\lambda$.

\begin{table}
\hspace{-0.2cm}
{\footnotesize
\begin{center}
\begin{tabular}{|r|r|r|r|r|r|r|r|r|r|}
  \hline
  $\mbox{Im}(j_3)$ & 0.0 & 0.1 & 0.2 & 0.3 & 0.4 & 0.5 & 0.6 & 0.7 & 0.8 \\
  \hline
  $0.05 i$ & -12.10 & -12.07 & -11.61 & -10.25 & -6.77 & 2.01 & 25.90 & 102.73 & 438.69 \\
  \hline
  $0.15 i$ & -36.48 & -36.59 & -35.67 & -32.59 & -24.76 & -5.99 & 39.81 & 157.27 & 457.34 \\
  \hline
  $0.25 i$ & -61.30 & -62.04 & -61.69 & -59.16 & -52.32 & -37.09 & -6.55 & 45.18 & 85.50 \\
  \hline
  $0.35 i$ & -86.60 & -88.49 & -89.72 & -89.72 & -87.72 & -83.10 & -77.42 & -81.46 & -131.85 \\
  \hline
  $0.45 i$ & -112.19 & -115.57 & -118.93 & -122.29 & -125.99 & -131.41 & -142.67 & -169.69 & -229.76 \\
  \hline
  $0.55 i$ & -137.72 & -142.67 & -148.20 & -154.68 & -162.95 & -174.81 & -193.71 & -225.15 & -274.84 \\
  \hline
  $0.65 i$ & -162.86 & -169.23 & -176.58 & -185.41 & -196.60 & -211.56 & -232.40 & -261.42 & -299.52 \\
  \hline
  $0.75 i$ & -187.31 & -194.83 & -203.54 & -213.90 & -226.58 & -242.52 & -262.73 & -287.87 & -317.31 \\
  \hline
  $0.85 i$ & -210.90 & -219.28 & -228.89 & -240.09 & -253.34 & -269.15 & -287.91 & -309.61 & -333.34 \\
  \hline
\end{tabular}
\end{center}
}
\caption{$\mbox{Im}(j_3)$ for several values of the real (horizontal) and 
imaginary part (vertical direction) of $\lambda$.}\label{tab:j3}
\end{table}

The new coefficient $j_3$ would cover several pages and is therefore not
given here in analytic form. Instead in table \ref{tab:j3} several
values for the imaginary part of $j_3$ as a function of $\lambda$
are given. The imaginary part is varied vertically and the real part
horizontally. Intermediate values can be obtained by interpolation. 
However, in the code that is used for the computation of the 
top cross section, the analytic expression for $j_3$ is used.

The infinite sums which appear in the formulae above as well as the sums in
$j_3$ can be evaluated numerically. At this point we mention some
issues which are relevant for these as well as all sums which appear 
in the other parts of the calculation. Single sums
which contain only $\Psi$-functions are done numerically without a
cutoff in Mathematica. Double sums and sums with $\Gamma$ functions
are calculated with a cutoff on the number of terms. 
This cutoff is chosen such that the
error of each individual sum is negligible in the threshold region.
For sums with poor convergence we evaluate the sums again up to some
cutoff, and then use an asymptotic expansion to the terms beyond the 
cutoff. The sum of the asymptotic expansion from the cutoff to infinity 
is then carried out analytically. 
The results have been checked with a numerical integration
in the region of $\lambda$ where (\ref{jder}) is convergent. 
We should note that it is often essential that the energy is 
complex, i.e. that the heavy quark has a finite width. When the 
width is taken to zero, $\lambda$ approaches the imaginary axis and 
some of the expressions above can no longer be evaluated in a meaningful 
way. Constructing a continuation of the sums applicable for stable 
quarks, for example, for the use of the Green function in 
non-relativistic bottom production sum rules, requires additional 
work \cite{Beneke:2014pta}. The sums appearing in the last line 
of (\ref{eq:j2}) and several others appearing below can in fact 
be expressed in terms of nested harmonic sums. These expressions 
as well as the corresponding ones for the  
${}_{q+1}F_q(1,1,1,\ldots;2,2,\ldots, 1-\lambda;1)$ function 
are used in the \texttt{QQbar\_threshold} code~\cite{Beneke:2016kkb}. 
The interested reader can find them in the source file 
\texttt{QQbarGridCalc.m} of the code that computes the grid for the 
\texttt{QQbar\_threshold} code.\footnote{Available from 
\texttt{https://qqbarthreshold.hepforge.org/downloads/}} 

We recall from the discussion of part a that we also need 
$J[1+a\epsilon;w(\epsilon)]$ in addition to $J^{(C)}[1;\ldots]$. 
Since the potential integrations in the Coulomb insertion are 
finite, we have the relation 
\begin{eqnarray}
J_b[1+\epsilon;\frac{w^{(1/\epsilon)}}{\epsilon}+w] &=&
J_b^{(C)}[1;w+w^{(1/\epsilon)} L_q]\,,
\label{eq:Jb(1+aeps)}
\end{eqnarray}
which completes the computation of the single Coulomb insertion.

\subsubsection{The $\lambda\rightarrow n$ limit of the single 
Coulomb insertion}
\label{sec:lambda2n}

To extract the correction to the energies and residues (wave functions 
at the origin squared) of the S-wave bound states, we have to determine 
the singular parts of the Laurent expansion of the correlation 
functions near the unperturbed bound state energies 
$E_n = -m(\alpha_s C_F)^2/(4 n^2)$, which corresponds to the 
expansion around positive integer $\lambda =n$. The exact correlation 
function has a single pole at the location of the exact bound state 
energy. Single insertions therefore have double poles, double insertions 
triple poles etc. at the location of the unperturbed 
bound state energy.

For the calculation of this limit we use different methods
depending on the structure of the expression. The simplest case 
involves $\Psi_i(1-\lambda)$ functions, which have poles at 
positive integer values of $\lambda$. Here we need the expansion 
of the $\Psi$-functions for 
negative integers, which is given by 
\begin{eqnarray}
\label{eq:psiexpansion}
\Psi_i(-n+\epsilon)&=&\frac{i!}{(-\epsilon)^{i+1}}
+\sum_{k=i}^{\infty}\frac{(-1)^k\epsilon^{k-i}}{(k-i)!}\,
\Big(\Psi_k(n+1)+((-1)^k-1)\,\Psi_k(1)\Big)\,.
\end{eqnarray}
The other, more complicated case refers to expressions such as in 
the last line of (\ref{eq:j2}) with a single sum involving 
$\Gamma$- and $\Psi_i$-functions. 
The parts with summations over the variable $k$ have additional
poles from expressions such as $\Psi(k-\lambda)$ in the sum over $k$. 
To extract these poles, we 
separate the terms into a) $k \leq n$ and b) $k>n$. There are 
no singularities from the terms b), while for the terms a) 
we apply (\ref{eq:psiexpansion}) to substitute the
$\Psi$-functions within the sum. The resulting sums
can be rewritten in terms of only 
harmonic and nested harmonic sums. In a few cases, 
namely parts d and e of the single insertion of the delta
potential discussed later in this section, the pole structure is 
more complicated and we did not succeed to write the Laurent 
expansion in terms of harmonic sums right away. However, we
can obtain the singularities for any specific value of $n$. By
assuming then a specific basis of harmonic sums for arbitrary $n$,
we obtain the rational coefficients in front of the assumed basis 
functions and check the result for further specific values of 
$n$.

After these general remarks we return to the single insertion of 
the Coulomb potential. Only part b of this insertion has poles in the
$\lambda\rightarrow n$ limit. Applying the procedure described above, 
we find 
\begin{eqnarray}
\hat{J}^{(C)}[1;w^{(c)}+w^{(L)} L_q+w^{(L^2)}L_q^2+w^{(L^3)}L_q^3]
&=&\frac{m^2}{16\pi^2}\Bigg\{\frac{n}{(n-\lambda)^2}\,
\bigg\{w^{(c)}+2w^{(L)}[L_n+S_1]
\nonumber\\ 
\nonumber&&\hspace{-6.5cm}
+\,w^{(L^2)}\bigg[\pi^2+4L_n^2-\frac{8}{n}S_1+8L_n S_1+4 S_1^2+4S_2\bigg]
+24w^{(L^3)}\bigg[ - 2 {S_{  2,1  }} + 2 {S_{  1,1,1  }}
\\ 
\nonumber&&\hspace{-6.5cm}+ \frac{{\pi }^2 {L_n}}{4} + \frac{{{L_n}}^3}{3} 
+ {{S_1}}^2  {L_n}  + {S_2}  {L_n}   +
  {S_1} \bigg( \frac{{\pi }^2}{4} - \frac{2 {L_n}}{n}+ {{L_n}}^2
  \bigg)\bigg]\bigg\}
  \\ 
\nonumber&&\hspace{-7cm}
-\frac{2}{(n-\lambda)}\,
\bigg\{w^{(L)}\bigg[1+\frac{n\pi^2}{3}+2S_1-2nS_2\bigg]+8w^{(L^2)}\bigg[-
n {S_3}  + n {S_{  2,1  }} - n {{\zeta }_3} + \frac{{L_n}}{2}
\\ 
\nonumber&&\hspace{-6.5cm} 
+\,\frac{n {\pi }^2 {L_n}}{6}+ {S_1} \bigg(
\frac{1}{2} + \frac{n {\pi }^2}{6} - n {S_2} + {L_n} \bigg)  + {S_2}
\bigg( 1 - n {L_n} \bigg)\bigg]
\\ \nonumber&&\hspace{-6.5cm} 
+\,24w^{(L^3)}\bigg[\frac{{\pi }^2}{8}
+ \frac{23 n {\pi }^4}{360} - \frac{{{S_1}}^3}{6} + n {{S_2}}^2
+ {{S_1}}^2 \bigg(
\frac{1}{2} + \frac{n {\pi }^2}{6} - n {S_2} \bigg) + 4 n {S_{3,1}}
\\ 
\nonumber&&\hspace{-6.5cm}  + \,3 {S_{ 1,1,1  }}-
  4 n {S_{  2,1,1  }} - 2 n {{\zeta }_3} {L_n} + \frac{{{L_n}}^2}{2} + \frac{n {\pi }^2 {{L_n}}^2}{6}+ {S_3} \bigg( - \frac{1}{3} - 2 n {L_n} \bigg)
\\
\nonumber&&\hspace{-6.5cm} + \,{S_{  2,1  }} \bigg( -2 + 2 n {L_n} \bigg) +
{S_2} \bigg( \frac{1}{2} - \frac{n {\pi }^2}{12} + 2 {L_n} - n {{L_n}}^2 \bigg)
  +
  {S_1} \bigg( -\frac{1}{n} - \frac{{\pi }^2}{12}
\\ 
&&\hspace{-6.5cm}-\,  2 n {S_3}+ 2 n {S_{  2,1  }} - 2 n {{\zeta }_3} + {L_n}
   + \frac{n {\pi }^2 {L_n}}{3} + {{L_n}}^2
  -{S_2} \bigg( \frac{1}{2}  + 2 n {L_n} \bigg)
     \bigg)\bigg]\bigg\}\Bigg\}\,,\qquad\quad
\end{eqnarray}
where $L_n=\ln\left(n\mu/(m C_F \alpha_s)\right)$. The harmonic
sums are defined as 
\begin{eqnarray}
&& S_a(n)=\sum_{k=1}^n \frac{1}{k^a}, \qquad 
S_{a,b}(n) \equiv \sum_{k=1}^n \frac{1}{ {k}^a }\, S_b(k), \qquad
S_{a,b,c}(n)\equiv \sum_{k=1}^{n}\frac{1}{k^a} \, S_{b,c}(k)\,.
\qquad
\end{eqnarray}
To shorten the notation we omit the argument $n$ (principal quantum number) 
of the harmonic sums and write $S_{a,b}\equiv S_{a,b}(n)$ etc.

\subsubsection{$1/r^2$ potential}
\label{sec:singleNA}

\begin{figure}[t]
\begin{center}
\makebox[0cm]{ \scalebox{0.4}{\rotatebox{0}{
     \includegraphics{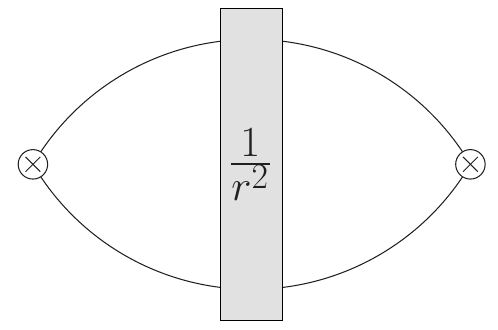}}}
\scalebox{0.4}{\rotatebox{0}{
     \includegraphics{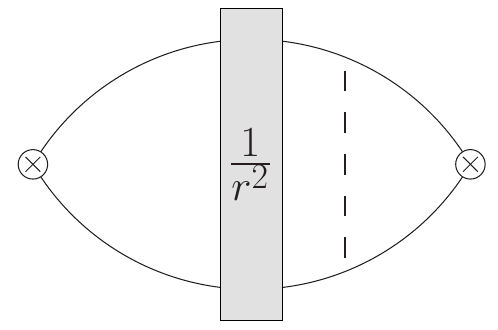}}}
\scalebox{0.4}{\rotatebox{0}{
     \includegraphics{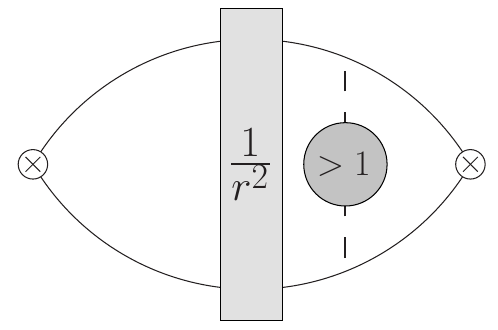}}}
\scalebox{0.4}{\rotatebox{0}{
     \includegraphics{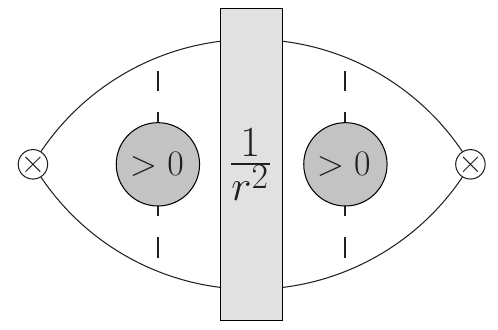}}}
     }
\end{center}
\hspace{2.35cm}(a)\hspace{3cm}(b)\hspace{3.05cm}(c)\hspace{3.05cm}(d)
 \caption{\label{fig:NAsubdiagrams} The four parts of the
$\frac{1}{r^2}$ potential insertion.}
\end{figure}

The single insertion of the $1/r^2$ potential, see (\refI{eq:vb1}), 
(\refI{eq:vb2}), generates ultraviolet $1/\epsilon$ poles from the integration 
over the potential loop momenta, which are related to the 
singularities in the dimensionally regulated hard current matching 
coefficients, and which have to be properly factorized. 
Power counting shows that the insertion has an overall
divergence coming from diagrams with less than two gluon exchanges,
and a vertex subdivergence, when there is no gluon exchange between 
the external vertex and the potential insertion. To accomplish the
correct factorization, we divide the integral into four different parts, 
according to their divergence structure:
\begin{eqnarray}
I[\frac{1}{2}+a\epsilon] &=& I_a[\frac{1}{2}+a\epsilon] + 2
I_b[\frac{1}{2}+a\epsilon] + 2 I_c[\frac{1}{2}+a\epsilon] +
I_d[\frac{1}{2}+a\epsilon],
\end{eqnarray}
Similar notation applies to the counterterm-including insertion 
function $J$. The diagrammatic representations of the parts 
$I_x[\frac{1}{2}+a\epsilon]$ are shown in 
figure~\ref{fig:NAsubdiagrams}. The first two diagrams have an overall
divergence and a divergence in the vertex subgraph(s) without gluon exchanges,
the third one has only a divergence in the left vertex subgraph and the
fourth one is finite. The calculation of the first two parts is
straightforward, since it involves only ordinary, solvable, dimensionally 
regularized two- and three-loop diagrams.

We shall now explain what we mean by ``properly factorizing'' the divergent 
parts on the example of part a. In analogy with the first line of 
(\ref{eq:i1u}) we define 
\begin{eqnarray}
&& I_a[\frac{1}{2}+u]=(\mu^2)^u\!\int\prod_{i=1}^4\Bigg[\frac{d^{d-1}{\bf
p}_i}{(2\pi)^{d-1}}\Bigg]\frac{\tilde{G}_0^{(0ex)}({\bf p}_1,{\bf
p}_2;E)\tilde{G}_0^{(0ex)}({\bf p}_3,{\bf p}_4;E)}{({\bf
q}_{23}^2)^{\frac{1}{2}+u}}
\nonumber\\
&& \hspace*{0.3cm} 
=\,\frac{m^2 \sqrt{-m E}}{64\pi^3} \left(-\frac{m E}{\mu^2}\right)^{\!-u} 
\!\left(-\frac{m E}{4\pi}\right)^{\!-2\epsilon}\,
\frac{\Gamma(1-\epsilon-u)\Gamma(\epsilon+u)^2
\Gamma(-\frac{1}{2}+u+2\epsilon)}{\Gamma(\frac{3}{2}-\epsilon)
\Gamma(2\epsilon+2 u)}\,,\hspace*{1.1cm}
\label{eq:i1halfu}
\end{eqnarray}
from which $J_a[\frac{1}{2}+a\epsilon;w(\epsilon)]$ follows according to 
the definition (\ref{eq:J-definition1}). The overall 
$\sqrt{-m E}$ factor indicates that the divergence of this 
integral persists in the imaginary part of the correlation function and 
is proportional to $G_0^{(0ex)}(0,0;E)$. Since the divergent part of 
the hard matching coefficient $c_v$ multiplies the $d$-dimensional 
correlation function, we must write the pole part 
of $J_a[\frac{1}{2}+a\epsilon;w(\epsilon)]$ in such a way that it 
multiplies the $d$-dimensional expression 
for $G_0^{(0ex)}(E)\equiv G_0^{(0ex)}(0,0;E)$, 
which is given by
\begin{eqnarray}
&& G_0^{(0ex)}(E) = (\tilde\mu^2)^\epsilon\!
\int\frac{d^{d-1}\bff{p}}{(2\pi)^{d-1}}
\frac{-1}{E -\frac{\bff{p}^2}{m}}  
= \frac{m\sqrt{-m E}}{8\pi^{3/2}} 
\left(-\frac{m E}{\mu^2}\right)^{\!-\epsilon} 
\!e^{\epsilon\gamma_E}\,\Gamma(-\frac{1}{2}+\epsilon)
\nonumber\\
&&\hspace*{1cm}
 =\,-\frac{m\sqrt{-m E}}{4\pi} \left[1+ 2(L_\lambda+1)\epsilon+
4\left(\frac{L_\lambda^2}{2}+L_\lambda+1+\frac{\pi^2}{16}\right)\epsilon^2
+O(\epsilon^3)\right].\qquad
\end{eqnarray} 
This results in 
\begin{eqnarray}
J_a[\frac{1}{2}+a\epsilon;w(\epsilon)] &=&-\frac{a m
w^{(1/\epsilon)}}{2\pi^2(a+1)\epsilon^2}\,G_0^{(0ex)}(E)
+\frac{m\Big(2a(\ln 2 -1) w^{(1/\epsilon)}+w\Big)}
{2\pi^2(a+1)\epsilon}\,G_0^{(0ex)}(E)
\nonumber \\
&&\hspace{-2.5cm}+ \,\frac{m^3C_F\alpha_s}{8\pi^3(a+1)\lambda} \Bigg[a
w^{(1/\epsilon)}\bigg(\ln^22-2\ln2-\frac{\pi^2}{24}(2a+5)-2a-2(a+1)L_{\lambda}
 \nonumber\\ 
&&\hspace{-2.5cm}
-\,(a+1)L_{\lambda}^2\bigg)+w\bigg(\ln
2-(a+1)L_{\lambda}-2-a\bigg)-\frac{1}{2}w^{(\epsilon)}\Bigg]
\label{eq:Jab}
\end{eqnarray}
for part a. We note that once again the pole part multiplying 
$G_0^{(0ex)}(E)$ as it should be. 
The three-loop contribution part b can be computed in the same way, where 
now one has to factorize the $d$-dimensional expression of 
the leading-order one-exchange Coulomb Green function $G_0^{(1ex)}(E)$. 
We find 
\begin{eqnarray}
2J_b[\frac{1}{2}+a\epsilon;w(\epsilon)] &=&-\frac{am
w^{(1/\epsilon)}}{ 2\pi^2(a+1)\epsilon^2}\,G_0^{(1ex)}(E)
+\frac{m\Big(2a(\ln 2 -1) w^{(1/\epsilon)}+w\Big)}
{2\pi^2(a+1)\epsilon}\,G_0^{(1ex)}(E) 
\nonumber \\ 
&&\hspace{-2.5cm}
+\,\frac{m^3 C_F \alpha_s}
{8\pi^3(a+1)}\Bigg\{aw^{(1/\epsilon)}\Bigg[\bigg(2(2a+2\ln2-\ln^22)
-\frac{2a+1}{4}\pi^2\bigg)L_{\lambda}+2(a+1)L_{\lambda}^2
\nonumber \\
&&\hspace{-2.5cm} 
+\,\frac{2}{3}(a+1)L_{\lambda}^3\Bigg]+w\bigg[2(2+a-\ln2)L_{\lambda}+(a+1)L_{\lambda}^2\bigg]+w^{(\epsilon)}
L_{\lambda}\Bigg\}.
\end{eqnarray}

The third part of the $1/r^2$-potential is the most complicated one, 
since it contains an infinite sum of Coulomb exchanges to the right of 
the potential insertion, for which an explicit expression is known only 
in four dimensions, while there is a logarithmic subdivergence in the left 
vertex subgraph, which must be computed in $d$ dimensions. We therefore 
first isolate the divergent part of this subgraph, such that it 
multiplies the formal $d$-dimensional expression of $G_0^{(>1ex)}(E)$, from 
a finite remainder. The pole part multiplying $G_0^{(>1ex)}(E)$ does 
not need to be evaluated further, since it cancels with an expression 
of opposite sign from the hard matching coefficients. In the finite 
remainder we insert the explicit four-dimensional representation of  
$G_0^{(>1ex)}(E)$ and proceed with the calculation. The
details of this calculation are given in appendix~\ref{app:details1}. 
The result is
\begin{eqnarray}\label{eq:Jc} \nonumber
2J_c[\frac{1}{2}+a\epsilon;w(\epsilon)] &=& -\frac{am
w^{(1/\epsilon)}}{2\pi^2(a+1)\epsilon^2}\,G_0^{(>1ex)}(E)
+\frac{m\Big(2a(\ln 2 -1)
w^{(1/\epsilon)}+w\Big)} {2\pi^2(a+1)\epsilon}\,G_0^{(>1ex)}(E)
\\ \nonumber
&&\hspace{-2.5cm}+\,\frac{m^3C_F\alpha_s}{4\pi^3 }\Bigg\{a
w^{(1/\epsilon)} \Bigg[\PG(1-\lambda)\bigg(\Psi_1(1-\lambda)
-2L_{\lambda}+\frac{2}{\lambda}L_{\lambda}-L_{\lambda}^2+\frac{2}{\lambda}
\\ \nonumber 
&&\hspace{-2.5cm}+\,\frac{1}{8(1+a)}(-16a-(3+2a)\pi^2-16\ln2+8\ln^2
2)\bigg)-\frac{1}{3}\PG(1-\lambda)^3-\frac{1}{3}\Psi_2(1-\lambda)
\\ \nonumber 
&&\hspace{-2.5cm}+\,\bigg(\PG(1-\lambda)^2
-\Psi_1(1-\lambda)\bigg)\bigg(1-\frac{1}{\lambda}+ L_{\lambda}\bigg)
-\frac{\pi^2}{6\lambda}+\frac{\pi^2}{2}
L_{\lambda}\Bigg]
\\ \nonumber 
&&\hspace{-2.5cm}+\,w\Bigg[\PG(1-\lambda)\bigg(\frac{1}{\lambda}
-\frac{2+a-\ln2}{(1+a)}-L_{\lambda}\bigg)+
\frac{1}{2}\bigg(\PG(1-\lambda)^2-\Psi_1(1-\lambda)\bigg)\Bigg]
\\
&&\hspace{-2.5cm}-\,\frac{w^{(\epsilon)}}{2(1+a)}\PG(1-\lambda)\Bigg\}\,.
\end{eqnarray}
The last part d is finite, so it can be done in four dimensions, where it 
is convenient to go to coordinate space. We then use on both sides of the 
potential insertion the integral representation (\refI{eq:greenint}) of the 
Coulomb Green function (with the zero-exchange term subtracted) 
and proceed in analogy to the Coulomb part. Again the
details of the calculation are presented in the appendix. The result
is 
\begin{eqnarray}\label{eq:Jd}
J_d[\frac{1}{2}+a\epsilon;w(\epsilon)]&=&
\frac{m^3 C_F \alpha_s}{8\pi^3\lambda}\Bigg\{-2a
w^{(1/\epsilon)}\Bigg[\bigg(\lambda\frac{\pi^2}{3}+\PG(1-\lambda)
-\lambda\Psi_1(1-\lambda)\bigg)L_{\lambda}
\nonumber\\ 
\nonumber &&\hspace{-2.5cm}
+\,\sum_{n=0}^{\infty}\big(\PG(1+n)
-\PG(1+n-\lambda)\big)\bigg(\PG(1+n)\PG(1+n-\lambda)-\PG(1+n-\lambda)^2
\\ 
&&\hspace{-2.5cm}-\,\Psi_1(1+n)+\Psi_1(1+n-\lambda)\bigg)\Bigg]
+w\bigg[-\lambda\frac{\pi^2}{3}-\PG(1-\lambda)+
\lambda\Psi_1(1-\lambda)\bigg]\Bigg\}\,.\qquad\quad
\end{eqnarray}

Note that the coefficients of the divergent parts of $J_a$, $2J_b$ and
$2J_c$ are the same. This is necessary for the sum of all divergent 
contributions to add to a term proportional to the full Green function
$G_0(E)=G_0^{(0ex)}(E)+G_0^{(1ex)}(E)+G_0^{(>1ex)}(E)$. That is, 
for the sum of all parts we have 
\begin{eqnarray}
J[\frac{1}{2}+a\epsilon;w(\epsilon)]&=&\left[
-\frac{am w^{(1/\epsilon)}}{
2\pi^2(a+1)\epsilon^2} +\frac{m\Big(2a(\ln 2 -1)
w^{(1/\epsilon)}+w\Big)}
{2\pi^2(a+1)\epsilon}\right] G_0(E)+O(\epsilon^0) 
\nonumber\\[-0.2cm]
\end{eqnarray}
as is required for cancelling the divergent part with the hard matching 
coefficient multiplying $G_0(E)$.

Finally we give the result for the singular terms in the limit
$\lambda\rightarrow n$. Only the all-order parts c and d can have such poles.
They were calculated with the methods described in section
\ref{sec:lambda2n} and read
\begin{eqnarray}\label{eq:polesJcd}
2\hat{J}_c[\frac{1}{2}+a\epsilon;w(\epsilon)]&=&\left[-\frac{am
w^{(1/\epsilon)}}{ 2\pi^2(a+1)\epsilon^2}+
\frac{m\Big(2a(\ln 2 -1) w^{(1/\epsilon)}+w\Big)}
{2\pi^2(a+1)\epsilon}\right]G_0^{\lambda\rightarrow n}(E)
 \nonumber\\ 
\nonumber
&&\hspace{-2.7cm}+\,\frac{m^3 C_F \alpha_s}{4\pi^3(n-\lambda)}\Bigg\{
 a w^{(1/\epsilon)}\Bigg[\frac{1}{(1+a)}\,
\bigg(2a+\frac{5\pi^2}{24}+\frac{\pi^2}{12}a+2\ln2-\ln^22\bigg)-
2S_1+S_1^2- S_2
\\&&\hspace{-2.7cm}
+\,2L_n(1-S_1)+L_n^2 \Bigg] +
w\bigg[L_n+\frac{(2+a-\ln2)}{(1+a)}-S_1\bigg] + w^{(\epsilon)}
\frac{1}{2(1+a)} \Bigg\}\,,
\\[0.1cm]
\hat{J}_d[\frac{1}{2}+a\epsilon;w(\epsilon)]&=&\frac{m^3 C_F
\alpha_s}{4\pi^3(n-\lambda)^2}\Bigg\{ a w^{(1/\epsilon)}
\bigg[1+L_n-S_1\bigg]+\frac{w}{2} \Bigg\}
 \nonumber \\ 
&&\hspace{-2.7cm}-\,\frac{m^3 C_F
\alpha_s}{4\pi^3(n-\lambda)}\Bigg\{ a w^{(1/\epsilon)}
\bigg[\frac{\pi^2}{3}+\frac{1}{n}(S_1-L_n)\bigg]-\frac{w}{2 n}
\Bigg\} \, .
\end{eqnarray}
Here $G_0^{\lambda\rightarrow n}(E)$ denotes the singular part of 
the $d$-dimensional leading-order Coulomb Green function. 
Recall that we suppress the argument $n$ of the harmonic sums.

\subsubsection{Delta potential}

The single insertion of the delta potential is the most complicated
part of the calculation. Naively one might think that a delta
potential is easy to calculate, because due to the momentum independence 
the left and right side of the potential insertion factorize. 
This is indeed the case at second order, since the tree-level coefficient 
function has no $1/\epsilon$ divergence and no $1/[\bff{q}^2]^\epsilon$ 
dependence. In this case, we can write 
\begin{eqnarray}
\int\prod_{i=1}^4\Bigg[\frac{d^{d-1}{\bf
p}_i}{(2\pi)^{d-1}}\Bigg]\tilde{G}_0({\bf p}_1,{\bf p}_2;E)
 \tilde{G}_0({\bf p}_3,{\bf
p}_4;E)&=&G_0(E)^2  \equiv I[0]\, ,
\end{eqnarray}
where the formal $d$-dimensional expression for $G_0(E)$ must be used. 
Then the required imaginary part of the insertion function is 
given by 
\begin{equation}
\mbox{Im}\,I[0] = 2\,\mbox{Re} \,G_0(E) \cdot 
\mbox{Im} \,G_0(E)\,.
\end{equation}
Recalling
\begin{eqnarray}
G_0(E)=\frac{m^2\,C_F\alpha_s}{4\pi} \bigg[ \frac{1}{4\epsilon}
+L_\lambda +\frac{1}{2} -\frac{1}{2\lambda} -\PG(1-\lambda) \bigg]
+O(\epsilon)\,,
\label{eq:zerodistancegreen}
\end{eqnarray}
we see that the divergent part already has the factorized form 
\begin{equation}
\mbox{Im}\,I[0]_{|\rm \,div} = \frac{m^2\,C_F\alpha_s}{16\pi\epsilon}
\cdot 2\,\mbox{Im} \,G_0(E)\,,
\end{equation}
which cancels with an infrared divergence in the hard two-loop matching 
coefficient, 
while the remaining piece of $\mbox{Im}\,I[0]$ can be evaluated in 
four dimensions.

\begin{figure}[t]
\begin{center}
\makebox[0cm]{ \scalebox{0.4}{\rotatebox{0}{
     \includegraphics{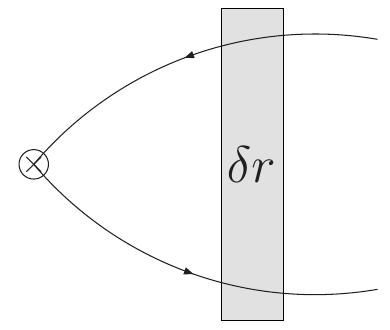}}}
\hspace{0.3cm} \scalebox{0.4}{\rotatebox{0}{
     \includegraphics{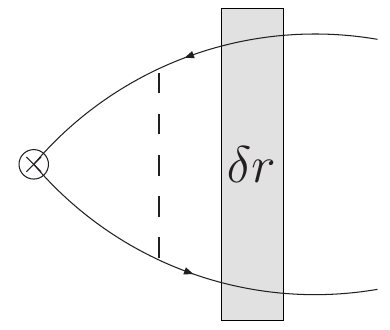}}}
\hspace{0.3cm}\scalebox{0.4}{\rotatebox{0}{
     \includegraphics{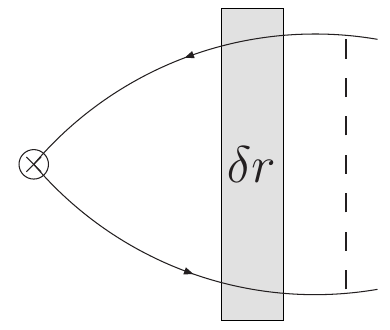}}}
     }
\end{center}
\hspace{4.8cm}(1)\hspace{2.8cm}(2)\hspace{2.7cm}(3)
 \caption{\label{fig:deltasinglediv} The three divergent vertex 
subdiagrams for the single insertion of a delta potential.}
\end{figure}

\begin{figure}[t]
\begin{center}
\makebox[0cm]{ \scalebox{0.4}{\rotatebox{0}{
     \includegraphics{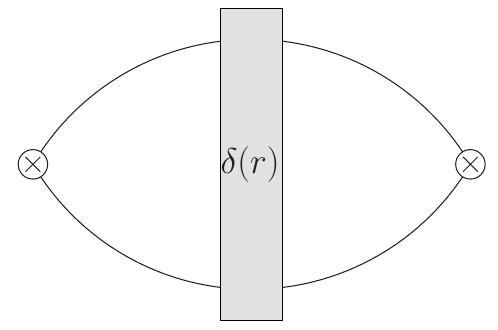}}}
\hspace{0.3cm} \scalebox{0.4}{\rotatebox{0}{
     \includegraphics{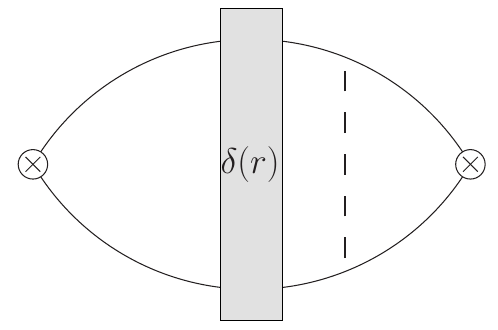}}}
\hspace{0.3cm}\scalebox{0.4}{\rotatebox{0}{
     \includegraphics{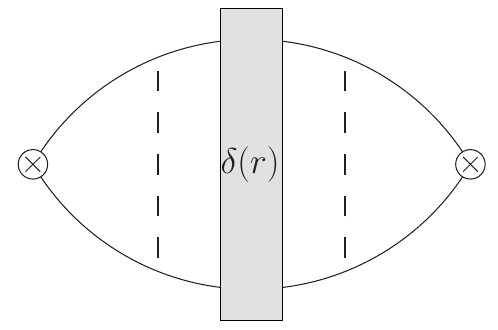}}}
     }
\end{center}
\hspace{3.75cm}(a)\hspace{3.5cm}(b)\hspace{3.35cm}(c)
\begin{center}
\makebox[0cm]{ \scalebox{0.4}{\rotatebox{0}{
     \includegraphics{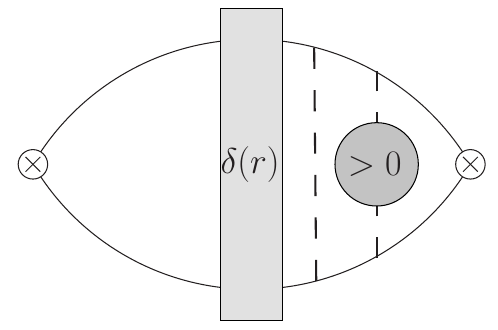}}}
\hspace{0.3cm} \scalebox{0.4}{\rotatebox{0}{
     \includegraphics{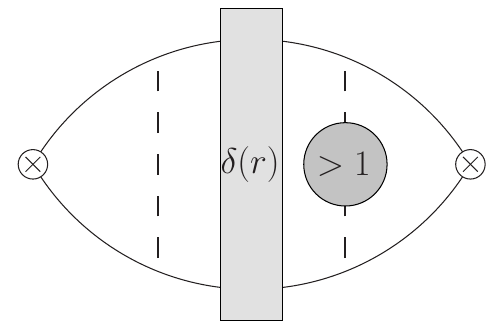}}}
\hspace{0.3cm}\scalebox{0.4}{\rotatebox{0}{
     \includegraphics{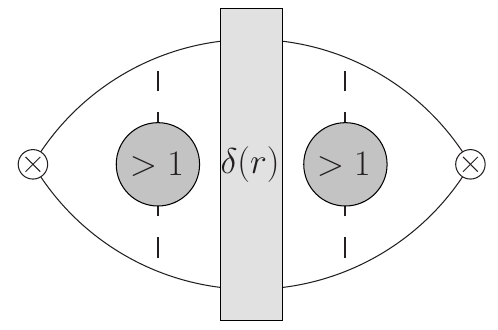}}}
     }
\end{center}
\hspace{3.75cm}(d)\hspace{3.5cm}(e)\hspace{3.35cm}(f)
 \caption{\label{fig:deltasubdiagrams} The delta potential single insertion 
is divided into six different parts.}
\end{figure}

However, the single insertion at NNNLO needs the one-loop corrected 
delta potential, which comes with the factor $(\mu^2/\bff{q}^2)^\epsilon$, 
a divergent coefficient function, and a counterterm. 
We therefore need the counterterm-including 
insertion function $\mbox{Im} \,J[\epsilon, w(\epsilon)]$. Due to the 
extra $\bff{q}^2$ dependence the calculation does not simplify as at second
order; instead is has to be performed along similar lines as
for the $1/r^2$ potential. The divergence structure is as follows. 
The diagram with no gluon exchange to the left and right of the 
potential insertion (part a below) has an overall divergence. Since the 
delta potential is more singular at short distances than the $1/r^2$ 
potential, the vertex subdiagram is now 
divergent with zero or one gluon exchanges in 
the vertex. The zero-exchange diagram has a linear subdivergence, the  
others are logarithmic. Note that the one gluon can be exchanged to the 
left or right of the potential insertion. The divergent vertex 
subgraphs are shown in figure~\ref{fig:deltasinglediv}. Therefore, we 
divide the insertion into six different parts, as shown in 
figure~\ref{fig:deltasubdiagrams}.

The first three parts are ordinary dimensionally regulated multi-loop 
integrals, which can be calculated with standard methods. We briefly discuss 
the method for the four-loop integral part c, which reads 
\begin{eqnarray}
I_c[a\epsilon]&=& \int\prod_{i=1}^{4}\Bigg[\frac{d^{d-1}{\bf
p}_i}{(2\pi)^{d-1}}\Bigg]\,
\frac{C_F^2 g_s^4
m^4}{({\bf p}_1^2-mE)({\bf p}_2^2-mE)({\bf p}_3^2-mE)({\bf
p}_4^2-mE)}
\nonumber\\ 
&& \times\,\frac{1}{({\bf p}_1-{\bf p}_2)^2[({\bf p}_2-{\bf
p}_3)^2]^{a\epsilon}({\bf p}_3-{\bf p}_4)^2}\, .
\label{eq:ideltac}
\end{eqnarray}
Inserting $\partial/\partial{\bf
p}_1\cdot ({\bf p}_1-{\bf p}_2)$ into the integrand (which makes the integral 
vanish) and performing an integration by parts yields the relation
\begin{eqnarray}
I_c[a\epsilon]&=& \frac{1}{d-4}\,(1^{+}5^{-}-1^{+}2^{-})\,I_c[a\epsilon]\, 
\end{eqnarray}
where $a^+$ ($a^-$) means that the power of the $a$th propagator 
in (\ref{eq:ideltac}) is raised (lowered) by one. In the term 
$1^{+}5^{-}I_c[a\epsilon]$ the gluon line to the left of the insertion is 
removed, which factorizes the remaining integral into a trivial one-loop 
integral over $\bff{p}_1$ and a three-loop integral that has the topology 
of part b (see figure~\ref{fig:deltasubdiagrams}). Applying a similar 
integration-by-parts (IBP) relation again, removes the gluon line to 
the right of the vertex. The resulting two-loop integral can be 
expressed in terms of Gamma functions.  In the other term 
$1^{+}2^{-}I_c[a\epsilon]$ the $\bff{p}_2$ integration can be 
trivially done in terms of Gamma functions, since only two massless 
propagators containing $\bff{p}_2$ remain. The result has  
the topology of part b, but with different propagator powers. 
Applying the IBP relation to this case works again, such that 
finally the entire four-loop diagram is expressed in terms of 
Gamma functions. Explicitly, the results for part a to c, setting $a=1$, are
\begin{eqnarray}
J_a[\epsilon;w(\epsilon)]&=&-\frac{E
m^3w^{(1/\epsilon)}}{48\pi^2\epsilon}+\frac{m^4 C_F^2
\alpha_s^2}{48\pi^2\lambda^2}\,\big(w^{(1/\epsilon)}(2+3L_\lambda)+w\big) \, ,
\label{eq:Jadelta}\\
2J_b[\epsilon;w(\epsilon)]&=&\left[-\frac{m^2C_F\alpha_sw^{(1/\epsilon)}}
{24\pi\epsilon^2} -
\frac{m^2C_F\alpha_s(2w^{(1/\epsilon)}-w)}{12\pi\epsilon}\right]G_0^{(0ex)}(E)
 \nonumber \\
&&\hspace{-1.5cm}+\, \frac{m^4C_F^2
\alpha_s^2}{96\pi^2\lambda}\Bigg\{w^{(1/\epsilon)}\bigg[19+\frac{7\pi^2}{12}+6L_\lambda-6L_\lambda^2\bigg]-w[1+6L_\lambda]-w^{(\epsilon)}\Bigg\}\,
,\\
J_c[\epsilon;w(\epsilon)]&=&\left[-\frac{m^2C_F\alpha_sw^{(1/\epsilon)}}
{24\pi\epsilon^2}-
\frac{m^2C_F\alpha_s(w^{(1/\epsilon)}-w)}{12\pi\epsilon}\right]G_0^{(1ex)}(E)
 \nonumber \\
&&\hspace{-1.5cm}+\,\frac{m^4
C_F^2
\alpha_s^2}{48\pi^2}\Bigg\{w^{(1/\epsilon)}\bigg[\bigg(\frac{5\pi^2}{12}-8\bigg)L_\lambda+2L_\lambda^3\bigg]
+w\bigg[2L_\lambda+3L_\lambda^2\bigg]+w^{(\epsilon)}L_\lambda\Bigg\}\,.
\qquad
\end{eqnarray}
In these expressions we omitted terms that do not contribute to the 
imaginary part. Note that $J_a[\epsilon;w_i(\epsilon)]$ has a divergent part
proportional to $E$ from the quadratic overall divergence, which produces 
a ``finite-width'' divergence $i\Gamma/\epsilon$, when the imaginary part 
is taken for the correlation function of unstable quarks. These divergences 
remain uncancelled in the pure QCD calculation and cancel instead with 
electroweak non-resonant terms as evidenced from the fact that 
$\Gamma$ is proportional to electroweak couplings. 

The parts d and e have both a divergence in the
left vertex subgraph including one gluon exchange from the Coulomb Green 
function and an all-order summation to the right of 
the insertion. The starting expression for part d is 
\begin{eqnarray}
I_d[a\epsilon]&=&\int\prod_{i=1}^{4}\Bigg[\frac{d^{d-1}{\bf
p}_{i}}{(2\pi)^{d-1}} \Bigg]\frac{C_F g^2 m^2}{({\bf p}^2_1-mE)
[({\bf p}_1-{\bf p}_2)^2]^{a\epsilon}({\bf p}^2_2-mE)({\bf
p}_2-{\bf p}_3)^2}
\nonumber\\
&&\times \,\tilde{G}_0^{(>0ex)}({\bf p}_3,{\bf p}_4;E)\,.
\label{eq:Ideltad}
\end{eqnarray}
We proceed as for part c of the $1/r^2$ potential and
calculate the divergent two-loop vertex integral over 
$\bff{p}_1$ and $\bff{p}_2$ in $d$ dimensions. 
We then factorize $G_0^{(>0ex)}(E)$, expand in $\epsilon$, 
and calculate the remaining part in $d=4$ dimensions. Similarly 
for part e, but in this case we must factorize  $G_0^{(>1ex)}(E)$,
see figure~\ref{fig:deltasubdiagrams}. The
details of the calculation are presented in appendix~\ref{app:details2}. 
The results read 
\begin{eqnarray}
2J_d[\epsilon;w(\epsilon)]&=&
-\frac{m^2C_F\alpha_sw^{(1/\epsilon)}}{12\pi\epsilon}\,G_0^{(>0ex)}(E)
 \nonumber\\ \nonumber
&&\hspace{-1.5cm}+\,\frac{m^4 C_F^2
\alpha_s^2}{8\pi^2}\,\Bigg\{w^{(1/\epsilon)}\bigg[C_{NInt}^{Log}[-C_F^2/(2
\lambda)^2]\ln(m^2 \alpha_s^2/\mu^2)-C_{NInt}^{const}[-C_F^2/(2
\lambda)^2]
\\ \nonumber
&&\hspace{-1.5cm}-\,\frac{1}{2}\bigg(\frac{11}{6}+3L_{\lambda}-\ln
2\bigg)(1+2L_{\lambda}-2\PG(1-\lambda))\bigg]
\\
&&\hspace{-1.5cm}+\,\frac{w}{6}\bigg[\frac{\pi^2}{2}+
\PG(1-\lambda)\bigg(\frac{3}{\lambda}+1\bigg)-L_{\lambda}\bigg]\Bigg\}\,,
\\
\label{eq:Je}
2J_e[\epsilon;w(\epsilon)]&=&
\left[-\frac{m^2C_F\alpha_sw^{(1/\epsilon)}}{24\pi\epsilon^2}-
\frac{m^2C_F\alpha_s(w^{(1/\epsilon)}-w)}{12\pi\epsilon}\right]G_0^{(>1ex)}(E)
 \nonumber\\ \nonumber 
&&\hspace{-1.5cm}+\frac{m^4 C_F^2
\alpha_s^2}{8\pi^2}\,\Bigg\{w^{(1/\epsilon)}\bigg[-\frac{\pi^2}{6}+\zeta_3
+\bigg(\frac{4}{3}-\frac{5\pi^2}{72}\bigg)\PG(1-\lambda)-L_\lambda^2\PG(1-\lambda)
\\ \nonumber &&\hspace{-1.5cm}
+\frac{1}{2\lambda}\bigg(\frac{\pi^2}{6}-2\PG(1-\lambda)
+\PG(1-\lambda)^2-\Psi_1(1-\lambda)\bigg)
-\lambda\sum_{n=1}^{\infty}\frac{\PG(n)^2+n \Psi_1(n)}{n(n-\lambda)}\bigg]
\\ &&\hspace{-1.5cm}
-\frac{1}{3}\bigg[w(1+3L_\lambda)+\frac{w^{(\epsilon)}}{2}\bigg]
\PG(1-\lambda)\Bigg\}\,,
\end{eqnarray}
where the symbols $C_{NInt}^X$ stand for integrals to be done numerically, 
as defined in the appendix. 
For $C_{NInt}^{Log}[-C_F^2/(2\lambda)^2]$ we find an analytic expression 
by making an ansatz for the function basis and fitting the rational 
coefficients with high accuracy, see~\eqref{eq:CNintLogAnalytic}. 
Finally, the last part f is finite. It can be
done by using the four-dimensional integral representation of the 
Green function with zero- and one-exchange subtracted. We obtain 
\begin{eqnarray}
J_f[\epsilon;w(\epsilon)]&=&
\frac{m^4C_F^2\alpha_s^2}{16\pi^2}\,\Bigg\{
2w^{(1/\epsilon)}\Bigg[\PG(1-\lambda)\bigg(\Psi_1(1-\lambda)-\PG(1-\lambda)^2
-\frac{\pi^2}{2}\bigg)
\nonumber\\
&&\hspace{-1.5cm}+\,\bigg(1-\frac{2}{\lambda}+L_\lambda\bigg)
\PG(1-\lambda)^2+\frac{1}{\lambda^2}
\sum_{k=1}^{\infty}\frac{1}{k}\,\bigg((\lambda-k)\Big(\PG(k)-\PG(k-\lambda)\Big)
+k\lambda\Psi_1(k)\bigg)^{\!2}\,\Bigg]
\nonumber\\
&&\hspace{-1.5cm}+\,w\,\PG(1-\lambda)^2\Bigg\}\,.
\end{eqnarray}
Note that the $1/\epsilon$ poles of the sum of all six parts precisely 
add to the full Coulomb Green function except for the finite-width divergence, 
which is necessary for a consistent pole cancellation as was explained in the 
previous section on the $1/r^2$ single insertion. 

The last three parts d, e, f 
have bound state poles for $\lambda \rightarrow n$. The
calculation of the poles for the parts d and e are more complicated
than for other parts and we had to apply the procedure of guessing 
and checking the form of the harmonic sums for general value of $n$ as 
explained in section~\ref{sec:lambda2n}. We find 
\begin{eqnarray}
2\hat{J}_d[\epsilon;w(\epsilon)]&=&
-\frac{m^2C_F\alpha_sw^{(1/\epsilon)}}{12\pi\epsilon}\,G_0^{\lambda
\rightarrow n}(E)
 \nonumber\\
&&\hspace{-2cm}-\,\frac{m^4 C_F^2 \alpha_s^2}
{8\pi^2(n-\lambda)}\,\Bigg\{w^{(1/\epsilon)}\bigg[\frac{11}{6}
+\frac{1}{2n}-S_1-\frac{S_1}{n}+\bigg(1+\frac{1}{n}\bigg)L_n\bigg]
+w\,\frac{3+n}{6n}\Bigg\}\,,
\\
2\hat{J}_e[\epsilon;w(\epsilon)]&=&
-\frac{m^2C_F\alpha_sw^{(1/\epsilon)}}{24\pi\epsilon^2}\,G_0^{\lambda
\rightarrow n}(E)-
\frac{m^2C_F\alpha_s(w^{(1/\epsilon)}-w)}{12\pi\epsilon}\,G_0^{\lambda
\rightarrow n}(E)
 \nonumber\\ 
\nonumber &&\hspace{-2cm}-\,\frac{m^4 C_F^2 \alpha_s^2}
{8\pi^2(n-\lambda)}\,\Bigg\{w^{(1/\epsilon)}
\bigg[\frac{4}{3}-\frac{5\pi^2}{72}+\frac{n\pi^2}{6}-\frac{S_1}{n}+
S_1^2-nS_2-L_n^2\bigg]
\\
&&\hspace{-2cm}-\,\frac{1}{3}w(1+3L_n)-\frac{1}{6}w^{(\epsilon)}\Bigg\}\,,
\\
\hat{J}_f[\epsilon;w(\epsilon)]&=&\frac{m^4 C_F^2 \alpha_s^2}
{8\pi^2}\Bigg\{
w\bigg[\frac{1}{2(n-\lambda)^2}+\frac{\frac{1}{n}-S_1}{n-\lambda}\bigg]+w^{(1/\epsilon)}\bigg[\frac{\frac{1}{n}-1+2L_n-2S_1}{2(n-\lambda)^2}
 \nonumber\\
&&\hspace{-2cm}+\,\frac{1}{n-\lambda}\bigg(\frac{1}{n^2}-\frac{1}{n} +
\frac{2}{n}L_n - \frac{\pi^2}{6} + \frac{n\pi^2}{6} - 2L_nS_1 -
\frac{3}{n}S_1 + 2S_1^2 -(1+n) S_2\bigg)\bigg] \Bigg\}.\,\;\;
\qquad
\end{eqnarray}

\subsubsection{Contact potential}

The momentum-space contact potential insertion arises from applying the 
equation-of-motion relation to the insertion of the kinetic energy 
correction, see I, section~\ref{sec:eqofmotion}. It is finite and 
multiplied by finite coefficient functions. Hence, we can directly 
evaluate the four-dimensional expression
\begin{equation}
I[\delta] = \int \prod_{i=1}^3\Bigg[\frac{d^{3}{\bf
p}_i}{(2\pi)^{3}}\Bigg] \,\tilde{G}_0({\bf p}_1,{\bf p}_2;E) 
\tilde{G}_0({\bf p}_2,{\bf p}_3;E)
= \int d^3\bff{r} \,G_0(0,\bff{r};E)^2 \,.
\label{eq:Ideldef}
\end{equation}
The $\bff{r}$-integral is reminiscent of (\ref{eq:pb1}) 
for $u=1/2$ and without the zero-exchange subtraction. Therefore 
we arrive at a parametric representation similar to $j(1/2)$ 
in (\ref{eq:defj}):
\begin{eqnarray}
I[\delta] &=&\frac{m^2}{2\pi}\left(-4 m E\right)^{-1/2} 
\int_0^{\infty}dz\,(1+z)^{\lambda}\int_0^1dy\,\frac{y(1-y)}{(1+y z)^2}
\nonumber\\
&=&\frac{m \lambda}{2\pi C_F \alpha_s}\int_0^{\infty}dz\,
(1+z)^{\lambda}\,\frac{(2+z)\ln(1+z)-2z}{z^3}
\nonumber \\ 
&=&\frac{m \lambda}{2\pi
C_F \alpha_s}\bigg(\frac{1}{2} + \lambda +\lambda^2\Psi_1(1-\lambda)
\bigg)\, .
\end{eqnarray}
The counterterm-including insertion function is identical, 
\begin{equation}
J[\delta;w] = w I[\delta]\,,
\end{equation}
and the singular part for $\lambda\to n$ reads: 
\begin{eqnarray}
\hat{J}[\delta;w]&=&\frac{m w}{2\pi C_F \alpha_s}\Bigg[
\frac{n^3}{(n-\lambda)^2} -\frac{3n^2}{(n-\lambda)} \Bigg]\, .
\end{eqnarray}
We note that $I[\delta]$ is actually related to the zero-distance 
Green function (\ref{eq:zerodistancegreen}) by 
\begin{equation}
I[\delta] = \frac{dG_0(E)}{dE}\,,
\label{eq:simpleIdelta}
\end{equation}
which provides a simpler way to derive the result. This relation 
follows from the fact that the product $\tilde{G}_0({\bf p}_1,{\bf p}_2;E) 
\tilde{G}_0({\bf p}_2,{\bf p}_3;E)$ integrated over $\bff{p}_2$ is 
the concatenation of two infinite sums of ladder diagrams which 
is a single ladder sum of diagrams with any one of the 
quark-antiquark propagators raised to the second power. Alternatively, 
we note that 
\begin{equation}
\frac{dG_0(E)}{dE} = \langle\bff{0}|\frac{1}{(H_0-E)^2}|\bff{0}
\rangle\,,
\end{equation}
where $|\bff{0}\rangle$ is the $\bff{r}=0$ position eigenstate 
and $H_0$ the unperturbed Hamiltonian, from which (\ref{eq:Ideldef}) 
follows by inserting a complete set of position eigenstates.

\subsection{Double insertions}

\subsubsection{Coulomb potential}
\label{sec:doublecoul}

The double insertion of two Coulomb potentials is finite and 
therefore we only need to calculate the insertion function 
$J^{(C)}[1,1;w^{(c)}+w^{(L)}]$ with logarithms 
defined in (\ref{eq:doubleJc}). Similar to part b of the 
single insertion of the Coulomb potential, we generate the 
logarithmic insertions from derivatives of the 
expression
\begin{equation}
I[1+u_1,1+u_2]=(\mu^2)^{u_1+u_2}\!
\int\prod_{i=1}^6\Bigg[\frac{d^{3}{\bf
p}_i}{(2\pi)^3}\Bigg]\frac{\tilde{G}_0({\bf p}_1,{\bf
p}_2;E)\tilde{G}_0({\bf p}_3,{\bf p}_4;E)
\tilde{G}_0({\bf p}_5,{\bf p}_5;E)}
{({\bf q}_{23}^2)^{1+u_1}({\bf q}_{45}^2)^{1+u_2}}\,,
\label{eq:i2u}
\end{equation}
The transformation to coordinate space results in 
\begin{eqnarray}
I[1+u_1,1+u_2]&=&
\frac{1}{4\pi \Gamma(1+2 u_1) \cos(\pi u_1)}\,
\frac{1}{4\pi \Gamma(1+2 u_2) \cos(\pi u_2)}\qquad
\nonumber\\
&&\hspace*{-2.5cm}\times\,(\mu^2)^{u_1+u_2}\!
\int d^3\bff{r}_1 d^3\bff{r}_2\,
G_0(0,\bff{r}_1;E)r_1^{2 u_1-1} G_0(\bff{r}_1,\bff{r}_2;E)
r_2^{2 u_2-1}G_0(\bff{r}_2,0;E)\,.\qquad
\end{eqnarray}
Now we observe that $G_0(0,\bff{r}_1;E)$ and $G_0(\bff{r}_2,0;E)$
depend only on $r_1=|\bff{r}_1|$ and $r_2$, respectively, which 
in turn implies that only the $l=0$ term in the 
partial-wave expansion (\refI{eq:gplexpand}) of 
$G_0(\bff{r}_1,\bff{r}_2;E)$ contributes. Thus, the angular 
integrals are trivial and give factors of $4\pi$. We then use the 
Laguerre representation (\refI{eq:gpartial}) of the $S$-wave 
Green function and the integral representation 
(\refI{eq:greenint}) for $G_0(0,\bff{r}_1;E)$ and
$G_0(\bff{r}_2,0;E)$. This factorizes the two $r$-integrals at the 
expense of a summation from the Laguerre representation 
such that 
\begin{equation}
I[1+u_1,1+u_2] = \frac{mp}{2\pi} \,\bigg(\frac{m}{8\pi p}\bigg)^{\!2}
\,\sum_{k=1}^\infty\frac{H(u_1,k) H(u_2,k)}{k (k-\lambda)}\,,
\end{equation}
where  $p=\sqrt{-m E}$, and\footnote{In \cite{Beneke:2005hg} we used a
slightly different definition for $H(u,k)$ with the argument $k$ 
shifted by one.} 
\begin{eqnarray}
H(u,k)&=&\frac{8\pi p}{m}
\frac{1}{\Gamma(1+2 u) \cos(\pi u)}\,
(\mu^2)^u\int_0^\infty dr\,r^{1+2 u}\,e^{-pr} L_{k-1}^{(1)}(2pr)\,G_0(r,0;E)
\nonumber\\
&=& \frac{1}{\Gamma(1+2u)\cos\pi
u}\left(\frac{\mu^2}{4p^2}\right)^{\!u}
 \int_0^{\infty}\!dt\left(\frac{1+t}{t}\right)^{\!\lambda}
\int_0^{\infty}\!ds \, e^{-(1+t)s}s^{2u+1}L_{k-1}^{(1)}(s) \nonumber
\\&=&
\frac{1}{\cos(\pi u)}
\left(\frac{\mu^2}{-4m E}\right)^{\!u}\,
\sum_{j=0}^{k-1}\frac{(-1)^j k!}{j!(k-1-j)!}
\frac{\Gamma(2+j+2u)\Gamma(1+j+2u)\Gamma(1-\lambda)}
{\Gamma(2+j)\Gamma(1+2u)\Gamma(2+j+2u-\lambda)}\,.
\nonumber\\[-0.2cm]
\label{eq:defH}
\end{eqnarray}
To obtain the last equality we use 
\begin{equation}
\int_0^{\infty}\!ds \, e^{-(1+t)s}s^{2u+1}L_{n}^{(1)}(s) = 
(n+1) \Gamma(2+2 u) (1+t)^{-2-2u} 
{}_2F_1\!\left(-n,2+2 u,2;\frac{1}{1+t}\right)\,,
\end{equation}
expand the hypergeometric function into its series 
representation and perform the $t$-integration term by term, see
 the appendix of \cite{Beneke:2005hg} for some more details.
Now we define the derivatives 
\begin{eqnarray}
H^{(n)}(k)&=& \frac{\partial^n}{\partial
u^n}\,H(u,k)_{|u=0}\,.
\end{eqnarray}
The zeroth and first derivative were already given in
\cite{Beneke:2005hg}, but since the NNLO Coulomb potentials 
involves $L_q^2$, we now also need the second derivative. Following
the method outlined in \cite{Beneke:2005hg}, we obtain 
\begin{eqnarray}
H^{(0)}(k)&=&\frac{k}{k-\lambda}\, ,
\\
H^{(1)}(k)&=&\frac{2}{k-\lambda}
\Bigg[kL_\lambda-k\PG(k-\lambda)+\lambda\Big(
\PG(1-\lambda)-\PG(k+1-\lambda)\Big)\Bigg]\, ,
\\ \nonumber
H^{(2)}(k)&=&\frac{4}{k-\lambda}\Bigg[k L_\lambda^2-\frac{2\lambda
L_\lambda}{k-\lambda}+\frac{5k\pi^2}{12}
+\frac{2\lambda}{(k-\lambda)^2}
\nonumber\\
&&-\,\frac{2k(k-1)_4F_3(1,1,1,2-k;2,2,2-\lambda;1)}{1-\lambda}
-k\Psi_1(1-\lambda)
\nonumber\\ 
\nonumber &&+\PG(1-\lambda)\bigg(-\frac{2\lambda}{k-\lambda}+
2\lambda L_\lambda+k\PG(1-\lambda)-2(k+\lambda)\PG(k-\lambda)\bigg)
\\ &&+\PG(k-\lambda)\bigg(\frac{4\lambda}{k-\lambda}
-2(k+\lambda)L_\lambda+2(k+\lambda)\PG(k-\lambda)\bigg)\Bigg]\, .
\label{eq:Hder}
\end{eqnarray}

At this point we can assemble the final result for the double-Coulomb
insertion function, which then reads:
\begin{eqnarray}
\label{eq:couldouble}
J^{(C)}[1,1;w^{(c)}+w^{(L)}
L_q+w^{(L^2)}L_q^2]&=&\frac{m^2\lambda}{64\pi^3 C_F
\alpha_s}\Bigg\{a_1
w^{(c)}\bigg[\Psi_1(1-\lambda)-\frac{\lambda}{2}\Psi_2(1-\lambda)\bigg]
\nonumber
\\ \nonumber &&\hspace{-6cm}+\,(a_1 w^{(L)}+w^{(c)}
\beta_0)\bigg[-\lambda\PG(1-\lambda)\Psi_2(1-\lambda)
-\frac{\lambda}{3}\Psi_3(1-\lambda)
-2\sum_{k=1}^{\infty}\frac{(k+\lambda)\PG(k-\lambda)}{(k-\lambda)^3}
\\ \nonumber
&&\hspace{-6cm}+\,L_\lambda\big(2\Psi_1(1-\lambda)
-\lambda\Psi_2(1-\lambda)\big)\bigg]+a_1
w^{(L^2)}\sum_{k=1}^{\infty}\frac{H^{(0)}(k)H^{(2)}(k)}{k(k-\lambda)}
\\  &&\hspace{-6cm}
+\,w^{(L)}
\beta_0\sum_{k=1}^{\infty}\frac{H^{(1)}(k)H^{(1)}(k)}{k(k-\lambda)}
+w^{(L^2)}\beta_0\sum_{k=1}^{\infty}\frac{H^{(1)}(k)H^{(2)}(k)}
{k(k-\lambda)}\Bigg\}\,.
\end{eqnarray}
The singular terms near $\lambda=n$ poles of this 
expression are:
\begin{eqnarray}
\hat{J}^{(C)}[1,1;w^{(c)}+w^{(L)}
L_q+w^{(L^2)}L_q^2]&=&\frac{m^2}{(4\pi)^3C_F\alpha_s}\,
\Bigg\{\frac{n^2}{(n-\lambda)^3}\,\Bigg[a_1 w^{(c)}
+2(a_1 w^{(L)}
\nonumber\\ 
\nonumber &&\hspace{-5.5cm}
+\,w^{(c)} \beta_0)(L_n+S_1)+4a_1 w^{(L^2)}\bigg(\frac{{\pi }^2}{4} + {{S_1}}^2 + {S_2} + {{L_n}}^2 + 2{S_1} \bigg( -\frac{1}{n} +  {L_n} \bigg)\bigg)
\\ \nonumber &&\hspace{-5.5cm} 
+\,4w^{(L)} \beta_0({{S_1}}^2 + 2 {S_1} {L_n} +
{{L_n}}^2)+8 w^{(L^2)}\beta_0\bigg({{S_1}}^3 + \frac{{\pi }^2
{L_n}}{4} + {S_2} {L_n} + {{L_n}}^3 
\\ \nonumber &&\hspace{-5.5cm} 
+ \,{S_1}^{\!2}\, \bigg(\! -\frac{2}{n}
+ 3 {L_n} \bigg)+ {S_1} \bigg( \frac{{\pi }^2}{4} + {S_2} - \frac{2
{L_n}}{n} + 3 {{L_n}}^2 \bigg)\bigg)\Bigg]
\\ \nonumber &&\hspace{-5.5cm} 
-\,\frac{n}{(n-\lambda)^2}\,\Bigg[a_1 w^{(c)}+2(a_1 w^{(L)}+w^{(c)}
\beta_0)(1+L_n+\frac{n\pi^2}{6}+2S_1-nS_2)
\\ \nonumber &&\hspace{-5.5cm} 
+\,4a_1 w^{(L^2)}\,\bigg(\frac{{\pi }^2}{4} + {{S_1}}^2 + 2 n {S_{  2,1  }} - 2 n {S_3}- 2 n {{\zeta }_3} + 2 {L_n} + \frac{n {\pi }^2}{3} {L_n} + {{L_n}}^2
\\
&&\hspace{-5.5cm}+ {S_1} \bigg( 2 - \frac{2}{n} + \frac{n {\pi }^2}{3}
- 2 n {S_2} + 4 {L_n} \bigg)  +
  {S_2} \bigg( 3 - 2 n {L_n} \bigg)\bigg)
\nonumber\\ 
\nonumber &&\hspace{-5.5cm} 
+\,4w^{(L)} \beta_0\,\bigg(\frac{{\pi }^2}{6} + 3 {{S_1}}^2 + n {S_3} - n {{\zeta }_3} + 2 {L_n}+ \frac{n {\pi }^2 {L_n}}{3} + {{L_n}}^2 
\\ \nonumber &&\hspace{-5.5cm} 
+ \,{S_1} \bigg( 2 - \frac{1}{n} + \frac{n {\pi }^2}{3} - 2 n {S_2} + 4 {L_n} \bigg)  +
  {S_2} \bigg( -1 - 2 n {L_n} \bigg)\bigg)
  \\ \nonumber &&\hspace{-5.5cm} +4w^{(L^2)}\beta_0\,\bigg(\frac{{\pi }^2}{2} + \frac{17 n {\pi }^4}{180} + 4 {{S_1}}^3 - 2 n {{S_2}}^2 - 4 {S_3} + 4 n {S_4} - 12 n {S_{  3,1  }} + \frac{7 {\pi }^2 {L_n}}{6}
  \\ \nonumber &&\hspace{-5.5cm}   + \,6 {{L_n}}^2+ n {\pi }^2
  {{L_n}}^2 + 2 {{L_n}}^3 + {{S_1}}^2 \bigg( 6 - \frac{6}{n} + n {\pi
  }^2 - 6 n {S_2} + 14 {L_n} \bigg)    
\\ \nonumber &&\hspace{-5.5cm}  + \, {{\zeta }_3} \bigg(\!-4 - 8 n {L_n} \bigg)
+ {S_{  2,1  }} \bigg( 4 + 4 n {L_n} \bigg)  +
  {S_2} \bigg( 2 + \frac{2}{n} + \frac{n {\pi }^2}{2} + 2 {L_n} - 6 n
  {{L_n}}^2 \bigg)     
\\ \nonumber &&\hspace{-5.5cm}  + \,{S_1}
   \bigg(\!- \frac{4}{n} + \frac{{\pi }^2}{3} + 4 n {S_{  2,1  }}
- 8 n {{\zeta }_3} + 12 {L_n} - \frac{8 {L_n}}{n} + 2 n {\pi }^2 {L_n}
+ 12 {{L_n}}^2 
\\ \nonumber &&\hspace{-5.5cm} + \,{S_2} \Big( 12 - 12 n {L_n} \Big)
   \bigg)\bigg)\Bigg]
\\ \nonumber &&\hspace{-5.5cm} +\,\frac{1}{(n-\lambda)}\,
\Bigg[(a_1 w^{(L)}+w^{(c)}
\beta_0)\bigg(1+\frac{\pi^2n}{3}+2S_1-2nS_2\bigg)  +4a_1
w^{(L^2)}\,\bigg(1 + \frac{n {\pi }^2}{3}
\\ \nonumber &&\hspace{-5.5cm} -\,2 n {S_3}  + 2 n {S_{  2,1  }}  - 2 n {{\zeta }_3} + {L_n} + \frac{n {\pi }^2}{3} {L_n}+
  {S_1} \bigg( 3 + \frac{n {\pi }^2}{3} - 2 n {S_2} + 2 {L_n} \bigg)
\\ \nonumber &&\hspace{-5.5cm}
+ \,{S_2} \bigg( 2 - 2 n - 2 n {L_n} \bigg)\bigg)
+4w^{(L)} \beta_0\,\bigg(1 + \frac{n {\pi }^2}{3} + \frac{n^2 {\pi
  }^4}{36} + 2 {{S_1}}^2 + n^2 {{S_2}}^2 + 6 n {S_3}   
\\ \nonumber&&\hspace{-5.5cm}
- \,5 n^2 {S_4} - 2 n {S_{  2,1  }} + 4 n^2 {S_{  3,1  }} 
- 2 n {{\zeta }_3} + {L_n} + \frac{n {\pi }^2 {L_n}}{3} 
  \\ \nonumber &&\hspace{-5.5cm}
+ \, {S_1} \bigg( 3 + n {\pi }^2 - 6 n {S_2} + 2 {L_n} \bigg)  
+ {S_2} \bigg(\!-2 n - \frac{n^2 {\pi }^2}{3} - 2 n {L_n} \bigg)\bigg)
 \\ \nonumber &&\hspace{-5.5cm} +\,8w^{(L^2)}\beta_0\bigg(\frac{11 {\pi }^2}{24} + \frac{73 n {\pi }^4}{360} + {{S_1}}^3 - 12 n^2 {S_5} + \frac{20 {S_{  1,1  }}}{n} + 20 n^2 {S_{  3,2  }} + 28 n^2 {S_{  4,1  }}
 \\ \nonumber &&\hspace{-5.5cm}+ \,4 n {S_{  2,1,1  }} + 4 n^2 {S_{  2,2,1  }} - 24 n^2 {S_{  3,1,1  }} -
  2 n^2 {{\zeta }_5} + 3 {L_n} + n {\pi }^2 {L_n} + \frac{n^2 {\pi }^4 {L_n}}{18} + \frac{3 {{L_n}}^2}{2}
  \\ \nonumber &&\hspace{-5.5cm} + \,\frac{n {\pi }^2 {{L_n}}^2}{2} +
  {{S_1}}^2 \bigg( \frac{11}{2} - \frac{10}{n} + \frac{7 n {\pi }^2}{6} - 7 n {S_2} + 4 {L_n} \bigg)  + {{\zeta }_3} \bigg(\! -4 n - \frac{2 n^2 {\pi }^2}{3} - 6 n {L_n} \bigg)
  \\ \nonumber &&\hspace{-5.5cm}+\,
  {S_{  2,1  }} \bigg( 2 n + \frac{2 n^2 {\pi }^2}{3} - 2 n {L_n} \bigg)  + {S_3} \bigg(\!\!-4 - \frac{2 n^2 {\pi }^2}{3} + 10 n {L_n} \bigg)  + {S_4} \bigg( 11 n - 10 n^2 {L_n} \bigg)
  \\ \nonumber &&\hspace{-5.5cm}+\,
  {{S_2}}^2 \bigg( \!-6 n + 2 n^2 {L_n} \bigg)  + {S_{  3,1  }} \bigg(\! -24 n + 8 n^2 {L_n} \bigg)  +
  {S_2} \bigg( \,\frac{1}{2} - \frac{10}{n} + \frac{n {\pi }^2}{4} - 4 n^2 {S_{  2,1  }}
  \\ \nonumber &&\hspace{-5.5cm}+ \,2 {L_n} - 6 n {L_n} - \frac{2 n^2 {\pi }^2 {L_n}}{3} - 3 n {{L_n}}^2 \bigg)  +
  {S_1} \bigg( 3 - \frac{3}{n} - \frac{3 {\pi }^2}{4} + n {\pi }^2 + \frac{n^2 {\pi }^4}{18} + 2 n^2 {{S_2}}^2
  \\ \nonumber &&\hspace{-5.5cm}- \,2 n {S_3} - 10 n^2 {S_4} + 6 n {S_{  2,1  }} + 8 n^2 {S_{  3,1  }} - 10 n {{\zeta }_3} + 9 {L_n} +
     \frac{7 n {\pi }^2 {L_n}}{3} + 3 {{L_n}}^2
\\ &&\hspace{-5.5cm}+ \,{S_2} \bigg( 9 - 6 n - \frac{2 n^2 {\pi }^2}{3} - 14 n {L_n} \bigg)
     \bigg)\bigg)\Bigg]\Bigg\}\, .
\end{eqnarray}

\subsubsection{Coulomb and $1/r^2$ potential}
\label{sec:doubleNA}

\begin{figure}[t]
\begin{center}
\makebox[0cm]{ \scalebox{0.4}{\rotatebox{0}{
     \includegraphics{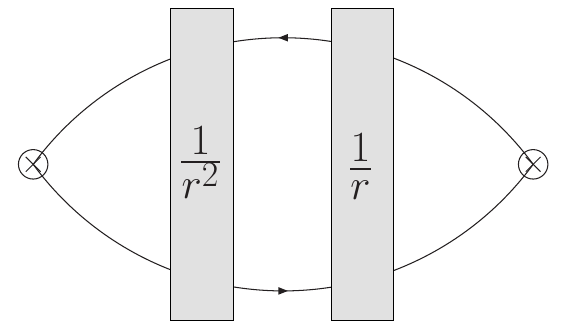}}}
\hspace{0.3cm} \scalebox{0.4}{\rotatebox{0}{
     \includegraphics{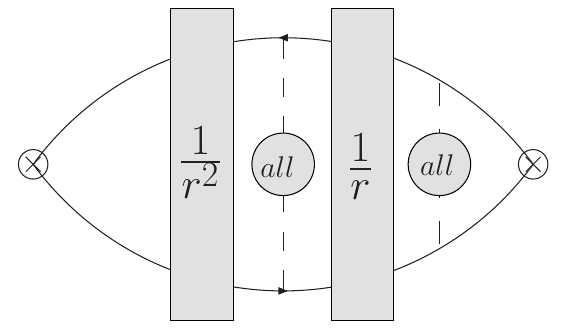}}}
\hspace{0.3cm}\scalebox{0.4}{\rotatebox{0}{
     \includegraphics{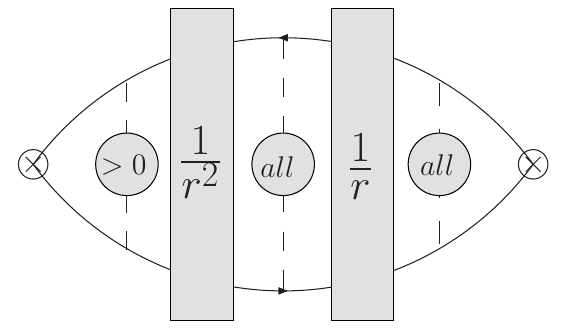}}}
     }
\end{center}
\hspace{3.3cm}(a)\hspace{3.65cm}(a+b)\hspace{3.55cm}(c)
 \caption{\label{fig:NACsubdiagrams} The three parts of the
$\frac{1}{r^2}$ and Coulomb potential double insertion.}
\end{figure}

The double insertion calculation of the Coulomb and the $1/r^2$ potential is 
more involved, because the more singular short-distance behaviour of 
the  $1/r^2$ potential causes ultraviolet divergences in the 
momentum integrals.  There is an overall divergence from the
diagram with no gluon exchanges between the potential insertions, 
which disappears in the imaginary part. In addition there are 
logarithmic divergences from diagrams with no gluon exchange 
between the left vertex and the $1/r^2$ potential insertion. These are 
the same subgraphs that appeared in the single insertion of the 
$1/r^2$ potential; the divergence is related to the renormalization 
of the external current as before. According to this 
divergence structure, we divide the insertion function 
$I[\frac{1}{2}+\epsilon,1+\epsilon]$ into
the three parts shown in figure~\ref{fig:NACsubdiagrams}. The 
calculation uses a combination of techniques employed for the 
single insertion of the $1/r^2$ potential and for the double insertion 
of the Coulomb potential. We sketch it here and present the results. 
Further details can be found in appendix~\ref{app:detailsNAC}.

Part a is an ordinary three-loop Feynman integral, which has a 
logarithmic, energy-independent overall divergence and a logarithmic 
divergence in the left one-loop vertex subgraph. It can be done with Feynman
parameters and Mellin-Barnes techniques. Dropping terms that do 
not contribute to the imaginary part, the result for 
the coun\-ter\-term-including insertion function is
\begin{eqnarray}
\nonumber J_a[\frac{1}{2}+\epsilon,1+\epsilon;w(\epsilon)]&=&
\frac{m w}{8 \pi^2
\epsilon}\bigg[\frac{\delta^{(1)}G_0^{(0ex)}(E)}{4\pi C_F
\alpha_s}\bigg]
+\frac{m^3}{64\pi^4}\Bigg\{w\beta_0\bigg[(3-\ln2)L_{\lambda}^2+\frac{4}{3}
L_{\lambda}^3\bigg]
\\ && \hspace{-3cm} +w
a_1\bigg[(3-\ln2)L_{\lambda}+L_{\lambda}^2\bigg]+
\frac{1}{2}w^{(\epsilon)}\beta_0 L_{\lambda}^2
+\frac{1}{2}w^{(\epsilon)}a_1 L_{\lambda} \Bigg\}\,.
\end{eqnarray}
Here we made use of the fact that at third-order the $1/r^2$ insertion 
is needed only with its tree-level coefficient, so there is no 
$w^{1/\epsilon}/\epsilon$ term in the function $w(\epsilon)$. 
We note the appearance of the expression 
\begin{equation}
\bigg[\frac{\delta^{(1)}G_0^{(0ex)}(E)}{4\pi C_F\alpha_s}\bigg]
=\langle G_0^{(0ex)}\left[\frac{1}{\bff{q}^2}\left\{
\left(\frac{\mu^{2\epsilon}}{\bff{q}^{2\epsilon}}-1\right)
\frac{\beta_0}{\epsilon} 
+\frac{\mu^{2\epsilon}}{\bff{q}^{2\epsilon}}a_1(\epsilon)\right\}
\right]G_0^{(0ex)}\rangle\,,
\label{eq:doubleNACa}
\end{equation}
which corresponds to the formal, $d$-dimensional expression for the 
the first-order correction to the correlation function without gluon
exchanges on both sides of the Coulomb potential insertion. This is 
precisely the quantity that arises when the divergent vertex subgraph 
in part a is contracted to a point and precisely what needs to be 
factorized if the $1/\epsilon$ pole in the first term on the 
right-hand side of (\ref{eq:doubleNACa}) is to cancel with the 
divergence in the two-loop hard matching coefficient of the current 
multiplying the NLO Green function.

Part b is more complicated, since it contains the divergent vertex 
subgraph and an all-order summation, similar to part c of the
$1/r^2$ single insertion. As by now familiar, we first calculate the 
vertex integral in $d$ dimensions. The remaining parts can then be 
done in four dimensions using some of the auxiliary functions from the 
calculation of the double insertion of the Coulomb
potential. The final result reads
\begin{eqnarray}
\nonumber J_b[\frac{1}{2}+\epsilon,1+\epsilon;w(\epsilon)]&=&\frac{m
w}{8 \pi^2 \epsilon}\bigg[\frac{\delta^{(1)}G_0^{(>0ex)}(E)}{4\pi
C_F \alpha_s}\bigg]
\nonumber\\ 
&&\hspace{-3.0cm}+\,\frac{m}{8 \pi^2}\bigg[w^{(\epsilon)}+2(1-\ln
2)w\bigg]J_b[1+\epsilon;\beta_0/\epsilon+a_1]
\nonumber\\ 
\nonumber&&\hspace{-3.0cm}
+\,\frac{m^3w}{64\pi^4}\,\Bigg\{\beta_0\Bigg[4L_{\lambda}^2
\bigg(\lambda\Psi_1(1-\lambda)-\PG(1-\lambda)\bigg)-
4L_{\lambda}\bigg(-\frac{\pi^2}{6}+\PG(1-\lambda)
\\ \nonumber
 &&\hspace{-3.0cm}
-\,\PG(1-\lambda)^2+ (1-\lambda)\Psi_1(1-\lambda)-\lambda\Psi_2(1-\lambda)+2\lambda\sum_{k=1}^{\infty}\frac{\PG(k-\lambda)}{(k-\lambda)^2}\,\bigg)
\\ \nonumber
 &&\hspace{-3.0cm}-\,
\frac{8}{3}\zeta_3-\frac{\pi^2}{3}
-2 \bigg(\frac{\pi^2}{3}+\frac{2}{\lambda^2}\bigg)\PG(1-\lambda)+
2  \bigg(1-\frac{2}{\lambda}\bigg)\bigg(\PG(1-\lambda)^2+\Psi_1(1-\lambda)\bigg)
\\ \nonumber
 &&\hspace{-3.0cm}+\,2\lambda\Psi_2(1-\lambda)-2\lambda\Psi_1^2(1-\lambda)-
 \frac{4}{3}\bigg(\PG(1-\lambda)^3+\Psi_2(1-\lambda)\bigg)
+\lambda\Psi_3(1-\lambda)
 \\ \nonumber
 &&\hspace{-3.0cm}-\,4\bigg(1+\lambda(-1+\PG(1-\lambda))\bigg)\PG(1-\lambda)
\Psi_1(1-\lambda)
+8\lambda\sum_{k=1}^{\infty}\frac{\PG(k)\PG(k-\lambda)}{(k-\lambda)^2}
 \\ \nonumber
 &&\hspace{-3.0cm}
 +\,4\sum_{k=1}^{\infty}\frac{(k+\lambda-2k\lambda)\PG(k-\lambda)}
{k(k-\lambda)^2}\,\Bigg]
\\ \nonumber
 &&\hspace{-3.0cm}+\,a_1\bigg[\bigg(-2L_{\lambda}-2+\PG(1-\lambda)-2\lambda\Psi_1(1-\lambda)\bigg)\PG(1-\lambda)
 \\
  &&\hspace{-3.0cm}+\,\bigg(2\lambda-3+2\lambda
  L_{\lambda}\bigg)\Psi_1(1-\lambda)+\lambda\Psi_2(1-\lambda)+\frac{\pi^2}{2}\,
\bigg]\Bigg\}\,,
\label{eq:doubleNAb}
\end{eqnarray}
where the finite expression 
$J_b[1+\epsilon;\beta_0/\epsilon+a_1]$ is given in (\ref{eq:JbC}), 
(\ref{eq:Jb(1+aeps)}), 
and 
\begin{equation}
\bigg[\frac{\delta^{(1)}G_0^{(>0ex)}(E)}{4\pi C_F
\alpha_s}\bigg]
= \langle G_0\left[\frac{1}{\bff{q}^2}\left\{
\left(\frac{\mu^{2\epsilon}}{\bff{q}^{2\epsilon}}-1\right)
\frac{\beta_0}{\epsilon} 
+\frac{\mu^{2\epsilon}}{\bff{q}^{2\epsilon}}a_1(\epsilon)\right\}
\right]G_0\rangle 
-\bigg[\frac{\delta^{(1)}G_0^{(0ex)}(E)}{4\pi C_F \alpha_s}\bigg]
\end{equation}
is the first-order correction to the Green
function with more than one gluon exchange. As it should be, this 
combines with (\ref{eq:doubleNACa}) to the full NLO Green function.

Finally the last part c is finite and can be done reusing some
results from the Coulomb double insertion. We obtain  
\begin{eqnarray}
\nonumber J_c[\frac{1}{2}+\epsilon,1+\epsilon;w(\epsilon)] &=&
\frac{m^3\beta_0 w}{32\pi^4}\Bigg[-2L_{\lambda}\bigg(\frac{\pi^2}{6}
-\Psi_1(1-\lambda)+\frac{\lambda}{2}\Psi_2(1-\lambda)\bigg)
\\ \nonumber &&\hspace{-3cm}
-\,\frac{2}{\lambda}\PG(1-\lambda)^2- \PG(1-\lambda)\,
\bigg(2\Psi_1(1-\lambda)+\lambda
\Psi_2(1-\lambda)\bigg) -2\PG(1-\lambda)^3
\\ \nonumber &&\hspace{-3cm}
+\,\lambda\Psi_1(1-\lambda)^2- \frac{\lambda}{2}\Psi_3(1-\lambda)
+2\lambda\,\bigg(1+2\lambda\PG(1-\lambda)\bigg)
\sum_{k=1}^{\infty}\frac{\PG(k-\lambda)}{k(k-\lambda)^2}
\\ \nonumber  &&\hspace{-3.3cm}
+ \,2\sum_{k=1}^{\infty}
\frac{(k+\lambda)\PG(k-\lambda)}{k-\lambda}\bigg(\frac{\PG(k-\lambda)}{k}-\frac{\PG(k)}{k-\lambda}\bigg)\Bigg]
\\[0.2cm] && \hspace{-3cm}+\frac{m^3 a_1 w}{32\pi^4}\,
\bigg[-\frac{\pi^2}{6}+\Psi_1(1-\lambda) -\frac{\lambda}{2}
\Psi_2(1-\lambda)\bigg]\, .
\label{eq:doubleNAc}
\end{eqnarray}

The last two parts are singular near the leading-order bound state 
poles. The singular terms for $\lambda \rightarrow n$ are
\begin{eqnarray}
\nonumber
\hat{J}_b[\frac{1}{2}+\epsilon,1+\epsilon;w(\epsilon)]&=&\frac{m
w}{8 \pi^2 \epsilon}\bigg[\frac{\delta^{(1)}G_0(E)}{4\pi C_F
\alpha_s}\bigg]^{\lambda\rightarrow n}
\\\nonumber &&\hspace{-3cm}
+\,\frac{m^3}{32\pi^4}\Bigg\{\frac{n}{(n-\lambda)^2}\bigg[w a_1
\bigg(\frac{3}{2}+L_n -\frac{1}{2}\ln 2 - S_1\bigg)
\\ \nonumber &&\hspace{-3cm}+ \,w\beta_0
(3+2L_n-\ln2-2S_1)(L_n+S_1) +\frac{\beta_0
w^{(\epsilon)}}{2}(L_n+S_1)+\frac{a_1 w^{(\epsilon)}} {4}\bigg]
\\ \nonumber &&\hspace{-3cm}+\,\frac{1}{n-\lambda}\bigg[-w a_1- w\beta_0\bigg(3+n\pi^2-\ln2-\frac{n\pi^2}{3}\ln2+L_n\bigg(4+\frac{2n\pi^2}{3}\bigg)
\\ \nonumber &&\hspace{-3cm}+ \,S_1\bigg(6-\frac{2n\pi^2}{3}+4L_n-2\ln2-4S_1 + 4 n
S_2\bigg)+(2-6n-4nL_n+2n\ln2)S_2
\\ &&\hspace{-3cm}-\,4nS_3+4nS_{2,1}-4n\zeta_3\bigg)
-\beta_0 w^{(\epsilon)}\bigg(\frac{1}{2}+n\frac{\pi^2}{6}+S_1-n
S_2\bigg)\bigg]\Bigg\}\, ,
\\
\hat{J}_c[\frac{1}{2}+\epsilon,1+\epsilon;w(\epsilon)]&=&\frac{m^3}
{32 \pi^4}\Bigg\{\frac{a_1 w n+2\beta_0 w n(L_n+S_1)}
{(n-\lambda)^3}
\nonumber \\ &&\hspace{-3cm}- \,\beta_0 w \,\bigg[\frac{1} {
(n-\lambda)^2}\,\bigg(2+\frac{2n\pi^2}{3}+2S_1-4 n S_2\bigg) +
\frac{\frac{1}{n}+4n\zeta_3+2 S_2-4n S_3} {n-\lambda}\bigg]\Bigg\}\,.
\qquad\quad
\end{eqnarray}

\subsubsection{Coulomb and delta potential}

The double insertion of the Coulomb potential and delta potential is
easy to calculate, because the integral factorizes into two parts,
which have already been calculated in the single insertion of the
Coulomb potentials:
\begin{eqnarray}
I[0,1+a\epsilon] &=&G_0(E)I[1+a\epsilon].
\end{eqnarray}
This equation is correct in $d$ dimensions, because we consider the 
double insertion of the Coulomb potential with the {\em tree level} 
delta potential, which does not contain factors of 
$(\mu^2/{\bf q}^2)^\epsilon$  (in contrast to the single insertion of 
the one-loop delta potential). 

The divergences in the imaginary part of $I[0,1+a\epsilon]$ come
from the $1/\epsilon$ poles in the real parts of $I[1+a\epsilon]$ and
$G_0(E)$. In the $J$-function they must factorized in such a way, 
that they multiply $d$-dimensional Coulomb Green functions (including 
the single Coulomb insertion) to allow for the cancellation with the 
divergences from the hard matching coefficient of the external current. 
The required result is
\begin{eqnarray}
J[0,1+\epsilon;w(\epsilon)] &=&\frac{m^2 C_F\alpha_s
w}{16\pi\epsilon}\bigg[\frac{\delta^{(1)}G_0(E)}{4\pi C_F
\alpha_s}\bigg]-\frac{m^2 w \beta_0}{192 \pi^2 \epsilon^2}\,G_0(E)
\nonumber \\ 
\nonumber &&\hspace{-2.5cm}+ \,\frac{m^2}{192
\pi^2 \epsilon} \,G_0(E)\,\big(2a_1w -2\beta_0 w - \beta_0 w^{(\epsilon)}\big)
+\frac{m^2 C_F\alpha_s
w^{(\epsilon)}}{16\pi}J[1+\epsilon;\beta_0/\epsilon+a_1(\epsilon)]
\\ \nonumber &&\hspace{-2.5cm}+\,\frac{m^2}{192
\pi^2} G_0(E)\,\big(2a_1w^{(\epsilon)} -2\beta_0 w^{(\epsilon)} - \beta_0
w^{(\epsilon^2)}+2 a_1^{(\epsilon)} w\big)
\\ \nonumber &&\hspace{-2.5cm}+\,\frac{m^4 C_F \alpha_s
w}{64\pi^3}\Bigg\{\beta_0\Bigg[L_\lambda^3+L_\lambda^2\bigg(\frac{1}{2}-\frac{1}{2\lambda}-3\PG(1-\lambda)+2\lambda\Psi_1(1-\lambda)\bigg)
\\ \nonumber &&\hspace{-2.5cm}+\,L_\lambda\bigg(\frac{5\pi^2}{72}
-\frac{1}{3}+(\lambda-4)\Psi_1(1-\lambda)+\lambda\Psi_2(1-\lambda)
\\ \nonumber &&\hspace{-2.5cm}+\,\PG(1-\lambda)\bigg(\frac{1}{\lambda}-1+3\PG(1-\lambda)-4\lambda\Psi_1(1-\lambda)\bigg)+4\,{}_4F_3(1,1,1,1;2,2,1-\lambda;1)\bigg)
\\ \nonumber &&\hspace{-2.5cm}+\,\bigg(1-\frac{1}{\lambda}\bigg)\bigg(\frac{5\pi^2}{144}-\frac{1}{6}\bigg)+\PG(1-\lambda)\,
\bigg(\frac{1}{3}-\frac{5\pi^2}{72}+(4-\lambda)\Psi_1(1-\lambda)-\lambda\Psi_2(1-\lambda)\bigg)
\\ \nonumber &&\hspace{-2.5cm}
+\,\PG(1-\lambda)^2\bigg(\frac{1}{2}-\frac{1}{2\lambda}+2\lambda\Psi_1(1-\lambda)\bigg)-\PG(1-\lambda)^3
+\frac{3}{2\lambda}(1-\lambda)\Psi_1(1-\lambda)
\\ \nonumber &&\hspace{-2.5cm}-\,\frac{1}{2}(1-\lambda)\Psi_2(1-\lambda) 
+2\,{}_4F_3(1,1,1,1;2,2,1-\lambda;1)\,\bigg(1-\frac{1}{\lambda}-2\PG(1-\lambda)\bigg)\Bigg]
\\ \nonumber &&\hspace{-2.5cm}+\,a_1\bigg[L_\lambda^2+L_\lambda\bigg(\frac{5}{6}-\frac{1}{2\lambda}-2\PG(1-\lambda)+\lambda\Psi_1(1-\lambda)\bigg)
+\bigg(\frac{1}{2\lambda}-\frac{1}{2}+\PG(1-\lambda)\bigg)
\\ &&\hspace{-2.5cm}
\times\,\bigg(\!-\frac{1}{3}+\PG(1-\lambda)-\lambda\Psi_1(1-\lambda)\bigg)\bigg]
\Bigg\}\, .
\end{eqnarray}
Here the first term in the first line stems from the $1/\epsilon$ pole of 
$G_0(E)$. The other $1/\epsilon$ divergences are related 
to those in the expansion of the Coulomb potential single insertion 
function $J[1+\epsilon;w(\epsilon)]$ given by the sum of 
(\ref{eq:Ja(1+aeps)}) and (\ref{eq:Jb(1+aeps)}).

The singular part for $\lambda\to n$ is:
\begin{eqnarray}
\hat{J}[0,1+\epsilon;w(\epsilon)] &=&\frac{m^2 C_F\alpha_s
w}{16\pi\epsilon}\bigg[\frac{\delta^{(1)}G_0(E)}{4\pi C_F
\alpha_s}\bigg]^{\lambda\rightarrow n}-\frac{m^2 w \beta_0}{192
\pi^2 \epsilon^2}\,G_0^{\lambda\rightarrow n}(E)
\nonumber \\ 
\nonumber &&\hspace{-2.5cm}+ \,\frac{m^2}{192
\pi^2 \epsilon} \,G_0^{\lambda\rightarrow n}(E)\,\big(2a_1w -2\beta_0 w -
\beta_0 w^{(\epsilon)}\big)+\frac{C_F\alpha_s
m^2w^{(\epsilon)}}{16\pi}\hat{J}[1+\epsilon;\beta_0/\epsilon+a_1(\epsilon)]
\\ \nonumber &&\hspace{-2.5cm}+\,\frac{m^2}{192
\pi^2} \,G_0^{\lambda\rightarrow
n}(E)\,\big(2a_1w^{(\epsilon)}+2a_1^{(\epsilon)}w -2\beta_0 w^{(\epsilon)}
- \beta_0 w^{(\epsilon^2)}\big)
\\ \nonumber &&\hspace{-2.5cm}+\,\beta_0 w\frac{m^4 C_F
\alpha_s} {32\pi^3}\Bigg[ \frac{n L_n+n S_1}{(n-\lambda)^3} +
\frac{-1-\frac{n\pi^2}{3}+ \frac{1+n}{2}L_n+nL_n^2
+\frac{n-3}{2}S_1-nS_1^2+2nS_2} {(n-\lambda)^2}
\\ \nonumber &&\hspace{-2.5cm}+\,\frac{1}{2(n-\lambda)}\Bigg(\!-\frac{4}{3}-\frac{24n+7}{72}\pi^2-\frac{2}{n}+S_1\bigg(-2-\frac{1}{n}+\frac{2n\pi^2}{3}+5S_1-4nS_2\bigg)
\\ \nonumber &&\hspace{-2.5cm}+\,S_2\bigg(2n-5\bigg)+8nS_3-4nS_{2,1}
+L_n\bigg(\frac{1}{n}-4-\frac{2n\pi^2}{3}-6S_1+4nS_2+L_n\bigg)\Bigg)\Bigg]
\\  &&\hspace{-2.5cm}+\,a_1w\frac{m^4 C_F
\alpha_s}{64\pi^3}\Bigg[ \frac{n}{(n-\lambda)^3}+\frac{\frac{1+n}{2}
+n L_n- n S_1} {(n-\lambda)^2} +\frac{-\frac{2}{3} +\frac{1}{2n}+
L_n- S_1} {(n-\lambda)} \Bigg]\,.
\end{eqnarray}

\subsubsection{Coulomb and contact potential}

The double insertion of the Coulomb potential with the contact potential,
which arises from applying  the equation of motion to the double insertion 
of the Coulomb potential with the kinetic energy 
correction, is completely finite. Therefore, it can be
easily calculated in coordinate space using a combination of techniques for 
the result for the Coulomb double insertion and the contact potential 
single insertion. This leads to
\begin{eqnarray}
I[\delta,1+a\epsilon] &=&\int \prod_{i=1}^5\Bigg[\frac{d^{3}{\bf
p}_i}{(2\pi)^{3}}\Bigg] \frac{G_0({\bf p}_1,{\bf p}_2)G_0({\bf p}_3,{\bf
p}_4)}{[{\bf q}_{23}^2]^{1+a\epsilon}} G_0({\bf p}_4,{\bf p}_5)
\nonumber 
\\ &=&
\frac{m^3}{(4\pi)^3}\,\frac{\lambda}{m C_F
\alpha_s}\sum_{k=1}^{\infty}\frac{H_{\delta}(k)[H^{(0)}(k)+a\epsilon
H^{(1)}(k)+O(\epsilon^2)]}{k(k-\lambda)}\,,
\end{eqnarray}
where ${\bf q}_{23}={\bf p}_2-{\bf p}_3$, and
\begin{eqnarray}
H_{\delta}(k) &=&\frac{-4\pi \lambda}{\sqrt{-mE}}\,\frac{k}{
(k-1-\lambda)(k-\lambda)(k+1-\lambda)}\, .
\end{eqnarray}
The remaining sum can be done partially except for a single sum that is
left:
\begin{eqnarray}
I[\delta,1+a\epsilon]&=&\frac{m\lambda^2}{16\pi^2C_F^2\alpha_s^2}\,\bigg(1+2\lambda\Psi_1(1-\lambda)-\lambda^2
\Psi_2(1-\lambda)\bigg)
 \nonumber\\ 
&&\hspace{-2cm} \nonumber +\,a\epsilon\frac{m\lambda^3}{8\pi^2C_F^2\alpha_s^2}
\,\Bigg[-\PG(1-\lambda)\bigg(\frac{1}{\lambda}+\lambda\Psi_2(1-\lambda)\bigg)-\frac{1}{3}\lambda\Psi_3(1-\lambda)+\Psi_1(1-\lambda)
\\ &&\hspace{-2cm} -\,\sum_{k=1}^{\infty}
\frac{2(k+\lambda)\PG(k-\lambda)}{(k-\lambda)^3}
+L_\lambda \bigg(\frac{1}{\lambda}+2\Psi_1(1-\lambda)-\lambda
\Psi_2(1-\lambda)\bigg)\Bigg]\, .
\end{eqnarray}
The $a$-dependent $O(\epsilon)$ terms are kept here, 
since they multiply the 
$\beta_0$-dependent part of the Coulomb potential (\refI{eq:vcoulombNLO}). 
They originate only from the potential, not from the integration measure,  
where we can therefore set $d=4$ from the start. 
The counterterm-including insertion function reads
\begin{eqnarray}
J[\delta,1+\epsilon;w]&=&\frac{m\beta_0
w\lambda^3}{8\pi^2C_F^2\alpha_s^2}\,
\bigg[-\PG(1-\lambda)\bigg(\frac{1}{\lambda}+\lambda\Psi_2(1-\lambda)\bigg)
-\frac{1}{3}\lambda\Psi_3(1-\lambda)
\nonumber \\ 
\nonumber &&\hspace{-2cm}+\,\Psi_1(1-\lambda)-
\sum_{k=1}^{\infty}\frac{2(k+\lambda)\PG(k-\lambda)}{(k-\lambda)^3}
+L_\lambda \bigg(\frac{1}{\lambda}+2\Psi_1(1-\lambda)-\lambda
\Psi_2(1-\lambda)\bigg)\bigg]
\\&&\hspace{-2cm}+\,\frac{m a_1
w\lambda^2}{16\pi^2C_F^2\alpha_s^2}\,\bigg(1+2\lambda\Psi_1(1-\lambda)
-\lambda^2 \Psi_2(1-\lambda)\bigg)\,,
\end{eqnarray}
where $w$ is assumed to be an $\epsilon$-independent constant.
The remaining sum shows good convergence and is easily 
calculated numerically.\footnote{
For instance, with Mathematica's  {\tt NSum} function.}

The singular parts of the Laurent expansion around the leading-order 
bound state poles $\lambda=n$ are easy to calculate and read:
\begin{eqnarray}
\hat{J}[\delta,1+\epsilon;w]&=&
\frac{m w}{8\pi^2 C_F^2 \alpha_s^2}\Bigg[\,
\beta_0 \,\bigg( \frac{2n^4(L_n+S_1)}{(n-\lambda)^3}
- \frac{n^3(3+18L_n+n\pi^2+24S_1-6 n S_2)}{3(n-\lambda)^2}
\nonumber\\  
&&\hspace{-2.5cm}
+\,\frac{n^2(3+6L_n+n\pi^2+12S_1-6 n S_2)}{n-\lambda} \bigg)
+ a_1\bigg(
\frac{n^4}{(n-\lambda)^3}-\frac{3n^3}{(n-\lambda)^2}+\frac{3n^2}{n-\lambda}\bigg)
\Bigg].\qquad\;
\end{eqnarray}

\subsection{Triple Coulomb potential insertion}

The triple insertion of the NLO Coulomb potential defined in 
(\ref{eq:tripledef}) is finite and therefore done in $d=4$
dimensions. Details of this calculation have already been given in the 
appendix of \cite{Beneke:2005hg}, but not the final result for 
the insertion function, which we provide here. The triple insertion 
involves a product of four Coulomb Green functions, two of which 
are adjacent to the external current vertex. For these we use the 
integral representation (\refI{eq:greenint}) of $G_0(0,r;E)$, 
while for the other two between the potential insertions we use the 
Laguerre polynomial representation (\refI{eq:gpartial})  of 
$G_{[l]}(r,r^\prime;E)$ for the $l=0$ partial wave. In this way all 
integrations can be performed at the expense of summations. 
The result is:
\begin{eqnarray}
J^{(C)}[1,1,1]&=&\frac{m^2}{(4\pi)^4 C_F^2\alpha_s^2}\,
\Bigg[\beta_0^3\sum_{j=1}^{\infty}\sum_{k=1}^{\infty}
\frac{H^{(1)}(j)K^{(1)}(j,k)H^{(1)}(k)}{j(j-\lambda)k(k-\lambda)}
\nonumber\\ 
\nonumber && \hspace{-1.5cm}+\,\beta_0^2
a_1\sum_{j=1}^{\infty}\sum_{k=1}^{\infty}\frac{2H^{(0)}(j)K^{(1)}(j,k)H^{(1)}(k)+H^{(1)}(j)K^{(0)}(j,k)H^{(1)}(k)}{j(j-\lambda)k(k-\lambda)}
\\ \nonumber && \hspace{-1.5cm}+\,\beta_0
a_1^2\sum_{j=1}^{\infty}\sum_{k=1}^{\infty}\frac{2H^{(0)}(j)K^{(0)}(j,k)H^{(1)}(k)+H^{(0)}(j)K^{(1)}(j,k)H^{(0)}(k)}{j(j-\lambda)k(k-\lambda)}
\\ &&
\hspace{-1.5cm}+\,a_1^3\,\frac{\lambda^2}{2}\bigg(\!-\Psi_2(1-\lambda)+\frac{\lambda}{3}\Psi_3(1-\lambda)\bigg)\Bigg]\,.
\end{eqnarray}
where the derivatives of the $H$-function, $H^{(n)}$, are the same as they 
appear in the double insertion of Coulomb potentials, see (\ref{eq:Hder}), 
and\footnote{We note the change in the normalization and arguments in the 
definition of $K^{(n)}(j,k)$ compared to~\cite{Beneke:2005hg}. The relation 
reads 
\[
K^{(n)\,\rm here}(j,k) = (m C_F\alpha_s)^2\, 
K^{(n)\,\rm Ref.\cite{Beneke:2005hg}}(j-1,k-1)
\]} 
\begin{eqnarray}
K^{(0)}(j,k)=\lambda^2j\delta_{jk}\, ,
\end{eqnarray}
and
\begin{eqnarray}
K^{(1)}(j,k)=2\lambda^2\left\{
\begin{array}{cl}
  1+j[\PG(j)+L_{\lambda}] &
  \qquad\textrm{if} \; j=k\hskip0.8cm\\[0.2cm]
  -\frac{\displaystyle \min (j,k)}{\displaystyle |k-j|} &
  \qquad\textrm{if} \; j\neq k \hskip0.1cm .
\end{array}\right.
\end{eqnarray}
The double sums can in principle be evaluated numerically. However,
we performed some parts of the sum analytically to speed up the
calculation. The result for the partly analytic result is too
long to be given here.

The singular part near the leading-order bound state poles $\lambda=n$ 
starts with a fourth-order pole as expected for a triple insertion. The 
result can be expressed in terms of (nested) harmonic sums as
\begin{eqnarray}
\hat{J}[1,1,1]&=&\frac{ m^2}{256\pi^4C_F^2
\alpha_s^2}\,\Bigg\{\frac{n^3}{(n-\lambda)^4}
\,\Bigg[a_1+2\beta_0(L_n+S_1)\Bigg]^3
\nonumber\\ 
&& \hspace{-1.6cm} \nonumber -\,\frac{n^2}{(n-\lambda)^3}\,
\Bigg[2a_1^3+2a_1^2\beta_0\bigg(3+6L_n+\frac{n\pi^2}{3}+8 S_1-2 n S_2\bigg)
+8a_1\beta_0^2\bigg(\frac{{\pi }^2}{6} + 3 {L_n}
\\ && \hspace{-1.6cm} \nonumber +\,
\frac{n {\pi }^2 {L_n}}{3} + 3 {{L_n}}^2 + 3 {S_1} - \frac{{S_1}}{n}
+ \frac{n {\pi }^2 {S_1}}{3} + 8 {L_n} {S_1} + 5 {{S_1}}^2 - {S_2} -
2 n {L_n} {S_2} - 2 n {S_1} {S_2}
\\ && \hspace{-1.6cm} \nonumber + \,n {S_3} - n
{{\zeta }_3}\bigg)+16\beta_0^3\bigg(\frac{{\pi }^2 {L_n}}{6} +
\frac{3 {{L_n}}^2}{2} + \frac{n {\pi }^2 {{L_n}}^2}{6} + {{L_n}}^3 +
\frac{{\pi }^2 {S_1}}{6} + 3 {L_n} {S_1} - \frac{{L_n} {S_1}}{n}
\\ && \hspace{-1.6cm} \nonumber +\,
\frac{n {\pi }^2 {L_n} {S_1}}{3} + 4 {{L_n}}^2 {S_1} + \frac{3
{{S_1}}^2}{2} -
  \frac{{{S_1}}^2}{n} + \frac{n {\pi }^2 {{S_1}}^2}{6} + 5 {L_n} {{S_1}}^2 + 2 {{S_1}}^3 - {L_n} {S_2} - n {{L_n}}^2 {S_2}
  \\ && \hspace{-1.6cm} \nonumber - \,{S_1} {S_2} - 2 n {L_n} {S_1} {S_2} - n {{S_1}}^2 {S_2} + n {L_n} {S_3} + n {S_1} {S_3} -
  n {L_n} {{\zeta }_3} - n {S_1} {{\zeta }_3}\bigg)\Bigg]
  \\ && \hspace{-1.6cm} \nonumber +\,\frac{n}{(n-\lambda)^2}
\,\Bigg[a_1^3+a_1^2\beta_0\bigg(9 + \frac{4\,n\,{\pi }^2}{3} + 6\,{L_n} + 14\,{S_1} - 8\,n\,{S_2}\bigg)
  +4a_1\beta_0^2\bigg(3 + \frac{{\pi }^2}{2} + \frac{2 n {\pi }^2}{3}
  \\ && \hspace{-1.6cm} \nonumber + \,\frac{n^2 {\pi }^4}{36} + 9 {L_n} + \frac{4 n {\pi }^2 {L_n}}{3} + 3 {{L_n}}^2 + 13 {S_1} - \frac{3 {S_1}}{n} + 2 n {\pi }^2 {S_1} + 14 {L_n} {S_1} + 11 {{S_1}}^2 - 3 {S_2}
  \\ && \hspace{-1.6cm} \nonumber - \,4 n {S_2} -
  \frac{n^2 {\pi }^2 {S_2}}{3} - 8 n {L_n} {S_2} - 12 n {S_1} {S_2} + n^2 {{S_2}}^2 + 9 n {S_3} - 5 n^2 {S_4} - 2 n {S_{ 2,1 }} + 4 n^2 {S_{3,1 }}
  \\ && \hspace{-1.6cm} \nonumber - \,5 n {{\zeta }_3}\bigg)
  +8\beta_0^3\bigg(\frac{{\pi }^2}{3} + \frac{5 n {\pi }^4}{72} + \bigg(
\frac{9}{2} + \frac{2 n {\pi }^2}{3} \bigg)  {{L_n}}^2 + {{L_n}}^3 +
5 {{S_1}}^3 + \bigg( \frac{5 n}{2} + n^2 {L_n} \bigg) {{S_2}}^2
\\ && \hspace{-1.6cm} \nonumber +\,
  {{S_1}}^2 \bigg( \frac{17}{2} - \frac{3}{n} + \frac{4 n {\pi }^2}{3} + 11 {L_n} - 8 n {S_2} \bigg)  + \bigg( 2 n + \frac{5 n^2 {\pi }^2}{6} + 9 n {L_n} \bigg)  {S_3}
  \\ && \hspace{-1.6cm} \nonumber -\,5n \bigg( \frac{1}{2} + n {L_n} \bigg)  {S_4} +
  3 n^2 {S_5} + \bigg( 1 - 2 n {L_n} \bigg)  {S_{  2,1  }} + \bigg( 2 n + 4 n^2 {L_n} \bigg)  {S_{  3,1  }} - 2 n^2 {S_{  3,2  }}
  \\ && \hspace{-1.6cm} \nonumber - \,6 n^2 {S_{  4,1  }} + \bigg( -2 - 2 n - \frac{5 n^2 {\pi }^2}{6} \bigg)  {{\zeta }_3} +
  {S_1} \bigg( 3 - \frac{2}{n} + \frac{{\pi }^2}{6} + \frac{2 n {\pi }^2}{3} + \frac{n^2 {\pi }^4}{36}
  \\ && \hspace{-1.6cm} \nonumber + \,\bigg( 13 - \frac{3}{n} + 2 n {\pi }^2 \bigg)  {L_n} + 7 {{L_n}}^2 +
     \bigg( -1 - 4 n - \frac{n^2 {\pi }^2}{3} - 12 n {L_n} \bigg)  {S_2} + n^2 {{S_2}}^2 + 11 n {S_3}
     \\ && \hspace{-1.6cm} \nonumber -\, 5 n^2 {S_4} - 2 n {S_{  2,1  }} + 4 n^2 {S_{  3,1  }} - 7 n {{\zeta }_3} \bigg)  +
  {L_n} \bigg( 3 + \frac{{\pi }^2}{2} + \frac{2 n {\pi }^2}{3} + \frac{n^2 {\pi }^4}{36} - 5 n {{\zeta }_3} \bigg)
  \\ && \hspace{-1.6cm} \nonumber +\,
  {S_2} \bigg(\! -2 + \frac{1}{n} - \frac{5 n {\pi }^2}{6} + \bigg( -3 - 4 n - \frac{n^2 {\pi }^2}{3} \bigg)  {L_n} - 4 n {{L_n}}^2 - n^2 {S_3} + n^2 {{\zeta }_3} \bigg) 
\\ && \hspace{-1.6cm} \nonumber + \,6 n^2 {{\zeta}_5}\bigg) \Bigg]
\\ 
&& \hspace{-1.6cm} \nonumber-\,\frac{1}{n-\lambda}\,
\Bigg[2a_1^2\beta_0\bigg(1 + \frac{n {\pi }^2}{3} + 2 {S_1} - 2 n {S_2}\bigg)+4a_1\beta_0^2\bigg(3 + n {\pi }^2 + \frac{n^2 {\pi }^4}{18} + \bigg( 2 + \frac{2 n {\pi }^2}{3} \bigg)  {L_n}
  \\ && \hspace{-1.6cm} \nonumber + 4 {{S_1}}^2 + \bigg(\! -6 n - \frac{2 n^2 {\pi }^2}{3} - 4 n {L_n} \bigg)  {S_2} + 2 n^2 {{S_2}}^2 +
  {S_1} \bigg( 8 + 2 n {\pi }^2 + 4 {L_n} - 12 n {S_2} \bigg)
  \\ && \hspace{-1.6cm} \nonumber + 12 n {S_3} - 10 n^2 {S_4} - 4 n {S_{  2,1  }} + 8 n^2 {S_{  3,1  }} - 4 n {{\zeta }_3}\bigg)
  +8\beta_0^3\bigg(1 + \frac{{\pi }^2}{3} + \frac{n {\pi }^2}{3} + \frac{8 n {\pi }^4}{45} + \frac{n^2 {\pi }^4}{36}
  \\ && \hspace{-1.6cm} \nonumber - \frac{n^3 {\pi }^6}{210} + \bigg( 1 + \frac{n {\pi }^2}{3} \bigg)  {{L_n}}^2 + 2 {{S_1}}^3 +
  \bigg( 4 n + n^2 + 2 n^2 {L_n} \bigg)  {{S_2}}^2 + {{S_1}}^2 \bigg( 7 - \frac{8}{n} + \frac{5 n {\pi }^2}{3}
  \\ && \hspace{-1.6cm} \nonumber + 4 {L_n} - 10 n {S_2} \bigg)  + 6 n^3 {{S_3}}^2 +
  \bigg(\! -6 n - 5 n^2 - \frac{8 n^3 {\pi }^2}{3} - 10 n^2 {L_n} \bigg)  {S_4} + 24 n^2 {S_5}
  \\ && \hspace{-1.6cm} \nonumber - 14 n^3 {S_6} + \frac{16 {S_{  1,1  }}}{n} + \bigg(\! -2 n - \frac{2 n^2 {\pi }^2}{3} - 4 n {L_n} \bigg)  {S_{  2,1  }} +
  \bigg( 4 n^2 + \frac{4 n^3 {\pi }^2}{3} + 8 n^2 {L_n} \bigg)  {S_{  3,1  }}
  \\ && \hspace{-1.6cm} \nonumber - 14 n^2 {S_{  3,2  }} - 34 n^2 {S_{  4,1  }} + 28 n^3 {S_{  4,2  }} + 40 n^3 {S_{  5,1  }} + 8 n^2 {S_{  3,1,1  }} - 8 n^3 {S_{  3,1,2  }} -
  8 n^3 {S_{  3,2,1  }}
  \\ && \hspace{-1.6cm} \nonumber - 24 n^3 {S_{  4,1,1  }} + \bigg(\! -4 n - \frac{7 n^2 {\pi }^2}{3} \bigg)  {{\zeta }_3} + 2 n^3 {{{\zeta }_3}}^2 +
  {S_1} \bigg( 5 - \frac{2}{n} - \frac{2 {\pi }^2}{3} + \frac{5 n {\pi }^2}{3} + \frac{n^2 {\pi }^4}{9}
  \\ && \hspace{-1.6cm} \nonumber + \bigg( 8 + 2 n {\pi }^2 \bigg)  {L_n} + 2 {{L_n}}^2 + \bigg( 4 - 10 n - \frac{4 n^2 {\pi }^2}{3} - 12 n {L_n} \bigg)  {S_2} +
     4 n^2 {{S_2}}^2 + 16 n {S_3}
     \\ && \hspace{-1.6cm} \nonumber - 20 n^2 {S_4} - 4 n {S_{  2,1  }} + 16 n^2 {S_{  3,1  }} - 8 n {{\zeta }_3} \bigg)  + {L_n} \bigg( 3 + n {\pi }^2 + \frac{n^2 {\pi }^4}{18} - 4 n {{\zeta }_3} \bigg)
     \\ && \hspace{-1.6cm} \nonumber -2
  {S_2} \bigg( 1 + \frac{4}{n} + n + \frac{2 n {\pi }^2}{3} + \frac{n^2 {\pi }^2}{6} + \bigg( 3 n + \frac{n^2 {\pi }^2}{3} \bigg)  {L_n} + n {{L_n}}^2 +6 n^2 {S_3} -2 n^2 {S_{  2,1  }}
  \\ && \hspace{-1.6cm} -2 n^2 {{\zeta }_3} \bigg)  +
  {S_3} \bigg(\! -2 + 8 n + \frac{11 n^2 {\pi }^2}{3} + 12 n {L_n} - 4 n^3 {{\zeta }_3} \bigg)  + 19 n^2 {{\zeta
  }_5}\bigg)\Bigg]\Bigg\}\, .
\label{eq:triplecoulombpolepart}
\end{eqnarray}

\subsection{Summary of results}
\label{sec:subsummary}

In this section we summarize the result for the expansion of the 
Green function up to the third order in terms of the insertions 
functions $J$, which have been calculated in the previous sections. 
The Green function expanded in the strong coupling and/or $v$ (treating 
$\alpha_s/v$ as $O(1)$) is written as
\begin{eqnarray}\label{eq:greensummary}
G(E)&=&G_0(E)+\sum_{n=1} \delta_n G(E)\,.
\end{eqnarray}
The leading-order Green function has been discussed in
paper~I, section~\ref{sec:CoulGreen} (see also (\ref{eq:zerodistancegreen})), 
and reads
\begin{eqnarray}
G_0(E)&=&\frac{m^2\alpha_s C_F}{4\pi}
\bigg(L_\lambda+\frac{1}{2}-\frac{1}{2\lambda}-\gamma_E-\Psi(1-\lambda)\bigg).
\end{eqnarray}

\subsubsection{First-order correction}

This correction arises entirely 
from the single insertion of the Coulomb potential
\begin{eqnarray}
-\frac{4\pi C_F \alpha_s}{{\bf q}^2}\frac{\alpha_s}{4\pi}(a_1+\beta_0
\ln\frac{\mu^2}{{\bf q}^2}).
\end{eqnarray}
Expressed in terms of the insertion functions, this correction is
\begin{eqnarray}
\delta_1 G(E)&=&4\pi \alpha_s C_F \,\frac{\alpha_s}{4\pi}\, 
J^{(C)}[1;a_1+\beta_0 L_q].
\label{eq:dG1}
\end{eqnarray}
We recall from paper I, section~5, that every insertion produces a 
factor $i\delta V i \hat{G}_0(E)$, hence single and triple insertions 
receive a factor of $(-1)$ from the factors of $i$ not included in the 
definition of the $J$-functions. 

\subsubsection{Second-order correction}

In this order in the non-relativistic power counting, we have to
include the single and double insertion of the Coulomb potential
(first and second line of (\ref{eq:GNNLO}) below), the single insertion
of the one-loop $1/m$ potential and tree-level $1/m^2$ potential 
(third and forth line in (\ref{eq:GNNLO})) with coefficient as given 
in paper I, section~\ref{sec:potentials}, as well as the
kinetic correction (last line in (\ref{eq:GNNLO})). The result is 
expressed in terms of the previously calculated insertion function as 
\begin{eqnarray} \label{eq:GNNLO}
\delta_2 G(E) &=& 
4\pi \alpha_s C_F \left(\frac{\alpha_s}{4\pi}\right)^2
J^{(C)}[1;a_2+(2a_1\beta_0+\beta_1)L_q+\beta_0^2 L_q^2]
\nonumber \\ 
&& \hspace*{-1cm} 
+ \, (4\pi \alpha_s C_F)^2 \left(\frac{\alpha_s}{4\pi}\right)^2 
J^{(C)}[1,1;a_1+\beta_0 L_q]
\nonumber \\ 
&&\hspace*{-1cm}
-\,\frac{\pi^2 (4\pi \alpha_s C_F)}{m}\,\frac{\alpha_s}{4\pi}\,
J[\frac{1}{2}+\epsilon;b_1(\epsilon)] 
+ \frac{4\pi \alpha_s C_F}{m^2} \, J[0;v_0(\epsilon)]
\nonumber \\ 
&&\hspace*{-1cm}
+\, 4\pi \alpha_s C_F 
\bigg(\frac{E}{m} J[1;1] + \frac{\pi\alpha_s C_F}{2 m} 
J[\frac{1}{2}+\epsilon;k(0)]\bigg)
\nonumber\\ 
&&\hspace*{-1cm}
+\,\frac{1}{4}\,\bigg(\frac{E^2}{m} J[\delta;1]
+\frac{2 E}{m}G_0(E)+ 8\pi \alpha_s C_F \,\frac{E}{m} J[1;1]
+\frac{2 (\pi\alpha_s C_F)^2}{m} J[\frac{1}{2}+\epsilon;k(0)]\bigg)\,.
\nonumber\\[-0.1cm]
&&
\end{eqnarray}
The last two lines arise after applying the equation-of-motion identities 
described in paper I, section
\ref{sec:eqofmotion} to the $\bff{p}^2/m^2$ potential and kinetic 
energy correction. 
The function $k(0)$ comes from an integration after using the
equation of motion and the definition is given in
(\refI{eq:def-k(u)}). 
 . 
\subsubsection{Third-order correction}

We split the third-order potential correction into 
three terms 
\begin{equation}
\delta_3G(E) = [\delta_3G(E)]_{\rm C} +[\delta_3G(E)]_{\rm nC} 
+ [\delta_3G(E)]_{\rm mixed}\,.
\end{equation}
The first term is associated with higher-order corrections 
and multiple insertions of the Coulomb potential and corresponds 
to the calculation already performed in \cite{Beneke:2005hg}. 
The other two terms represent the single-insertion of loop 
corrections to the non-Coulomb potential and the mixed double insertion 
of leading-order non-Coulomb potentials with the one-loop 
correction to the Coulomb potential.

In terms of insertion functions, the first two terms read
\begin{eqnarray}
[\delta_3G(E)]_{\rm C} &=& 
4\pi\alpha_s C_F\left(\frac{\alpha_s}{4\pi}\right)^3\times
\nonumber\\ 
&&\hspace{-1.2cm}\bigg\{
J^{(C)}[1;a_3+(2a_1\beta_1+\beta_2+3a_2\beta_0+8\pi^2C_A^3)L_q
+(\frac{5}{2}\beta_0\beta_1+3a_1\beta_0^2)L_q^2+\beta_0^3L_q^3]
\nonumber\\
&&\hspace{-1.2cm} +\,2\, (4\pi \alpha_s
C_F) \,J^{(C)}[1,1;a_2+(2a_1\beta_0+\beta_1)L_q+\beta_0^2 L_q^2]
\nonumber \\ 
 &&\hspace{-1.2cm} +\,(4\pi \alpha_s
C_F)^2\,J^{(C)}[1,1,1]\,\bigg\}\,,
\label{eq:GNNNLOC} \\[0.2cm]
[\delta_3G(E)]_{\rm nC}&=& 
-\,\frac{\pi^2 (4\pi \alpha_s C_F)}{m}\left(\frac{\alpha_s}{4\pi}\right)^{\!2}
\,\bigg\{J[\frac{1}{2}+2\epsilon;-\frac{8}{3\epsilon}(2 C_FC_A+C_A^2)
+\frac{2\beta_0}{\epsilon}b_1(\epsilon)+4 b_2(\epsilon)]
\nonumber \\ 
&&\hspace{-0.6cm} 
-J[\frac{1}{2}+\epsilon;\frac{2\beta_0}{\epsilon} b_1(\epsilon)]\bigg\}
\nonumber\\
&&\hspace{-1.2cm}
+ \,\frac{4\pi \alpha_s C_F}{m^2} \,\frac{\alpha_s}{4\pi}\, 
\bigg\{J[\epsilon;\frac{1}{\epsilon}\bigg(\frac{7}{3}C_F
-\frac{11}{6}C_A+\beta_0v_0(\epsilon)\bigg)+v_q^{(1)}(\epsilon)]
\nonumber \\ 
&&\hspace{-0.6cm}
+\,J[0;-\frac{1}{\epsilon}\beta_0 v_0(\epsilon)
+\bigg[\bigg(\frac{\mu^2}{m^2}\bigg)^{\!\epsilon}-1\bigg]
\frac{1}{\epsilon}\bigg(\frac{C_F}{3}+\frac{C_A}{2}\bigg)
+\bigg(\frac{\mu^2}{m^2}\bigg)^{\!\epsilon}v_m^{(1)}(\epsilon)]\bigg\}
\nonumber \\ 
&&\hspace{-1.2cm}
+\,4\pi \alpha_s C_F\,\frac{\alpha_s}{4\pi}\,
\bigg\{\frac{E}{m}J[1+\epsilon;\frac{1}{\epsilon}\bigg(\frac{8}{3}C_A
+\beta_0\bigg)+v_p^{(1)}(\epsilon)] + \frac{\pi\alpha_s C_F}{2m}\times
\nonumber \\ 
&&\hspace{-0.6cm}
\bigg(\!J[\frac{1}{2}+2\epsilon;\frac{k(\epsilon)}{\epsilon}
\bigg(\frac{8}{3}C_A+\beta_0\!\bigg)+k(\epsilon)v_p^{(1)}(\epsilon)]
-J[\frac{1}{2}+\epsilon;
\frac{k(0)}{\epsilon}\bigg(\frac{8}{3}C_A+\beta_0\!\bigg)]\bigg)\bigg\}.
\nonumber\\[-0.1cm]
\label{eq:GNNNLOnC}
\end{eqnarray}
Note that there is no contribution from the kinetic correction to
the single insertion part at third order, since the kinetic energy term 
in the Lagrangian is not renormalized. The contribution from  the 
${\bf p}^2/m^2$ potential (last two lines in previous equation) 
was again reduced to other insertions with the equation-of-motion 
identities. 

The two-loop $d$-dimensional $1/m$ potential contains momentum dependence 
of the form
\begin{eqnarray}
\label{eq:NAform-eps2}
\Bigg[\bigg(\frac{\mu^2}{\bff{q}^2}\bigg)^{\!2\epsilon}-
\bigg(\frac{\mu^2}{\bff{q}^2}\bigg)^{\!\epsilon}\,\Bigg].
\end{eqnarray}
Such terms do not correspond directly to the definition of 
$J[1/2+a\epsilon]$, because there we always subtract 1 from
$(\mu^2/\bff{q}^2)^{a\epsilon}$, see (\ref{eq:J-definition1}). We therefore 
use the combination
$(J[1/2+2\epsilon,...]-J[1/2+\epsilon,...])$ in the result for the 
non-Coulomb contributions to recover the form 
of (\ref{eq:NAform-eps2}).

Finally, the mixed Coulomb-non-Coulomb contributions are given by
\begin{eqnarray}\label{eq:GNNNLOmixed}
[\delta_3G(E)]_{\rm mixed}&=&
2\,(-4\pi\alpha_s C_F)\,\frac{\alpha_s}{4\pi}\,\times \Bigg\{
\nonumber\\ 
&&\hspace{-1.2cm}
+\,\frac{\pi^2 (4\pi \alpha_s C_F)}{m}\,\frac{\alpha_s}{4\pi}\,
J[\frac{1}{2}+\epsilon,1+\epsilon;b_1(\epsilon)] 
- \frac{4\pi \alpha_s C_F}{m^2} \,J[0,1+\epsilon;v_0(\epsilon)]
\nonumber \\
&&\hspace{-1.2cm} -\,4\pi\alpha_s C_F \,
\bigg\{\frac{E}{m} J[1,1+\epsilon;1]
+\frac{\pi\alpha_sC_F}{2m} J[\frac{1}{2}+\epsilon,1+\epsilon;k(0)]
\nonumber\\ 
&&
+\,\frac{1}{16 m} \bigg(\!
J[\frac{1}{2}+2\epsilon;k(\epsilon)\left(\frac{\beta_0}{\epsilon}+
a_1(\epsilon)\right)]
-J[\frac{1}{2}+\epsilon;k(0)\frac{\beta_0}{\epsilon}]\bigg)\bigg\}
\nonumber \\ 
&&\hspace{-1.2cm}
-\,\frac{1}{4} \,\bigg\{\frac{E^2}{m} J[\delta,1+\epsilon;1]
+\frac{2 E}{m} J[1+\epsilon;\frac{\beta_0}{\epsilon}+a_1(\epsilon)]
\nonumber \\ 
&&
+\,\frac{\pi\alpha_s C_F}{2 m} 
\left(J[\frac{1}{2}+2\epsilon;k(\epsilon)\left(\frac{\beta_0}{\epsilon}
+a_1(\epsilon)\right)]-J[\frac{1}{2}+\epsilon;k(0)\frac{\beta_0}{\epsilon}]
\right)
\nonumber \\ 
&&+ 4\pi\alpha_s C_F \,\frac{2 E}{m} J[1,1+\epsilon;1]
+ 4\pi\alpha_s C_F \,\frac{\pi \alpha_s C_F}{2 m} 
J[\frac{1}{2}+\epsilon,1+\epsilon;k(0)]\bigg\}\Bigg\}.
\nonumber\\[-0.1cm]
\end{eqnarray}
Here the last three lines originate from the double insertion 
of the one-loop Coulomb potential with the kinetic energy corrections 
after using the equation-of-motion identity (\refI{eq:eomkindouble}). 
We note that $J[1,1+\epsilon;1]$ equals $J^{(C)}[1,1;1]$ 
given in (\ref{eq:couldouble}) up to $O(\epsilon)$ terms, which can 
be dropped.

\subsubsection{Divergent part}
\label{sec:divpart}

The second- and third-order results are still divergent. Only the
combination with the matching coefficients of the external currents 
(from second order), the ultrasoft correction (from third order) 
is finite in the $\Gamma\rightarrow 0$ limit. For non-vanishing 
width, that is, imaginary part of $E$, the divergences cancel only 
together with the non-resonant contribution to the $e^+ e^- \to 
W^+ W^- b\bar b$ cross section. 

The divergent part of the second-order Green function reads
\begin{eqnarray}
\label{eq:GNNLOdivpart}
\mbox{Im}[\delta_2 G]_{\rm div} &=& \frac{\alpha_s^2
C_F}{6\epsilon}\left(C_F+\frac{3}{2}C_A\right) \mbox{Im}[G_0(E)]
+\frac{m \alpha_s C_F\Gamma}{8\pi \epsilon} \, .
\end{eqnarray} 
The coefficient of $\mbox{Im}[G_0(E)]$ is related to the 
anomalous dimension of the non-relativistic vector current 
\cite{Beneke:1997jm}. The pole proportional to the width indicates 
that the combination with the non-resonant contribution is required 
for consistency at NNLO. 
The divergent part of the third-order Green function is given by
\begin{eqnarray}\label{eq:GNNNLOdivpart}
\mbox{Im}[\delta_3 G]_{\rm div} &=& 
\Bigg[ -\frac{1}{\epsilon^2}\,
  \bigg(\,
       \frac{7}{72}C_F^{2}
       +\frac{2}{9}C_A^{2}
       +\frac{23}{48}C_A C_F
       +\beta_0\,\bigg(\frac{C_A}{24}+\frac{C_F}{36}\bigg)
  \bigg)
\nonumber
\\
&& \hspace*{-1.5cm}- \, \frac{1}{\epsilon}\, \Bigg\{
 \left(\frac{11}{24}
      -\frac{L_m}{12}
 \right)C_F^{2}
+
 \left(\frac{427}{324}
      -\frac{4\ln{2}}{3}
      -\frac{L_m}{8}
 \right)C_AC_F
-
 \left(\frac{5}{216}
      +\frac{2\ln{2}}{3}
 \right)C_A^{2}
\nonumber
\\
&&\hspace*{-1.5cm} +\,
 \left(\frac{C_A}{24}
      +\frac{C_F}{54}
 \right)\beta_0
-
 \left(\frac{1}{30}
      -\frac{29\,n_f}{162}
 \right)C_F T_F
+ \frac{49}{216}C_A T_F n_f \Bigg\} \Bigg]\, 
\frac{\alpha_s^3 C_F}{\pi}\, \mbox{Im}[G_0(E)]
\nonumber
\\
&& \hspace*{-1.5cm}+\,\frac{\alpha_s^2
C_F}{6\epsilon}\left(C_F+\frac{3}{2}C_A\right) \mbox{Im}[\delta_1 G(E)]
\nonumber\\ 
&& \hspace*{-1.5cm}-
\frac{m\alpha_s^2 C_F\Gamma}{32\pi^2}\,
\Bigg[\frac{4C_A+3\beta_0}{9\epsilon^2}+\frac{42 C_F-71 C_A+40 n_f
T_F+6\beta_0}{27\epsilon}\Bigg]\, ,
\nonumber\\ 
&& \hspace*{-1.5cm}+\,\mbox{Im}[\delta^{us}G(E)]_{\rm div}\end{eqnarray}
with $L_m=\ln(\mu/m)$. The divergent part of the third-order ultrasoft 
contribution has already been given in \cite{Beneke:2008cr} and includes 
further divergent parts proportional to the top width $\Gamma$.

An additional source of finite-width divergence to the cross section  
arises from the multiplication of the non-relativistic Green function with 
the short-distance matching coefficients. Up to the third order, the 
relevant expressions is
\begin{equation}
\label{eq:sigma}
\sigma_{t\bar t} = \sigma_0 \times 12\pi e_t^2 \,\mbox{Im}\left[
\frac{N_{c}}{2m^{2}}\left(c_v 
\left[c_v-\frac{E}{m}\,\left(c_v+\frac{d_v}{3}\right)\right] G(E)
+\ldots\right)\right],
\end{equation}
see (\refI{R1}), which should be expanded out order by order. The 
following observation needs to be made 
here.\footnote{See also (\refI{eq:pitoNRQCD}) and the corresponding 
footnote there, as well as (\ref{eq:piv}) below for the distinction 
between complex and real energy variables.} The term $E d_v$ involving 
the matching coefficient $d_v$ of the higher-derivative current 
(\refI{eq:currentspinprojection}) arises from an exact equation-of-motion 
relation for the Green function. When $E$ is taken to be complex to 
include the finite decay width of the quark, $E$ must be assumed to 
be complex in this term. On the other hand, the term involving 
$E c_v$ in square brackets arises from the expansion of the kinematic 
factor $1/s = 1/(2 m+E)^2 =1/(4 m^2) (1-E/m + \ldots)$ and, since 
$s$ is always real, $E$ must always be understood as a real quantity in 
the product $E c_v$.

This being said, the additional contribution to the finite-width 
divergence of the round bracket in (\ref{eq:sigma}) at second order 
is given by 
\begin{equation}
\label{eq:finitewidthdivextcurrent2ndorder}
-\frac{d_v^{(0)}}{3 m} \,\mbox{Im}[E G_0(E)]_{\rm div} = 
-\frac{m\alpha_s C_F\Gamma}{48\pi\epsilon}\,,
\end{equation}
where the $1/\epsilon$ pole originates from the real part of $G_0(E)$.
At third order there is a trivial term arising from multiplying out 
$c_v^2 G(E)$ given by $2 c_v^{(1)} \alpha_s/(4\pi) 
\,\mbox{Im}[\delta_2 G]_{\rm div}$. The additional finite-width divergences 
involving the higher-derivative current read
\begin{eqnarray}
&&-\frac{d_v^{(0)}}{3 m} \,\mbox{Im}[E \delta_1 G(E)]_{\rm div} 
-\frac{1}{3 m}\frac{\alpha_s}{4\pi} \left[d_v^{(1,\rm fin)}+
d_v^{(0)} c_v^{(1)}\right]\,\mbox{Im}[E G_0(E)]_{\rm div} 
\nonumber\\
&& = 
-\frac{m\alpha_s^2 C_F\Gamma}{192\pi^2}\,\left[
-\frac{\beta_0}{3\epsilon^2}+\frac{1}{\epsilon}
\left(-\frac{2\beta_0}{3}+\frac{2 a_1}{3}+c_v^{(1)}+d_v^{(1,\rm fin)}\right)
\right].
\label{eq:finitewidthdivextcurrent3rdorder}
\end{eqnarray}
The (unsubtracted) one-loop matching coefficient $d_v^{(1)}$ has 
itself a $1/\epsilon$ pole. Only the finite part $d_v^{(1,\rm fin)}$ should be kept  
here, as indicated above, since the divergent part is already 
included as a counterterm in the definition (\refI{eq:defUS}) of the 
ultrasoft correction $\delta^{us}G(E)$.

The cancellation of the divergent parts not proportional to $\Gamma$ 
with the vertex correction and the ultrasoft correction has been 
checked explicitly. What needs to be checked is that $Z_J^{-2} \,\mbox{Im}
\,[G(E)]$ is finite as $\epsilon\to 0$, where $Z_J$ is the 
renormalization constant of the non-relativistic 
current $\psi^\dagger \sigma^i\chi$, which also relates 
the bare matching coefficient to the renormalized one 
through $c_v^{\rm bare} = Z_J^{-1} c_v$. We can construct 
$Z_J^{-1}$ from the $\mu$-dependence of $c_v$ by imposing 
that $c_v^{\rm bare}$ must be independent of $\mu$, which 
implies
\begin{equation}
\frac{d}{d\ln\mu} \,\ln Z_J^{-1} = 
-\frac{d}{d\ln\mu} \,\ln c_v\,.
\label{eq:ZJeq}
\end{equation}
We then compute the right-hand side from the renormalized, 
$\epsilon$-independent coefficients 
(\refI{eq:cv1}) -- (\refI{eq:cv3}) to the three-loop order. 
We make an ansatz for the $\overline{\rm MS}$ renormalization 
factor in terms of an expansion in $\alpha_s$ in terms of 
pure poles of maximal order $n$ at $O(\alpha_s^n)$ and up 
to $n-1$ powers of $L_m=\ln(\mu/m)$, and 
compute the $\mu$-derivative, now with the $d$-dimensional 
QCD beta function for $\alpha_s$. Matching both sides of 
(\ref{eq:ZJeq}), 
we obtain 
\begin{eqnarray}
Z_J^{-1} &=& 1+ \left(\frac{\alpha_s}{4\pi}\right)^{\!2} 
\frac{1}{\epsilon}\,C_F\pi^2 \left[-\frac{4}{3} C_F-2 C_A\right]
\nonumber\\
&&+\,  \left(\frac{\alpha_s}{4\pi}\right)^{\!3}
C_F\pi^2 \,\Bigg\{\,
\frac{1}{\epsilon^2}
\left[-\frac{20}{9}C_F^2-\frac{10}{3}C_F C_A-\frac{8}{9}C_A^2 
+\frac{2\beta_0}{3}\left(\frac{4}{3} C_F+2 C_A)\right)\right]
\nonumber\\
&&\hspace*{1cm}+\,\frac{1}{\epsilon}
\,\bigg[\left(-\frac{40}{3}C_F^2-20C_F C_A-\frac{16}{3}C_A^2 
\right)\ln\frac{\mu}{m}
+\left(-\frac{172}{9}+32\ln{2}\right) C_F^2
\nonumber \\
&&\hspace{2cm}
+\left(-\frac{1808}{81}-16\ln{2}\right) C_F C_A
+\left( -\frac{128}{27}-16\ln{2}\right)  C_A^2
\nonumber \\
&&\hspace{2cm}
+\,\frac{400}{81} C_F T_F n_f
+\frac{148}{27} C_A T_F n_f
-\frac{16}{15} C_F T_F
\bigg]\,\Bigg\}
\label{eq;ZJinv}
\end{eqnarray}
The absence of an $O(\alpha_s)$ term reflects the 
well-known fact that the non-relativistic current is renormalized 
first at the two-loop order. The 
$O(\alpha_s^2)$ was first given in \cite{Beneke:1997jm},
and the third order terms agrees with \cite{Marquard:2014pea}. 
It is then straightforward to verify that $Z_J^{-2} \,\mbox{Im}
\,[G(E)]$ is indeed finite as $\epsilon\to 0$.\footnote{
In \cite{Beneke:2007pj} it was stated that the pole part of 
the third-order correction to the wave function at the origin, 
$|\psi(0)|^2$, as given in Eq.~(14) of that paper, cancels with twice 
the divergent part of the third-order correction to $c_v$, 
while the correct statement should be that it ``cancels with twice 
the divergent part of the third-order correction to $Z_J^{-1}$".}

The remaining finite-width divergences cancel upon including higher-order 
finite-width effects not captured by making the energy in the QCD 
spectral function complex, the $P$-wave contribution, and the 
non-resonant contribution to the process $e^+ e^-\to W^+ W^- b\bar b$, 
as will be discussed further in the following section. The cancellation 
of finite-width divergences in the sum of resonant and non-resonant 
contributions has been shown in \cite{Jantzen:2013gpa} at NNLO. 
At NNNLO this check is not possible at present. 

\subsubsection{Checks of renormalization scale independence}

In addition to the cancellation of the $1/\epsilon$ poles, 
we checked that for real energy $E$, the $\mu$-dependence 
of the cross section, resp. $R$-ratio (\refI{R1}), cancels up to terms formally of 
higher order, i.e. 
\begin{equation}
\frac{d\sigma}{d\ln \mu}{}_{|t\bar t} = 
{\cal O}(v\alpha_s^4)\qquad \mbox{(N$^4$LO)}\,.
\end{equation}
This check includes the logarithms from the running of 
the strong coupling and the matching coefficients of the 
non-relativistic currents and effective Lagrangian. 
When all pieces are combined, the remaining explicit 
dependence on $\ln\mu$ must be cancelled by the scale 
dependence of $\alpha_s(\mu)$. The confirmation of this 
fact is non-trivial, especially at NNNLO, 
as it requires expanding the 
insertion functions in $\lambda$, which contains implicit 
dependence on $\alpha_s(\mu)$. With the exception of 
a few insertion-function derivatives, we verify the 
scale independence analytically. For the numerical 
evaluations, we achieve relative $5\cdot 10^{-7}$ 
accuracy. 

The restriction to real energy $E$ is necessary, as for 
non-vanishing top width, there is uncancelled scale 
dependence associated with the finite-width divergences 
discussed above. This scale dependence cancels with the 
non-resonant contributions. In section~\ref{sec:refine} 
we explain how the finite-width scale dependence can 
be separated from the $\ln\mu$ terms related to the 
strong coupling. 


\section{Refinements}
\label{sec:refine}

Up to now we considered the perturbative expansion of the non-relativistic 
vector current two-point function in PNRQCD perturbation theory in 
dimensional regularization. In this 
section we discuss several modifications of the perturbative expansion, 
which are useful or required to achieve an accurate result for the 
top pair production threshold. First we elaborate on finite-width effects 
and separate the scale dependence due to the factorization 
of resonant and non-resonant effects from the renormalization scale 
dependence of the strong coupling. We then consider  
the need to resum PNRQCD perturbation theory for energy values near 
the bound state poles and the elimination of the pole mass in terms 
of renormalized mass parameters, which lead to a better convergence 
of the expansion. Some subtleties that arise with the implementation 
of these better-suited mass parameters in PNRQCD perturbation 
theory near the continuum threshold are discussed subsequently.

\subsection{Top quark width effects}
\label{sec:widtheffects} 

The width of the top quark is given at tree-level by the expression 
\begin{equation}\label{eq:widthSM}
\Gamma_t=\frac{G_F m_t^3}{8\pi\sqrt{2}}
\bigg(1-\frac{M_W^2}{m_t^2}\bigg)^2\bigg(1+2\,\frac{M_W^2}{m_t^2}\bigg)\,.
\end{equation}
Including the effect from the finite bottom quark mass, 
NLO \cite{Jezabek:1988iv} and NNLO \cite{Blokland:2005vq,Gao:2012ja,Brucherseifer:2013iv} QCD, 
as well as one-loop electroweak corrections \cite{Denner:1990ns,Eilam:1991iz}, 
this evaluates to $\Gamma_t \approx 1.36$~GeV for $m_t=173.3$~GeV. 
Within the accuracy considered here the top quark decays exclusively 
to $b W^+$. Despite the fact that $\Gamma_t\ll m_t$, the top width 
has a large effect on the production threshold, since $\Gamma_t$ 
is of the same order as the energy scale $m_t\alpha_s^2$ of the bound state 
dynamics and the kinetic energy of the top quark.\footnote{The ratio 
$\Gamma_t/m_t$ is therefore counted as 
$\alpha_s^2\sim v^2$ in non-relativistic power counting. This is 
consistent with the counting of the finite-width divergence in 
(\ref{eq:GNNLOdivpart}) etc.} 
The toponium states acquire a large width such that 
only a single broad resonance from the original $n=1$ bound state remains. Since $\Gamma_t$ is larger 
than the strong interaction scale, the top quark does not form 
top-hadrons before it decays, making it behave like a free 
quark in first approximation \cite{Bigi:1986jk,Kuhn:1980gw}. Because 
of this the ``line shape'' of the production threshold can be 
computed locally with perturbative effective field theory methods.

\subsubsection{Resonant finite-width corrections}
\label{sec:higherwidth}

The effective non-relativistic Lagrangians presented in paper I 
literally apply only to the dynamics of a pair of stable heavy 
quarks. The effective Lagrangian accounts for  quarks close to their 
mass-shell. When the propagator of an unstable heavy quark field 
is expanded around the position of its complex resonance pole 
$m_t^2-i m_t\Gamma_t$, one finds that the terms 
\begin{eqnarray}
{\cal L}_{\Gamma} &=& \psi^\dag \Bigg( i \frac{
\Gamma_t}{2}+i\frac{\Gamma_t {\bf D}^2}{4 m^2}-\frac{\Gamma_t^2}{8
m}\Bigg)\psi -\chi^\dag \Bigg( i \frac{ \Gamma_t}{2}+i\frac{\Gamma_t
{\bf D}^2}{4 m^2}-\frac{\Gamma_t^2}{8 m}\Bigg)\chi\, .
\label{eq:width-lagrangian}
\end{eqnarray}
have to be added to the NRQCD 
Lagrangian~\cite{Beneke:2007zg,Beneke:2004xd} to achieve 
NNNLO accuracy. The first term in both brackets is a leading-order effect, 
since $\Gamma_t$ counts as $i D^0$. This term modifies the 
non-relativistic propagator to  
\begin{eqnarray}
\frac{i\delta^{ij}}{p^0-\frac{{\bf p}^2}{2m}+i\frac{\Gamma_t}{2}}, 
\end{eqnarray}
and the one of the quark-antiquark pair with energy $E=\sqrt{s}-2 m$ 
by $E\to E+i\Gamma_t$. Since we {\em defined} the QCD contribution to 
the production cross section by continuing the correlation function 
to complex energy according to this prescription, the leading 
finite-width effect is exactly accounted for in our PNRQCD calculation.

The remaining terms lead to additional finite-width corrections to 
the cross section, which become relevant first at NNLO. There are no NNNLO 
corrections to the Lagrangian, since the width counts as ${\cal O}(v^2)$ and 
since the bilinear terms in the Lagrangian are not renormalized by QCD 
loop corrections, but the second-order correction terms in the Lagrangian 
generate third-order corrections to the cross section, when combined with 
the single insertion of the first-order Coulomb potential.

The effect of the additional terms can be included to all orders by a 
simple redefinition of the leading-order non-relativistic Lagrangian. 
The lifetime dilatation term 
$i \Gamma_t\bff{D}^2/(4 m^2)$ amounts to a rescaling of the 
kinetic term $\bff{D}^2/(2m)$, while the term quadratic in the 
width is a correction to the energy shift. If $G_0(m,E)$ denotes 
the leading-order Coulomb Green function, where we have now made 
the dependence on the mass through the kinetic term explicit, 
the finite-width terms (\ref{eq:width-lagrangian}) are included 
exactly in the expression 
\begin{equation}
G_{\Gamma}(m,E) = G_0\left(\frac{m}{1+i\Gamma/(2m)},
{\cal E}-\frac{\Gamma^2}{4 m}\right)\,,
\label{eq:gammaeffectsexact}
\end{equation} 
where here $E$ is real and ${\cal E}=E+i\Gamma$.
Up to NNNLO only a single insertion of the higher-order finite-width 
terms in (\ref{eq:width-lagrangian}) is required. The single insertion 
is obtained by expanding (\ref{eq:gammaeffectsexact}) to first order, 
which results in the NNLO terms 
\begin{equation}
-\frac{\Gamma^2}{4 m} \,\mbox{Im}\left[\frac{\partial G_0}{\partial E}
(m,{\cal E})\right] - 
\frac{\Gamma}{2} \,
\mbox{Re}\left[\frac{\partial G_0}{\partial m}(m,{\cal E})\right]
\label{eq:gammaeffectssingle}
\end{equation}
to the imaginary part of the Green function. The derivatives are 
straightforward to compute. The explicit expressions can be found in 
\cite{Beneke:2017rdn}. The second term, which 
originates from the lifetime dilatation term, causes an additional 
finite-width divergence 
\begin{equation}
-\frac{m\alpha_s C_F\Gamma}{16\pi\epsilon}
\end{equation}
in the imaginary part of the Green function, 
$\mbox{Im}\,G_{\Gamma}(m,E)$. 

Since the Coulomb potential is independent of $m$ to all orders 
in $\alpha_s$, (\ref{eq:gammaeffectsexact}) holds for the Green function, 
when corrections to the Coulomb potential are included, but not 
in general. However, this is sufficient to obtain the NNNLO correction 
from the higher-order finite-width terms in the Lagrangian by 
simply replacing $G_0$ with $\delta_1 G$ in (\ref{eq:gammaeffectsexact}) 
and (\ref{eq:gammaeffectssingle}), since only the Coulomb 
potential contributes to $\delta_1 G$. The derivatives can be computed 
using (\ref{eq:dG1}) and the results from section~\ref{sec:singleC}. 

While the extra terms are straightforward to obtain, we do 
not include them in this paper. Rather we focus on the QCD contribution  
alone defined through the pure-QCD correlation functions continued to 
complex energy $E+i\Gamma$ and refer to \cite{Beneke:2017rdn} for the 
computation of NNLO electroweak effects including the NNLO finite-width 
corrections discussed above.\footnote{In the framework 
of unstable particle effective theory 
\cite{Beneke:2004xd,Beneke:2003xh} the matching coefficient 
of the non-relativistic $t\bar t$ production currents must  
be computed by ``on-shell'' matching on the complex-mass pole 
position $p^2=m^2-i m\Gamma$ (rather than $p^2=m^2$), which 
introduces another finite-width correction to the QCD on-shell 
matching for stable quarks. This correction is included at NNLO in 
\cite{Beneke:2017rdn} through the 
term $\sigma_{C^{(k)}_{{\rm Abs},Z_t}}$, which also 
includes the corresponding correction for the $P$-wave contribution.  
}

In the following we adopt the following convention. 
Whenever we discuss the Green function $G(E)$ or $G_0(E)$ itself, 
we regard it as an analytic function of the complex variable $E$. 
However, when the Green function appears in the formula for 
the top cross section or the current correlation function, such 
as in (\ref{eq:piv}) below, then $E$ refers to the real quantity 
$\sqrt{s}-2 m$, and we use $\mathcal{E}=E+i\Gamma$ for the complex 
argument of the Green function including the top decay width.

\subsubsection{Non-resonant corrections and 
finite-width scale dependence}

The uncancelled finite-width divergences make it clear that 
the non-relativistic description of the resonant $t\bar t$ process 
is not complete. A consistent treatment can only be given for 
the  process $\sigma_{e^+ e^-\to W^+ W^- b \bar b}$ involving 
the final state of decayed top quarks, which, however, can be 
produced also non-resonantly, through off-shell internal top quark lines, 
or without internal top quarks at all. ``Unstable particle effective 
theory'' \cite{Beneke:2004xd,Beneke:2003xh} provides a systematic 
framework for combining resonant and non-resonant contributions 
in an expansion in $\Gamma/m$, in which the finite-width divergences 
of the resonant cross section are consistently cancelled with 
divergences of the non-resonant contribution. The framework has 
been used to study the $W^+ W^-$ pair production threshold with 
high accuracy \cite{Beneke:2007zg,Actis:2008rb}. For the case of 
the top pair threshold non-resonant production of the $W^+ W^- b \bar b$ 
contributes from NLO \cite{Beneke:2010mp}, but the NLO contribution 
is finite in dimensional regularization, and divergences appear first 
at NNLO, consistent with the non-relativistic computation of the 
resonant cross section with the same regularization.

The result for the third-order resonant top pair production cross section 
derived in this paper depends on the scale of dimensional regularization 
through $\alpha_s=\alpha_s(\mu)$ and explicit logarithms of $\mu$. It is 
useful to separate the explicit scale dependence into the part that 
cancels the implicit scale dependence of $\alpha_s$, and the part that is 
associated with the finite-width divergences for the following 
reason.\footnote{There is no remaining dependence on $\mu$ due to factorizing 
the QCD expression into hard, potential, soft and ultrasoft 
contributions after summing all such contributions at a given order.}
Denoting by $\mu_w$ the scale related to the finite-width scale 
logarithms, and $\mu$ the standard coupling renormalization scale, 
the physical cross section can be represented as the sum of two terms, 
\begin{equation} 
\sigma_{e^+ e^-\to W^+ W^- b \bar b} = 
\sigma_{t\bar t, \rm resonant}(\mu,\mu_w) + 
\sigma_{\rm non-resonant}(\mu,\mu_w)\,,
\end{equation}
The dependence on $\mu$ from the scale dependence of the strong coupling 
cancels {\em within} the two contributions  
separately up to dependence of higher order, hence at NNNLO, we have
\begin{equation}
\mu\frac{d\sigma}{d\mu}{}_{|t\bar t, \rm resonant} = 
{\cal O}(v\alpha_s^4)\qquad \mbox{(N4LO)}\,.
\end{equation}
On the other hand, $\mu_w$ is a factorization scale and the dependence on it 
cancels order-by-order 
exactly, but only {\em between} the two terms. Since the non-resonant 
contribution is not included in the present calculation, and known 
fully only to NNLO~\cite{Beneke:2017rdn}, the NNNLO result discussed here 
has an uncancelled dependence on $\mu_w$ of the form 
\begin{equation}
\mu_w\frac{d\sigma}{d\mu_w}{}_{|t\bar t, \rm resonant} = 
{\cal O}(\alpha_s \Gamma)\qquad \mbox{(NNLO)}\,,
\end{equation}
which is formally of second order. If we did not separate the two scales, 
the scale variation of the third-order cross section would not 
parametrically represent the improvement of the calculation of the 
resonant QCD cross section when going from NNLO to NNNLO, which is the 
main result of this work.\footnote{However, as will be seen in the later 
analysis, the dependence on $\mu_w$ is numerically smaller than the $\mu$ 
dependence at the same order except below threshold.}

In the following we provide the expressions that have to be added 
to the resonant NNNLO cross section to convert the previous result 
with a single scale to the one with the dependence on $\mu$ and $\mu_w$ 
separated. Here we discuss the full third-order cross section and not 
only the Green function with potential insertions, more precisely, we 
consider the imaginary part of the vector current correlator 
\begin{equation}
\label{eq:piv}
\Pi^{(v)}(q^2) = \frac{N_{c}}{2m^{2}}\,c_v 
\left[c_v-\frac{E}{m}\,\left(c_v+\frac{{\cal E}}{E}\frac{d_v}{3}\right)\right] 
G({\cal E}) \,.
\end{equation}
To this end we reanalyzed the ultrasoft 
calculation \cite{Beneke:2008cr}, and extracted the logarithms of $\mu$ due to  
the finite-width divergent parts. Note, 
however, that we do not include the finite-width scale 
dependence generated by the higher-order width effects discussed 
in section~\ref{sec:higherwidth}, since these are not included in 
our definition of the QCD contributions.

The second order finite-width scale dependence is in direct correspondence 
with the two divergent terms (\ref{eq:GNNLOdivpart}), 
(\ref{eq:finitewidthdivextcurrent2ndorder}). The expression 
\begin{equation}
\mbox{Im}[\delta_2\Pi^{(v)}(q^2)]_{\mu_w} = 
 \frac{N_{c}}{2m^{2}}\,\frac{m \alpha_s C_F\Gamma}{4\pi} 
\left[2-\frac{1}{3}\right]\ln\frac{\mu_w}{\mu}\,.
\label{eq:2ndorderlogmuw}
\end{equation}
has to be added to the vector current spectral function to convert the 
corresponding logarithms of $\mu$ into logarithms of $\mu_w$. By 
construction it vanishes for $\mu_w=\mu$. The third-order 
contribution to be added is 
\begin{eqnarray}
\mbox{Im}[\delta_3\Pi^{(v)}(q^2)]_{\mu_w} &=& 
 \frac{N_{c}}{2m^{2}}\,\frac{m \alpha_s C_F\Gamma}{4\pi}\,
\frac{\alpha_s}{4\pi} \,\Bigg\{\,
\left[\frac{28}{3}C_F+2 C_A-\beta_0\left(2-\frac{1}{3}\right)\right]
\ln^2\frac{\mu_w}{\mu}
\nonumber\\
&&\hspace*{-2cm} 
+\,\bigg[\left(-\frac{56}{3}C_F-4 C_A\right)\ln\frac{m}{\mu}
+\left(-\frac{278}{9}-8\ln 2\right) C_F 
+\left(-\frac{49}{27}+12\ln 2\right)C_A 
\nonumber\\
&&\hspace*{-2cm} 
-\,\frac{88}{27} T_F n_f\bigg]\,\ln\frac{\mu_w}{\mu}
\,\Bigg\}\,.
\label{eq:3rdorderlogmuw}
\end{eqnarray}
We sketch the derivation of these results in the following subsection.

\subsubsection{Derivation of the finite-width scale dependence}
\label{sec:finitewidth}

The NNLO finite-width scale dependence is straightforward to obtain. 
Only three terms in (\ref{eq:GNNLO}) contain a finite-width divergence, 
which lead to a corresponding scale dependence:
\begin{eqnarray} 
\label{eq:GNNLOmuw}
\delta_2 G(E) &\supset& 
4\pi \alpha_s C_F 
\,\frac{E}{m} J[1;1] + \frac{1}{4}\,\bigg(\frac{2 E}{m}G_0(E)+ 
8\pi \alpha_s C_F \,\frac{E}{m} J[1;1]\bigg)\,,
\end{eqnarray}
where the first term arises from the $\bff{p}^2/(m^2 \bff{q}^2)$ 
potential and the second from the kinetic energy correction 
$\bff{p}^4/(4 m^3)$. The finite-width divergence and scale 
dependence of $E J[1;1]$ and $EG_0(E)$ can be inferred from 
(\ref{eq:Ja(1+aeps)}) and (\ref{eq:zerodistancegreen}), respectively, 
and are given by
\begin{eqnarray} 
\mbox{Im}\left[E G_0(E)\right]_{\rm div} &=& 
\Gamma \times \frac{m^2\alpha_s C_F}{4\pi} \left(\frac{1}{4\epsilon} 
+\ln \mu_w\right)\,
\\
\mbox{Im}\left[E J[1;1]\right]_{\rm div} &=& 
\Gamma \times \frac{m^2}{16\pi^2} \left(\frac{1}{4\epsilon} 
+\ln \mu_w\right).
\end{eqnarray}
Including the piece from the derivative current 
(see (\ref{eq:finitewidthdivextcurrent2ndorder})), we obtain 
\begin{equation}
\mbox{Im}\left[\delta_2 G(E)\right]_{\mu_w}-
\frac{d_v^{(0)}}{3 m} \,\mbox{Im}[E G_0(E)]_{\mu_w} = 
\frac{m\alpha_s C_F\Gamma}{4\pi} \left(1+1-\frac{1}{3}\right)
\ln\frac{\mu_w}{\mu}\,,
\label{eq:2ndordermuwresult}
\end{equation}
from which (\ref{eq:2ndorderlogmuw}) follows. The three terms in 
brackets arise, in order, from the  $\bff{p}^2/(m^2 \bff{q}^2)$ 
potential, kinetic energy correction and derivative current. 

Extracting the finite-width logarithms at the third order is considerably 
more involved, hence we adopt the following more systematic approach. In 
the expressions for $\Pi^{(v)}(q^2)$ and $G(E)$ we first replace 
$\alpha_s(\mu)$ by 
\begin{equation}
\alpha_s(\mu) = \alpha_s(\mu_r) + \frac{2\beta_0}{4\pi}\alpha_s(\mu_r)^2 
\ln\frac{\mu_r}{\mu}+\ldots,
\label{eq:couplingreplace}
\end{equation}
and re-expand them to NNNLO in $\alpha_s(\mu_r)$ in non-relativistic 
perturbation theory. The logarithms of $\mu_r/\mu$ convert the logarithms 
of $\mu$ in the original expression into logarithms 
of $\mu_r$, except for those not related to the running of the coupling. 
These left-over logarithms are precisely the finite-width scale dependent 
logarithms, hence in these we rename $\mu \to\mu_w$. In practice it is 
difficult to carry out this expansion in the insertion 
functions with a complicated dependence on $\lambda$. However, since 
the $\mu_w$ dependence must cancel with the non-resonant cross section, 
which is computed by an expansion in $E$, it must be polynomial in $E$ 
before taking the imaginary part. 
Specifically, at NNLO and NNNLO the 
dependence on $E$ and $\alpha_s$ can only be $\alpha_s E$ (NNLO) and 
$\alpha_s^2 E, E^2$ (NNNLO), respectively. The NNNLO non-resonant terms proportional 
to $E^2$ do not have a logarithmic divergence and hence do not 
generate $\mu_w$ dependence. It follows that the terms in 
$\Pi^{(v)}(q^2)$ or $G(E)$, which can contain finite-width scale dependence 
can arise at NNLO only in two-loop diagrams of order ${\cal O}(m E \alpha_s)$ 
and at NNNLO in three-loop diagrams of  ${\cal O}(m E \alpha_s^2)$. This 
is consistent with the fact that the NNLO non-resonant contribution 
is obtained from electroweak two-loop diagrams with one gluon 
exchange \cite{Beneke:2017rdn}.

The task of determining the NNLO finite-width scale dependence now reduces 
to identifying all terms in (\ref{eq:GNNLO}), which contain two-loop 
contributions, which gives
\begin{eqnarray}
[\delta_2 G(E)]_{\rm 2-loop} &=& 
\bigg[\frac{4\pi \alpha_s C_F}{m^2} \, J[0;v_0(\epsilon)]
+ 4\pi \alpha_s C_F \frac{E}{m} \,J[1;1]
\nonumber\\ 
&&\hspace*{-2cm}
+\,\frac{1}{4}\,\bigg(\frac{E^2}{m} J[\delta;1]
+\frac{2 E}{m}G_0(E)+ 8\pi \alpha_s C_F \,\frac{E}{m} J[1;1]\bigg)
\,\bigg]_{\rm 2-loop}
\nonumber\\
&& \hspace*{-2cm}=\,
4\pi \alpha_s C_F \frac{E}{m}\left(-\frac{m^2}{16\pi^2}\,v_0(0) + 
I_a[1]+\frac{1}{4}\,\bigg[E\,\frac{dI_a[1]}{dE} +(2+2)I_a[1]\bigg]\,
\right)\,\quad
\label{eq:GNNLO2loop}
\end{eqnarray}
where $v_0(0) = -2/3$ as defined in (\refI{eq:defv0}) and 
\begin{equation}
I_a[1] = J_a[1;1] = \frac{G_0(E)^{\rm (1ex)}}{4\pi\alpha_s C_F} 
= \frac{m^2}{16\pi^2}\left(L_\lambda+\frac{1}{2}\right).
\label{eq:Ia1}
\end{equation}
In this equation we dropped the $1/\epsilon$ pole term, since the 
procedure of identifying the finite-width scale dependence is carried 
out for the minimally subtracted expressions. None of the $\ln\mu$ 
contained in (\ref{eq:GNNLO2loop}) through $L_\lambda$ will be converted 
to  $\ln\mu_r$ by the substitution (\ref{eq:couplingreplace}), hence 
$\mu\to\mu_w$ is the correct interpretation. After having made the 
substitution $\mu\to\mu_w$ in the logarithms not converted to 
$\ln\mu_r$, we rename $\mu_r$ back to $\mu$.
It follows that the 
expression that must be added to the single scale result in order to 
separate the finite-width dependence is 
\begin{equation}
\left[\delta_2 G(E)\right]_{\mu_w} = 
\frac{m\alpha_s C_F E}{4\pi}\,2 \ln\frac{\mu_w}{\mu}\,,
\label{eq:d2Gmuw}
\end{equation}
which reproduces the corresponding terms in (\ref{eq:2ndordermuwresult}) 
after taking the imaginary part.

To extract the finite-width scale dependence from the NNNLO result, 
we must examine the expression
\begin{eqnarray} 
&& \bigg[\underbrace{[\delta_3 G(E)]_{\rm nC+mixed}}_{C),D)} 
+ \underbrace{\delta^{us}G(E) 
-\frac{\alpha_s}{4\pi}\frac{d_v^{(1,\rm fin)}}{3 m} E G_0(E)}_{D)} 
-\underbrace{\frac{d_v^{(0)}}{3 m} E \delta_1G(E)}_{B)}\bigg]_{\rm 3-loop}
\nonumber\\
&&+\,\underbrace{\frac{\alpha_s}{4\pi} c_v^{(1)} 
\bigg[2\delta_2G(E)-\frac{d_v^{(0)}}{3 m} E G_0(E)
\bigg]_{\rm 2-loop}}_{A)}\nonumber\\
&&+\,
\frac{2\beta_0}{4\pi}\alpha_s(\mu_r)
\ln\frac{\mu_r}{\mu}
\bigg[\underbrace{\delta_2G(E)}_{C)}
-\underbrace{\frac{d_v^{(0)}}{3 m} E G_0(E)}_{B)}\bigg]_{\rm 2-loop}
\label{eq:3rdordermuwterms}
\end{eqnarray}
after eliminating $\alpha_s(\mu)$ in favour of $\alpha_s(\mu_r)$. 
The term in the last line arises from the substitution 
(\ref{eq:couplingreplace}) in the second order result, exploiting that 
it is proportional to a single power of $\alpha_s$. We already dropped 
third-order terms such as $[\delta_3 G(E)]_{\rm C}$ and others than 
cannot yield linear terms in $E$. For the following discussion we group 
the terms as indicated by the underbraces.

Term A) is straightforward to deal with, since it can be related to 
(\ref{eq:2ndordermuwresult}) and (\ref{eq:d2Gmuw}). We then find that 
it contributes $11 c_v^{(1)}/3 \ln(\mu_w/\mu)$ to the curly bracket 
of (\ref{eq:3rdorderlogmuw}).

For terms B) we calculate
\begin{eqnarray}
&& - \frac{d_v^{(0)}}{3 m} \left(
\Big[E \delta_1G(E)\Big]_{\rm 3-loop} 
+ \frac{2\beta_0}{4\pi}\alpha_s(\mu_r)
\ln\frac{\mu_r}{\mu}
\Big[E G_0(E)\Big]_{\rm 2-loop}\right)
\nonumber\\
&& \hspace*{1cm} = \,
- \frac{d_v^{(0)}}{3 m} \,4\pi\alpha_s C_F E 
\left(\frac{\alpha_s}{4\pi}\,J_a[1+\epsilon;
\frac{\beta_0}{\epsilon}+a_1(\epsilon)]
+ \frac{2\beta_0}{4\pi}\alpha_s(\mu_r)
\ln\frac{\mu_r}{\mu}\,I_a[1]\right). \qquad
\end{eqnarray}
Keeping only the logarithmic terms from (\ref{eq:Ja(1+aeps)}) and 
(\ref{eq:Ia1}), we obtain, recalling that 
$L_\lambda=-\frac{1}{2}\ln\left(-4 mE/\mu^2\right)$,
\begin{eqnarray}
&& - \frac{d_v^{(0)}}{3 m} \,
\frac{m^2 \alpha_s C_F E}{4\pi}\frac{\alpha_s}{4\pi}
\left(\beta_0\left[L_\lambda^2+\ln\frac{\mu_r^2}{\mu^2}
\left(L_\lambda +\frac{1}{2}\right)\right]+a_1 L_\lambda\right)
\nonumber\\
&& \hspace*{1cm} = \,
 - \frac{d_v^{(0)}}{3 m} \,
\frac{m^2 \alpha_s C_F E}{4\pi}\frac{\alpha_s}{4\pi}
\Bigg(\beta_0\left[\left(-\frac{1}{2}\ln\frac{-4 m E}{\mu_r^2}\right)^2
-\frac{1}{4}\ln^2\frac{\mu_r^2}{\mu^2} + \frac{1}{2}\ln\frac{\mu_r^2}{\mu^2}
\right]
\nonumber\\
&& \hspace*{1cm}
+\,a_1\left[-\frac{1}{2}\ln\frac{-4 m E}{\mu_r^2}
-\frac{1}{2}\ln\frac{\mu_r^2}{\mu^2} \right]\Bigg)\,.
\label{eq:termB}
\end{eqnarray}
The important point is that there is no term of the form 
$\ln (-4 mE) \ln\mu$. Hence, the finite-width scale dependence is ``local'' 
as it should be, and the terms B) contribute 
\begin{equation}
 - \frac{d_v^{(0)}}{3} \left(-\beta_0\ln^2\frac{\mu_w}{\mu} + 
(a_1-\beta_0)\ln\frac{\mu_w}{\mu} \right)
\end{equation}
to the curly bracket of (\ref{eq:3rdorderlogmuw}), which follows from 
subtracting (\ref{eq:termB}) with a single scale, i.e. $\mu \to \mu_r$, 
from the same expression with $\mu \to \mu_w$.

Turning to terms C) and D), we first note that there is again only a few 
terms in $\delta_3G(E)$ that can cause finite-width scale dependence. The 
pure Coulomb terms $[\delta_3 G]_{\rm C}$ can be dropped 
as already mentioned. Inspecting 
the remaining terms in (\ref{eq:GNNNLOnC}), (\ref{eq:GNNNLOmixed}), 
we identify
\begin{eqnarray}
[\delta_3G(E)]_{\rm nC+mixed}&\supset& 
\frac{4\pi \alpha_s C_F}{m^2} \,\frac{\alpha_s}{4\pi}\, 
\bigg\{J[\epsilon;\frac{1}{\epsilon}\bigg(\frac{7}{3}C_F
-\frac{11}{6}C_A+\beta_0v_0(\epsilon)\bigg)+v_q^{(1)}(\epsilon)]
\nonumber \\ 
&&\hspace{-2cm}
+\,J[0;-\frac{1}{\epsilon}\beta_0 v_0(\epsilon)
+\bigg[\bigg(\frac{\mu^2}{m^2}\bigg)^{\!\epsilon}-1\bigg]
\frac{1}{\epsilon}\bigg(\frac{C_F}{3}+\frac{C_A}{2}\bigg)
+\bigg(\frac{\mu^2}{m^2}\bigg)^{\!\epsilon}v_m^{(1)}(\epsilon)]\bigg\}
\nonumber \\ 
&&\hspace{-2.5cm}
+\,4\pi \alpha_s C_F\,\frac{\alpha_s}{4\pi}\,
\frac{E}{m}J[1+\epsilon;\frac{1}{\epsilon}\bigg(\frac{8}{3}C_A
+\beta_0\bigg)+v_p^{(1)}(\epsilon)]
\nonumber \\ 
&&\hspace{-2.5cm}
+\,2\,(-4\pi\alpha_s C_F)\,\frac{\alpha_s}{4\pi}\,
\left(-\frac{1}{4}\right)
\bigg\{\frac{E^2}{m} J[\delta,1+\epsilon;1]
+\frac{2 E}{m} J[1+\epsilon;\frac{\beta_0}{\epsilon}+a_1(\epsilon)]
\bigg\},\qquad
\label{eq:G3NNNLOmuw}
\end{eqnarray}
which originate from the single insertion of the delta potential, 
the single insertion of the $\bff{p}^2/(m^2 \bff{q}^2)$ 
potential, and the double insertion of the kinetic energy correction 
with the NLO Coulomb potential. Note that the corresponding double 
insertion with the  $\bff{p}^2/(m^2 \bff{q}^2)$ potential does not 
contribute, since $E J[1,1+\epsilon;1]$ is a finite expression, 
see (\ref{eq:couldouble}) and the remark after (\ref{eq:GNNNLOmixed}). 
In (\ref{eq:G3NNNLOmuw}) we assign all terms involving $\beta_0$ and 
$a_1$ to contribution C), and the remaining terms to D).

Contribution C) is then given by
\begin{eqnarray}
&& 
\bigg[\frac{4\pi \alpha_s C_F}{m^2} \,\frac{\alpha_s}{4\pi}\, 
\left(J[\epsilon;\frac{\beta_0v_0(\epsilon)}{\epsilon}]- J[0;\frac{\beta_0v_0(\epsilon)}{\epsilon}]\right)
+ 4\pi\alpha_s C_F\,\frac{\alpha_s}{4\pi}\,
\frac{E^2}{2m} J[\delta,1+\epsilon;1]
\nonumber\\
&&+ \,4\pi \alpha_s C_F\,\frac{\alpha_s}{4\pi}\,
\frac{E}{m} 
J[1+\epsilon;\frac{2 \beta_0}{\epsilon}+a_1(\epsilon)]\bigg]_{\rm 3-loop}
+
\frac{2\beta_0}{4\pi}\alpha_s(\mu_r)
\ln\frac{\mu_r}{\mu}
\Big[\delta_2G(E)\Big]_{\rm 2-loop}\!.\qquad
\end{eqnarray}
Keeping only the logarithmic terms, this evaluates to 
\begin{eqnarray}
&&
\frac{m\alpha_s C_F E}{4\pi}\frac{\alpha_s}{4\pi}
\left(-4\beta_0v_0(0) L_\lambda -\frac{1}{4}\beta_0L_\lambda +2\beta_0 L_\lambda^2+a_1 L_\lambda
+\beta_0\ln\frac{\mu_r^2}{\mu^2} \left[2 L_\lambda +\frac{7}{8}-v_0(0) 
\right]\right)\nonumber\\
&& \rightarrow -2\beta_0 \ln^2\frac{\mu_w}{\mu} + 
\left(-2 \beta_0 v_0(0)+a_1-2\beta_0\right)\ln\frac{\mu_w}{\mu}\,,
\label{eq:termC}
\end{eqnarray}
where the expression after the arrow is the term that contributes to 
the curly bracket of  (\ref{eq:3rdorderlogmuw}) to separate the finite-width 
scale dependence in the single-scale result.

Contribution D) follows from assembling the remaining pieces
\begin{eqnarray} 
&&\bigg[\frac{4\pi \alpha_s C_F}{m^2} \,\frac{\alpha_s}{4\pi}\, 
\bigg\{\,J[\epsilon;\frac{1}{\epsilon}\bigg(\frac{7}{3}C_F
-\frac{11}{6}C_A\bigg)+v_q^{(1)}(\epsilon)]
\nonumber \\ 
&&
+\,J[0;\bigg[\bigg(\frac{\mu^2}{m^2}\bigg)^{\!\epsilon}-1\bigg]
\frac{1}{\epsilon}\bigg(\frac{C_F}{3}+\frac{C_A}{2}\bigg)
+\bigg(\frac{\mu^2}{m^2}\bigg)^{\!\epsilon}v_m^{(1)}(\epsilon)]\,\bigg\}
\nonumber \\ 
&&
+\,4\pi \alpha_s C_F\,\frac{\alpha_s}{4\pi}\,
\frac{E}{m}J[1+\epsilon;\frac{8C_A}{3\epsilon}+v_p^{(1)}(\epsilon)]
+ \delta^{us}G(E) 
-\frac{\alpha_s}{4\pi}\frac{d_v^{(1,\rm fin)}}{3 m} E G_0(E)
\bigg]_{\rm 3-loop}\!\!\!\!.\qquad
\end{eqnarray}

To obtain the contribution from $ \delta^{us}G(E)$ we extracted the 
three-loop ultrasoft diagrams from the all-order summed 
result \cite{Beneke:2008cr}. With $d_v^{(1,\rm fin)} = (-C_F) 
\left(32 \ln(\mu/m)-16/3\right)$ from (\refI{eq:dv1loop}), 
$v_q^{(1)}(\epsilon)$ from (\refI{eq:epsterms}) and 
$v_p^{(1)}(0)=a_1$ from  (\refI{eq:vp1epsterms}), we obtain 
the contribution 
\begin{eqnarray}
&&
\left(\frac{28}{3}C_F+2 C_A\right) \ln^2\frac{\mu_w}{\mu} 
+  \ln\frac{\mu_w}{\mu}\,\Bigg\{
\left(-\frac{56}{3}C_F-4 C_A\right)\ln\frac{m}{\mu} 
\nonumber\\
&& + \left(-\frac{14}{9}-8\ln 2\right) C_F 
+ \left(-\frac{19}{3}+12\ln 2\right)C_A +a_1\Bigg\}
\label{eq:termD}
\end{eqnarray}
to the curly bracket of (\ref{eq:3rdorderlogmuw}). Note the presence of 
the logarithm $\ln(\mu/m)$, which originates from the ultrasoft and 
$d_v^{(1,\rm fin)}$ terms.  Adding terms A) to 
D) results in (\ref{eq:3rdorderlogmuw}).

\subsubsection{\boldmath Corrections from $O(\epsilon)$ 
hard matching coefficients}
\label{sec:opeshardmatching}

In Section~3.5 of paper~I we noted that one does not need the 
$d$-dimensional expressions of the hard matching coefficients 
of the non-relativistic currents as long as one computes the 
pair production cross section of stable heavy quarks. However, 
when the finite width is included the resonant cross section 
is divergent, as discussed above. The product of 
$O(\epsilon)$ terms in the hard matching coefficients 
with the finite-width divergences produces finite terms that 
must be consistently included in the sum of the resonant and 
non-resonant cross section. 

The finite terms are obtained from the products of the divergent 
part proportional to $\Gamma$ calculated in 
Section~\ref{sec:divpart} and the $O(\epsilon)$ 
terms of the matching coefficients $c_v$, $d_v$ given in 
Section~3.5 of paper~I. The contribution to the 
cross section (\ref{eq:sigma}) is
\begin{eqnarray}
\sigma_{t\bar t}{}_{|\,O(\epsilon)\,c_v, d_v}
&=& \sigma_0 \times 12\pi e_t^2 \,\frac{N_{c}}{2m^{2}}
\,\bigg\{-\frac{d_v^{(0,\epsilon)}}{3 m} 
\,\mbox{Im}[E G_0(E)]_{\Gamma/\epsilon}
\nonumber\\
&& -\,\frac{1}{3m}\,d_v^{(0,\epsilon)}\,
\mbox{Im}[E \delta_1G(E)]_{\Gamma/\epsilon^2}
\nonumber\\
&& +\,2\frac{\alpha_s}{4\pi} c_v^{(1,\epsilon)}\,
\mbox{Im}[\delta_2 G]_{\Gamma/\epsilon} 
-\frac{1}{3m} \frac{\alpha_s}{4\pi} 
\left(c_v^{(1,\epsilon)}+c_v^{(1)}d_v^{(0,\epsilon)}+
d_v^{(1,\epsilon)}\right)\,\mbox{Im}[E G_0(E)]_{\Gamma/\epsilon}
\nonumber\\
&&-\,\frac{1}{3m}\left(d_v^{(0,\epsilon^2)}\,
\mbox{Im}[E \delta_1G(E)]_{\Gamma/\epsilon^2} 
+d_v^{(0,\epsilon)}\,
\mbox{Im}[E \delta_1 G(E)]_{\Gamma/\epsilon}\right) 
\bigg\}\,,
\label{eq:oepsfrommatchabstract}
\end{eqnarray}
where the superscript $(n,\epsilon^m)$ refers to the $O
(\epsilon^m)$ coefficient of the $n$-loop coefficient, and 
the subscript $\Gamma/\epsilon$ (or similar) on the square 
brackets means that the coefficient of $\Gamma/\epsilon$ (or similar) 
of the expression should be used.  The given 
expression makes use of the fact that the tree-level value of 
$c_v$ equals 1 in $d$ dimensions, i.e.~$c_v^{(0)}=1$ and 
$c_v^{(0,\epsilon^m)} = 0$. On the other hand $d_v^{(0)}=1$ but 
$d_v^{(0,\epsilon^m)} \not= 0$, see paper~I. The first line of 
(\ref{eq:oepsfrommatchabstract}) is NNLO, while the remaining ones 
contribute to the NNNLO cross section. The second term in 
the curly bracket is divergent and provides another contribution 
to the finite-width divergence.

Inserting the results for the finite-width divergences of the 
Green function (\ref{eq:GNNLOdivpart}), 
(\ref{eq:finitewidthdivextcurrent2ndorder}), 
(\ref{eq:finitewidthdivextcurrent3rdorder}) and 
for the $d$-dimensional matching coefficients 
(\refI{eq:cv1}), (\refI{eq:dv1loop}), we find
\begin{eqnarray}
\sigma_{t\bar t}{}_{|\,O(\epsilon)\,c_v, d_v}
&=& \sigma_0 \times\alpha_s C_F \, \frac{\Gamma}{m}\,
\bigg\{-\frac{1}{9} \left(1+\frac{2\beta_0\alpha_s}{4\pi} \,
\ln\frac{\mu}{\mu_w}\right)
\nonumber\\
&& \hspace*{-2cm}+\, 
\frac{\alpha_s}{4\pi} \,\bigg[
\frac{\beta_0}{27\epsilon} 
+C_F \left(\frac{4}{3}  \ln^2 \frac{m^2}{\mu_w^2}
+ \frac{124}{9} \ln \frac{m^2}{\mu_w^2}
+ \frac{428}{27}+\frac{2\pi^2}{9}\right)
+\frac{8\beta_0}{81}-\frac{2 a_1}{27}
\bigg]\bigg\}\,,\qquad
\label{eq:oepsfrommatchinserted}
\end{eqnarray}
where the assignment of $\mu_w$ has been determined by the 
procedure discussed above. The NNNLO term in the second line 
includes a finite-width divergence, which must cancel with 
the yet unknown NNNLO non-resonant correction. We define the 
resonant part by minimal subtraction of the finite-width 
divergence and consequently drop this pole.
Note that (\ref{eq:oepsfrommatchinserted}) 
is an energy-independent constant. 
Numerically, it is very small. Choosing $m_t=171.5$~GeV, 
$\Gamma=1.36$~GeV,  
$\mu_w =350\,$GeV 
(50~GeV) and $\alpha_s=\alpha_s(80~\mbox{GeV}) = 0.1208$ 
we obtain $-0.00014$ for the NNLO and $5\cdot 10^{-5}$ 
$(9.8\cdot 10^{-4})$ for the NNNLO contribution to the $R$-ratio 
$\sigma_{t\bar t}/\sigma_0$, which amounts to a sub-permille 
correction to the cross section.

The corresponding finite-width related terms 
from the $O(\epsilon)$ terms in the 
P-wave matching coefficient to the P-wave contribution to the 
cross section, not discussed in this paper, have already been given 
in Eqs.~(4.3) and (4.4) of \cite{Beneke:2013kia}. 

\subsubsection{\boldmath Corrections from the $d$-dimensional top decay 
width}
\label{sec:ddimwidth}

In the implementation of the cross section in 
the \texttt{QQbar\_threshold} code~\cite{Beneke:2016kkb} as well as 
in the numerical analysis of Section~\ref{sec:results} below,  the top 
width is treated as a {\em numerical} parameter. This implies that 
higher-order corrections to the tree-level width $\Gamma^{(0)}$ are 
treated non-perturbatively through the replacement
$E\rightarrow E+i\Gamma$. As pointed out in \cite{Beneke:2017rdn}, a 
subtlety arises when this result is combined with the non-resonant 
contribution, which is computed in dimensional regularization. 
For example, the pole part of the NNLO non-resonant contribution is 
proportional to the {\em algebraic} expression for $\Gamma^{(0)}$, 
calculated in the SM, and the finite part follows from expanding 
the divergent hard top-self-energy diagrams up to 
${\cal O}(\epsilon^0)$. For consistency, the top width in 
the effective Lagrangian (\ref{eq:width-lagrangian}) for the 
resonant part of the physical cross section must also be treated 
as a $d$-dimensional hard matching coefficient, given by the algebraic 
expression, calculated order by order. In order to correct for 
the use of the width as a numerical parameter, we must 
replace 
\begin{equation}
\label{eq:widthepssub}
\Gamma \to \Gamma + \Gamma_\epsilon\,,
\end{equation}
where now on the right-hand side the $d$-dimensional width 
separated into the four-dimensional part $\Gamma$ and the remaining 
$O(\epsilon)$ terms $\Gamma_\epsilon$ appears. The latter 
contribute additional finite-width related terms to the cross section, 
when the $O(\epsilon)$ terms multiply the finite-width 
divergence of the Green function, 
which will be calculated in the following. 

Essentially we must treat the width as a $d$-dimensional hard 
matching coefficient on the same footing as the other 
coefficients of the non-relativistic effective Lagrangian. 
We expand $\Gamma_\epsilon$ in $\epsilon$ and the strong 
coupling by writing 
\begin{equation}
\label{eq:epstopwidthexp}
\Gamma_\epsilon = \Gamma^{(0,\epsilon)} + 
 \Gamma^{(0,\epsilon^2)} + \ldots 
+\frac{\alpha_s}{4\pi}\left(\Gamma^{(1,\epsilon)} + \ldots
\right) + O(\alpha_s^2)\,,
\end{equation}
where the meaning of the superscript is as in the previous 
subsection. We will not need electroweak loop corrections 
to the top width or any other terms not indicated explicitly 
to compute the correction terms up to NNNLO. For example, 
the $d$-dimensional generalization of the expression 
(\ref{eq:widthSM}) for the tree-level top decay width 
reads \cite{Beneke:2017rdn}
\begin{equation}
[\Gamma+\Gamma_\epsilon]^{(0)} =\frac{G_F m_t^3}{8\pi\sqrt{2}}
\,(1-x_W)^2\,(1+2(1-\epsilon)x_W)\,\frac{\sqrt{\pi}}
{2\Gamma(3/2-\epsilon)}
\left(\frac{4\mu_w^2e^{\gamma_E}}{m_t^2(1-x_W)^2}\right)^{\!\epsilon},
 \label{eq:Gamma0d}
\end{equation}
where $x_W=M_W^2/m_t^2$, from which $\Gamma^{(0,\epsilon^n)}$ 
is obtained by expansion in $\epsilon$. We are not aware of 
a published result of the $d$-dimensional one-loop QCD 
correction to the width, nor its expansion to 
$O(\epsilon)$. In the following, we use an expansion 
of the $O(\epsilon)$ term $\Gamma^{(1,\epsilon)}$ 
in $x_W$ to order $x_W^8$, which provides a good 
approximation.\footnote{We thank J.~Piclum for supplying us 
with this unpublished result.} 

Including (\ref{eq:epstopwidthexp}) in the substitution 
$E\to E+i\Gamma$ in the Green function to account for finite-width 
effects, produces the following additional terms from 
the $d$-dimensional width:
\begin{eqnarray}
\sigma_{t\bar t}{}_{|\,\Gamma_\epsilon}
&=& \sigma_0 \times 12\pi e_t^2 \,\frac{N_{c}}{2m^{2}}
\,\bigg\{
\Gamma^{(0,\epsilon)}\,\mbox{Im}[\delta G_2]_{\Gamma/\epsilon}
-\frac{1}{3 m}\, \Gamma^{(0,\epsilon)}
\,\mbox{Im}[E G_0(E)]_{\Gamma/\epsilon}
\nonumber\\
&&+\, \Gamma^{(0,\epsilon)}\,
\mbox{Im}[\delta G_3]_{\Gamma/\epsilon^2}
-\frac{1}{3 m}\, \Gamma^{(0,\epsilon)}
\,\mbox{Im}[E \delta G_1(E)]_{\Gamma/\epsilon^2}
\nonumber\\[0.1cm]
&&+\, \Gamma^{(0,\epsilon^2)}\,
\mbox{Im}[\delta G_3]_{\Gamma/\epsilon^2}
+ \Gamma^{(0,\epsilon)}
\left(\mbox{Im}[\delta G_3]_{\Gamma/\epsilon} 
+ 2 \frac{\alpha_s}{4\pi}\,
c_v^{(1)} \,\mbox{Im}[\delta G_2]_{\Gamma/\epsilon}
\right) 
\nonumber \\
&&+ \,\frac{\alpha_s}{4\pi}\,
\Gamma^{(1,\epsilon)} \,\mbox{Im}[\delta G_2]_{\Gamma/\epsilon}
\\
&& -\,\frac{1}{3m} \frac{\alpha_s}{4\pi} 
\Big((c_v^{(1)}+d_v^{(1)})\Gamma^{(0,\epsilon)}
+\Gamma^{(1,\epsilon)}\Big)\,\mbox{Im}[E G_0(E)]_{\Gamma/\epsilon}
\nonumber\\
&&-\,\frac{1}{3m}\left((d_v^{(0,\epsilon)}\Gamma^{(0,\epsilon)} 
+\Gamma^{(0,\epsilon^2)})\,
\mbox{Im}[E \delta_1G(E)]_{\Gamma/\epsilon^2} 
+\Gamma^{(0,\epsilon)}\,
\mbox{Im}[E \delta_1 G(E)]_{\Gamma/\epsilon}\right) 
\bigg\}\,.
\nonumber
\label{eq:oepsfromwidthabstract}
\end{eqnarray}
Here the first two terms in the curly bracket are NNLO terms, 
while all others are NNNLO. The third 
and fourth terms are divergent and 
provide another contribution to the finite-width divergence. 
The remaining third-order terms are finite terms, which are 
required to obtain a consistent, scheme-independent combination 
with the yet unknown NNNLO non-resonant contribution. 

We next insert the expressions for the finite-width divergences of the 
Green function from (\ref{eq:GNNLOdivpart}), (\ref{eq:GNNNLOdivpart}) 
(\ref{eq:finitewidthdivextcurrent2ndorder}), 
(\ref{eq:finitewidthdivextcurrent3rdorder}) and recall 
that $\mbox{Im}[\delta G_3]$ in  (\ref{eq:GNNNLOdivpart}) 
includes the divergent part of the third-order ultrasoft 
contribution, $\mbox{Im}[\delta^{us}G(E)]$. The piece proportional 
to $\Gamma$ reads  \cite{Beneke:2008cr} 
\begin{eqnarray}
\label{eq:usfinitewidthdiv}
\mbox{Im}[\delta^{us}G(E)]_{\Gamma/\epsilon^{2,1}} &=& 
\frac{m\alpha_s^2 C_F \Gamma}{6\pi^2}\,\bigg(
\left[\frac{C_F}{4}+\frac{C_A}{8}\right]\frac{1}{\epsilon^2} 
+ \frac{1}{\epsilon}\,\bigg[
-\left(\frac{C_F}{4}+\frac{C_A}{8}\right)\ln\frac{m^2}{\mu_w^2}
\nonumber\\
&&
+\,C_F\left(-\frac{1}{2}\ln 2+\frac{1}{12}\right)
+C_A\left(\frac{3}{4}\ln 2-\frac{11}{24}\right)\bigg]\,\bigg)\,.
\end{eqnarray}
The extra contribution to the cross section is then given by
\begin{eqnarray}
\sigma_{t\bar t}{}_{|\,\Gamma_\epsilon}
&=& \sigma_0 \times\frac{\alpha_s C_F}{m}\,
\bigg\{\Gamma^{(0,\epsilon)}\,
\frac{5}{6} \left(1+\frac{2\beta_0\alpha_s}{4\pi} \,
\ln\frac{\mu}{\mu_w}\right)
\nonumber\\
&&+\, 
\frac{\alpha_s}{4\pi} \,\bigg[\Gamma^{(0,\epsilon)}
\,\bigg(\frac{1}{\epsilon}\left[\frac{4}{3} C_F+\frac{2}{9}C_A
-\frac{5\beta_0}{18}\right] 
-\left(4 C_F+\frac{2}{3} C_A\right)\ln\frac{m^2}{\mu_w^2} 
\nonumber\\
&&+\,C_F\left(-\frac{8}{3}\ln 2-\frac{134}{9}\right)
+C_A\left(4\ln 2+\frac{5}{27}\right)-\frac{40 \,T_F n_f}{27}
-\frac{2\beta_0}{27}-\frac{a_1}{9}
\bigg)
\nonumber\\
&&+\, 
\Gamma^{(0,\epsilon^2)} \left(\frac{4}{3} C_F+\frac{2}{9}C_A
-\frac{5\beta_0}{18}\right)
+ \frac{5}{6} \,\Gamma^{(1,\epsilon)}\bigg]\,\bigg\}\,.
\label{eq:oepsfromwidthinserted}
\end{eqnarray}
The first line constitutes the NNLO contribution (plus the NNNLO 
term from converting $\alpha_s(\mu_w)$ to $\alpha_s\equiv 
\alpha_s(\mu)$), while the remaining ones are NNNLO. The NNNLO 
term includes a finite-width divergence, which arises from 
the substitution $\Gamma \to \Gamma + \Gamma_\epsilon$ in
(\ref{eq:GNNNLOdivpart}), (\ref{eq:finitewidthdivextcurrent3rdorder}),
including the ultrasoft contribution (\ref{eq:usfinitewidthdiv}) 
to (\ref{eq:GNNNLOdivpart}). This divergence must cancel with 
the yet unknown 
NNNLO non-resonant correction and is minimally subtracted 
here. The NNLO term was already given in \cite{Beneke:2017rdn} 
together with further NNLO terms of similar origin which arise from 
the finite width divergence produced by the 
higher-order finite-width terms $\Gamma_t \,{\bf D}^2/(4 m^2)$, 
$\Gamma_t^2/(8 m)$ in the effective Lagrangian 
(\ref{eq:width-lagrangian}). We count these terms as electroweak 
corrections and therefore do not include them in the analysis 
of QCD corrections below.

Numerically, this contribution is again very small. Choosing 
the same parameters as above,  $m_t=171.5$~GeV, 
$\Gamma=1.36$~GeV and $\alpha_s=\alpha_s(80~\mbox{GeV}) = 0.1208$, 
we obtain $0.0041$ ($-0.0003$) for the NNLO and  $-0.0020$ 
($-0.0002$) for the NNNLO contribution to the $R$-ratio 
$\sigma_{t\bar t}/\sigma_0$, with $\mu_w =350\,$GeV 
(50~GeV). The sum amounts to a permille correction to the cross 
section.

\subsection{Pole resummation}
\label{sec:poleresummation}

The correlation function $G(E)$ in the pole scheme is an analytic function 
with poles corresponding to bound states at negative $E$ and a cut at 
$E>0$. The exact Green function has only single poles, 
\begin{eqnarray}
G(E) &\stackrel{E\to E_n}{=}&
\frac{|\psi_n(0)|^2}{E_n -E}+ \mbox{non-singular}
\nonumber\\
&=&
\frac{|\psi_n^{(0)}(0)|^2 \big[1 +F_n^{(1)}+F_n^{(2)}+\cdots \big] 
      }{ \big[  E_n^{(0)}+E_n^{(1)} +E_n^{(2)} +\cdots \big]  -E} 
+\mbox{non-singular},
\label{eq:GFsinglepole}
\end{eqnarray}
where $n$ enumerates the $S$-wave bound state poles. The bound state energy 
and the residue of the pole have expansions in powers of $\alpha_s$ 
(and $\ln\alpha_s$), as indicated in the second equation, with 
leading order expressions
\begin{equation}
E_n^{(0)} = -\frac{m (\alpha_s C_F)^2}{4 n^2}, \qquad\quad 
|\psi_n^{(0)}(0)|^2 = \frac{(m \alpha_s C_F)^3}{8\pi n^3}\,.
\label{eq:zerothorderboundstateparameters}
\end{equation} 

The calculation of $G(E)$ in resummed non-relativistic perturbation 
theory does not produce the correct singular behaviour 
(\ref{eq:GFsinglepole}). At the N3LO we compute 
(see also (\ref{greenexpandII}), (\ref{masterthirdorderII}))
\begin{eqnarray}
G_{\rm N^3LO} &=&
G_0+\delta_1 G+\delta_2 G+\delta_3 G,
\label{eq:insertion_formula_N3LO}
\\
\delta_1 G &=& - \langle 0| \hat{G}_0 \delta V_1 \hat{G}_0 |0\rangle,
\label{eq:insertion_formula_1}
\\
\delta_2 G &=&  -\langle 0| \hat{G}_0 \delta V_2 \hat{G}_0 |0\rangle +
   \langle 0| \hat{G}_0 \delta V_1 \hat{G}_0 \delta V_1 \hat{G}_0|0\rangle,
\label{eq:insertion_formula_2}
\\
\delta_3 G&=&  -\langle 0| \hat{G}_0 \delta V_3 \hat{G}_0 |0\rangle +
  2 \langle 0| \hat{G}_0 \delta V_1 \hat{G}_0 \delta V_2 \hat{G}_0|0\rangle
\nonumber\\
&& 
 - \langle 0| \hat{G}_0 \delta V_1 \hat{G}_0 \delta V_1 \hat{G}_0 \delta V_1 
\hat{G}_0|0\rangle +\delta^{us}G(E)\,.  
\label{eq:insertion_formula_3}   
\end{eqnarray}
The expansion around $G_0$ generates poles of higher and higher order 
at the location of the lowest order bound state energies $E_n^{(0)}$ 
rather than a single pole at the exact bound state energy. The 
singularities are given by
\begin{eqnarray}
G_{\rm N^3LO}(E) 
&\stackrel{E\to E_n^{(0)}}{=}&
\frac{|\psi_n^{(0)}(0)|^2}{E_n^{(0)}-E}+ \delta_1 {\cal G}(E) 
+\delta_2 {\cal G}(E)+\delta_3 {\cal G}(E)
+\mbox{non-singular},
\label{eq:GFexpanded}
\end{eqnarray}
with 
\begin{eqnarray}
\delta_1 {\cal G}(E) 
&\equiv&
-\frac{ |\psi_n^{(0)}(0)|^2 E_n^{(1)}}{(E_n^{(0)}-E)^2} 
+\frac{ |\psi_n^{(0)}(0)|^2 F_n^{(1)} }{E_n^{(0)}-E},
\\
\delta_2 {\cal G}(E)
&\equiv&
\frac{ |\psi_n^{(0)}(0)|^2 ( E_n^{(1)})^2 }{(E_n^{(0)}-E)^3}
-\frac{ |\psi_n^{(0)}(0)|^2 \left(E_n^{(2)}+E_n^{(1)} F_n^{(1)} \right)}
{(E_n^{(0)}-E)^2} 
+\frac{ |\psi_n^{(0)}(0)|^2 F_n^{(2)} }{E_n^{(0)}-E},
\qquad\\
\delta_3 {\cal G}(E)
&\equiv&
-\frac{|\psi_n^{(0)}(0)|^2 (E_n^{(1)})^3 }{(E_n^{(0)}-E)^4} 
+ \frac{ |\psi_n^{(0)}(0)|^2 (2E_n^{(2)} E_n^{(1)}+(E_n^{(1)})^2 F_n^{(1)})}
{(E_n^{(0)}-E)^3}
\nonumber\\
&&
-\frac{ |\psi_n^{(0)}(0)|^2 (E_n^{(2)} F_n^{(1)} +E_n^{(1)} F_n^{(2)}
+E_n^{(3)})}{(E_n^{(0)}-E)^2} 
+\frac{ |\psi_n^{(0)}(0)|^2 F_n^{(3)} }{E_n^{(0)}-E}\,.
\end{eqnarray}
It is evident that non-relativistic perturbation theory breaks down in the 
vicinity of the lowest order bound state locations. For example, 
the third-order correction to the Green function contains a fourth order 
pole, whose coefficient is determined by the first-order energy correction. 
Since this arises from the one-loop correction to the Coulomb potential 
only, the most singular contribution can arise only from the triple 
insertion of the Coulomb potential. This can indeed be seen from the 
leading $1/(n-\lambda)^4$ pole in 
(\ref{eq:triplecoulombpolepart}), since the singularity at 
$E=E_n^{(0)}$ corresponds to positive integer $\lambda=n$. 
As an aside we note that at any order $k$ the correction $F_n^{(k)}$ 
to the squared wave function at the origin and the correction 
$E_n^{(k)}$ to the bound state energy can be determined from the single 
and double pole of $\delta_k {\cal G}(E)$, respectively.

The top pair production cross section near threshold needs the 
evaluation of $G(E)$ on a line parallel to the real axis with imaginary 
part $\Gamma$, hence $|E_n^{(0)}-E| \geq \Gamma_t$. The convergence of 
(\ref{eq:GFexpanded}) near the bound state poles therefore requires
\begin{equation}
\frac{|E_n^{(1)}|}{\Gamma_t}\ll 1\,.
\label{eq:poleresumcondition}
\end{equation}
Parametrically, $E_n^{(1)} \sim m \alpha_s^3\sim m v^3$, $\Gamma \sim 
m \alpha_{\rm EW} \sim m v^2$, so (\ref{eq:poleresumcondition}) is 
formally always satisfied whenever the non-relativistic expansion is 
justified ($v\ll 1$). However, numerically,
\begin{equation}
|E_1^{(1)}| = \frac{m \alpha_s^3(\mu) C_F^2}{16 \pi}
\left(2 a_1+4\beta_0\Big[\ln\frac{\mu}{m\alpha_s C_F}+S_1\Big]\,\right) 
\approx 0.8\,\mbox{GeV}
\label{eq:e11shift}
\end{equation}
in the pole scheme and $\Gamma_t\approx 1.36\,$GeV, 
resulting in an effective expansion parameter of about 0.6 near the peak of 
the cross section.

This problem was first noticed in \cite{Beneke:1999qg} and remedied by 
subtracting the singular terms (\ref{eq:GFexpanded}) from the Green 
function calculated in non-relativistic perturbation theory and adding 
back the exact form (\ref{eq:GFsinglepole}) of the single pole with 
bound state energy and residue computed in the corresponding order 
of perturbation theory. This results in an improved,  ``pole resummed'' (PR) 
Green function $G^{PR}_{\rm N^3LO}$ defined by
\begin{eqnarray}
G_{\rm N^3LO}^{PR}(E)
&\equiv &
G_{\rm N^3LO}(E) - {\Delta}^{PR}(E),
\\
\Delta^{PR}(E)&=&
\bigg[ 
\frac{|\psi_n^{(0)}(0)|^2}{E_n^{(0)}-E}
+\sum_{i=1}^{3} \delta_i {\cal G}(E) 
\bigg]
-\frac{|\psi_n^{(0)}(0)|^2 \big[1+ \sum_{i=1}^{3} F_n^{(i)} \big] }
{ \big[ \sum_{i=0}^{3} E_n^{(i)}\big]-E}\,.
\end{eqnarray}
The numerical effect of pole resummation on the top pair cross section 
will be studied in sections~\ref{sec:insertionproblems} and~\ref{subsec:poleresum} 
below. Elsewhere 
pole resummation applied to the vector current correlator $\Pi^{(v)}(q^2)$ 
will always be understood without explicit mentioning. 

Since the residues 
of the excited bound states decrease quickly with $n$, in practice it is 
sufficient to carry out the pole resummation procedure up to some 
$n_{\rm max}$, and we choose $n_{\rm max}=6$. The formulas for the 
expansion of the S-wave bound state energy and wave function in the 
origin are collected in appendix~\ref{sec:enpsicorr}. The various 
third-order corrections to the energy level and wave function at the 
origin that make up the complete result have originally been obtained in 
\cite{Kniehl:1999ud,Kiyo:2000fr,Beneke:2005hg,Penin:2005eu} and 
\cite{Beneke:2005hg,Penin:2005eu,Beneke:2007gj,Beneke:2007pj,Beneke:2007uf}, 
respectively.\footnote{Expressed in terms of some hard
matching coefficients and some order $\epsilon$ parts of the
potentials that were not known at the time.}

\subsection{Conversion of the top mass scheme}
\label{sec:massdef}

Up to now the top pair production cross section has been expressed in 
terms of the top quark pole mass. The pole mass is gauge invariant and 
infrared-finite in perturbation theory~\cite{Kronfeld:1998di}, but 
it is intrinsically ambiguous by an amount of order 
$\Lambda_{\rm QCD}$ \cite{Bigi:1994em,Beneke:1994sw,Beneke:2021lkq}, the strong interaction 
scale, due to a strong infrared renormalon divergence of the 
perturbative series that relates the pole mass to a short-distance 
mass, such as the $\overline{\rm MS}$ mass. No matter to what accuracy 
the  top pair production cross section is computed and measured, one must 
find that the top pole mass cannot be determined with an uncertainty 
less than ${\cal O}(\Lambda_{\rm QCD})$.\footnote{Available results 
on the low-order and asymptotic high-order perturbative series 
coefficients suggest that the intrinsic ambiguity is only 70~MeV 
when all five light quark masses are neglected, and 110 MeV accounting 
for the bottom and charm mass \cite{Beneke:2016cbu}. This is still a factor 
2-3 larger than 
the accuracy goal for the experimental threshold scan.}

The threshold cross section by itself is less sensitive to the infrared 
regime of QCD than the top quark pole mass. The large shifts in the 
cross section prediction that must occur in the pole mass scheme to 
prevent that the pole mass can be determined with accuracy better than 
its intrinsic ambiguity, are therefore spurious and can be eliminated 
by expressing the cross section in another, less infrared sensitive 
mass renormalization scheme \cite{Beneke:1998rk,Beneke:1998jj}. In order 
not to spoil the 
non-relativistic expansion, the new mass parameter must not differ 
from the pole mass by an amount parametrically larger than $m v^2$, 
which rules out the direct use of the $\overline{\rm MS}$ mass, for 
which the difference is of order $m \alpha_s \sim m v$.

\subsubsection{The potential-subtracted mass scheme}
\label{sec:PSscheme}

Since the leading infrared sensitivity in the pole mass is closely 
related to the Coulomb potential \cite{Beneke:1998rk,Hoang:1998nz} 
and since the Coulomb potential dominates the threshold dynamics, 
it is most natural to define a short-distance mass compatible with 
non-relativistic power counting by subtracting the infrared integral 
of the Coulomb potential from the pole mass \cite{Beneke:1998rk}. 
More precisely, the ``potential-subtracted'' (PS) mass is defined 
as 
\begin{equation}
\label{eq:ps-pole-mass}
m_{\rm PS}(\mu_f) = m_{\rm pole}-\delta m_{\rm PS}(\mu_f),
\end{equation}
with
\begin{equation}
\label{eq:deltam} 
\delta m_{\rm PS}(\mu_f) =
-\frac{1}{2}\int\limits_{|{\bf q}\,|<\mu_f} \!\!\!\frac{d^3{\bf
q}}{(2\pi)^3}\,\tilde{V}(\bff{q})\,.
\end{equation}
The infrared sensitivity precisely cancels in this combination, from 
which it follows that the relation between the PS mass and the 
$\overline{\rm MS}$ mass is free from the (leading) infrared renormalon 
divergence. When the top pair production cross section is expressed 
in terms of the PS mass, the large corrections disappear, as was 
explicitly demonstrated at NNLO in \cite{Beneke:1999qg}. While the 
PS mass scheme is particularly natural for the pair production 
threshold, any other scheme $X$ that subtracts the leading renormalon 
divergence from the pole mass can be used for the cross section 
calculation, as long as $\delta m_X \sim m v^2$, as shown, 
for example, in \cite{Hoang:1999zc}. 

The Coulomb potential in momentum space can be written in the form 
\begin{eqnarray}
\tilde{V}(\bff{q}) &=& -\frac{4\pi \alpha_s(\bff{q}) C_F}{\bff{q}^2} 
\Bigg[1+a_1 \frac{\alpha_s(\bff{q})}{4\pi} + 
a_2 \left(\frac{\alpha_s(\bff{q})}{4\pi}\right)^{\!2} 
+  \left(\frac{\alpha_s(\bff{q})}{4\pi}\right)^{\!3}
\left(a_3+8\pi^2 C_A^3 \ln\frac{\nu^2}{\bff{q}^2}
\right)
\nonumber\\
&& +\, {\cal O}(\alpha_s^4)\Bigg]\,.
\end{eqnarray}
When expressed in terms of the running coupling $\alpha_s(\bff{q})$, 
an explicit logarithm due to the factorization of soft and ultrasoft 
contributions appears only at ${\cal O}(\alpha_s^4)$.\footnote{See 
Section~\ref{sec:coulombpotential} of paper I for the Coulomb potential 
expressed in terms of $\alpha_s(\mu)$ and references.}
Performing the integral in (\ref{eq:deltam}), the expansion of 
$\delta m_{\rm PS}(\mu_f)$ in $\alpha_s\equiv\alpha_s(\mu)$ to 
NNNLO accuracy reads~\cite{Beneke:2005hg}
\begin{eqnarray}
\delta m_{\rm PS}(\mu_f)
&=&
\frac{\mu_f \alpha_s C_F}{\pi}\,
\Bigg[1 
+ \frac{\alpha_s}{4\pi} \Big(2 \beta_0 l_1 +a_1\Big) 
+ \bigg(\frac{\alpha_s}{4\pi}\bigg)^2 
     \bigg( 4 \beta_0^2 l_2+2\,\Big(2 a_1\beta_0+\beta_1\Big )l_1
     + a_2 \bigg) 
\nonumber \\
&& 
+ \,\bigg(\frac{\alpha_s}{4\pi}\bigg)^3 
  \bigg(   8 \beta_0^3 l_3 + 
   4 \Big(3 a_1\beta_0^2+\frac{5}{2}\beta_0\beta_1\Big) l_2 
   +2 \Big(3 a_2\beta_0+2 a_1\beta_1+\beta_2\Big) l_1 
\nonumber \\
&& \hspace{0cm}    
+ \,a_3 + 
   16 \pi^2 C_A^3 \Big[\ln\frac{\nu}{\mu_f}+1\Big]\bigg)
\Bigg],
 \label{PSmass}
\end{eqnarray}
where $l_1=\ln(\mu/\mu_f)+1$, $l_2=\ln^2(\mu/\mu_f)+2\ln(\mu/\mu_f)+2$,
$l_3=\ln^3(\mu/\mu_f)+3 \ln^2(\mu/\mu_f)+6\ln(\mu/\mu_f)+6$. 
Note that in addition to the subtraction scale $\mu_f$, the 
PS mass also depends on the scale $\nu$ in the ultrasoft logarithm 
in the third-order Coulomb potential. Different values of 
$\mu_f$ and $\nu$ correspond to different {\em definitions} 
of the PS mass.

The scale $\mu_f$ should be chosen of order $m \alpha_s$ in order 
not to violate the power counting of the  non-relativistic expansion, 
so the relation (\ref{PSmass}) is accurate to order $m \alpha_s^5$ just 
as the third-order bound state masses. The ``standard'' choice 
adopted in previous NNLO and NNNLO analyses of the 
top quark pair production threshold \cite{Beneke:1999qg,Beneke:2015kwa}
is $\mu_f=20\,$GeV. Unless mentioned otherwise, the term ``PS scheme'' 
will imply this choice. However, in the results section~\ref{sec:results} 
we will also analyze the effect of changing $\mu_f$. Contrary to the 
sensitivity of the PS mass definition to $\mu_f$, which can be 
${\cal O}(\mbox{GeV})$, the $\nu$-dependence 
from the last line in (\ref{PSmass}) is relatively small. With 
$\mu_f=20$~GeV and $\mu=80$~GeV, the PS mass changes by $+8.8$~MeV and 
$-3.5$~MeV when varying $\nu$ from 2~GeV to 50~GeV, respectively.
Unless explicitly said we set $\nu=\mu_f$ as done in 
previous applications. The motivation for choosing $\nu$ at the 
potential rather than ultrasoft scale stems from the fact that 
one would not like the subtracted mass to depend on ultrasoft 
dynamics, and $\nu=\mu_f$ minimizes the factorization logarithm 
that is sensitive to the ultrasoft scale.

In the following, we discuss two ways to convert the pair production 
cross section calculation in the pole mass scheme to the PS mass 
scheme, or related schemes.

\subsubsection{Insertion method}
\label{sec:insertionmethod}

We assume a generic relation of the form (\ref{eq:ps-pole-mass}), 
but drop the argument $\mu_f$ and label ``PS'' on $\delta m(\mu_f)$ for 
ease of notation. We also expand
\begin{equation}
\delta m = \delta m_0+\delta m_1 +\ldots\,,
\end{equation}
and note that with $\mu_f\sim m v$, $\delta m_n \sim m v^{n+2}$. We can 
implement the new mass scheme directly in the NRQCD Lagrangian by 
first replacing $m=m_{\rm pole} \to m_{\rm PS}+\delta m$, and then 
expanding systematically in $\delta m$. In addition, since a phase 
factor $e^{-i m t}$ is extracted from the relativistic heavy quark 
field when defining the non-relativistic field $\psi$, the rephasing with 
$m_{\rm PS}$ instead of $m$ results in a residual mass term $\delta m \psi^\dagger 
\psi$ in the effective Lagrangian. Keeping only terms up to NNNLO, 
the non-relativistic Lagrangian (\refI{eq:nrqcdqq}) is modified to
\begin{eqnarray}
\label{eq:nrqcdps}
{\cal L}_{\psi}&=&
\psi^{\dag} \Bigg([i D^0 -\delta m_0] - 
\sum_{i=1}^3 \delta m_i + 
\left[1-\frac{\delta m_0+\delta m_1}{m_{\rm PS}}\right]
\frac{{\bf D}^2}{2m_{\rm PS}} + 
\frac{{\bf D}^4}{8 m^3_{\rm PS}}\, \Bigg)\psi
\nonumber \\[0.1cm] 
&& -\,
\left[1-\frac{\delta m_0}{m_{\rm PS}}\right]
\frac{d_1 g_s}{2m_{\rm PS}}\psi^{\dag} {{\bm{ \sigma}} \cdot {\bf B}}\,\psi
+ O\bigg(\frac{1}{m_{\rm PS}^2}\bigg)\,,
\end{eqnarray}
where in all other terms of (\refI{eq:nrqcdqq}) not written here 
the $\delta m$ correction is already of higher order. We bracketed 
the combination $[i D^0 -\delta m_0]$ to emphasize that both, 
$i D^0, \delta m_0 \sim m v^2$, hence the leading-order mass correction 
$\delta m_0$ must be considered as part of the leading-order 
Lagrangian, while $\delta m_k \psi^\dagger\psi$ for $k\geq 1$ should be 
treated as a NkLO two-point interaction. In the coefficient of the 
kinetic term $\delta m_k/m$ counts as $m v^{k+2}$, hence the mass 
corrections given define NNLO and NNNLO Lagrangian corrections. 
By the same reasoning the $\delta m_0/m_{\rm PS}$ correction to 
the chromomagnetic interaction is a NNNLO term. Since one 
always needs two insertions of this interaction for a non-vanishing 
contribution to the pair production cross section, this correction 
can be dropped.

Matching to the PNRQCD Lagrangian implies the following three 
modifications (up to NNNLO). 1) The energy variable of the 
non-relativistic Green function, $E_{\rm PS}$ is now related to 
the cms energy by $\sqrt{s} = 2 m_{\rm PS} + E_{\rm PS}$. 2) The potential 
modification can be described by the substitution
\begin{equation}
V_k \to V_k +2 \delta m_k \,.
\end{equation}
which is precisely what is required by (\ref{eq:deltam}) to cancel 
the infrared part of the Coulomb potential contribution to the 
$i$th order potential. 3) There are additional single kinetic energy 
insertions $\delta m_k \,\bff{p}^2/m_{\rm PS}^2$ with $k=0,1$.
The PS scheme cross section can now be calculated 
according the standard rules for PNRQCD perturbation 
theory. For example, just as for the first-order correction to 
the Coulomb potential, one needs to consider up to three insertions 
of the $\delta m_1$ term, and the mixed insertion of one $\delta m_1$ 
together with the NNLO non-Coulomb potentials, and so on.

Given the third-order calculation in the pole scheme, there is 
a simpler way to obtain the result which avoids the explicit 
calculation of the new insertions. We can regard the dimensionless 
object 
\begin{equation}
\hat{G}(\lambda,\ln(m/\mu)) \equiv \frac{G(E)}{m^2}
\end{equation}
as a function of the scaling variable $\lambda$, and, since no 
other dimensionful variable is available, all energy dependence 
resides in the dependence on $\lambda$, $\ln(m/\mu)$. Recalling the 
origin of the corrections to the (P)NRQCD Lagrangian, the  
result of calculating insertions can also be obtained by 
substituting $m\to m_{\rm PS} + \delta m$ in $\hat{G}$, which 
amounts to first substituting systematically 
\begin{eqnarray}
\lambda &\to& \frac{\alpha_s C_F}
{\displaystyle 2 \, \sqrt{-\frac{(E_{\rm PS}-2\delta m)}
{m_{\rm PS}+\delta m}}}
= \lambda_{\rm PS} +\delta\lambda_1 +\delta\lambda_2 
+\delta\lambda_3 + \ldots 
\nonumber\\
&=&  \lambda_{\rm PS} - 
\underbrace{\frac{4\delta m_1}{(\alpha_s C_F)^2}
\,\frac{\lambda_{\rm PS}^3}{m_{\rm PS}}}_{-\delta\lambda_1} + \ldots
\\
\ln\frac{m}{\mu} &\to& \ln\frac{m_{\rm PS}}{\mu} 
+\delta\ln_2 + \delta\ln_3 + \ldots
= \ln\frac{m_{\rm PS}}{\mu} 
+\underbrace{\frac{\delta m_0}{m_{\rm PS}}}_{\delta\ln_2}+\ldots \,,
\label{eq:PSsubstitutions} 
\end{eqnarray}
where 
\begin{equation}
\lambda_{\rm PS} \equiv 
\frac{\alpha_s C_F}
{
2 \, \sqrt{-\frac{(E_{\rm PS}-2\delta m_0)}{m_{\rm PS}}}}\,,
\end{equation}
and then expanding in $\delta \lambda_k$ and $\delta\ln_k$ up 
the the third order. The NLO, NNLO ... cross section in the PS insertion 
scheme is then defined by collecting all NLO terms, NNLO terms, 
and so on.

We implemented this procedure by expanding all insertion functions 
given in this paper to the appropriate order (there is no need to 
expand functions that appear first at third order). In general this 
can be done analytically, but for the hypergeometric functions, as well 
as for some of the sum expressions, it turns out to be more efficient 
to calculate the derivative numerically. The same substitutions and 
expansion have been applied to the pole parts of the insertion 
functions. 

\subsubsection{Shift method}
\label{sec:shiftmethod}

The shift method works as follows: given a PS mass value, for the computation 
of the NkLO cross section in the PS shift scheme, first obtain the 
numerical, order-dependent value of the pole mass from 
(\ref{eq:ps-pole-mass}) with $\delta m_{\rm PS} = \sum_{i=0}^k \delta m_i$. 
Note the last term in this sum is of the same order $m v^{k+2}$ as 
the toponium energy level shifts in the NkLO cross section computation. 
Then evaluate the expressions for the cross sections, energy levels, 
wave functions at the origin (as required for pole resummation) 
in the {\em pole} scheme with the pole mass as determined above. 
Given that the pole scheme expressions have already been calculated, 
this amounts to a technically straightforward implementation of 
the conversion to the PS or any similar mass renormalization scheme. 

The large shifts of the cross section for fixed pole mass input are 
now compensated by the use of an order-dependent pole mass from a fixed 
PS mass input. The shift scheme differs from the insertion scheme 
by multiple insertions of higher order than considered. Unlike the 
PS insertion scheme, which achieves an exact leading renormalon 
cancellation by combining terms of the same order directly, the shift 
scheme is ``renormalon-free'' only up to order NkLO. However, 
at $k=3$ the cancellation is numerically already very precise and 
the residual renormalon divergence is no longer a practical problem. 
This conceptual disadvantage is outweighed by the technical simplicity 
of the shift scheme and the difficulties with the insertion scheme 
discussed in the next subsection.

The insertion method cannot be used for the conventional 
$\overline{\rm MS}$ scheme, since in this case $\delta m_0$ would 
be of order $m\alpha_s\sim m v$, which spoils the non-relativistic 
power counting. At the Lagrangian level, this can be seen from 
the fact that the non-dynamical term $\delta m_{\overline{\rm MS}} 
\psi^\dagger\psi$ would parametrically dominate over the kinetic 
terms. However, the $\overline{\rm MS}$ mass {\em can} be 
employed in the shift method, although combining the 
$k$-loop pole-$\overline{\rm MS}$ mass relation with the NkLO 
cross section calculation leads to a mismatch by one order 
in the mass vs toponium energy level shift. The  $\overline{\rm MS}$ 
shift scheme together with the present calculation of the third-order 
spectral function in PNRQCD perturbation theory has been employed in 
\cite{Beneke:2014pta} to determine the bottom quark mass from 
Upsilon sum rules, resulting in a slightly more precise value 
of the bottom $\overline{\rm MS}$ quark mass than the determination 
via the PS mass. A similar observation was made 
in \cite{Kiyo:2015ooa} for the toponium energy levels. 
We shall investigate the top pair production threshold in 
the $\overline{\rm MS}$ shift scheme in the results section. 

\subsection{Insertion schemes and pole resummation} 
\label{sec:insertionproblems}

In Section~\ref{sec:poleresummation} we discussed the necessity to 
perform a resummation on PNRQCD perturbation theory for $G(E)$ in 
the vicinity of the bound state poles, since the effective expansion 
parameter near the singularity is the first order shift of the pole 
divided by the distance to the pole, see 
(\ref{eq:poleresumcondition}). In the insertion scheme the 
leading order energy of the $n$th bound state pole, and its first order 
correction are given by (adopting $m_{\rm PS} = 171.5~$GeV and 
$\alpha_s(80\,\mbox{GeV})=0.120383$ as in the later 
section~\ref{sec:results})
\begin{equation}
E_{n,\rm PS}^{(0)} = 
-\frac{m_{\rm PS} (\alpha_s(\mu) C_F)^2}{4 n^2} + 2 \delta m_0 
\stackrel{n=1}{=} (-1.10 + 2.04)\,\mbox{GeV} 
\end{equation}
and
\begin{eqnarray}
E_{n,\rm PS}^{(1)} &=& - \frac{m_{\rm PS} \alpha_s^3(\mu) C_F^2}{16 \pi n^2}
\left(2 a_1+4\beta_0\Big[\ln\frac{n \mu}{m_{\rm PS}\alpha_s C_F}+S_n\Big]
\,\right) 
+  2 \delta m_1 
\nonumber\\
&\stackrel{n=1}{=}& (-0.77 + 0.81)\,\mbox{GeV} = 0.04\,\mbox{GeV}
\label{eq:e1corPS}
\end{eqnarray}
respectively, where the second equation generalizes 
(\ref{eq:e11shift}) to the PS scheme and arbitrary $n$. The accumulation 
point of bound state poles is now at positive energy $2 \delta m_0$. 
There is a large cancellation between the binding energy and the 
mass correction in the first-order correction to the $1S$ energy level, 
which persists at higher orders \cite{Beneke:1997jm,Beneke:2005hg}, 
and can be viewed as being due to the infrared subtraction of the 
Coulomb contribution, which implements the renormalon cancellation. 
As a consequence, the quantity $|E_n^{(1)}|/\Gamma_t$, which controls 
the need for pole resummation is now only 0.03 rather than 0.6 
and we do not expect a drastic breakdown of PNRQCD perturbation 
theory in the PS insertion scheme.

However, the cancellation becomes less effective as $n$ increases. 
Since the binding energy scales as $1/n^2$, the first-order energy 
correction asymptotes to the first order mass shift $\delta m_1$, 
implying that the expansion parameter quickly increases to 
$2\delta m_1/\Gamma_t\approx 0.6$ for the higher bound state poles. 
Even though the contribution of higher poles to $G(E)$ is suppressed 
by their decreasing residues $|\psi_n(0)|^2 \propto 1/n^3$, 
this suggests that pole resummation is still necessary in the insertion 
schemes to obtain precise results. Contrary to expectations we then 
find an obviously unphysical dip-peak structure in the cross section 
near $E=2 \delta m_0$. We will show this behaviour for the third-order 
cross section in the following results section. Here we shall demonstrate and 
explain its origin in an exactly solvable toy model.

To this end, we assume that the exact potential is given by 
\begin{eqnarray}
\tilde{V}(\bff{q}) &=& -\frac{4\pi \alpha_s C_F}{\bff{q}^2} 
\Bigg[1+\varepsilon a_1 \frac{\alpha_s}{4\pi}\Bigg]\,.
\end{eqnarray}
We adopt $a_1=43/9$, such that the potential corresponds to the 
one-loop Coulomb potential in the absence of running coupling 
effects. To enhance the features that we wish to demonstrate we 
enhance the one-loop correction by choosing $\varepsilon =10$ in 
the figures shown below. We treat the one-loop correction in 
PNRQCD perturbation theory, and employ the pole and the PS 
scheme. With the above potential the relation between the poles and PS 
mass is given exactly by 
\begin{eqnarray}
m_{\rm pole} &=& m_{\rm PS} +\frac{\mu_f \alpha_sC_F  }{\pi} 
\left[ 1+\delta_{a1}\right].
\label{eq:polePSrela1model}
\end{eqnarray}
with $\delta_{a1} = (\alpha_s/4\pi) \varepsilon a_1$. The exact $S$-wave 
bound state energy levels in the PS scheme read
\begin{eqnarray}
E_{n, \rm PS}&=& 
-\frac{(\alpha_s C_F)^2}{4 n^2} m_{\rm pole}(1+2\delta_{a1} +\delta_{a1}^2)
+\frac{2 \mu_f \alpha_sC_F  }{\pi} 
\left[ 1+\delta_{a1}\right]\,,
\end{eqnarray}
where for $m_{\rm pole}$ the expression (\ref{eq:polePSrela1model}) 
should be substituted. 
The model is exactly solvable, since it corresponds to the solvable Coulomb 
problem with an effective coupling $\tilde{\alpha}_s = 
\alpha_s (1+\delta_{a1})$. The exact solution for the Green function 
is the standard zero-distance Coulomb Green function 
(\refI{eq:G00MSbar})
\begin{eqnarray}
G(\sqrt{s},m) = \frac{m^2}{4\pi}
\left[-\sqrt{-\frac{E}{m}} -\tilde{\alpha}_s C_F 
\left\{
\frac{1}{2}\ln\left(\frac{-4 m E}{\mu^2}\right) -\frac{1}{2} 
+\gamma_E+\Psi(1-\tilde\lambda)\right\}\right]\quad
\label{eq:Ga1}
\end{eqnarray}
with $\sqrt{s} = 2 m_{\rm pole}+E$ and  
$\tilde{\lambda}=\lambda (1+\delta_{a_1})$. 
By expanding $G(\sqrt{s}+i \Gamma_t,m_{\rm pole})$ 
in $\delta_{a_1}$, we generate 
the PNRQCD perturbative approximation to the exact result to the 
desired order. The PS insertion scheme is obtained from 
$G(\sqrt{s}+i \Gamma_t,m_{\rm PS} +\delta m)$ and subsequent 
expansion, including $\delta m$ (except $\delta m_0$ in the combination 
$\sqrt{s}-2 (m_{\rm PS}+\delta m_0)$). 

\begin{figure}[p]
\begin{center}
\includegraphics[width=0.55\textwidth]{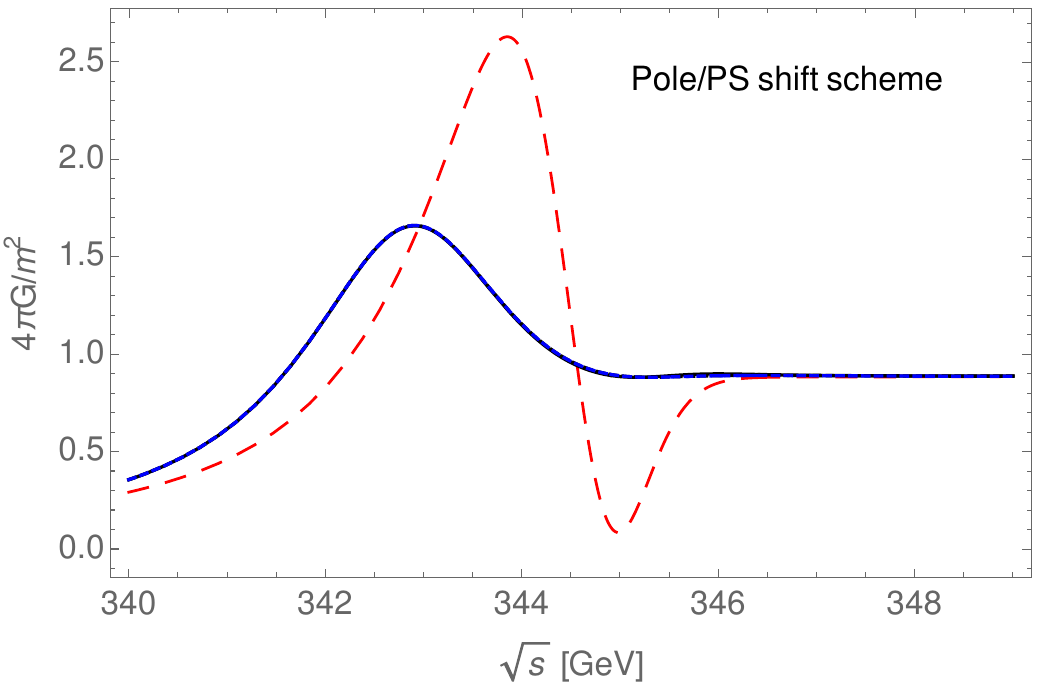}
\vskip0.4cm
\includegraphics[width=0.55\textwidth]{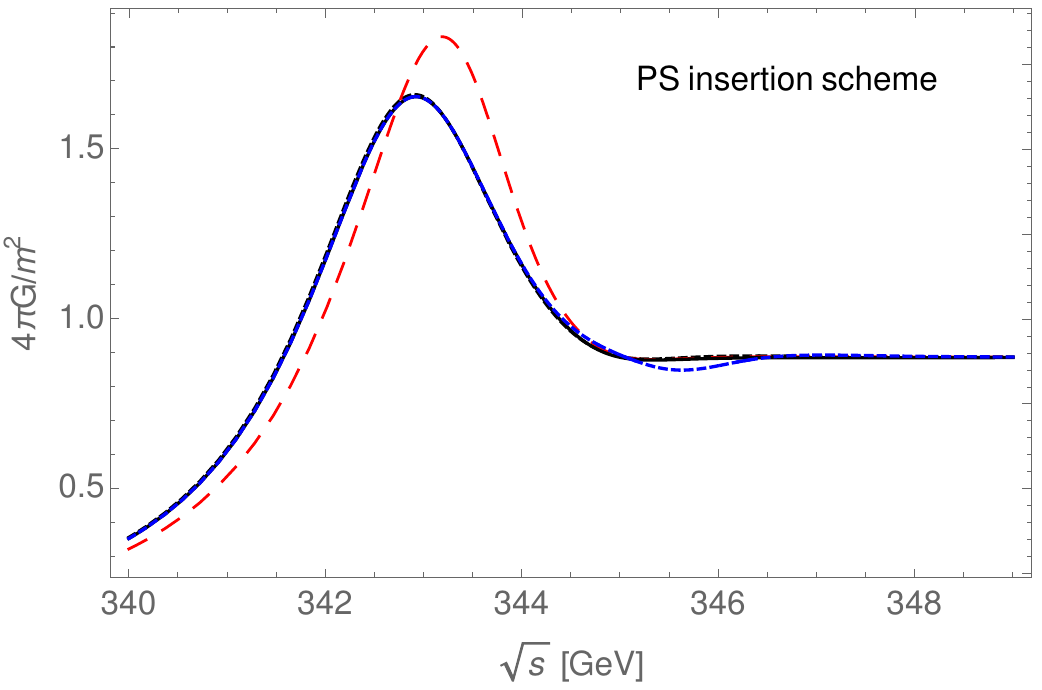}
\vskip0.4cm
\includegraphics[width=0.55\textwidth]{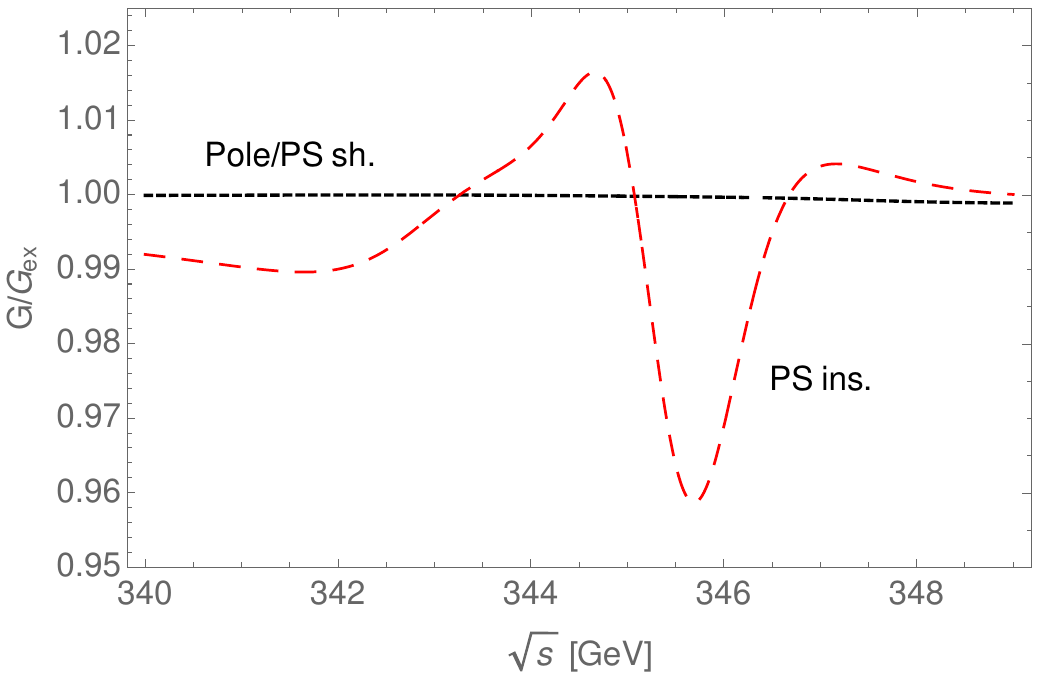}
\vskip0.4cm
\end{center}
\caption{\label{fig:a1modnopoleresum}
Upper panel: No (long-dashed/red), first pole (solid/black), 
first six poles (short-dashed/blue) poles resummed in the PS 
shift (= pole) scheme. Middle panel: as upper panel, but in the 
PS insertion scheme. Lower panel: ratio of first six poles resummed 
to exact in the PS shift (short-dashed/black) and PS insertion 
(long-dashed/red) scheme.
}
\end{figure}

In Figure~\ref{fig:a1modnopoleresum} we show $G(\sqrt{s}+i\Gamma_t,m)$ 
divided by $m^2/(4 \pi)$ without pole resummation (long-dashed/red) in 
the PS~shift (upper panel) and PS insertion (middle panel) at 
third order in non-relativistic perturbation theory. The parameters 
used are: $m_{\rm pole} = 173.208\,$GeV which implies 
$m_{\rm PS} = 171.387\,$GeV by (\ref{eq:polePSrela1model}), 
$\Gamma_t=1.36\,$GeV, and $\alpha_s=0.14$. Also shown in the 
plot is the result with pole resummation applied to the first bound 
state pole only (solid/black), to the first six bound state 
poles (short-dashed/blue), and the exact Green function 
(short-dashed/black). Note that in the toy model the PS shift 
and pole scheme give identical results from the third order 
on, since the mass relation (\ref{eq:polePSrela1model}) receives 
no higher-order corrections.

The following observations can now be made:
(1) The long-dashed curve (no pole resummation) is clearly a poor 
approximation, pole resummation is certainly necessary, at least for 
the first bound state pole. The effect of pole resummation is less 
important in the PS insertion scheme. In the toy model, 
(\ref{eq:e1corPS}) becomes 
\begin{eqnarray}
E_{1,{\rm PS} - a_1 {\rm model}}^{(1)} &=& - \frac{m_{\rm PS} 
\alpha_s^3 C_F^2}{16 \pi}
\left(2 \varepsilon a_1\right) 
+  \frac{2 \mu_f \alpha_sC_F  }{\pi} \delta_{a1} 
\nonumber\\
&=& (-1.59 + 1.27)\,\mbox{GeV} = -0.32\,\mbox{GeV}\,,
\end{eqnarray}
with a smaller first-order correction to the first bound-state pole 
than in the pole scheme. Since the cancellation is smaller than 
in the real case (\ref{eq:e1corPS}), we may expect the effect of pole 
resummation to be even smaller in that case. 
(2) With pole resummation applied to the first bound state pole 
only, the third-order result differs from the exact result by less 
than 1\% over the entire energy region in both schemes, and 
the corresponding curves are indistinguishable in both plots.  
(3) When further poles are resummed (here up to $n=6$) the 
approximation improves in the PS shift scheme (invisible on the 
scale of the upper panel). However---and this is the unexpected 
feature which motivates this discussion---the approximation 
becomes worse in the PS insertion scheme, where a dip develops 
slightly below the nominal continuum threshold at $2 m_{\rm pole} 
= 346.42\,$GeV, as seen in the middle plot. This effect is 
highlighted in the lower panel of Figure~\ref{fig:a1modnopoleresum}, 
which shows the third-order approximation with the first six poles 
resummed normalized to the exact result. While the PS shift result 
provides a perfect approximation to the exact result, the PS 
insertion scheme shows a peak-dip oscillation of more than 
5\%, which is larger than the target accuracy of the third-order 
approximation. The effect does not go away as the order of the 
approximation is increased and grows when more poles are resummed.

\begin{figure}[t]
\begin{center}
\includegraphics[width=0.6\textwidth]{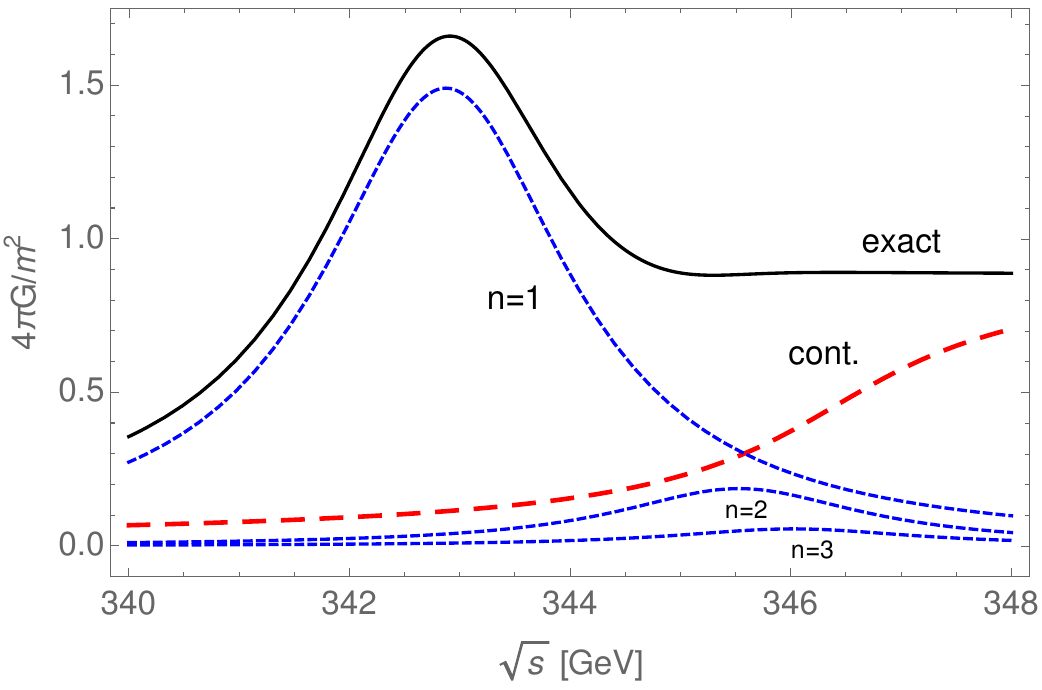}
\end{center}
\vspace*{-0.4cm}
\caption{\label{fig:continuumpoleseparated} 
Exact Green function separated into continuum and the first 
three bound state pole contributions according to 
(\ref{eq:Ga1separated}).
}
\end{figure}

To understand the origin of this effect we use the identity
\begin{eqnarray}
\mbox{Im}\,[\psi(1-\lambda)] &=& 
- \mbox{Im}\,[\psi(1+\lambda)] 
+\sum_{n=1}^\infty \left(-\frac{m\alpha_s^2 C_F^2}{2 n^3}\right)
\mbox{Im}\left[\frac{1}{\displaystyle 
-\frac{m\alpha_s^2 C_F^2}{4 n^2} - E}
\right],\quad
\end{eqnarray}
which allows us to separate the continuum and pole contribution 
to (\ref{eq:Ga1}) according to 
\begin{eqnarray}
\mbox{Im}\,[G(\sqrt{s},m)] &=& \frac{m^2}{4\pi}\,
\mbox{Im}\left[-\sqrt{-\frac{E}{m}} -\tilde{\alpha}_s C_F 
\left\{
\frac{1}{2}\ln\left(\frac{-4 m E}{\mu^2}\right) 
-\Psi(1+\tilde\lambda)\right\}\right]\
\nonumber\\
&&+\,
\sum_{n=1}^\infty \frac{m^3\tilde{\alpha}_s^3 C_F^3}{8 \pi n^3}\;
\mbox{Im}\left[\frac{1}{\displaystyle 
-\frac{m\tilde{\alpha}_s^2 C_F^2}{4 n^2} - E}
\right].\label{eq:Ga1separated}
\end{eqnarray}
The continuum contribution in the first line has no imaginary part 
for negative energy. The second line, on the contrary, is real for 
positive energy and represents the bound state poles for negative 
energy values. The separate contributions from the continuum and 
the first three bound state poles is shown in 
Figure~\ref{fig:continuumpoleseparated} (parameters as above). 
When the above expression is expanded in PNRQCD perturbation 
theory, the pole resummation procedure restores the form 
of the exact pole contribution in the second line with bound-state 
energy and residue at the appropriate order. Let us focus 
on the expansion in non-relativistic perturbation theory of 
the continuum contribution. In the pole and PS shift scheme 
this amounts to the expansion in $\delta_{a_1}$ in $\tilde{\alpha}_s$ 
and $\tilde{\lambda}$, but non-analytic energy dependence 
in $\sqrt{-E/m}$ and $\ln(-4 m E)$ remains untouched. In the 
PS insertion scheme, however, these terms are expanded around 
$E=0$ into a series of increasingly singular terms, similar to 
the energy denominator of the pole terms. For example, 
the term $\sqrt{-E/m}$ is expanded as
\begin{eqnarray}
\sqrt{-\frac{E}{m}} &=& 
\sqrt{-\frac{\sqrt{s}-2 (m_{\rm PS}+\delta m)}
{m_{\rm PS}+\delta m}}
\nonumber\\
&=& \sqrt{-\frac{E_{\rm PS}-2\delta m_0}{m_{\rm PS}}} \bigg[1-\frac{1}{2}
\frac{2\delta m_1}{E_{\rm PS}-2\delta m_0}-
\left(\frac{1}{8} \left(\frac{2\delta m_1}{E_{\rm PS}-2\delta m_0}\right)^2 +
\frac{\delta m_0}{2 m_{\rm PS}}\right)
\nonumber\\
&& +\,\ldots
\bigg]
\end{eqnarray}
For $E_{\rm PS} \approx 2 \delta m_0$, corresponding to 
$\sqrt{s}\approx 2 (m_{\rm PS}+\delta m_0)$, the parameter of this 
expansion is $2\delta m_1/\Gamma_t$, just as for the highly 
excited $S$-wave bound states in the insertion scheme, resulting in 
a poorly convergent series.

Let us emphasize the similar behaviour of the high-$n$ bound states 
and the continuum threshold: in the pole and shift schemes, the 
first-order energy shift decreases as $1/n^2$, hence 
$|E_n^{(1)}|/\Gamma_t \propto 1/n^2$, and the continuum threshold 
is not expanded at all, consistent with $n\to \infty$. In the 
insertion scheme the high-$n$ bound states {\em and} the continuum 
are uniformly shifted by $2\delta m$, resulting in a expansion 
parameter $2 \delta m_1/\Gamma_t$ in non-relativistic perturbation 
theory for the high-$n$ bound-states and the continuum near the zeroth 
order continuum threshold  $\sqrt{s}=2 (m_{\rm PS}+\delta m_0)$.
While the pole resummation procedure cures the deficiency of the 
non-relativistic expansion near the bound state poles, the singular 
expansion of the continuum threshold remains in the insertion scheme. 
This is the origin 
of the unphysical peak-dip near the continuum threshold seen in 
the threshold cross section in the insertion scheme in the middle 
and lower panel of Figure~\ref{fig:a1modnopoleresum}. The obvious 
solution to the problem appears to be a resummation of the singular 
terms at the continuum threshold. This is effectively done within 
the shift scheme, which resums all terms, not just the singular ones, 
which would indeed be difficult to separate in practice given the complexity 
of the insertion functions.

Does it follow that the insertion scheme must be abandoned? We recall 
from the discussion of Figure~\ref{fig:a1modnopoleresum} that the 
insertion method provides a very good approximation when only 
the first pole is resummed. That is, there seems to be a cancellation 
of the singular terms between the high-$n$ bound states and the 
continuum when no pole resummation is applied to either. Such a 
cancellation is consistent with the uniform behaviour of the 
singular expansion of  the highly excited states and the continuum. 
That a close analytic connection between the two exists has been 
noted in previous contexts \cite{Beneke:2014pta,Beneke:2016jpx}.
In particular, in the first reference, it was noted that the gluon 
condensate correction to the moments of the Green function 
are divergent for the continuum contribution and the bound state 
contribution separately. The divergence arises from the threshold 
of the continuum and the infinite sum $\sum_{n=1}^\infty$ over 
bound states. However, in $\sum_{n=n_{\rm min}}^\infty$ plus 
the continuum the divergence cancels.

\begin{figure}[t]
\begin{center}
\includegraphics[width=0.6\textwidth]{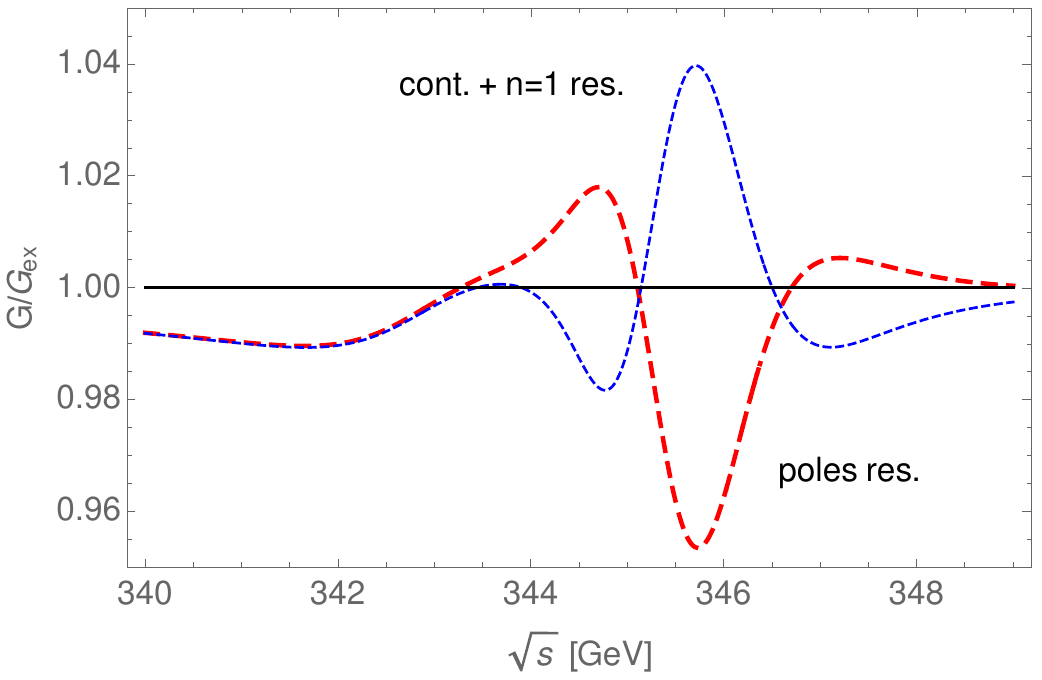}
\vskip0.4cm
\includegraphics[width=0.6\textwidth]{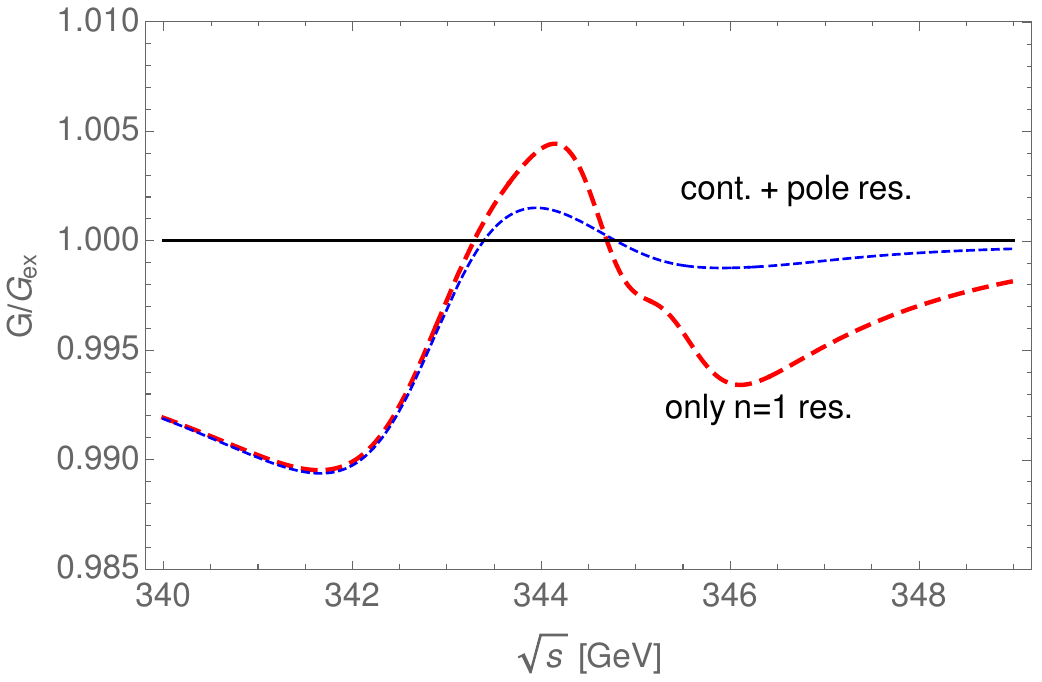}
\end{center}
\caption{\label{fig:a1modelpartialresums}
Upper panel: Pole resummation but continuum expanded (long-dashed/red) 
vs. continuum resummed but no pole resummation except for 
$n=1$ (short-dashed/blue). 
Lower panel:  No resummation except for the $n=1$ bound-state pole 
(long-dashed/red) vs. continuum and pole resummation (short-dashed/blue). 
See text for more explanations.
}
\end{figure}

We did not attempt to prove a cancellation of the singular terms 
in the expansion of the bound states and continuum threshold for 
the top pair threshold, but provide further numerical evidence in 
Figure~\ref{fig:a1modelpartialresums}. The curves in this figure 
refer to the PS insertion scheme in the toy model in the third-order 
in non-relativistic perturbation theory with parameters as above. 
The curves are normalized to the exact Green function. 
The upper panel shows the quality of two approximations for 
which the continuum and the high-$n$ bound state poles are 
not treated coherently: the long-dashed/red curve includes pole 
resummation, but the expansion of the continuum. This is the 
standard approximation with pole resummation and it exhibits the 
peak-dip artefact already seen in the lower panel of 
Figure~\ref{fig:a1modnopoleresum}. The short-dashed/blue curve on 
the other hand refers to the opposite case, where the continuum 
is resummed (which corresponds to the exact continuum in the toy model), 
but no pole except for $n=1$ is resummed. A dip-peak artefact of 
opposite sign and similar magnitude appears. When the poles and 
continuum are treated coherently, a much improved approximation 
is obtained and shown in the lower panel of 
Figure~\ref{fig:a1modelpartialresums} (note the different vertical 
scale of this panel). Here the long-dashed/red 
curve applies no resummations at all except for the first bound 
state pole, which achieves an accuracy of a few permille to 
to one percent (below threshold) as already 
mentioned. Resumming both, poles and continuum, is even closer to 
the exact result, but both curves are within the expected accuracy of 
the third-order approximation to the exact result.

It follows from this study that the insertion scheme must be used 
with care. The following schemes are found to be good approximations: 
(1) the shift scheme with pole resummation. It is sufficient to 
resum the first six poles. (2) the insertion scheme with pole 
resummation {\em only} for the first bound state pole. In the 
following presentation of results for the NNNLO top threshold 
in QCD we adopt the PS shift scheme with pole resummation to 
$n=6$ as the standard scheme.


\section{Top-quark cross section}
\label{sec:results}

We proceed to analyze the theoretical result for the  top 
pair production cross section around the threshold at the 
next-to-next-to-next-to-leading order in the resummed 
non-relativistic expansion. The results of the present paper 
have already been incorporated in~\cite{Beneke:2015kwa}, 
which reported the first complete NNNLO QCD calculation. 
Subsequent publications~\cite{Beneke:2015lwa,Beneke:2017rdn} 
adding important Higgs Yukawa coupling, electroweak and non-resonant 
effects also made use of the present results, and provided 
a dedicated discussion of these non-QCD effects. In the following, 
we focus on the detailed analysis of the third-order QCD 
correction, which was not included in the letter 
publication~\cite{Beneke:2015kwa}, but do not repeat the discussion of 
non-QCD effects. 

The cross section depends on a number of Standard Model 
parameters, and the renormalization and finite-width scale. 
We adopt 
\begin{eqnarray}
&&m_t^{\rm PS}(\mu_f=20\,\mbox{GeV}) = 171.5\,\mbox{GeV}\,,\\
&&\Gamma_t = 1.36\,\mbox{GeV}\,,\\
&&\alpha_s(m_Z) = 0.1180\,,\\
&&m_Z = 91.1876 \,\mbox{GeV}\,,\\
&&\sin^2\theta_w = 0.222897\,,\\
&&\mu=80\,\mbox{GeV}\,,\\
&&\mu_w=350\,\mbox{GeV}
\end{eqnarray}
as reference values. These settings imply the $\overline{\rm MS}$ and 
pole mass values $\overline{m}_t \equiv m_t^{\overline{\rm MS}}
(m_t^{\overline{\rm MS}}) = 163.395\,$GeV and 
$m_t^{\rm pole} = 173.208\,$GeV, respectively, using four-loop 
conversion formulas \cite{Beneke:2005hg,Marquard:2015qpa}.
The numerical results are generated with version 2.2 of the 
\texttt{QQbar\_threshold} code~\cite{Beneke:2016kkb} with setting 
that ignores non-QCD effects. The default call to evaluate 
the R-ratio $R(\sqrt{s})$ for given center-of-mass energy 
$\sqrt{s}$ at third order reads
\begin{verbatim}
TTbarRRatio[sqrts, {80 (*mu*), 350 (*muw*)}, {171.5, 1.36}, 
 "N3LO", alphaSmZ -> 0.1180, 
 BeyondQCD -> {"None"}, ResonantOnly -> True, Production -> "SWaveOnly",
 MassScheme -> {"PSshift", 20}]
\end{verbatim}
The PS shift mass scheme with $\mu_f=20~$GeV is our reference 
scheme.  
The above includes S-wave production from virtual $s$-channel 
$Z$-boson exchange, but not the small P-wave 
contribution~\cite{Beneke:2013kia}. The $R$-ratio 
can be converted to the cross section in picobarn through
\begin{equation}
\sigma(s) = \frac{4 \pi\alpha(m_Z)^2}{3 s}\,R(s) 
= 0.828977\,\left(\frac{344\,\mbox{GeV}}{\sqrt{s}}\right)^2
\,R(s) \,\mbox{pb}\,,
\end{equation} 
where the QED coupling $\alpha(m_Z)=1/128.944$ has been used. 
Except for an update of the strong coupling and top width 
input, this corresponds 
to the result published in letter form in \cite{Beneke:2015kwa}.

\begin{figure}[t]
\begin{center}
\includegraphics[width=0.65\textwidth]{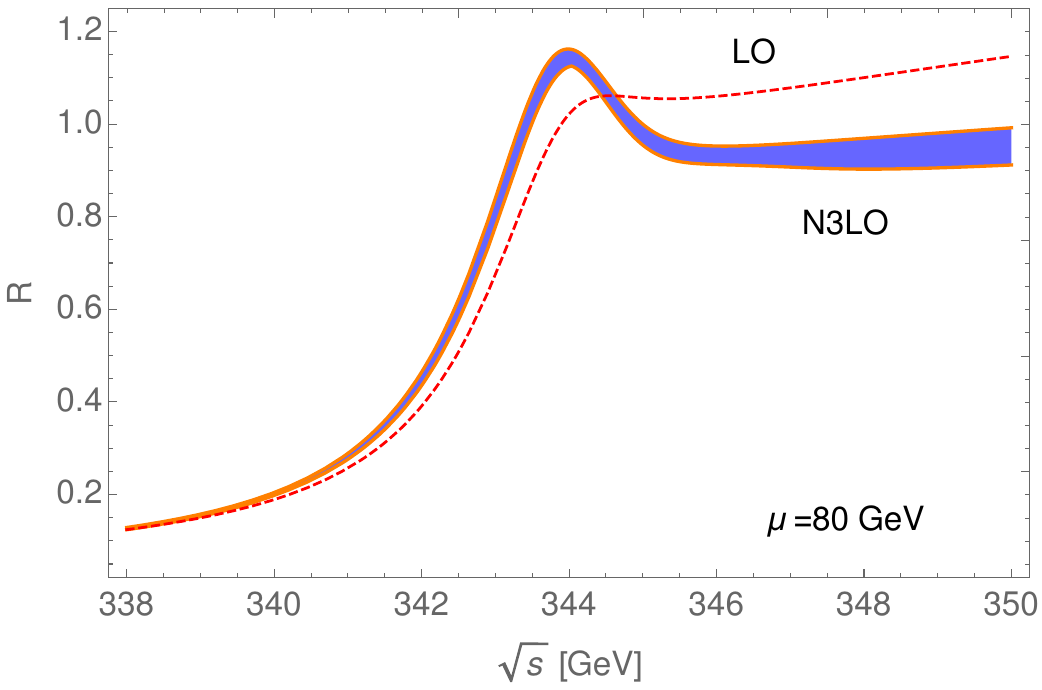}
\end{center}
\caption{\label{fig:mainresult}
Top-pair production cross section near threshold at NNNLO in 
QCD. The width of the band reflects the scale variation in 
the interval $[50,350]~$GeV. For comparison, the leading-order 
cross section (dashed) is depicted.}
\end{figure}

The third-order result for the cross section 
is shown in Figure~\ref{fig:mainresult} 
including the scale uncertainty from varying $\mu$ from 50 to 
350~GeV. For comparison we also show the (strongly 
scale-dependent) leading-order cross section (evaluated with 
$\mu=80$~GeV), which highlights 
the importance of QCD corrections.

\subsection{Scale dependence}

In this section we provide a more detailed assessment of the 
residual scale dependence than in \cite{Beneke:2015kwa}, including a 
discussion of finite-width scale dependence.

\subsubsection{Renormalization scale dependence}

\begin{figure}[t]
\begin{center}
\includegraphics[width=0.65\textwidth]{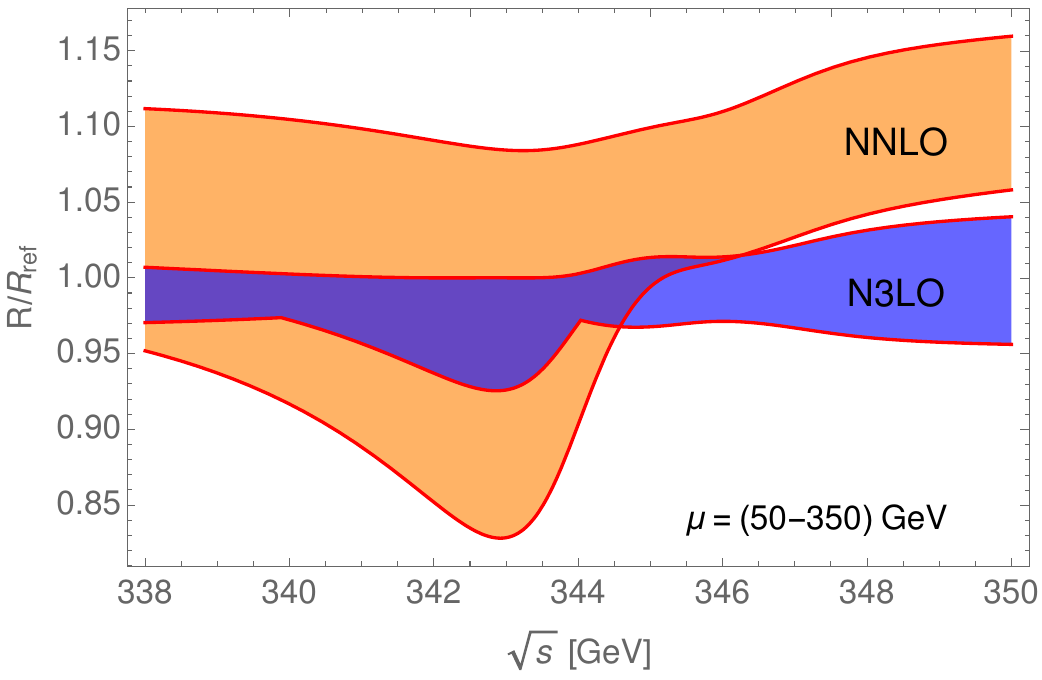}
\end{center}
\caption{\label{fig:relativescaledependence}
Relative scale-dependence of the energy-dependent 
second-order (NNLO) and 
third-order (N3LO) cross section when normalized to the 
third-order cross section at the reference scale 
$\mu=80\,$GeV.
}
\end{figure}

In order to display the change in the predicted cross section and the 
reduction of the scale uncertainty from second to third order we 
define the $R$-ratio normalized to a reference prediction for which 
we choose the third-order result at the default scale $\mu=80~$GeV.
Figure~\ref{fig:relativescaledependence} shows these normalized 
$R$-ratio results with uncertainty bands from the scale variation 
in the standard interval $[50,350]$~GeV, keeping $\mu_w=350~$GeV 
fixed. We will discuss the 
reasons for this interval choice below.

The comparison of NNLO and NNNLO shows a large reduction of the 
scale dependence in the vicinity and below the location of the 
cross section peak around 344~GeV. In contrast, there is a sizeable 
negative correction to the cross section above the threshold and 
the scale dependence is only slightly diminished in this region. 
This is caused by the large negative third-order correction $c_3$ to 
the matching coefficient of the vector current.

\begin{figure}[p]
\begin{center}
\includegraphics[width=0.55\textwidth]{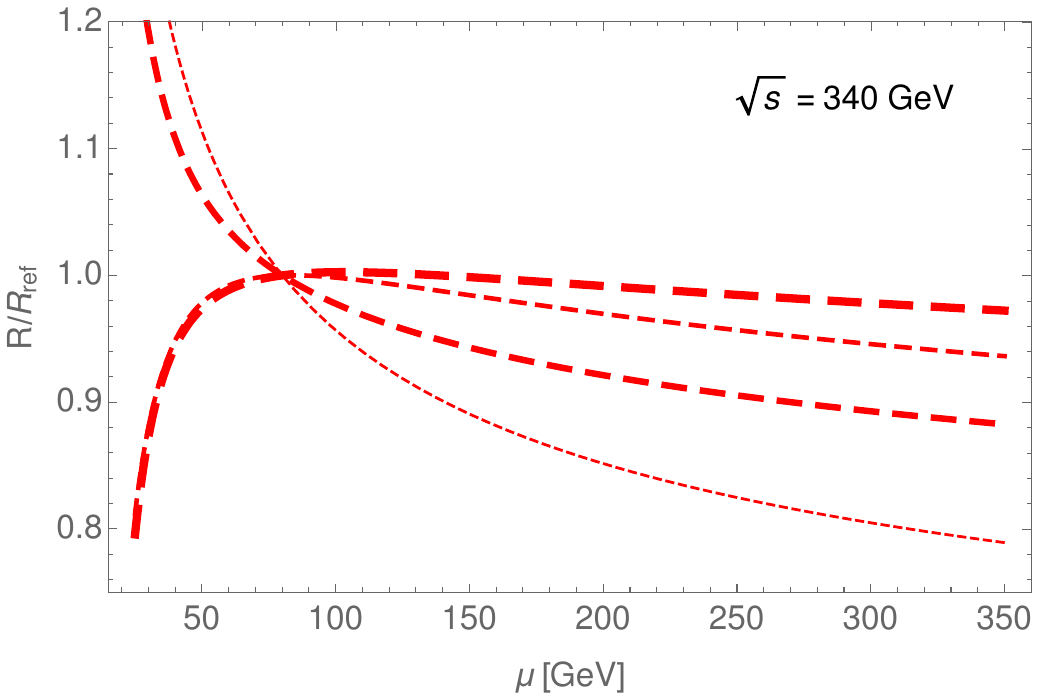}
\vskip0.2cm
\includegraphics[width=0.55\textwidth]{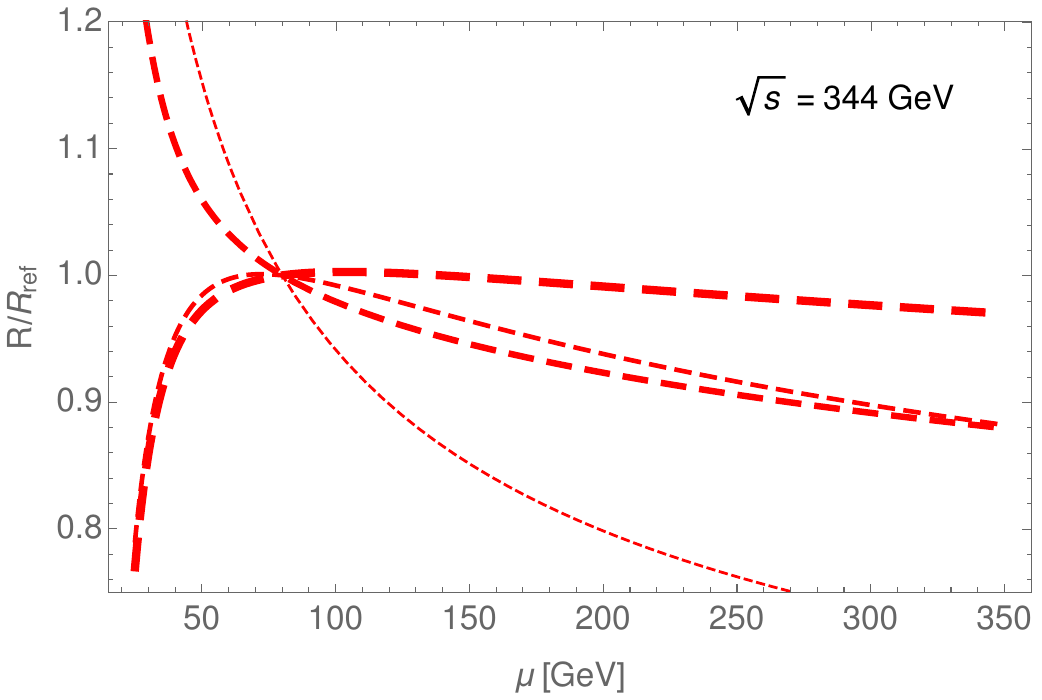}
\vskip0.2cm
\includegraphics[width=0.55\textwidth]{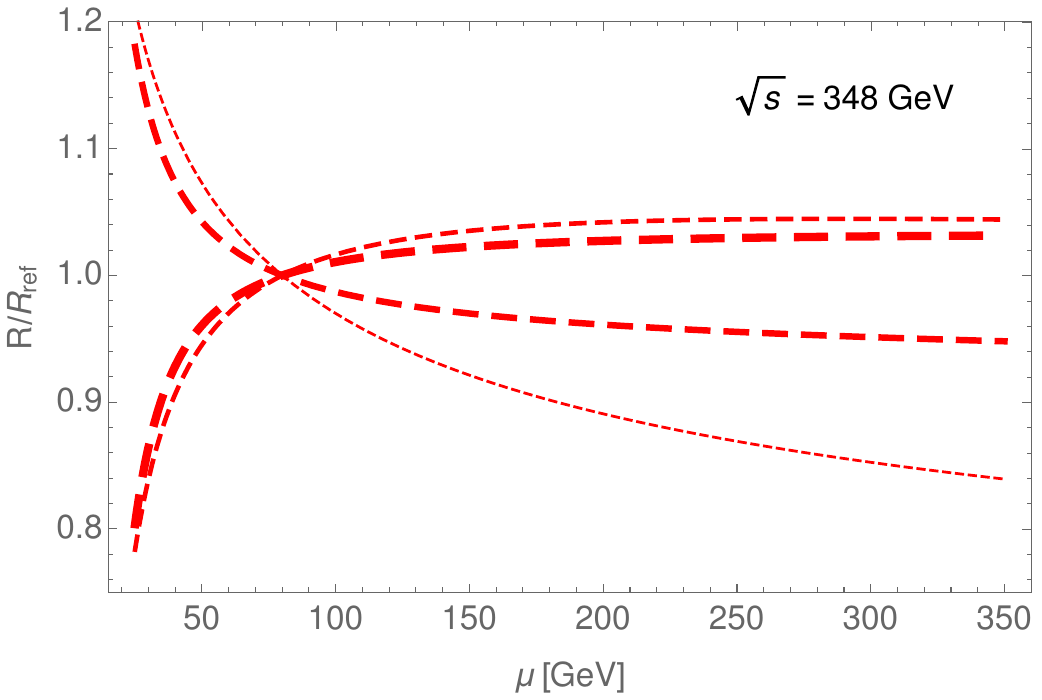}
\end{center}
\caption{\label{fig:relscaledepenergies}
Scale dependence of the cross section (normalized to the reference 
scale $\mu=80~$GeV) below, near and above the peak.  The order of 
approximation increases from LO to NNNLO as the line 
style changes from thin to thicker and short to longer dashes.
}
\end{figure}

In Figure~\ref{fig:relscaledepenergies} we show the scale dependence 
explicitly for three selected energy values, 340, 344, and 348~GeV, 
corresponding to $e^+ e^-$ cms energies below, near and above the 
top pair production threshold. The scale-dependent cross section at 
a given order is now normalized to its value at the reference scale 
$\mu=80\,$GeV in the {\em same} order of approximation, hence all 
lines cross at the point $(\mu=80~\mbox{GeV},1)$.  As the order of 
the theoretical calculation increases from LO to NNNLO the line 
style changes from thin to thicker and short to longer dashes. All 
three plots visualize the reduction of the scale dependence as the 
curves become flatter above $\mu\approx 50$~GeV with increasing 
approximation order. Another common feature is that perturbation 
theory breaks down at renormalization scales not far below 50~GeV 
with large sign-alternating variations from order to order. A 
similar observation has already been made in the analysis of the 
Coulomb potential contributions at third order \cite{Beneke:2005hg}, 
although in that case the breakdown 
of perturbation theory occurred at somewhat lower scales.

The breakdown of perturbation theory at such large scales for a 
process whose characteristic scales span from the ultrasoft 
scale of a few GeV to $2 m_t\approx 350$~GeV is surprising and 
troublesome. It could originate from large logarithms of the ratios 
of the involved scales or from some other source of systematically 
large corrections. Indeed the resummation of logarithms for the 
top threshold \cite{Hoang:2013uda} indicates that in the resummed 
result the various scales can be taken at their natural values. 
In the case of the pure Coulomb contributions it is the series of 
multiple insertions of the Coulomb potential, which converges 
slowly at small scales \cite{Beneke:2005hg}, while for larger scales 
the third order approximation is in excellent agreement with the 
numerical solution that sums these multiple insertions to all orders. 
Quite generally, sign-alternating diverging series in QCD are 
more efficiently summed by employing larger scales (making 
$\alpha_s(\mu)$ smaller) at the expense of larger logarithms 
\cite{Beneke:1992ea}. Given these observations, we defined the 
lower limit of the scale variation interval $[50,350]\,$GeV, also 
adopted in \cite{Beneke:2015kwa}, by the scale below which 
perturbation theory becomes unstable and unreliable. It should be 
kept in mind that the theoretical uncertainty 
estimate from scale variation depends on this choice.

\begin{figure}[t]
\begin{center}
\includegraphics[width=0.6\textwidth]{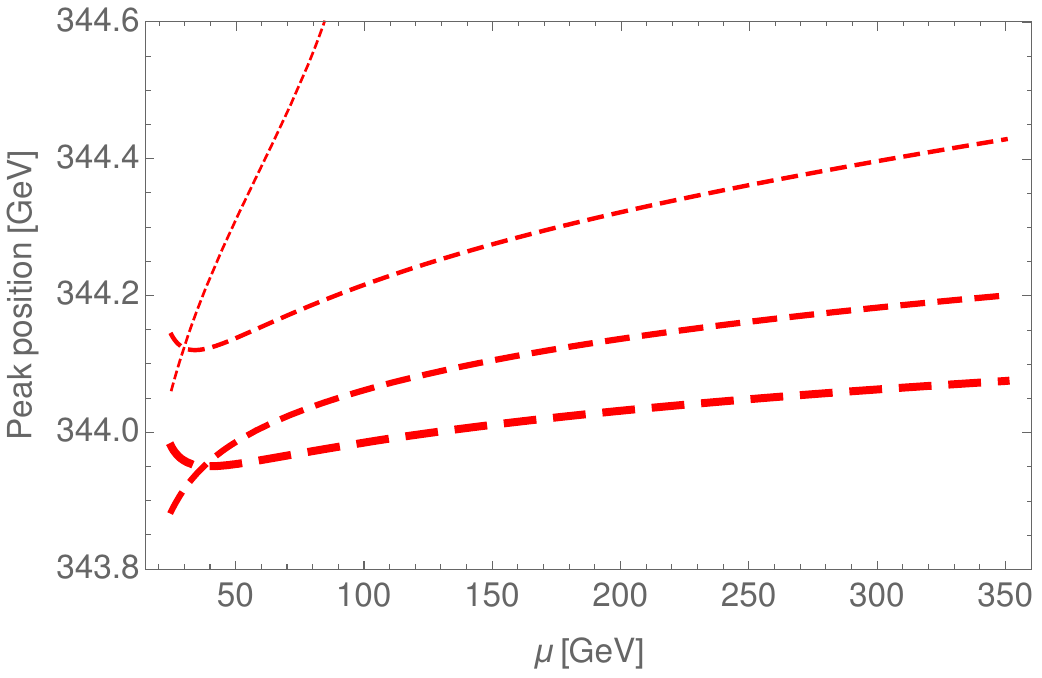}
\vskip0.2cm
\includegraphics[width=0.6\textwidth]{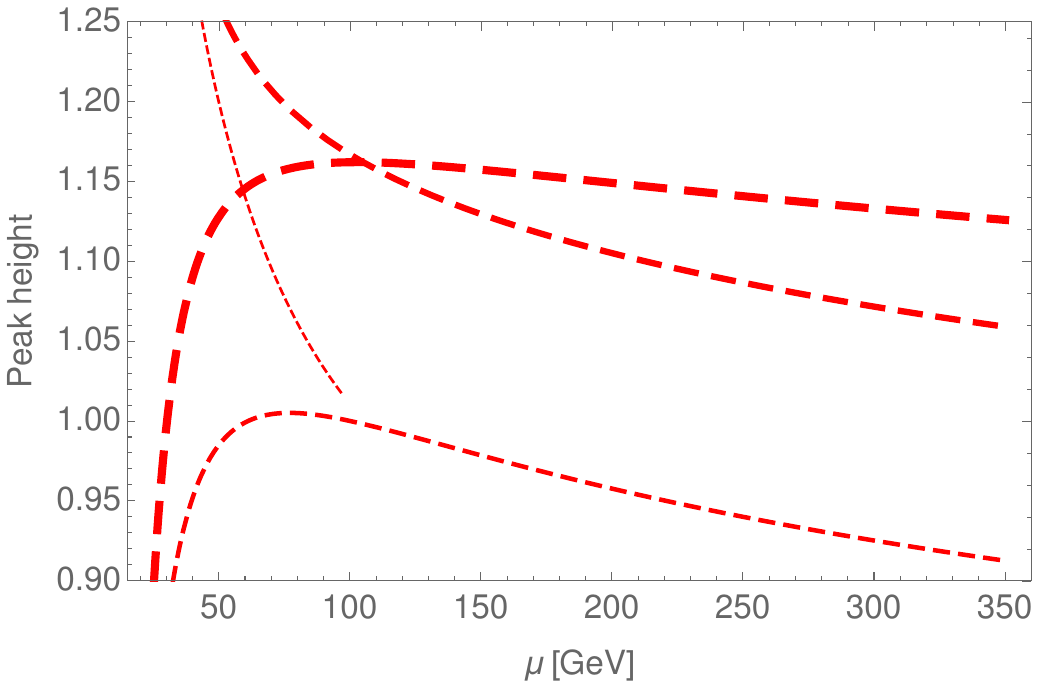}
\end{center}
\caption{\label{fig:peak}
Scale dependence of the peak (maximum of the $R$-ratio) 
location and height.  The order of 
approximation increases from LO to NNNLO as the line 
style changes from thin to thicker and short to longer dashes. 
The LO line ends near $\mu=98\,$GeV, since for larger scales 
the LO cross section does not exhibit a resonance peak.
}
\end{figure}

In Figure~\ref{fig:peak} we display the position and the height of 
the peak of the $R$-ratio as a function of the renormalization scale 
and different orders of approximation. The line style encodes the
order as in the previous figure. We observe that the resonance 
parameters stabilize as the orders increases, supporting the 
suitability of the threshold for the precision top mass determination.
Numerical values are given and compared to other mass 
renormalization schemes in Table~\ref{tab:peak} below.

\subsubsection{Finite-width scale dependence}

\begin{figure}[p]
\begin{center}
\includegraphics[width=0.55\textwidth]{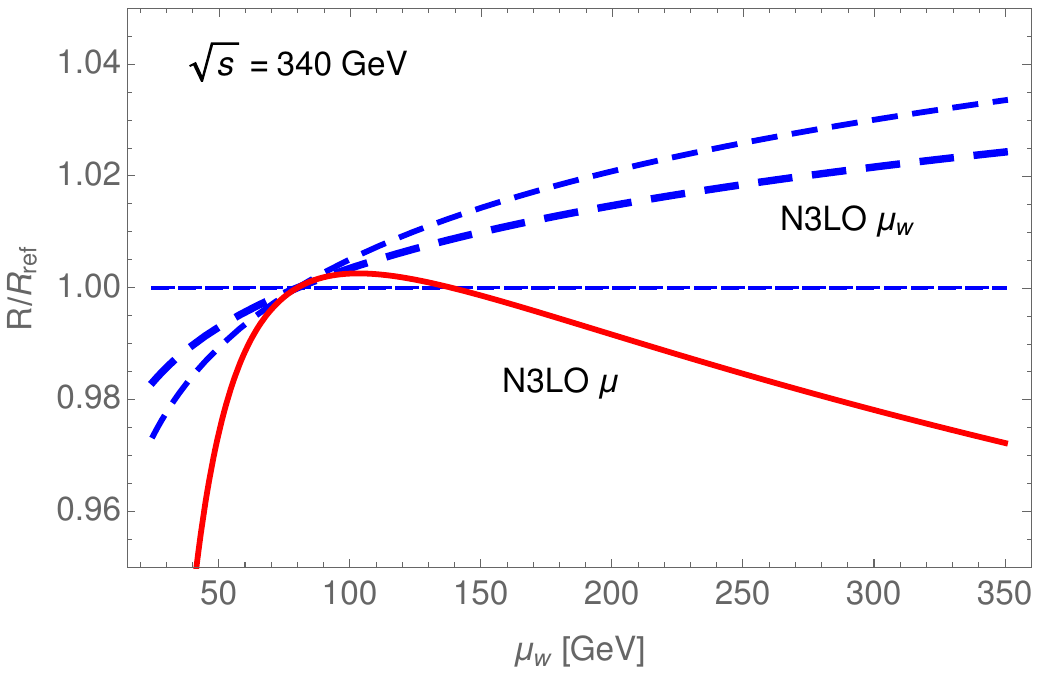}
\vskip0.3cm
\includegraphics[width=0.55\textwidth]{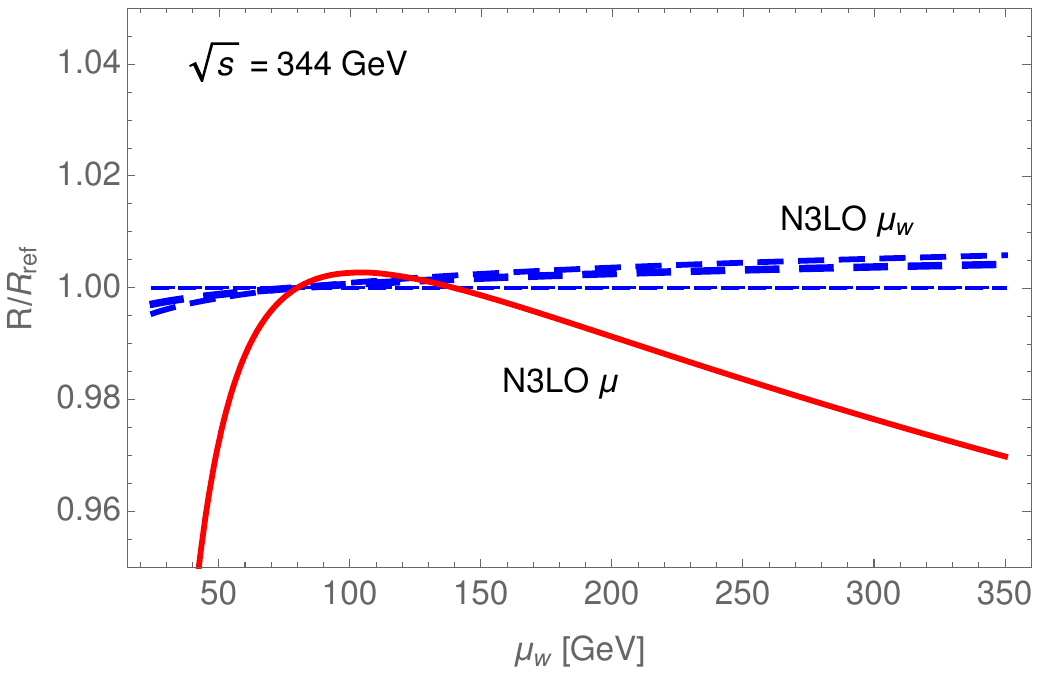}
\vskip0.3cm
\includegraphics[width=0.55\textwidth]{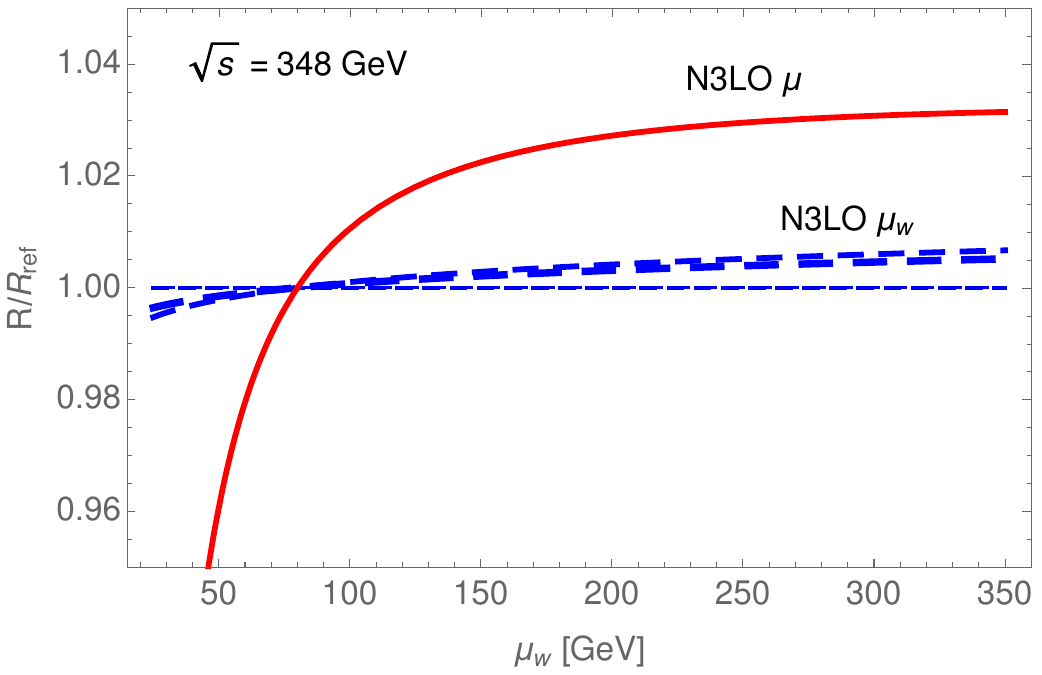}
\end{center}
\caption{\label{fig:relwidthscaledepenergies}
Finite-width ($\mu_w$) scale dependence of the cross section 
(normalized to the value at the scale $\mu_w=80~$GeV) below, 
near and above the peak (dashed/blue). The renormalization 
scale is fixed to $\mu=80~$GeV in all curves. 
The order of approximation increases from LO to NNNLO as the line 
style changes from thin to thicker and short to longer dashes. 
For comparison we also show the renormalization ($\mu$) scale 
dependence at NNNLO (solid/red).
}
\end{figure}

In addition to the renormalization scale dependence there is 
also a residual finite-width scale dependence starting at 
NNLO, which is cancelled only together with the non-resonant 
cross section not included here. 
Figure~\ref{fig:relwidthscaledepenergies} displays this 
finite-width scale dependence in the same 
format as for the renormalization scale dependence 
(Figure~\ref{fig:relscaledepenergies}). Even though the expected 
residual scale dependence is parametrically 
larger than the renormalization scale dependence, namely NNLO rather than 
N4LO, the 
$\mu_w$ variation is seen to be numerically much smaller (less than 1\%) 
than the $\mu$ dependence. The $\mu$ dependence at NNNLO is shown 
for comparison as the solid line. An exception occurs when $\sqrt{s}$ 
is sufficiently below threshold (upper panel), where the 
resonant cross section decreases and the non-resonant 
contribution becomes relatively more important. 

The actual $\mu_w$ dependence of the full cross section is 
smaller than shown in this Figure, since the NNLO 
non-resonant cross section is known. The cancellation and 
reduction of finite-width scale dependence upon summing 
resonant and non-resonant contributions has been discussed 
in \cite{Beneke:2017rdn}. 

\subsection{Mass schemes}

As discussed in the previous section, adopting a mass renormalization 
convention that eliminates the leading infrared sensitivity of the 
pole mass is crucial for the precise calculation of the top-antitop 
line shape. In this subsection we quantify this statement. We compare 
the reference PS shift scheme to the pole scheme, the MS shift scheme 
and the PS shift scheme with a different value of $\mu_f$. We also 
show the result in the PS insertion scheme and the effect of 
pole resummation.

\subsubsection{Pole vs PS shift}

\begin{figure}[t]
\begin{center}
\includegraphics[width=0.65\textwidth]{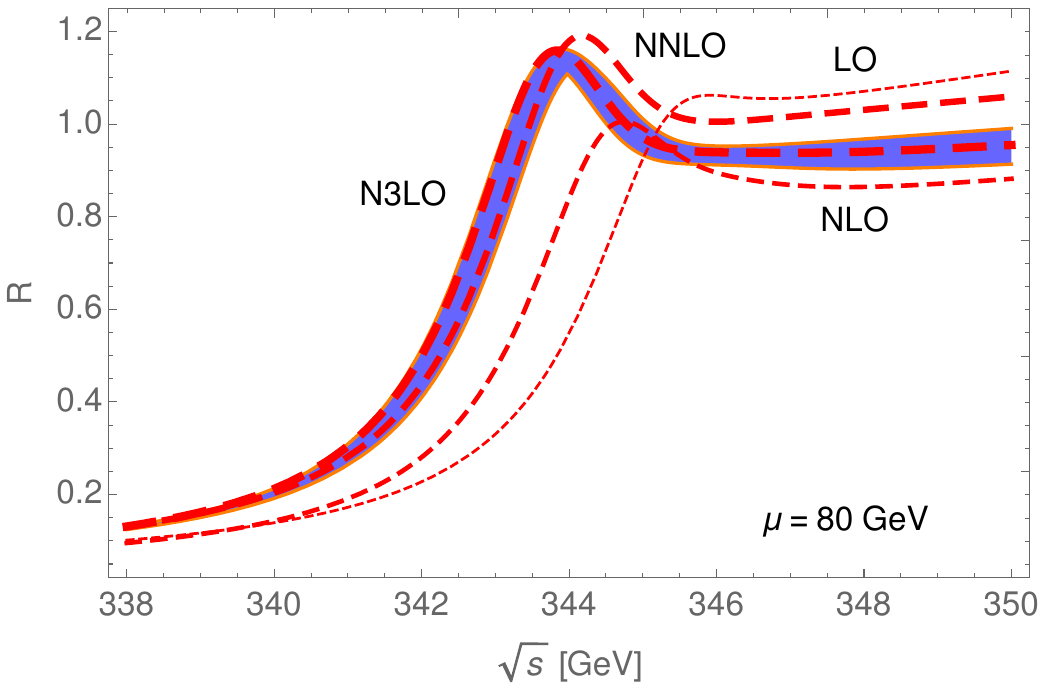}
\vskip0.4cm
\includegraphics[width=0.65\textwidth]{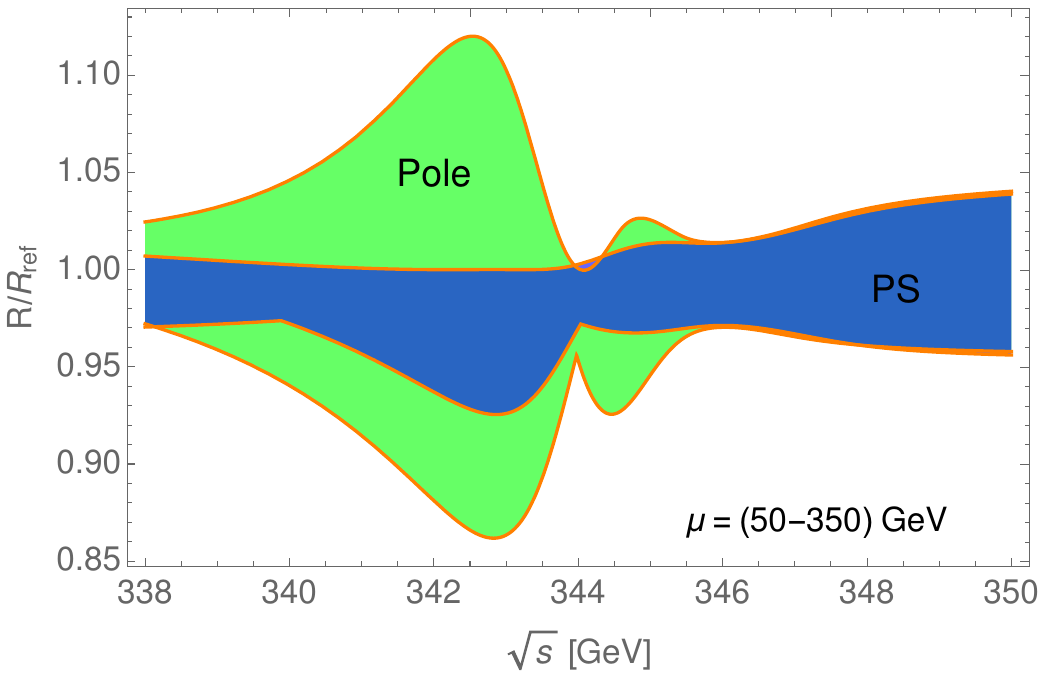}
\end{center}
\caption{\label{fig:polescheme}
Upper panel: cross section ($R$-ratio) in the pole scheme at LO, NLO, NNLO 
and NNNLO. The band shows the scale variation at NNNLO. 
Lower panel: comparison of the scale dependence in the pole vs. the 
PS scheme at NNNLO. The $R$-ratio is divided by its reference value in the 
PS scheme for both bands.
}
\end{figure}

We begin by showing the threshold cross section in the pole scheme 
in Figure~\ref{fig:polescheme} (upper panel). The width of the band 
represents the 
third-order scale dependence, which can be compared to the 
corresponding band in  Figure~\ref{fig:mainresult} for the PS scheme. 
The better performance of the potential subtraction scheme is 
clearly seen around the peak. The larger variation in the pole 
scheme is primarily a consequence of the large scale dependence 
of the peak location related to the systematic shifts from the 
uncancelled leading renormalon divergence. These shifts are also 
clearly visible by comparing the successive LO, ..., NNNLO 
approximations (dashed/red) lines in the Figure. Sufficiently above 
threshold where the energy dependence of the cross section becomes 
mild, both schemes give nearly identical results. These features 
are highlighted in the lower panel, which shows the relative scale 
dependence as function of cms energy $\sqrt{s}$ in both schemes, by 
normalizing the $R$-ratio to the reference result at $\mu=80~$GeV. 
In Table~\ref{tab:peak} we provide, in the second and third column for 
the PS shift and pole scheme, respectively, numerical values of the shift 
of the maximum of the cross section at a given order relative to 
the previous order, as well as the scale dependence form varying
$\mu$ within $[50,350]\,$GeV at the given order. At NNNLO, 
the scale dependence of the peak is more than a factor of three 
smaller in the PS scheme than in the pole scheme (61~MeV vs. 
228~MeV). We remark that as a matter of principle the 
shift of the peak position {\em in the pole scheme} should never become smaller than 
twice the intrinsic ambiguity of the top pole mass, which amounts to 
a peak shift larger than about 140~MeV~\cite{Beneke:2016cbu} when all other five quark 
flavours are massless, as assumed here.

\begin{table}[b]
\newcommand{\m}{\hphantom{$-$}}
\newcommand{\cc}[1]{\multicolumn{1}{c}{#1}}
\renewcommand{\tabcolsep}{0.8pc} 
\renewcommand{\arraystretch}{1.0} 
\caption{Shift of the peak position with respect to the previous 
order / $\pm$ scale variation of the peak position in the four 
mass schemes discussed in this section. All numbers in units of MeV. 
At NNNLO, the peak is located at $\sqrt{s}=$ 343.972 (PS shift) / 
343.841 (pole) / 343.985 (MSshift) / 343.972 (PS shift, $\mu_f=50~$GeV) 
GeV in the four schemes.
 }
\label{tab:peak}
\begin{center}
\begin{tabular}{@{}clllll}
\hline  \vspace{-4mm}\\  
\mbox{Order} &  \hspace{-5mm}   
     & \mbox{PS shift} 
     & \mbox{Pole}  
     & \mbox{MS shift}
     & \hspace{-7mm} \mbox{PS shift}, $\mu_f=50~$GeV  
\\
\hline   \\[-2mm]
NLO  & \hspace{-5mm} 
     & $-367 / \pm 145$
     & $-1167 / \pm 385$
     & $+1008 / \pm 482\;\;$
     & $+154 / \pm 44$
 \vspace{2mm} \\
NNLO  & \hspace{-5mm} 
     & $-149 / \pm 107$
     & $-571 / \pm 304$
     & $+46 / \pm 103$
     & $+2 / \pm 16$
 \vspace{2mm} \\
NNNLO  & \hspace{-5mm} 
     & $-65 / \pm 61$
     & $-333 / \pm 228$
     & $-81 / \pm 26$
     & $-9 / \pm 13$
 \vspace{2mm} \\
\hline
\vspace{-10mm}
\end{tabular}
\end{center}
\end{table} 


\subsubsection{MS shift vs PS shift}

\begin{figure}[t]
\begin{center}
\includegraphics[width=0.65\textwidth]{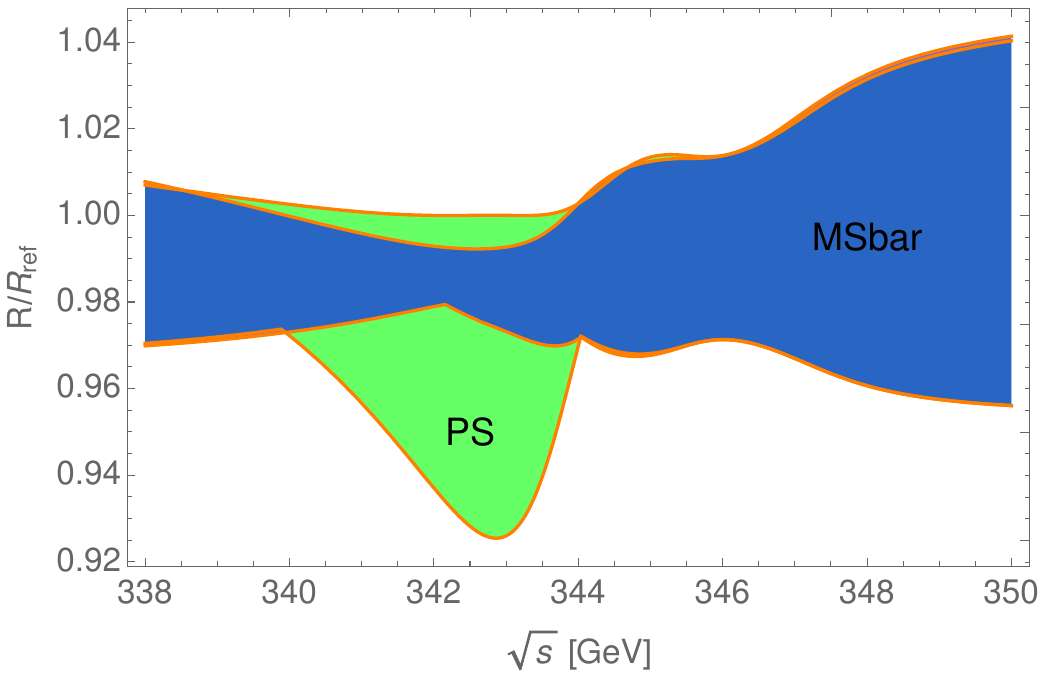}
\end{center}
\caption{\label{fig:masshiftscheme}
Scale dependence at NNNLO normalized to reference cross section 
in the MS and PS shift schemes. The reference cross section is the PS scheme 
cross section at $\mu=80~$ GeV for both bands. Note the different vertical 
scale compared to the analogous Figure~\ref{fig:polescheme} 
for the pole-to-PS scheme comparison. 
}
\end{figure}

In Figure~\ref{fig:masshiftscheme} we compare the scale dependence 
of the threshold cross section in the MS shift scheme, which 
employs the $\overline{\rm MS}$ mass $\overline{m}_t$ as input, 
to the reference PS shift scheme. 
The scale dependence is almost exactly the same in both schemes far below 
and above threshold. However, in the region most relevant to 
the top mass determination directly below the peak at $\sqrt{s}\approx 
344~$GeV, the MS shift scheme is much better behaved. This 
demonstrates that the observation made in \cite{Kiyo:2015ooa} for 
the would-be 1S toponium bound state energy also applies to the 
cross section itself.

The technical origin of this different behaviour of the 
PS and MS scheme is as follows. The scale variation band 
in the PS shift scheme exhibits two kinks at 
$\sqrt{s}=340~$GeV and 344~GeV (better visible in the 
lower panel of Figure~\ref{fig:polescheme}) and develops 
a downward ``nose'' with a maximal width of the band near 
343~GeV.  The kinks arise because in the 
energy region below 340~GeV and above 344~GeV the scale 
variation is determined by the maximal value of the 
cross section attained at some $\mu_*$ within the interval 
$[50,350]~$GeV and the minimal value at the lower 
boundary $\mu=50~$GeV. On the other hand between 
340 and 344~GeV, the minimal value is attained at the 
upper limit $\mu=350~$GeV of the scale variation interval, 
that is  $\sigma(\sqrt{s},\mu=350\,\mbox{GeV}) < 
\sigma(\sqrt{s},\mu=50\,\mbox{GeV})$.
In other words, in the most interesting energy region just below 
the peak, the uncertainty of the PS scheme cross section 
is determined by a comparatively sizeable scale dependence 
for {\em large} values of $\mu$. The cross section in 
the MS shift scheme does not exhibit this scale dependence 
at large $\mu$, and therefore the ``nose'' is absent, 
as seen in Figure~\ref{fig:masshiftscheme}.

We could not find a physics argument for this different behaviour 
and it may well be accidental (see also the following 
subsection). Given the opposite behaviour of the PS shift 
(scale dependence maximal) and MS shift (scale dependence minimal) 
behaviour in the region of most interest for the top mass 
determination, we caution that the scale uncertainty in 
the MS shift scheme may underestimate the theoretical uncertainty, 
while the one in the PS shift might be too conservative.
We find additional evidence for this conclusion in 
Table~\ref{tab:peak}, which provides the shift of the 
peak of the cross section upon increasing the order of approximation 
and its scale dependence. The small scale dependence of only 
$\pm26~$MeV reflects the discussion above, yet the shift of the 
peak of $-81~$MeV relative to NNLO is of the same size 
as in the PS shift scheme.

\subsubsection{PS shift with different $\mu_f$ or smaller 
$\nu$}

\begin{figure}[t]
\begin{center}
\includegraphics[width=0.65\textwidth]{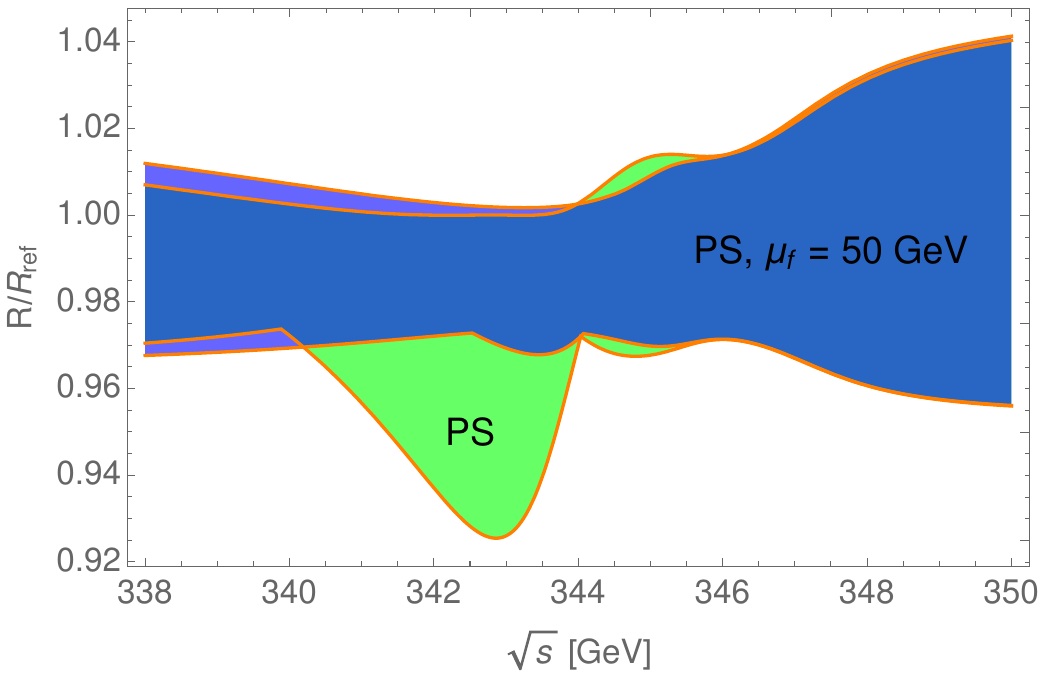}
\vskip0.4cm
\includegraphics[width=0.65\textwidth]{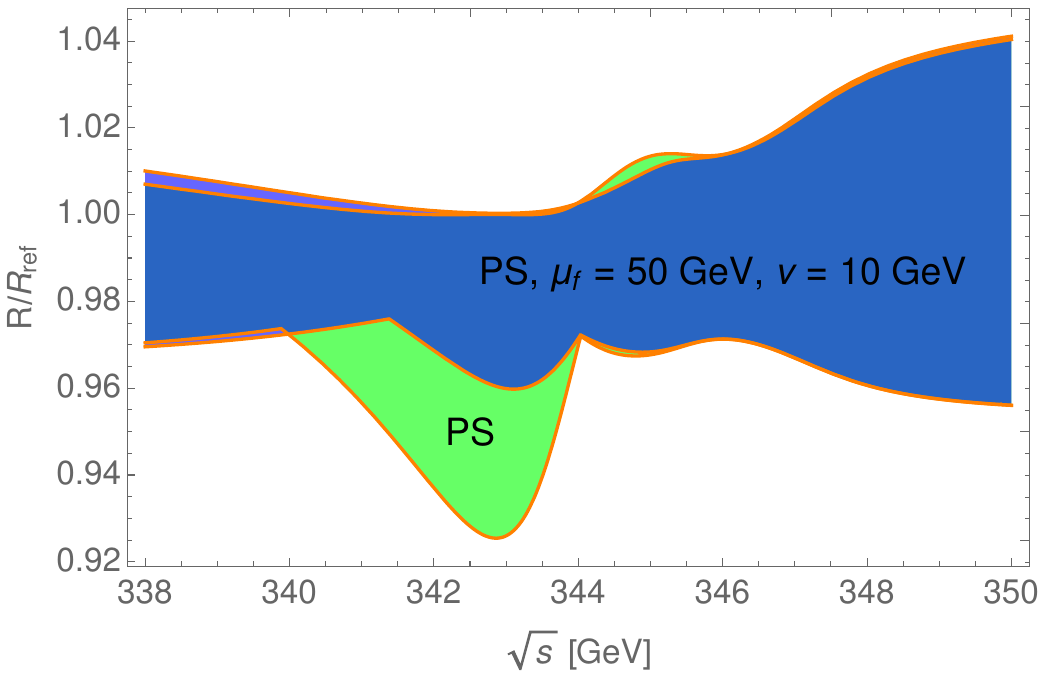}
\end{center}
\caption{\label{fig:psshiftschememufnu}
Scale dependence at NNNLO cross section normalized to the corresponding 
cross section value at $\mu=80$~GeV. Upper panel: PS shift scheme with 
$\mu_f=\nu=50$~GeV, 
lower panel:   $\mu_f=50$~GeV, $\nu=10$~GeV, both compared to the 
default  $\mu_f=\nu=20$~GeV.
}
\end{figure}

The scales $\mu_f=20~$GeV and $\nu=\mu_f$ (see Section~\ref{sec:PSscheme}) 
have been universally used in previous works to define the PS 
mass. Here we explore the effect of changing $\mu_f$ to 50~GeV 
while maintaining $\nu=\mu_f$, and of setting $\mu_f=50$~GeV 
together with $\nu=10$~GeV. For both cases we recompute the 
PS mass from the reference value $m_{\rm PS}(\mu_f=\nu=20\,
\mbox{GeV}) = 171.5$~GeV,  and find $m_{\rm PS}(\mu_f=\nu=50\,
\mbox{GeV}) = 169.600$~GeV and  $m_{\rm PS}(\mu_f=50\,\mbox{GeV},\nu=10\,
\mbox{GeV}) = 169.615$, respectively. The conversion is done 
with coupling renormalization scale $\mu=80$~GeV. There is an uncertainty 
in the conversion from this choice, which affects the PS mass value, 
but not the scale dependence of the cross section prediction for 
fixed input mass, which we discuss next. 

Figure~\ref{fig:psshiftschememufnu} displays the 
scale dependence of the cross section in the same format as for the 
MS shift scheme. Interestingly, the ``nose'' of the reference 
PS scheme disappears (upper panel). 
It seems that the PS shift scheme with $\mu_f=50$~GeV behaves similarly 
to the MS shift scheme.  For this choice of $\mu_f$, the peak position is 
particularly stable and its scale uncertainty is very small, 
as can be seen from Table~\ref{tab:peak}. 
The additional change of $\nu$ has only a small 
effect (lower panel), although we see a smaller ``nose'' reappearing. 

\subsubsection{PS insertion vs. PS shift}

\begin{figure}[t]
\begin{center}
\includegraphics[width=0.45\textwidth]{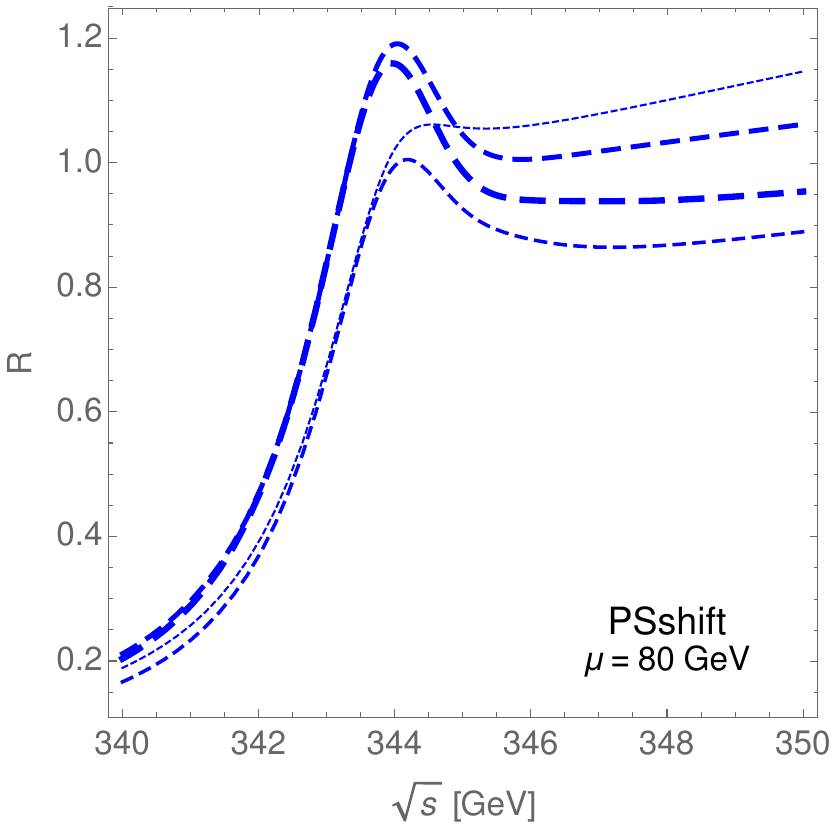}\hskip0.5cm
\includegraphics[width=0.45\textwidth]{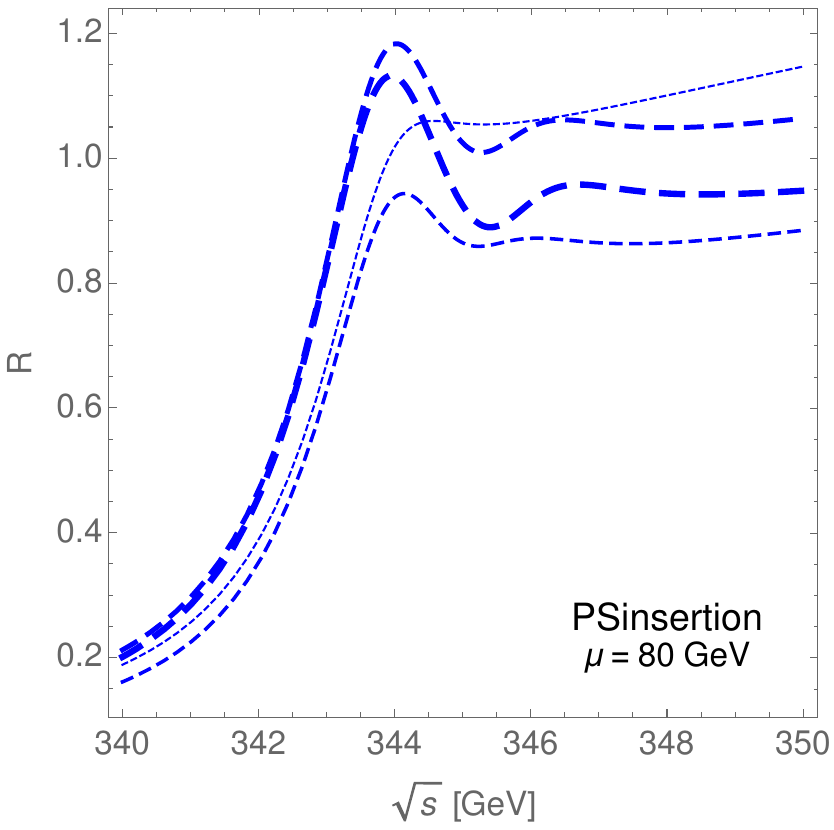}
\end{center}
\caption{\label{fig:psinsertion}
LO, NLO, NNLO, N3LO in PS shift (left) and PS insertion (right).
The order of approximation increases from LO to NNNLO as the line 
style changes from thin to thicker and short to longer dashes.}
\end{figure}

The treatment of the subleading mass corrections as perturbations 
in non-relativistic effective theory seems to be most natural from 
the EFT and renormalon-cancellation perspective. However, 
in Section~\ref{sec:insertionproblems} we pointed out a generic 
difficulty in the implementation of the insertion scheme related 
to local instabilities of the expansion near the continuum threshold, 
and discussed the issue in an exactly solvable approximation. Here 
we display this instability for the full implementation of the 
top line shape in QCD.

In Figure~\ref{fig:psinsertion} we show on the left the $R$-ratio 
in the standard PS shift scheme and on the right the PS insertion 
result for the same input PS mass 171.5~GeV. As in previous figures 
the order of approximation increases from LO to NNNLO as the line 
style changes from thin to thicker and short to longer dashes.
The left panel displays the smooth result obtained in the PS 
shift scheme, of which LO and NNNLO have already been 
shown in Figure~\ref{fig:mainresult}. The right figure clearly 
exhibits the unphysical peak-dip oscillation in the PS insertion 
scheme with an amplitude of about 10\%. The nominal threshold is at 
$2 m_{t,\rm pole} = 346.415\,$ GeV and the mass of the 
would-be $n=1$ toponium state is $343.831\,$ GeV 
(NNNLO, PS scheme). The oscillation occurs directly below the 
nominal threshold as expected. In both plots the pole resummation 
procedure is applied to the first six bound state poles. From 
the general discussion in Section~\ref{sec:insertionproblems} we 
know that there is a subtle connection between the singular terms 
from the expansion of the high-$n$ bound-state poles and the 
continuum. We therefore discuss next the effect of pole resummation 
on the third-order QCD cross section in the PS shift and 
insertion scheme. 

\subsubsection{Effect of pole resummation}
\label{subsec:poleresum}

\begin{figure}[t]
\begin{center}
\includegraphics[width=0.55\textwidth]{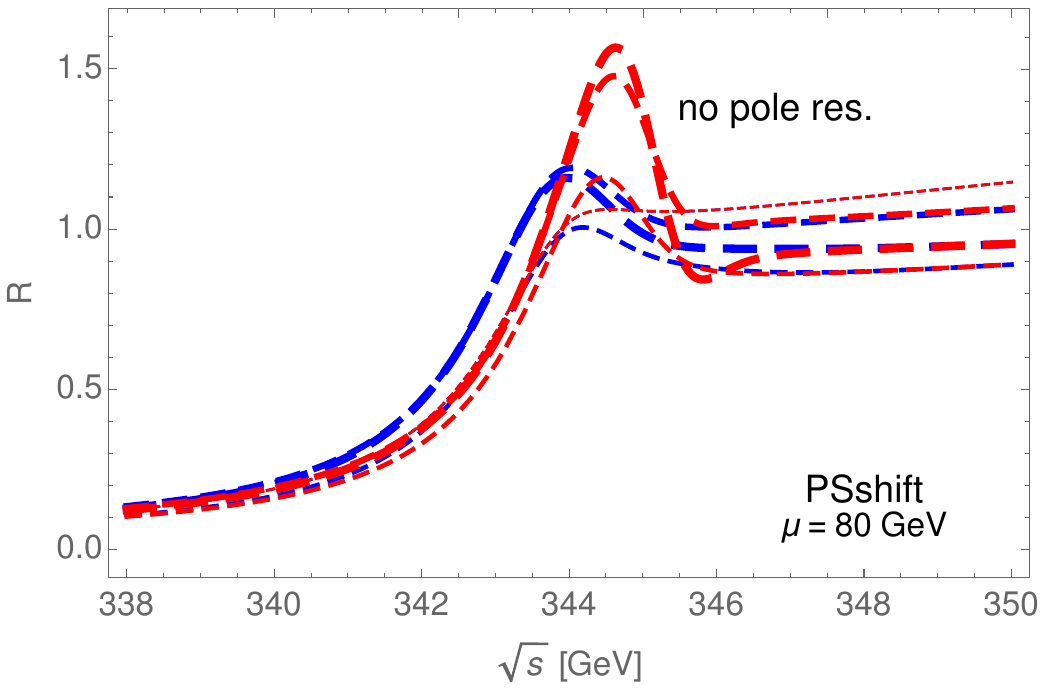}
\vskip0.4cm
\includegraphics[width=0.55\textwidth]{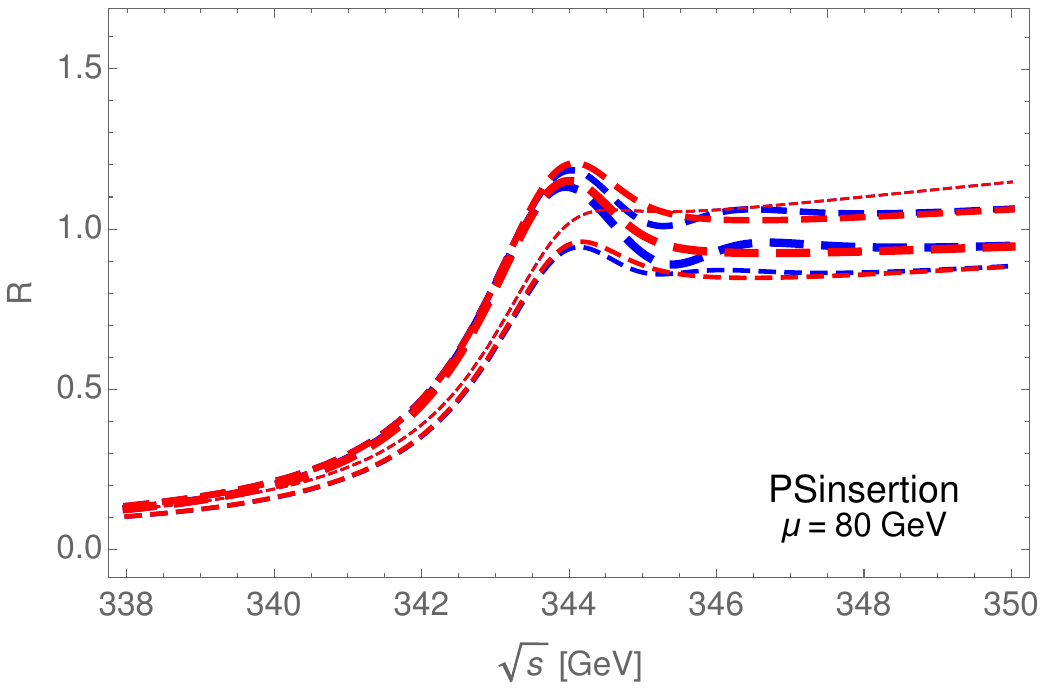}
\vskip0.4cm
\end{center}
\caption{\label{fig:nopoleresum}
Comparison of the cross section without pole resummation 
(lighter grey/red) and with pole resummation (darker grey/blue) 
for various orders in the PS shift scheme (upper panel) 
and the PS insertion scheme (lower panel). The order of 
approximation increases from LO to NNNLO as the line 
style changes from thin to thicker and short to longer dashes.
}
\end{figure}

In Figure~\ref{fig:nopoleresum} we show the effect of pole resummation 
in the PS shift and the PS insertion scheme. As expected and explained 
in Section~\ref{sec:poleresummation}, pole resummation is indispensable 
to obtain a reliable result in the PS shift scheme (upper panel), 
which in this 
respect is identical to the pole scheme. On the other hand, pole 
resummation is negligible in the peak region in the insertion scheme 
(lower panel), 
since the ratio of the first-order $n=1$ energy-level correction 
(\ref{eq:e1corPS}) to the top width is very small. However, the 
unphysical peak-dip structure discussed and already shown above 
appears. 

\begin{figure}[t]
\begin{center}
\includegraphics[width=0.55\textwidth]{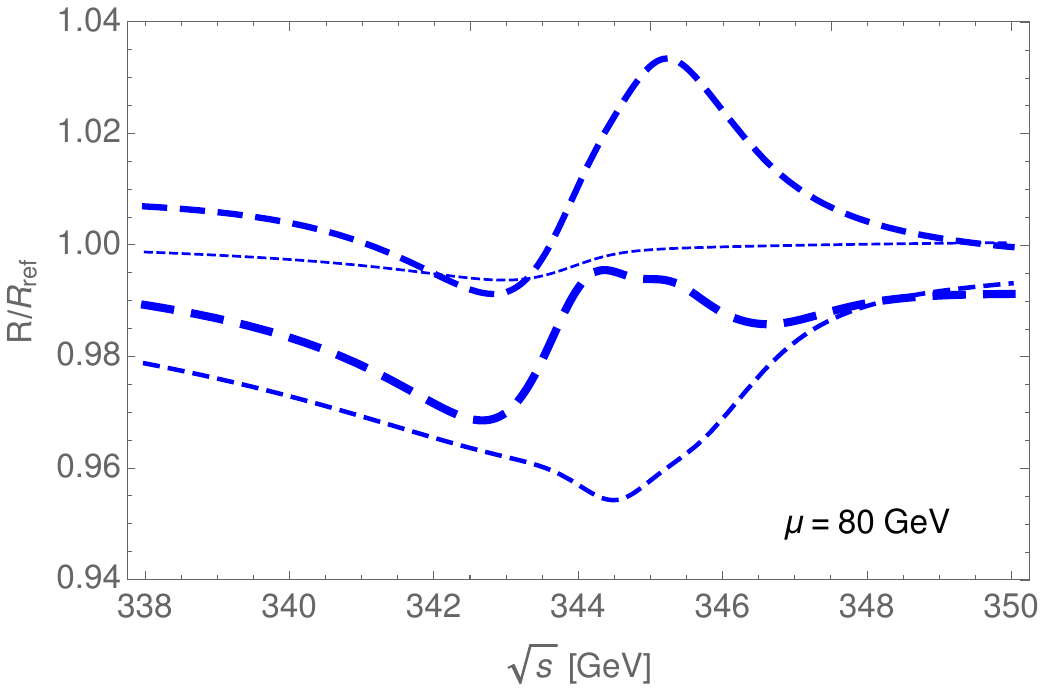}
\vskip0.4cm
\includegraphics[width=0.55\textwidth]{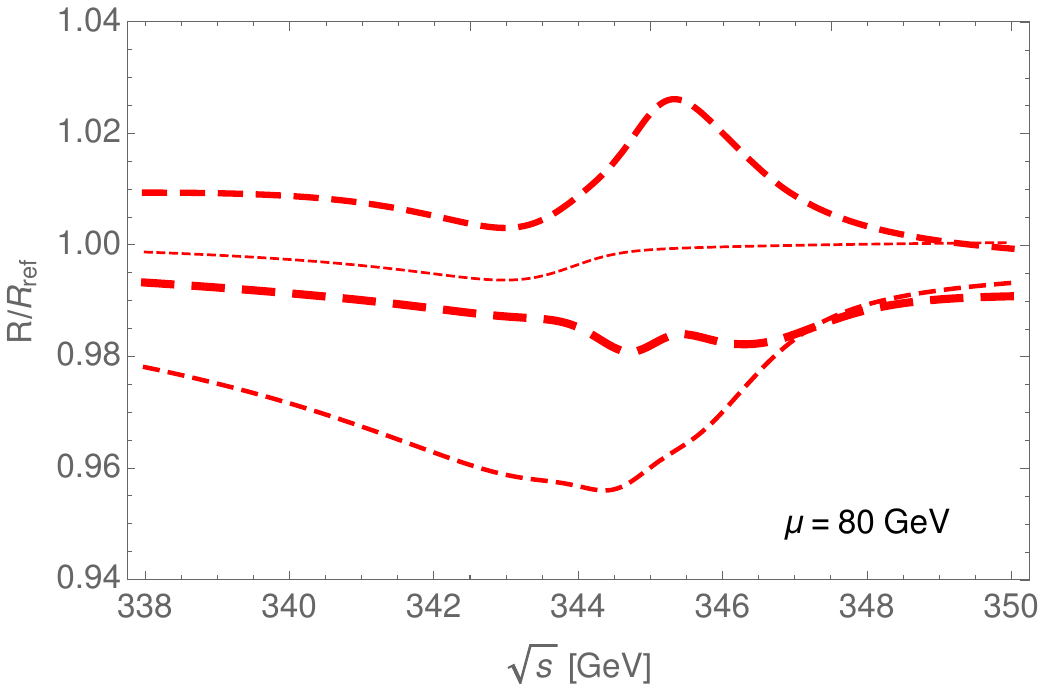}
\vskip0.4cm
\end{center}
\caption{\label{fig:PSinstoshoftoptimal}
Upper panel: ratio of PS insertion scheme without pole resummation 
to PS shift with pole resummation. The order of 
approximation increases from LO to NNNLO as the line 
style changes from thin to thicker and short to longer dashes. 
Lower panel: as above, but now the resummation of the first 
bound-state pole is included in the PS insertion scheme.
}
\end{figure}

In Section~\ref{sec:insertionproblems} we found that this problematic 
aspect of the insertion scheme results from treating 
the high-$n$ (in reality, $n>1$) bound-state poles and continuum 
threshold differently, since only the former are resummed, and 
showed that in an exactly solvable model the best approximation 
in the insertion scheme results from resumming only the most 
prominent $n=1$ pole. We see this confirmed for the full-QCD 
threshold cross section in Figure~\ref{fig:PSinstoshoftoptimal}. 
Here we show (upper panel) the ratio of the PS insertion result 
{\it without} pole resummation to the reference result in the 
PS shift scheme which includes pole resummation.  
The difference between these two approximations is never larger 
than 3\% at NNNLO in the energy range of interest. This improves 
when the first (and only the first) bound-state pole is resummed 
in the PS insertion scheme. The ratio to the PS shift scheme 
is shown in the lower panel to never deviate more than 1.5\% 
at NNNLO, which is within the scale uncertainty of the PS 
shift scheme result.

\subsection{Analysis of individual contributions}

Up to now we discussed the sensitivity of the top threshold 
cross section to the renormalization scale, the mass renormalization 
scheme and the pole resummation procedure. It is also of 
interest to look at the importance of the individual contributions 
that build up the third-order result, which consist of 
hard matching coefficients, the Coulomb and non-Coulomb potential 
insertions, where the latter are the main new technical result of 
this work, and the ultrasoft contribution. The separation of 
the individual terms is manifestly factorization scheme dependent. 
In the present case, all terms are defined by minimal subtraction 
($\overline{\rm MS}$) in dimensional regularization. Nevertheless, 
we find it instructive to show the size of the individual terms, 
their scale dependence, and the cancellation of scale dependence, 
as individual terms are summed.

We display the results at three values of the center-of-mass energy, 
below the peak at $\sqrt{s} = 340\,\mbox{GeV}$, above the peak at 
$\sqrt{s} = 348\,\mbox{GeV}$, and on the peak. In order that 
the contributions from individual potentials and other terms add 
up to their sum, we must turn off pole resummation (since this 
procedure resums multiple insertions of to all orders near the 
resonance energies). As discussed above, in the case the insertion 
scheme is more accurate and we therefore adopt the PS insertion 
scheme in this section. Furthermore, ``on the peak'', instead of 
the cross section, we show the individual contributions to the 
1S bound-state pole residue rather than the cross section itself, 
since the latter is sensitive to small corrections to the energy 
level near the peak, while we intend to show the individual 
contributions to the normalization. 

We show the scale variation of individual contributions within the 
standard interval $[50,350]\,\mbox{GeV}$. However, departing from 
the previous sections, we set the central value to 100~GeV rather 
than 80~GeV to avoid mostly one-sided scale variations.

\subsubsection{Pure Coulomb corrections vs exact solution}
\label{sec:coulombonly}

\begin{figure}[p]
\begin{center}
\vskip-0.7cm
\includegraphics[width=0.8\textwidth]{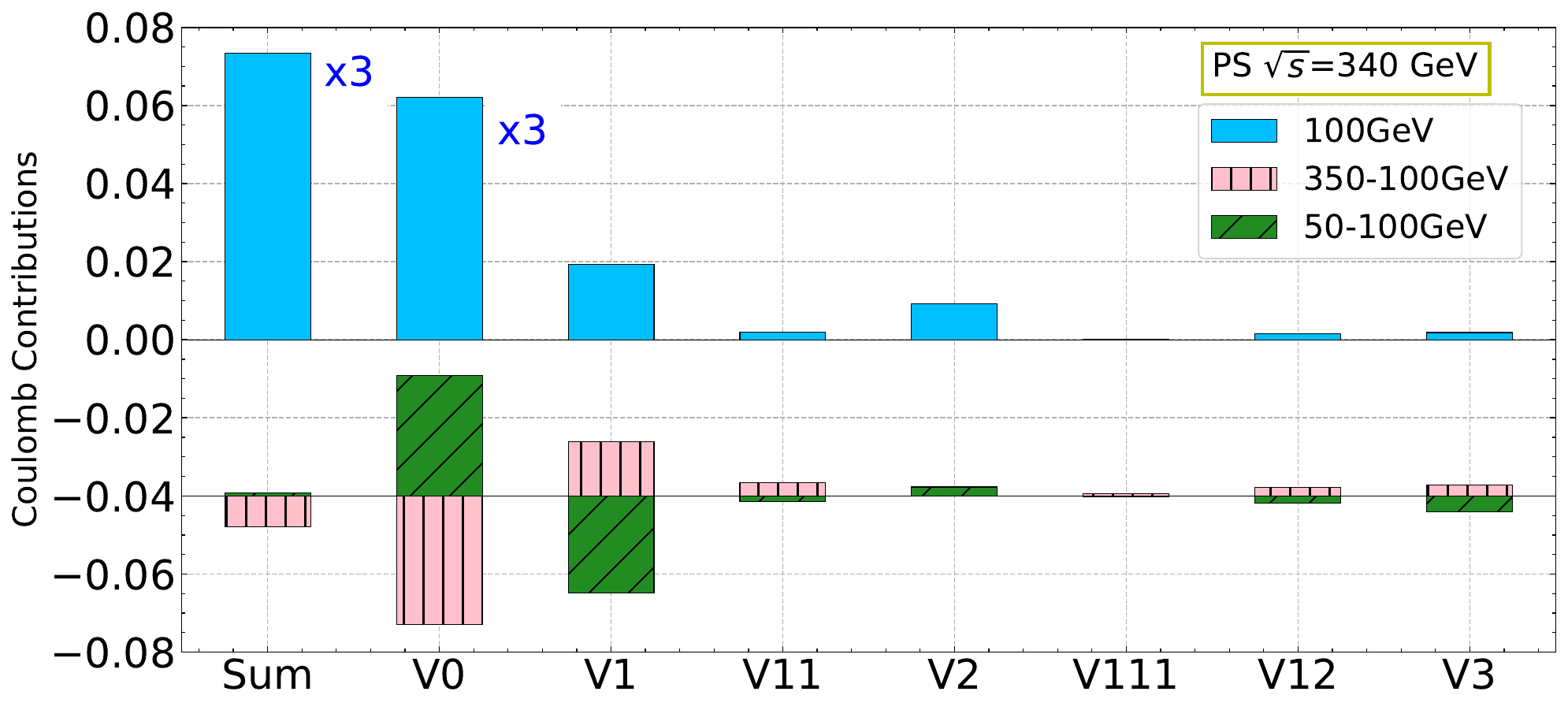}
\vskip0.5cm
\includegraphics[width=0.8\textwidth]{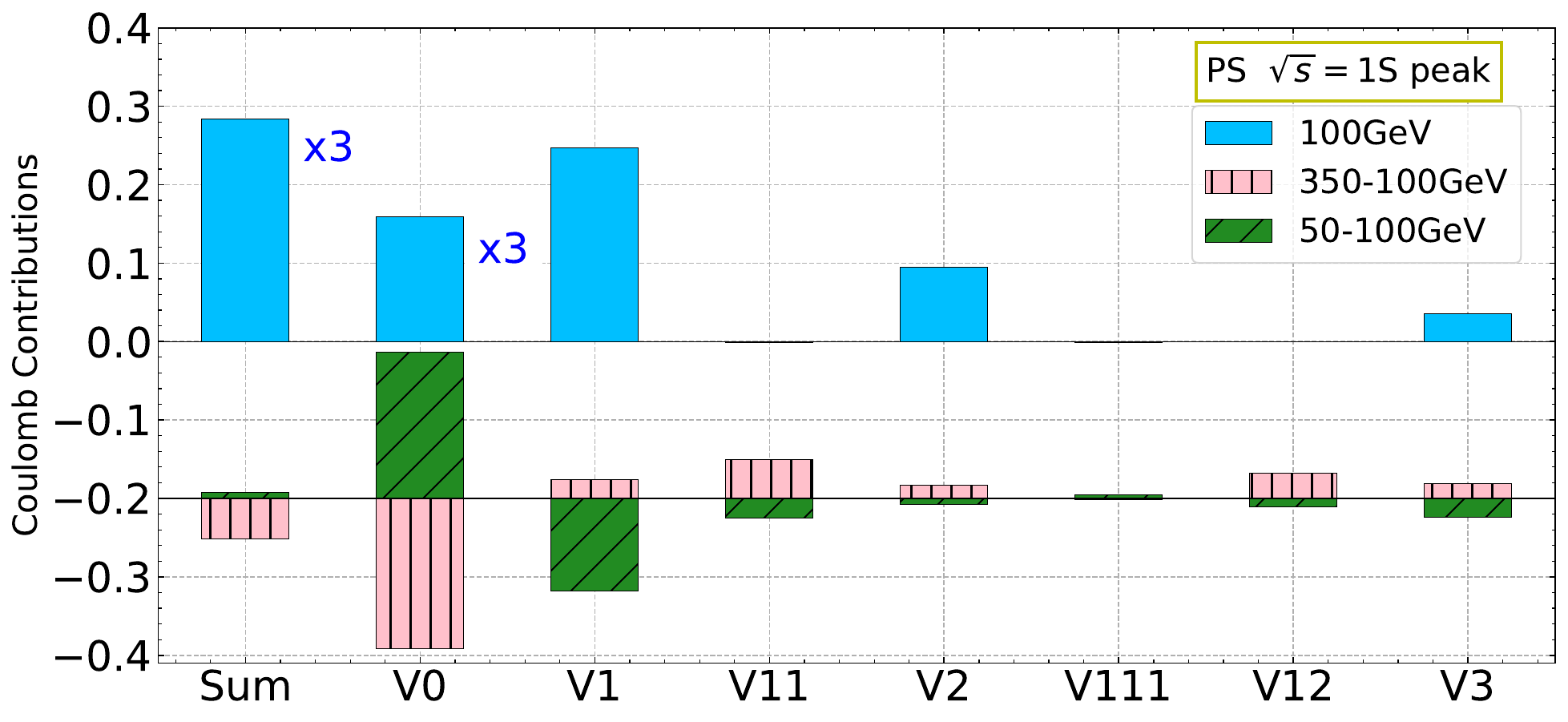}
\vskip0.5cm
\includegraphics[width=0.8\textwidth]{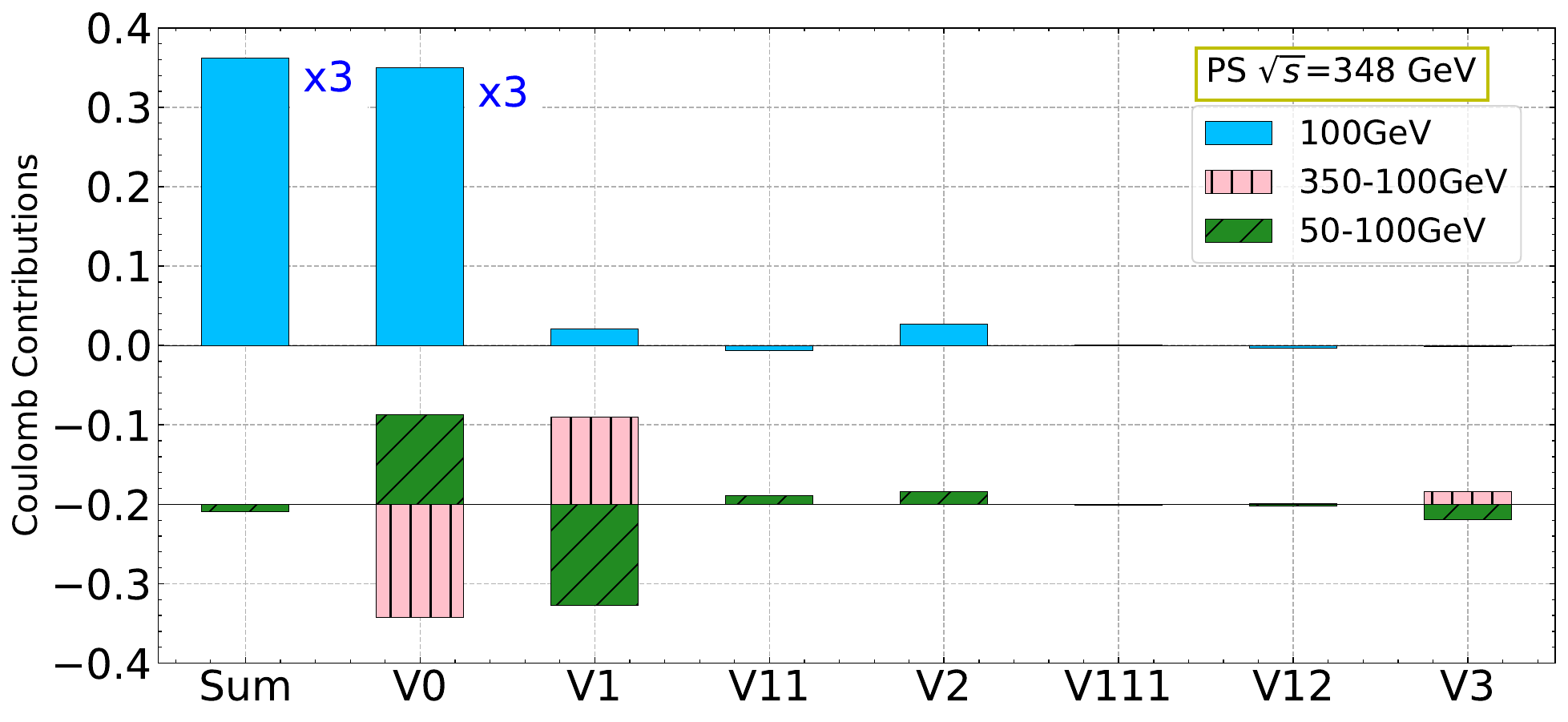}
\end{center}

\vskip-0.0cm
\caption{\label{fig:coulombpartsonly}
Coulomb potential contributions separated according to order and 
single, double and triple insertions for three values of 
$\sqrt{s}$. ``Vxy'' refers to a double insertion 
of the xth order with the yth order Coulomb potential, which 
contributes at the (x+y)th order to the cross section. 
``x3'' means that the height of the bar must be multiplied by three.
}
\end{figure}

We begin by showing in Figure~\ref{fig:coulombpartsonly} the 
contributions from the Coulomb potential only, originally computed  
already in \cite{Beneke:2005hg}. The left most bar represents the 
sum of all terms, from left to right the individual contributions 
are ordered according to order from LO to NNNLO, separated 
at every order into the single, double, and the triple insertions. 
The notation is such that, for example, 
``Vxy'' refers to a double insertion 
of the xth order with the yth order Coulomb potential, which 
contributes at the (x+y)th order to the cross section.

We observe that non-relativistic perturbation theory converges 
well. Higher-order Coulomb potential corrections are most 
important in the vicinity of the peak (middle panel). For 
the central scale $\mu=100$~GeV, the multiple insertions 
are always smaller than the single insertion at the same 
order. This may be partly accidental. A better estimate of the 
importance of a given term can sometimes be obtained from 
its scale dependence shown by the two hatched lower bars 
(off-set by the amounts $-0.04$, or $-0.2$). For example, at 
$\sqrt{s}=348$~GeV, the single insertion of the one-loop Coulomb 
potential is negligible at the central scale, but its scale 
dependence is large and opposite to the leading-order 
term, resulting in an almost complete cancellation. In general, 
it is evident that the addition of higher-order terms 
systematically removes scale dependence as it should be 
for a reliable perturbative approximation. 

\begin{figure}[t]
\begin{center}
\includegraphics[width=0.65\textwidth]{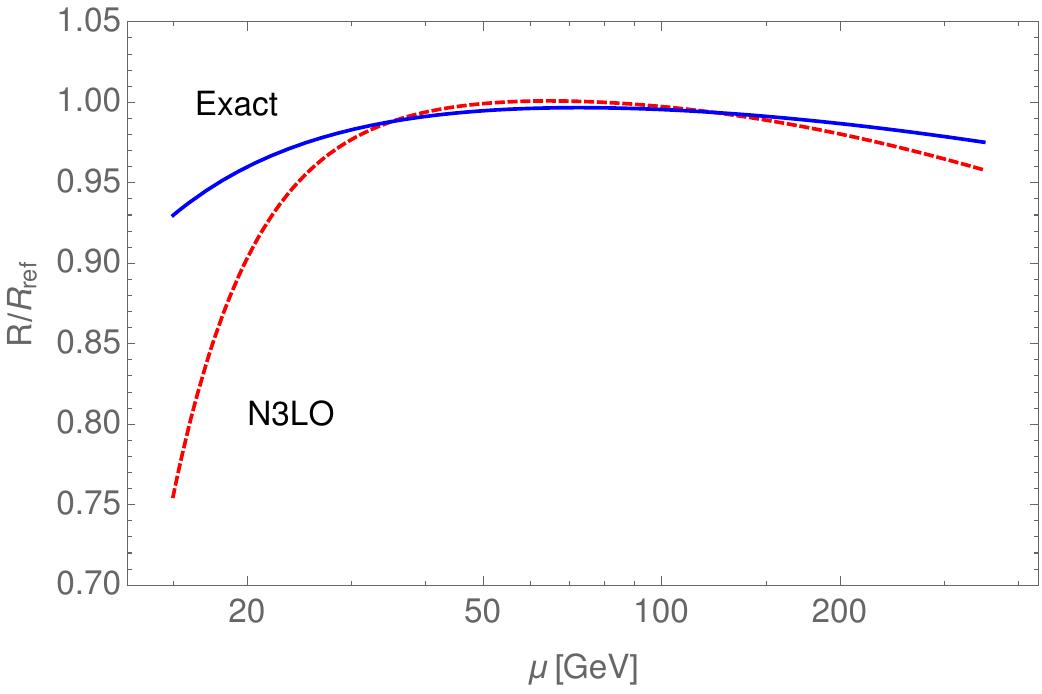}
\vskip0.4cm
\includegraphics[width=0.65\textwidth]{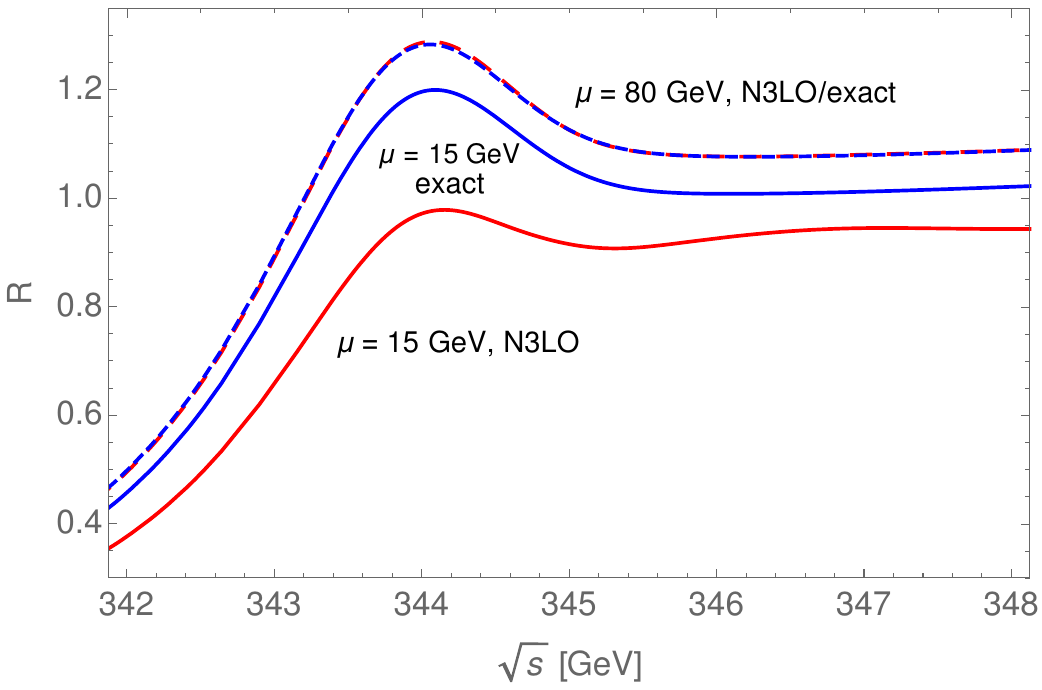}
\end{center}
\caption{\label{fig:exactvspert}
Upper panel: Scale-dependence of the exact solution of the 
Schr\"odinger equation with the Coulomb potential (to three loops) 
vs the perturbative NNNLO solution at $\sqrt{s}=344~$GeV.
Lower panel: Energy dependence of the exact and NNNLO solution 
for two values of the renormalization scale, $\mu=80$~GeV and 
$\mu=15$~GeV.
}
\end{figure}

The Schr\"odinger equation with the three-loop Coulomb potential 
can be solved numerically rather than expanding it order by 
order in non-relativistic perturbation theory into single, 
double and multiple insertions. The upper panel of 
Figure~\ref{fig:exactvspert} compares the scale dependence of 
the exact solution to the one of the NNNLO truncation at 
$\sqrt{s}=344$~GeV, where the 
latter corresponds to what is contained in the third-order 
computation of the cross section. The main conclusion from this 
comparison is that the NNNLO truncation is very close to the 
exact result when the scale is chosen between 30~GeV and 
200~GeV. This supports the reference choice of 80~GeV, which 
at first sight appears rather large compared to the natural 
toponium Bohr radius scale of about $(20-30)$~GeV, an 
observation already made in \cite{Beneke:2005hg}. The lower 
panel of the figure displays this fact for the energy 
dependence: at $\mu=80$~GeV the NNNLO and exact result are 
indistinguishable on the scale of the figure, while the curves 
for $\mu=15$~GeV are far apart. However, the scale dependence 
of the exact result remains moderate to $\mu=15$~GeV, while 
the truncated result is no longer accurate at such small 
scales. 

\subsubsection{Non-Coulomb contributions}
\label{sec:allnoncoulomb}

\begin{figure}[p]
\begin{center}
\vskip-1cm
\includegraphics[width=0.8\textwidth]{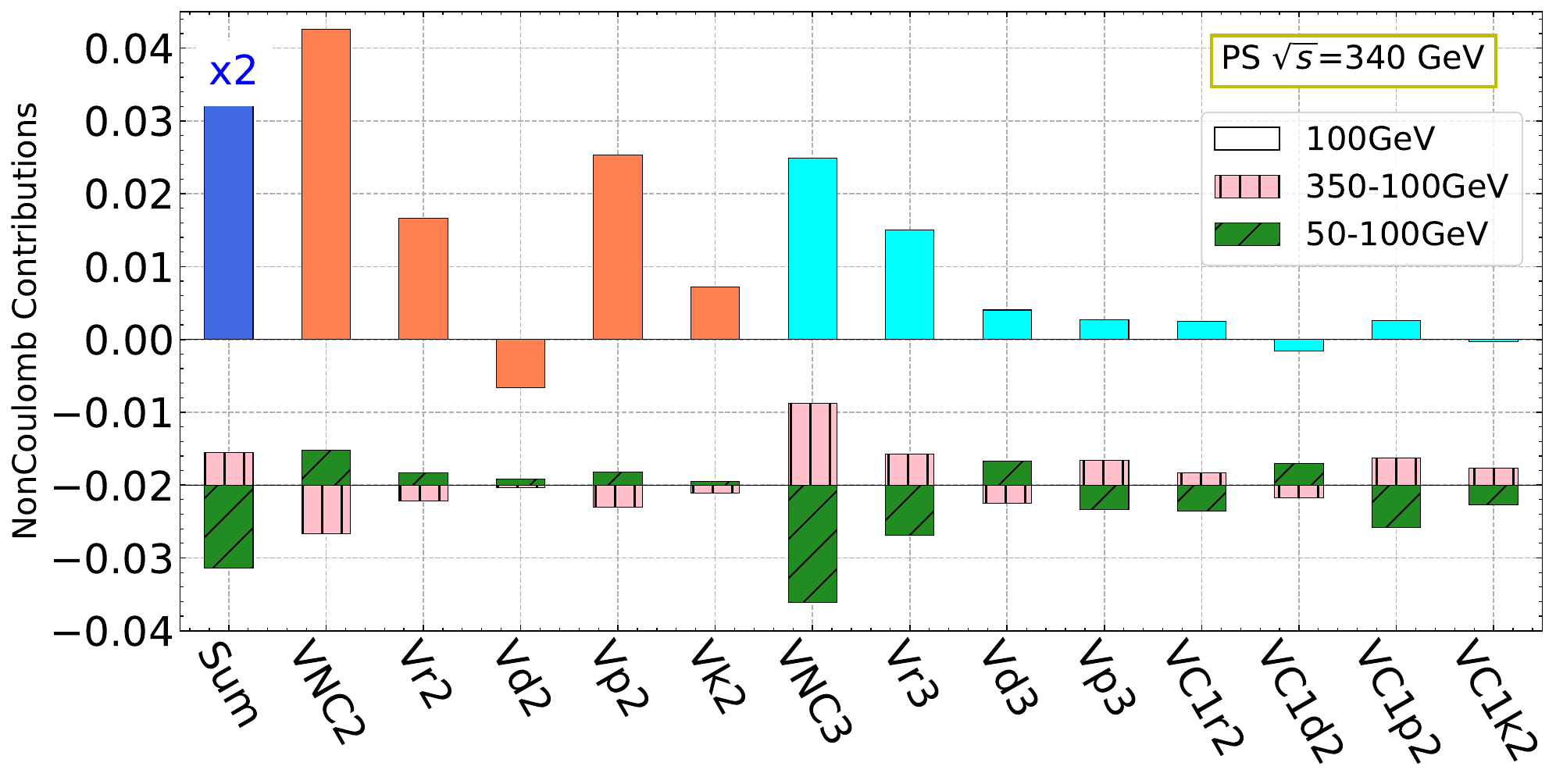}
\vskip-0.0cm
\includegraphics[width=0.8\textwidth]{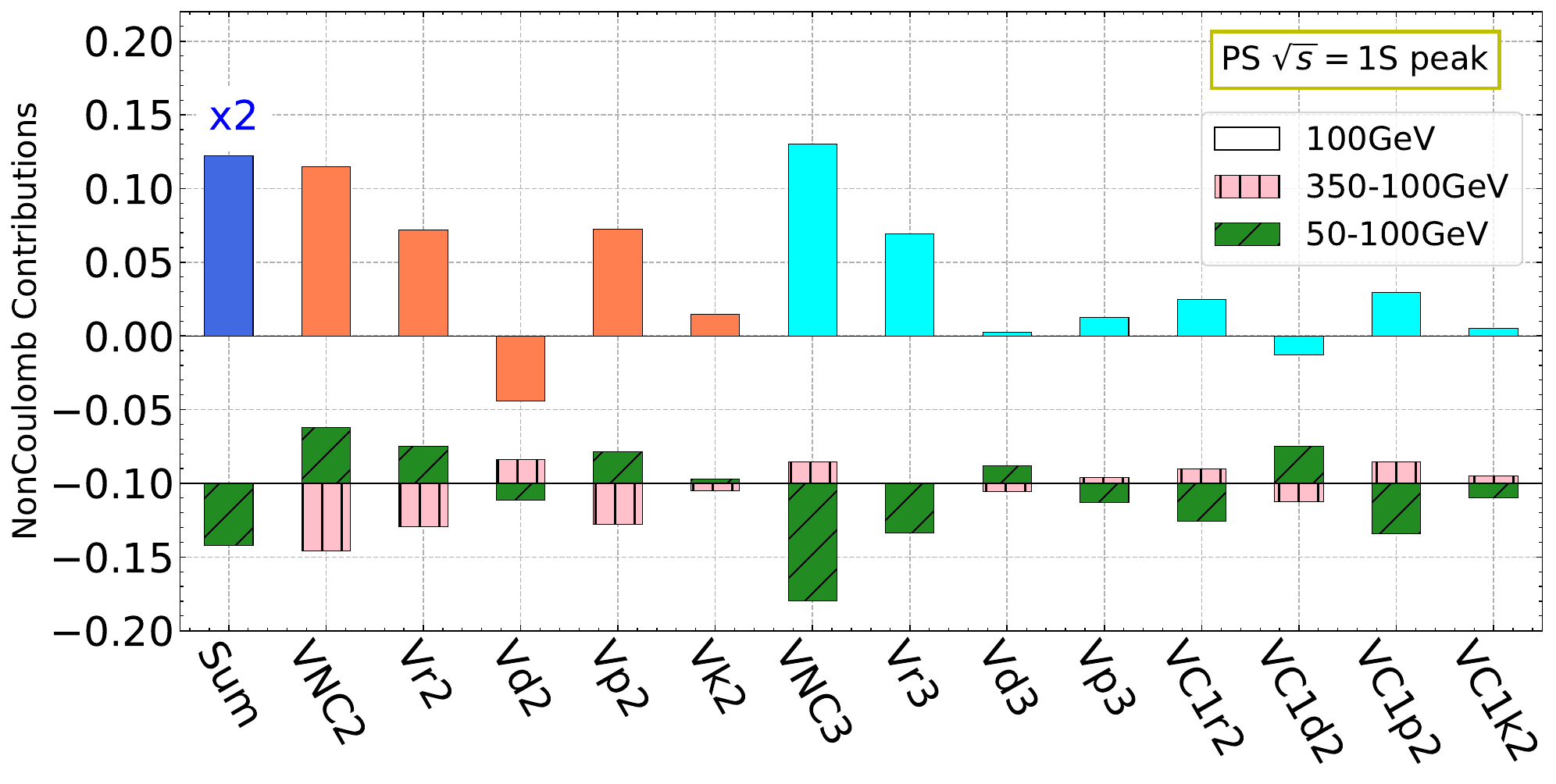}
\vskip-0.0cm
\includegraphics[width=0.8\textwidth]{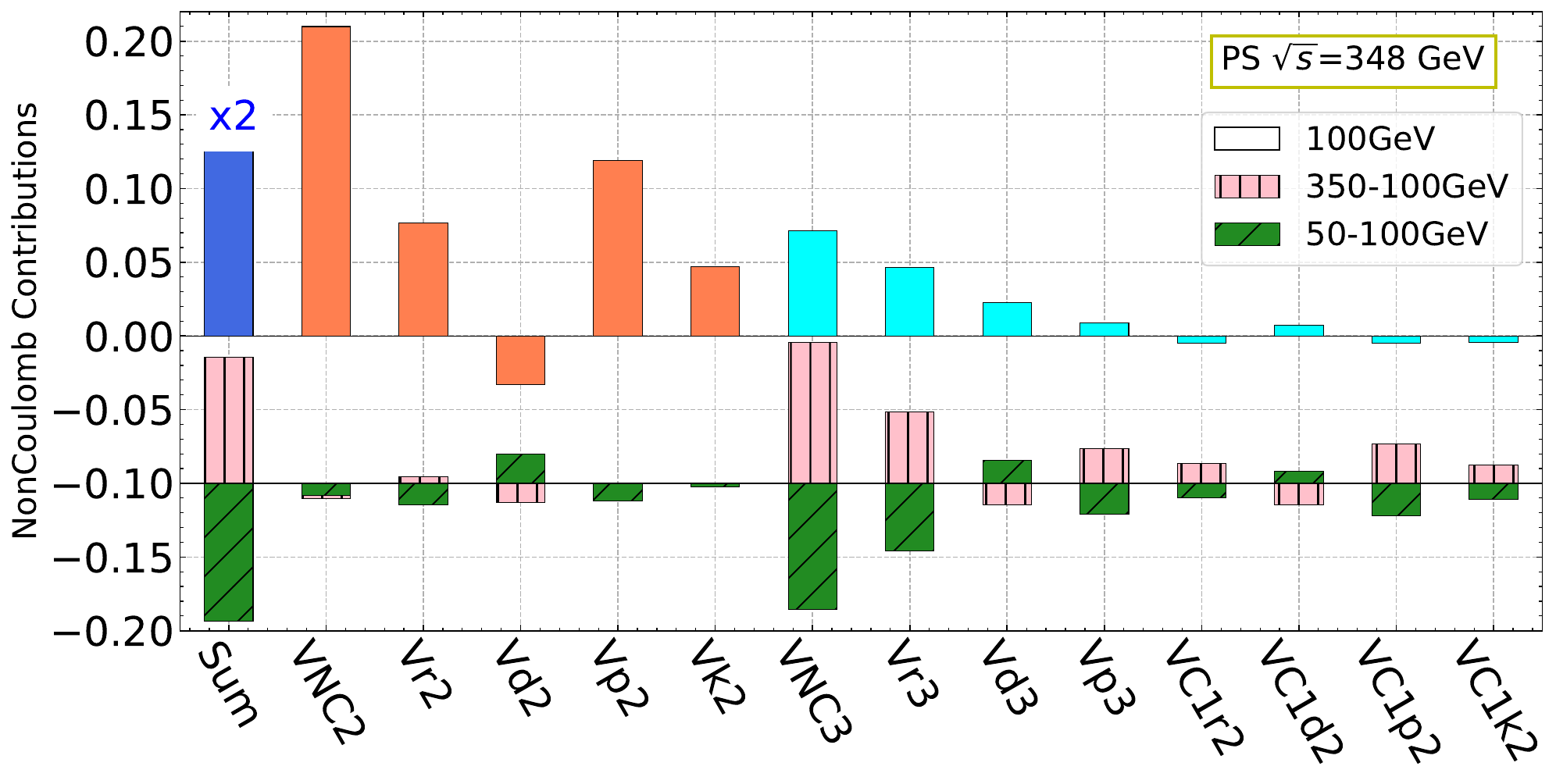}
\end{center}

\vskip-0.3cm
\caption{\label{fig:noncoulombparts}
Non-Coulomb potential corrections at NNLO and NNNLO separated into 
the contributions from the individual potentials, and in single 
insertions and double insertions. See text for further notation.}
\end{figure}

The new result described in this paper refers to the 
non-Coulomb potential contributions at third order, including 
the double insertions of the NNLO non-Coulomb potentials 
with the NLO Coulomb potential. The size of the new 
contributions is shown in Figure~\ref{fig:noncoulombparts}, 
together with the NNLO non-Coulomb potential 
contributions.\footnote{We recall that NNLO is the first order 
at which non-Coulomb contributions are present.} 

The non-Coulomb contribution constitutes a sizeable fraction 
of about 25\% (``Sum'' in the figures) of the total cross 
section. The sum of all third-order terms (``VNC3'') is 
not particularly small relative to the second-order sum 
(``VNC2'') and even exceeds the second-order sum on the 
peak. The individual second order terms are given by the 
four bars to the right of ``VNC2'', referring in sequence to 
the $1/(m|\bff{q})|$ potential (``Vr2''), the $1/m^2$ potential  
(``Vd2''), the $\bff{p}^2/\bff{q}^2$-potential (``Vp2''), and 
the relativistic kinetic energy correction (``Vk2''). 
The corresponding single insertions at third order are 
denoted with the same short-hand, replacing 2 by 3, to the 
right of ``VNC2'', followed by the mixed non-Coulomb-Coulomb 
double insertions with obvious notation. 

The plot shows that the single insertion of the two-loop 
$1/(m|\bff{q})|$ potential is the largest third-order 
non-Coulomb correction. Again it is instructive to focus on 
the scale-dependence of the various terms. This turns out 
to be large at third-order, in fact larger than at second 
order. The large scale-dependence and slow convergence of the 
series of non-Coulomb corrections does not by itself 
constitute a concern. Contrary to the Coulomb corrections, 
the non-Coulomb contributions depend on the factorization 
scale on top of the dependence on the scale of the strong 
coupling. The factorization scale dependence cancels together 
with hard matching coefficient contributions at the same 
order, and the ultrasoft contribution. Only the size and 
scale dependence of the sum of these, shown below, is indicative 
of the quality of the non-relativistic resummed perturbative 
expansion.

\subsubsection{All contributions}
\label{sec:allcontributions}

%
\begin{figure}[p]
\begin{center}
\hspace*{-0.5cm}
\includegraphics[width=0.9\textwidth]{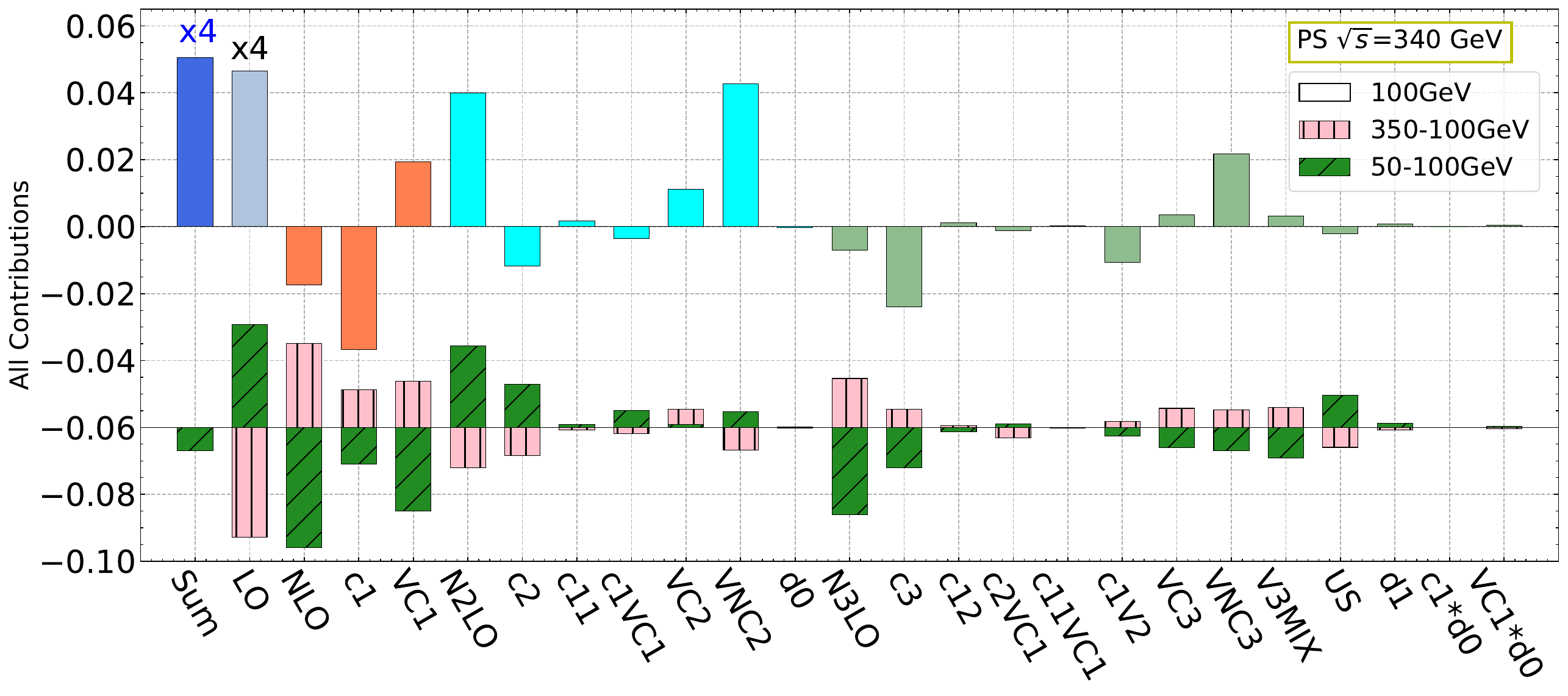}
\vskip0.4cm
\hspace*{-0.5cm}
\includegraphics[width=0.9\textwidth]{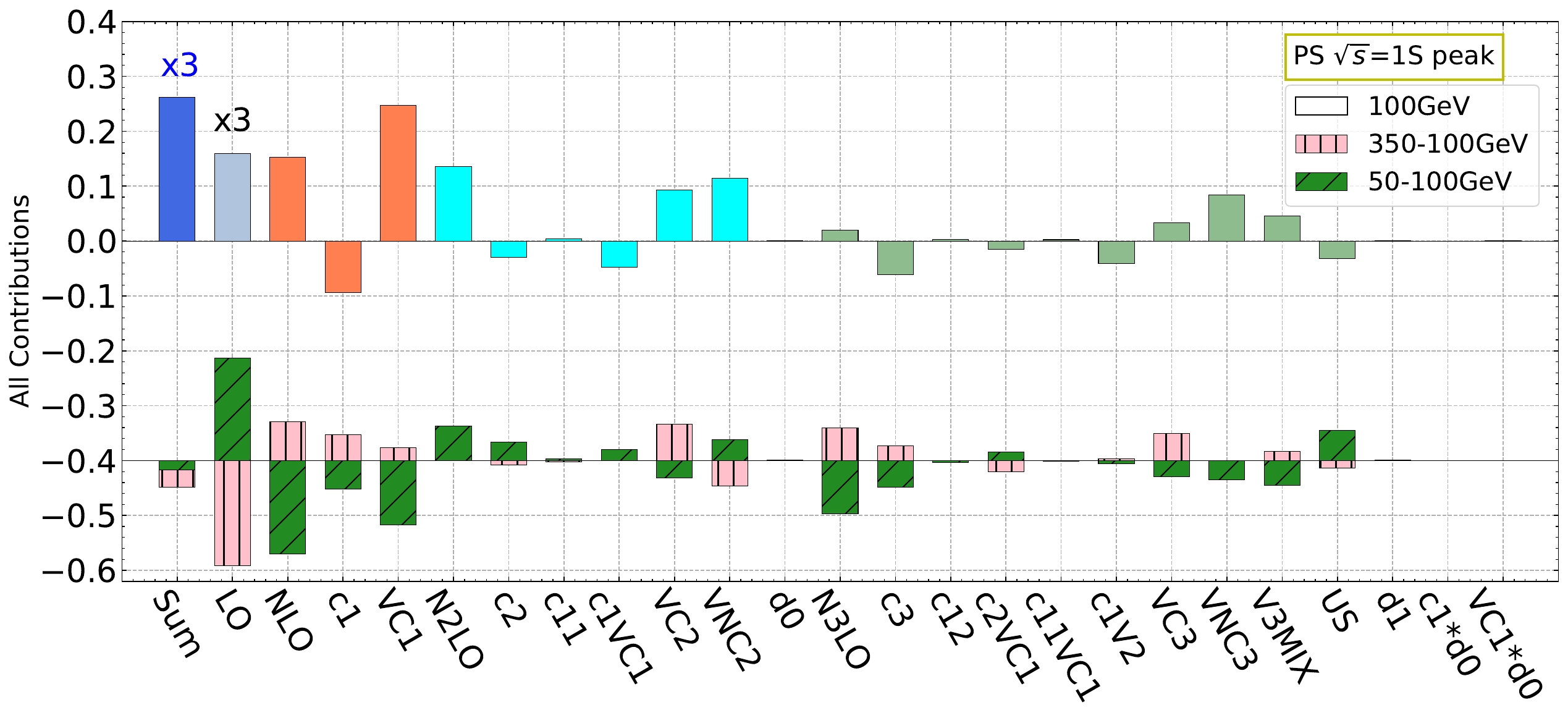}
\vskip0.4cm
\hspace*{-0.5cm}
\includegraphics[width=0.9\textwidth]{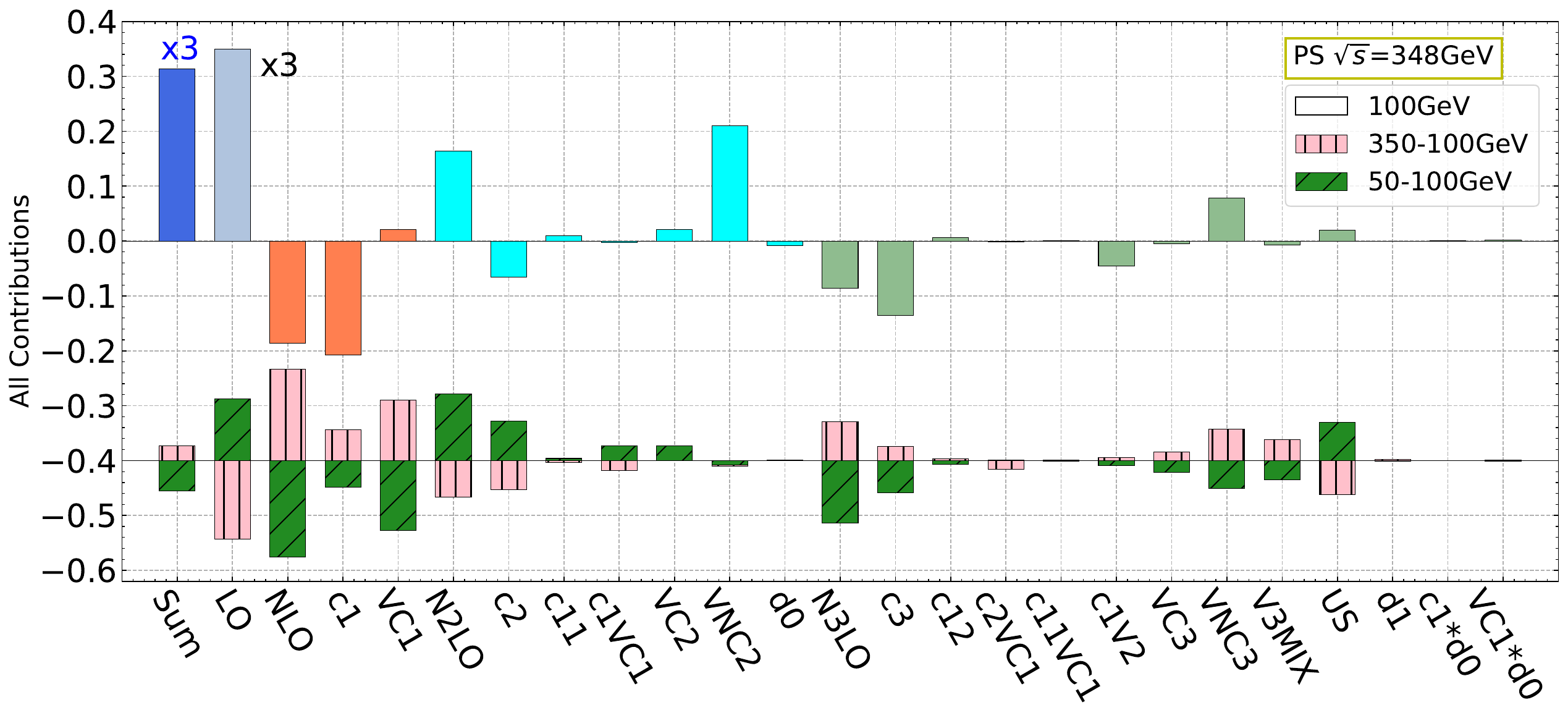}

\vspace*{-0.5cm}
\end{center}
\caption{\label{fig:allcontributions}
Overview of all contributions to the cross sections, including 
the short-distance coefficients (``ci'') and the ultrasoft 
contribution (``US''). See text for further notation.
}
\end{figure}

In Figure~\ref{fig:allcontributions} 
we finally give an overview 
of the size and scale dependence of all contributions. To 
shorten the list, we combine all Coulomb terms at a given 
order $n$ (that is, single, double, triple insertions) that were 
already shown separately before, into a 
single contribution ``VCn''. Similarly, for the non-Coulomb 
correction, we combine the single insertions into ``VNCn'' 
and the double insertions at third-order into ``V3MIX''. 
The plot shows in addition the ultrasoft contribution and 
products of the hard matching coefficients $c_v^{(n)}$ with the 
leading order (``c1", ``c2'', etc. in obvious notation, 
while, for example, ``c12'' denotes the terms proportional 
to $c_v^{(1)} c_v^{(2)}$ at NNNLO).  

It is worth noting that, at least within the adopted range 
of scale $\mu$, the hard matching coefficient $c_v$ is 
negative and counterbalances the mainly positive contribution 
from the potential insertions. This is particularly true 
at the third-order. We also notice that the scale dependence 
of the sum of third-order terms remains 
sizeable even after adding all 
third-order terms (see ``N3LO'' in the figure). In fact, 
one of the motivations for performing the third-order 
calculation has been the observation that the scale uncertainty 
of the NNLO computation of the cross section near threshold 
turned out to be unexpectedly 
large. The sizeable scale dependence of the third-order 
terms cancels this scale dependence to a large extent 
as seen by comparing to the scale dependence of the sum on 
the left (see also Figure~\ref{fig:relativescaledependence}, 
which displays this reduction over the entire energy range).


\section{Conclusion}
\label{sec:conclusion}

The present paper concludes a series of two papers, which 
presented the conceptual and computational details of the 
calculation of the top-quark pair production cross section 
at the third-order in QCD in non-relativistic, resummed 
perturbation theory. The main result has already been 
published in letter form \cite{Beneke:2015kwa}, and in 
addition to the results presented here relies on further 
pieces at the third-order, in particular the Coulomb 
contributions \cite{Beneke:2005hg}, the ultrasoft contribution 
\cite{Beneke:2008cr}, and the three-loop 
matching coefficient of the vector 
current \cite{Marquard:2014pea}. The 
complex organization of the full third-order computation 
in non-relativistic effective field theory, the 
unconventional nature of higher-order computations with 
Coulomb Green functions relative to standard multi-loop 
computations with free-theory propagators and the subtleties 
in implementing the third-order calculation in order to 
obtain the final result shown in \cite{Beneke:2015kwa} motivated the
present work. 

Part I \cite{Beneke:2013jia} summarized the effective field 
theory set-up together with all hard and potential 
matching coefficients required for the production of 
a heavy quark anti-quark pair near threshold at the third order. 
It can also be read in parts as a review of non-relativistic and 
potential non-relativistic QCD. The present part II first 
presented the calculation of the missing non-Coulomb potential 
contributions to the non-relativistic current 
correlation function to third order in PNRQCD perturbation 
theory, extending the methods based on dimensional regularization 
developed in \cite{Beneke:1999fe} and for third-order 
computations in \cite{Beneke:2005hg} to singular 
potential insertions at the third-order.

The literal use of the third-order calculation for the cross 
section near threshold faces several difficulties. The result 
exhibits uncancelled singularities and scale dependence, which 
is understood to be related to non-resonant production. 
Non-relativistic, resummed perturbation theory breaks down 
in the vicinity of the poles of the exact Green function. 
Curing this problem requires a further resummation. Finally, 
the pole mass is not a useful parameter due to its intrinsic 
ambiguity of order $\Lambda_{\rm QCD}$, and must be replaced 
by a more suitable definition, which in turn requires a careful 
reconsideration of the pole resummation procedure. These 
issues, which are crucial for the eventual precision determination 
of the top quark mass from the threshold, have been discussed in 
depth here for the first time.

In the final chapter of this paper we provided a detailed 
analysis of the result presented in \cite{Beneke:2015kwa}, 
analyzing its scale dependence, the use of different mass 
parameters, the effect of pole resummation, and the breakdown 
of the result into its many individual contributions. 

The third-order computation of the cross section in QCD is 
now complete. From the phenomenological point of view its main 
benefit is summarized in Figures~\ref{fig:mainresult} and 
\ref{fig:relativescaledependence}, which demonstrate the 
considerable reduction in theoretical uncertainty in going 
from NNLO to NNNLO. The development within QCD is matched by 
corresponding progress on electroweak and non-resonant 
effects \cite{Beneke:2017rdn}. Further improvement can be expected 
from merging the present NNNLO result with the summation 
of logarithms of $E/m$ \cite{Hoang:2013uda,Pineda:2006ri}, 
which could also shed led on the somewhat puzzling observation 
that non-relativistic perturbation theory for the 
top threshold becomes unstable already for renormalization 
scales smaller than 50~GeV.

\noindent
\subsubsection*{Acknowledgement}
We thank K.~Schuller for collaboration in an early stage of 
this work, and in particular, for collaboration on the calculation 
of the third-order insertion functions in 
section~\ref{sec:insertions}, see~\cite{Schuller:2008}.
We thank A.~Maier, C.~Peset and J. Piclum for comments on the text, 
and A.~Maier and J.~Piclum for their collaboration 
on and continuous support of the \texttt{QQbar\_threshold} code. 
YK thanks the DFG Excellence Cluster ``Origins and Structure 
of the Universe'' for hospitality. This work was supported in part 
by the Deutsche Forschungsgemeinschaft (DFG) 
Sonder\-for\-schungs\-be\-reich/Trans\-regio~9 
``Com\-puter\-ge\-st\"utzte
Theoretische Teil\-chen\-physik'', the DFG Graduiertenkolleg
``Elementar\-teil\-chen\-physik an der TeV-Skala'' and by JSPS KAKENHI 
Grant Number JP22K03602.

\appendix

\section{Glossary of symbols and functions}
\label{app:glossary}

For convenience we provide here a list of symbols and special functions 
that appear in the main text.

\begin{itemize}
\item  $\tilde\mu^2 = {\displaystyle 
 \mu^2 \frac{e^{\gamma_E}}{4\pi}}$
\item  $\gamma_E=0.577216...$ is the Euler-Mascheroni number.
\item  $\PG(x)=\gamma_E+\Psi(x)$, where $\Psi(x)$ is the Euler 
Psi function, $\Psi(x) = {\displaystyle \frac{d}{dx}}\ln \Gamma(x)$, 
$\Psi_n(x) = {\displaystyle \frac{d^n}{dx^n}}\Psi(x)$.
\item $E=\sqrt{s}-2 m$ 
\item $\lambda = {\displaystyle \frac{\alpha_s C_F}{2\sqrt{-\frac{E}{m}}}}$
\item $L_m ={\displaystyle \ln\frac{\mu}{m}}$  
\item $L_q={\displaystyle \ln\frac{\mu^2}{{\bf q}^2}}$
\item $L_{\lambda}={\displaystyle \ln\left(\frac{\lambda\mu}
{m \alpha_s C_F}\right)} = -{\displaystyle \frac{1}{2}}\,  {\displaystyle  
\ln\left(-\frac{4 m E}{\mu^2}\right)}$
\item $L_{n}={\displaystyle \ln\left(\frac{n \mu}
{m \alpha_s C_F}\right)} $
\item Harmonic sums: 
\[S_a(n)={\displaystyle \sum_{k=1}^n \frac{1}{k^a}}, \qquad 
S_{a,b}(n) \equiv {\displaystyle  \sum_{k=1}^n \frac{1}{ {k}^a }\, S_b(k)}, 
\qquad
S_{a,b,c}(n)\equiv {\displaystyle \sum_{k=1}^{n}\frac{1}{k^a} \, S_{b,c}(k)}\]
We usually omit the argument $n$ in the main text.
\end{itemize}


\section{More details on potential insertions}
\label{app:details}

\subsection{Single insertion of the $1/r^2$ potential}
\label{app:details1}

Here we give the details of the third and fourth part of this
calculation. 

\subsubsection{Part c} This part is the most
complicated one, because it has a subdivergence in the left vertex 
subgraph and contains the all-order Coulomb summation to the right of 
the potential insertion. The divergent part has to 
be factorized correctly to cancel divergences from the hard Wilson
coefficient. From its definition, see figure~\ref{fig:NAsubdiagrams}, 
part c is given by
\begin{eqnarray}
I_c[\frac{1}{2}+a\epsilon]&=&\int\prod_{i=1}^{4}\Bigg[\frac{d^{d-1}{\bf
p}_i}{(2\pi)^{d-1}}\Bigg] \frac{(2\pi)^{d-1} \delta^{(d-1)}({\bf
p}_1-{\bf p}_2)}{{\bf p}_1^2-mE}\frac{m\tilde{G}_0^{(>1ex)}({\bf
p}_3,{\bf p}_4;E)}{[({\bf p}_2-{\bf
p}_3)^2]^{\frac{1}{2}+a\epsilon}}.
\end{eqnarray}
The integration over $\bff{p}_2$ is trivial. After this integration 
the divergence arises from the integral over ${\bf p}_1$. We introduce 
the coordinate 
representation of the Green function through the Fourier transform 
of $\tilde{G}_0^{(>1ex)}({\bf p}_3,{\bf p}_4;E)$, which renders the 
$\bff{p}_4$ integral trivial. At this point we can perform the integration  
over $\bff{p}_1$ by introducing the Feynman parameter $x$ and 
obtain\footnote{For the $d$-dimensional integration over $\bff{p}_1$, 
we must add a factor $(\tilde\mu^2)^\epsilon$ for the $\overline{\rm MS}$ 
scale, but we then set $\mu=1$ to simplify the notation.}
\begin{eqnarray}
I_c[\frac{1}{2}+a\epsilon]&=& 
\frac{me^{\gamma_E \epsilon}\Gamma((a+1)\epsilon)}{(4\pi)^{\frac{3}{2}}
\Gamma(\frac{1}{2}+a\epsilon)}
\int_0^1 dx \,x^{-\frac{1}{2}+a\epsilon}\,\bar{x}^{-(a+1)\epsilon}
\int d^{d-1}{\bf r} \,G_0^{(>1ex)} (\bff{r},0;E) \label{eq:Ic}
\nonumber  \\
&& \times\,\int\frac{d^{d-1}{\bf p}_3}{(2\pi)^{d-1}} 
\frac{e^{i{\bf p}_3{\bf r}}}
{[x{\bf p}_3^2-mE]^{(a+1)\epsilon}}
\end{eqnarray}
with $\bar x=1-x$.
The $1/\epsilon$ pole comes only from the $\Gamma$-function in the prefactor, 
while the integrations are finite, 
hence we can expand the integrand in $\epsilon$. The important observation 
is that the leading term of the $\bff{p}_3$-integral is simply 
$\delta^{(d-1)}(\bff{r})$, so that the overall $1/\epsilon$ pole 
indeed multiplies the $d$-dimensional expression 
$G_0^{(>1ex)} (0,0;E)$ as required for the proper factorization 
of the pole part. Since $I_c[1/2+a\epsilon]$ can be multiplied 
by a $1/\epsilon$ term from the potential coefficients, we expand to 
$O(\epsilon)$ and obtain
\begin{eqnarray}
I_c[\frac{1}{2}+a\epsilon]&=&
\frac{m}{4\pi^2}\,\Bigg\{\frac{G_0^{(>1ex)}(E)}{a+1}
\Bigg[\frac{1}{\epsilon}+2(1-\ln2)
+\epsilon\bigg(4+\left(a^2+2a-\frac{1}{2}\right)\frac{\pi^2}{6}
\nonumber \\ 
\nonumber &&
-\,4\ln2+2\ln^22\bigg)\Bigg]
 +\frac{1}{2} \int d^{d-1}{\bf r} \,G_0^{(>1ex)}
 (r,0;E)\int_0^1 \frac{dx}{\sqrt{x}}\,\Bigg[L^{(1)}(r)
 \\ 
 &&+\,\epsilon
 \bigg(a\ln4x-(1+a)\ln(1-x)\bigg)L^{(1)}(r)
 +\epsilon\frac{1+a}{2} L^{(2)}(r)+O(\epsilon^2)\Bigg]\Bigg\}.
\qquad
\label{eq:i1halfa}
\end{eqnarray}
In the second term we introduced the abbreviations
\begin{equation}
L^{(n)}(r) \equiv (-1)^n\int\frac{d^{d-1}{\bf p}}{(2\pi)^{d-1}}
e^{i{\bf p}\cdot {\bf r}}\,[\ln(x{\bf p}^2-mE)]^n=\frac{\partial^n}{\partial
u^n} L(u,r)_{|_{u=0}}.
\end{equation}
where 
\begin{equation}
L(u,r) \equiv \int\frac{d^{d-1}{\bf p}}{(2\pi)^{d-1}}\frac{e^{i{\bf p}\cdot {\bf
r}}}{(x{\bf p}^2-mE)^u}\,.
\end{equation}
This definition is useful, because we can obtain an analytic expression 
for the generating function $L(u,r)$. Carrying out the angular integration, 
we obtain in $d=4$ dimensions 
\begin{eqnarray}
L(u,r) &=& 
\frac{1}{2\pi^2r}\int_{0}^{\infty}\!
dp\,\frac{p\sin(pr)}{(xp^2-mE)^u}
\nonumber \\ 
&=&\frac{1}{2^{u+\frac{1}{2}}}\frac{(-\frac{r^2 mE}{x})^{\frac{3+2u}{4}}}
{\pi^{\frac{3}{2}} r^3 (-mE)^u\Gamma(u)}\,
K_{\frac{3}{2}-u}\!\left(r\,\sqrt{\frac{-mE}{x}}\right),
\label{eq:lur}
\end{eqnarray}
where $K_n(x)$ is the modified Bessel function of the second kind.
Now we use the integral representation  (\refI{eq:greenint}) 
of the Green function $G_0^{(>1ex)}(\bff{r},0;E)$ 
and generate the $\bff{r}$-integrals 
$\int d^{3}{\bf r} \,G_0^{(>1ex)}(\bff{r},0;E) L^{(n)}(r)$ required for 
the evaluation of (\ref{eq:i1halfa}) from the $u$-derivatives of 
$\int d^{3}{\bf r} \,G_0^{(>1ex)}(\bff{r},0;E) L(u,r)$, which can be 
done analytically. Then we can perform the Feynman parameter integral 
over $x$ and finally the parameter integral that is contained in the 
integral representation  (\refI{eq:greenint}). 

Explicitly, for the $O(\epsilon^0)$ terms in (\ref{eq:i1halfa}) we 
obtain 
\begin{eqnarray}
&&\int_0^1 \frac{dx}{\sqrt{x}}
\int d^{d-1}{\bf r} \,G_0^{(>1ex)}(\bff{r},0;E)\,L^{(1)}(r)
\nonumber\\
&&\hspace{0cm}=\,-\frac{m\sqrt{-m E}}{\pi}
\int_0^\infty \!dt \left(\left(\frac{1+t}{t}\right)^{\!\lambda} 
-1-\lambda\ln\frac{1+t}{t}\right)
\big(\ln(-4 mE)+2 \ln(1+t)\big) + O(\epsilon)
\nonumber\\
&&\hspace{0cm}
= \,-2\ln(-4mE)\,G_0^{(>1ex)}(E) +
\frac{m\sqrt{-mE}}{2\pi}\bigg[4( 1-\lambda  )\PG(1-\lambda)
+2\lambda \PG(1-\lambda)^2
\nonumber \\ 
&&\hspace{0.5cm}
+ \,2\lambda \bigg( \frac{{\pi }^2}{2} - \Psi_1(1 - \lambda )
\bigg)\bigg] + O(\epsilon)\,.
\end{eqnarray}
Here the finite part is obtained by setting $d=4$ and using the 
four-dimensional representation  (\refI{eq:greenint}) 
of the Green function $G_0^{(>1ex)}(\bff{r},0;E)$. In order to compute 
$I_c[1/2+a\epsilon]$ to $O(\epsilon)$, 
we would need the $O(\epsilon)$ terms as well, but these 
cannot be obtained without the $O(\epsilon)$ terms of $G_0^{(>1ex)}
(\bff{r},0;E)$, which are not known. However, it is clear from 
(\ref{eq:i1halfa}) that the inaccessible $O(\epsilon)$ terms are 
independent of $a$, and hence they drop out from the counterterm-including 
insertion function $J_c[1/2+a\epsilon]$, which is all we 
need in the end.

For the $O(\epsilon)$ terms of (\ref{eq:i1halfa}) 
the calculation is performed in the
same way. For technical reasons it is useful to combine the parts
with $L^{(1)}$ and $L^{(2)}$:
\begin{eqnarray}
&&\hspace*{-0.5cm}
\int_0^1 \frac{dx}{\sqrt{x}} \int d^{3}{\bf r}\, G_0^{(>1ex)}(\bff{r},0;E)
\bigg([a\ln4x-(1+a)\ln(1-x)]L^{(1)}+\frac{1+a}{2}L^{(2)}\bigg)
\nonumber\\ 
&& = \frac{4 a m\sqrt{-mE}}{2\pi}\Bigg\{\frac{{\pi }^2}{6}( 3\lambda -1 )  
+ ( \lambda-1 ) \PG(1-\lambda) ^2 -
\frac{\lambda}{3}\PG(1-\lambda)^3
\nonumber\\ 
\nonumber && - \,( \lambda-1 ) \Psi_1(1 - \lambda)
+\PG(1-\lambda)\bigg( 2 -2\lambda- \frac{{\pi }^2 \lambda }{6} 
+ \lambda \Psi_1(1 - \lambda ) \bigg)
- \frac{1}{3}\lambda \Psi_2(1 - \lambda )
\\ 
&&  - \,\frac{8}{3}\lambda \zeta_3 
+\ln(-4mE)\Bigg[( \lambda-1  ) \PG(1-\lambda)  -
  \frac{\lambda}{2} \PG(1-\lambda)^2
  - \frac{\lambda}{2} \bigg( \frac{{\pi }^2}{2} - \Psi_1(1 - \lambda )
  \bigg)\Bigg]
\nonumber\\
&& -\, \ln^2(-4mE)\,\frac{\lambda}{4}\PG(1-\lambda)
\Bigg\}+\delta_{\not{a}}\, ,
\end{eqnarray}
where the $\Psi_n(x)$ are the polygamma functions and
$\delta_{\not{a}}$ is an $a$-independent term, which drops out in
the counterterm-including $J$-function. 
The final result is given in equation (\ref{eq:Jc}).

\subsubsection{Part d}  
This part is finite, so it can be done directly using the coordinate space 
representation  (\refI{eq:greenint}) of the Green function 
on both sides of the potential insertion. We then proceed as in the 
calculation of the single Coulomb insertion. 
We remain with the integrals over the variables
from the Green function representation, which can be converted into a
single sum:
\begin{eqnarray}
&&\hspace{-0.2cm}I_d[\frac{1}{2}+a\epsilon]=
\frac{m^{2}}{8\pi^{2}}\frac{a\epsilon(-4mE
)^{\frac{1}{2}-a\epsilon}}{\cos[\pi (a\epsilon-\frac{1}{2})]}
\int_{0}^{\infty}\!dt_{1} dt_{2}\,
\frac{\left(\left(\frac{1+t_{1}}{t_{1}}\right)^{\lambda}-1\right)
\left(\left(\frac{1+t_{2}}{t_{2}}\right)^{\lambda}-1\right)}
{(1+t_{1}+t_{2})^{1+2a\epsilon}}
\nonumber \\ 
&&\hspace{0.2cm}=\frac{m^{2}}{16\pi^{2}}\frac{(-4mE )^{\frac{1}{2}-a\epsilon}
\Gamma(2a\epsilon)}{\cos[\pi
(a\epsilon-\frac{1}{2})]}
\sum_{k=1}^{\infty}\frac{\Big(\Gamma(k+2a\epsilon)\Gamma(k-\lambda)
-\Gamma(k)\Gamma(k+2a\epsilon-\lambda)\Big)^2}
{\Gamma(k)\Gamma(k+2a\epsilon)\Gamma(k+2a\epsilon-\lambda)^2}\,.
\qquad \quad
\label{eq:i1halfd}
\end{eqnarray}
Then we can expand the result in $\epsilon$. In the expanded form
some parts of the sum can be done and we are left with
\begin{eqnarray}
I_d[\frac{1}{2}+a\epsilon]&=&\frac{m^2
\sqrt{-mE}}{4\pi^3}\bigg(-\lambda\frac{\pi^2}{3}-\PG(1-\lambda)+\lambda\Psi_1(1-\lambda)\bigg)
\\ \nonumber && \hspace{-1.5cm}-a\epsilon\frac{m^2
\sqrt{-mE}}{4\pi^3}\Bigg[\bigg(-\lambda\frac{\pi^2}{3}-\PG(1-\lambda)+\lambda\Psi_1(1-\lambda)\bigg)\ln(-4mE)
\\ \nonumber &&\hspace{-1.5cm}+2\sum_{k=1}^{\infty}(\PG(k)-\PG(k-\lambda))\bigg(\PG(k)\PG(k-\lambda)-\PG(k-\lambda)^2
-\Psi_1(k)+\Psi_1(k-\lambda)\bigg)\Bigg]\, .
\end{eqnarray}
The final result for $J_d[1/2+a\epsilon]$ can be found in equation 
(\ref{eq:Jd}). We should emphasize again that the $O(\epsilon)$ terms 
in the above expressions for  $I_d[1/2+a\epsilon]$ are not complete. 
In (\ref{eq:i1halfd}) the $\epsilon$-dependence stems from the insertion 
of $1/[\bff{q}^2]^{1/2+a\epsilon}$, but we used the four-dimensional 
representation of the Green function and ignored its  $O(\epsilon)$ terms.
This procedure yields the correct result for the counterterm-including 
$J_d[1/2+a\epsilon]$, since the ignored terms are independent of $a$. 
In general, this procedure works whenever the potential $\bff{p}_i$ 
integrations are finite, such as here or in part c after extraction of 
the divergent part, since then the integral can be expanded 
in $\epsilon$ before integration.

\subsection{Single insertion of the delta potential}
\label{app:details2}

\subsubsection{Part d}

The starting expression for part d is (\ref{eq:Ideltad}), which 
has a divergence in the left two-loop subgraph. In principle, one can
proceed in the same way as for part c of the $1/r^2$ potential:
calculate the divergent subdiagram in $d$ dimensions, factorize the 
zero-distance Green function $G_0^{(0ex)}(E)$, then
expand in $\epsilon$ and calculate the remainder in $d=4$
dimensions. However, it turns out that this integral is more
complicated. The reason is that the above strategy leads to
hypergeometric function, which cannot be expanded easily in 
$\epsilon$. We could use the summation representation
for hypergeometric functions and proceed, but this yields 
an additional sum. Unfortunately, this sum shows a slow
convergence, so that the actual numerical evaluation is also very
slow. A faster way is to do the integration numerically. This is
possible when using the momentum space representation of the Green
function. Similar numerical integrals appear in the ultrasoft 
contribution computed in~\cite{Beneke:2008cr}. The result for the 
counterterm-including $J$-function is found to be
\begin{eqnarray}
2J_d[\epsilon;w(\epsilon)]&=&
-\frac{m^2C_F\alpha_sw^{(1/\epsilon)}}{12\pi\epsilon}\,G_0^{(>0ex)}(E)
\\ \nonumber
&&\hspace{-1.5cm}+\,\frac{m^4 C_F^2
\alpha_s^2}{8\pi^2}\,\Bigg\{w^{(1/\epsilon)}\bigg[C_{NInt}^{Log}[-C_F^2/(2
\lambda)^2]\ln(m^2 \alpha_s^2/\mu^2)-C_{NInt}^{const}[-C_F^2/(2
\lambda)^2]
\\ \nonumber
&&\hspace{-1.5cm}-\,\frac{1}{2}\,\bigg(\frac{11}{6}+3L_{\lambda}-\ln
2\bigg)(1+2L_{\lambda}-2\PG(1-\lambda))\bigg]
\\ \nonumber
&&\hspace{-1.5cm}+\,\frac{w}{6}\bigg[\frac{\pi^2}{2}
+\PG(1-\lambda)\bigg(\frac{3}{\lambda}+1\bigg)-L_{\lambda}\bigg]\Bigg\}\,,
\end{eqnarray}
where $C_{NInt}^{Log}$ and $C_{NInt}^{const}$ represent numerical 
integrations given by:
\begin{eqnarray}
C_{NInt}^{Log}(e)&=&\mbox{real}+ i \, \mbox{Im}\,\Bigg\{\frac{3}{4}\ln(-e)
+\int_{0}^{\infty}\!dp\;
\bigg\{\frac{-\sqrt{-e}(2\ln(1/p)-1)}{\pi(p^2-e)}
\nonumber\\ 
\nonumber && \hspace{-1.5cm}
+\,\frac{9}{4}\Bigg[\frac{\sqrt{-e}}{9\pi(e-p^2)}
\bigg(8-3ip\ln\frac{\sqrt{-e}}{\sqrt{-e}-ip}
+3ip\ln\frac{\sqrt{-e}}{\sqrt{-e}+ip}+8\ln\frac{p}{2\sqrt{-e}}
\\ \nonumber && \hspace{-1.5cm}
-8\,\PG\!\left(1-\frac{2}{3\sqrt{-e}}\right)\bigg)
-\frac{iep}{\pi(e-p^2)^2(2\sqrt{-e}+3e)}
\\ \nonumber && \hspace{-1.5cm}
\times\bigg((e(\sqrt{-e}-2ip)+\sqrt{-e}p^2)\,
_2F_1(1,1-\frac{2}{3\sqrt{-e}},2-\frac{2}{3\sqrt{-e}};
\frac{ip-\sqrt{-e}}{ip+\sqrt{-e}})
\\ && \hspace{-1.5cm}-\,(e(\sqrt{-e}+2ip)+\sqrt{-e}p^2)\,
_2F_1(1,1-\frac{2}{3\sqrt{-e}},2-\frac{2}{3\sqrt{-e}};
\frac{ip+\sqrt{-e}}{ip-\sqrt{-e}})\bigg)\Bigg]\bigg\}\Bigg\}\qquad
\label{eq:CNintLogNum}\\
&=&\mbox{real}+ i \, \mbox{Im}\,\left[-\frac{5}{2} L_\lambda + 
\left(1 - \frac{1}{2\lambda}\right) \PG(1-\lambda) \right],
\label{eq:CNintLogAnalytic}\\
C_{NInt}^{const}(e)&=&
\frac{9i}{32\pi}\int_{0}^{\infty}\!dp\;
\mbox{Im}\,\Bigg\{\bigg(\!-i(p^2+e)\ln\frac{\sqrt{-e}+ip}{\sqrt{-e}-ip}+
2p\sqrt{-e}+2p\sqrt{-e}\ln\frac{1}{p^2-e}\bigg)
 \nonumber \\ \nonumber &&
\hspace{-1.5cm}
\times\Bigg[\frac{16(1-2\ln(1/p))}{9p(p^2-e)}
 +\frac{-32}{9 p(p^2-e)}\bigg(1+\ln\frac{p}{2\sqrt{-e}}
-\PG\!\left(1-\frac{2}{3\sqrt{-e}}\right)\bigg)
\\ \nonumber && \hspace{-1.5cm}
+\,\frac{4i}{3(p^2-e)}\bigg(\ln\frac{\sqrt{-e}}{\sqrt{-e}-ip}
-\ln\frac{\sqrt{-e}}{\sqrt{-e}+ip}\bigg)
\\ \nonumber && \hspace{-1.5cm}
+\,\frac{4e}{(e-p^2)^2(2\sqrt{-e}+3e)}
\\ \nonumber && \hspace{-1.5cm}
\times\bigg((-ie+p(2\sqrt{-e}-ip))_2F_1(1,1-\frac{2}{3\sqrt{-e}},
2-\frac{2}{3\sqrt{-e}};\frac{ip-\sqrt{-e}}{ip+\sqrt{-e}})
\\&& \hspace{-1.5cm}
+\,(ie+p(2\sqrt{-e}+ip))_2F_1(1,1-\frac{2}{3\sqrt{-e}},
2-\frac{2}{3\sqrt{-e}};\frac{ip+\sqrt{-e}}{ip-\sqrt{-e}})\bigg)\Bigg]\Bigg\}\,.
\end{eqnarray}
In the numerical integration in \eqref{eq:CNintLogNum} 
the imaginary part has been taken
before the integration to avoid a divergence that would appear in the 
integration of the real part.  
The analytic expression 
\eqref{eq:CNintLogAnalytic} for this integral has been obtained by making an 
ansatz for the appearing functions and checking the rational coefficients 
numerically to high accuracy.

\subsubsection{Part e}

This part e has also a divergence in the left two-loop 
vertex subgraph, but the starting expression now reads 
\begin{eqnarray}
I_e[a\epsilon]&=&\int\prod_{i=1}^{4}\Bigg[\frac{d^{d-1}{\bf
p}_{i}}{(2\pi)^{d-1}}\Bigg] \frac{C_F g^2 m^2}{({\bf p}^2_1-mE)({\bf
p}_1-{\bf p}_2)^2({\bf p}^2_2-mE) [({\bf p}_2-{\bf
p}_3)^2]^{a\epsilon}}
\nonumber\\
&&\times \,\tilde{G}_0^{(>1ex)}({\bf p}_3,{\bf p}_4)\,.
\end{eqnarray}
In contrast to the previous part, the strategy applied to 
the $1/r^2$ single insertion leads to a useful result. First, the left 
vertex subgraph is calculated in $d$ dimensions. 
This is done by using Feynman parameters and the 
integration-by-parts relation $[d-1-2a_2-a_1+a_11^+(3^--2^-)] \,I_e[a\epsilon]=0$, 
which expresses the original integral in terms of 
two simple ones: 
\begin{equation}
I_e[a\epsilon]=\frac{(1^+2^- -1^+3^-)}{d-4}\, I_e[a\epsilon]\,.
\end{equation}
After these manipulations we obtain 
\begin{eqnarray}
\nonumber I_e[a\epsilon]&=&\frac{C_F\alpha_s m^2}{(4\pi)^{d-2}\,(d-4)}
\int\prod_{i=1}^{2}\Bigg[\frac{d^{d-1}{\bf
p}_{i}}{(2\pi)^{d-1}}\Bigg]\frac{\tilde{G}_0^{(>1ex)}({\bf p}_1,{\bf
p}_2)}{\Gamma(a\epsilon)}
\nonumber\\
&& \times \int_0^1 dx\,
\Bigg[\frac{\Gamma(2-\frac{d-1}{2})\Gamma(1+a\epsilon-\frac{d-1}{2})}
{(-mE)^{2-\frac{d-1}{2}}}
\frac{x^{a\epsilon-1}\bar{x}^{\frac{d-1}{2}-a\epsilon-1}}{(x{\bf
p}^2_1-mE)^{1+a\epsilon-\frac{d-1}{2}}}
\nonumber \\  
&& 
-\, \frac{\Gamma(\frac{d-1}{2}-1)\Gamma(\frac{d-1}{2}-a\epsilon)
\Gamma(4-d+a\epsilon)}
{\Gamma(d-2-a\epsilon)}
\frac{x^{a\epsilon-\frac{d-1}{2}}\bar{x}^{d-3-a\epsilon}}{(x{\bf
p}^2_1-mE)^{4-d+a\epsilon}}\Bigg]\, .\qquad
\end{eqnarray}
At this point we can do the $x$-integral. Writing the resulting hypergeometric
function as a sum, we then perform the remaining integrals using the
Laguerre representation (\refI{eq:gpartial}) of the Green function. 
The resulting expression can be simplified to a single remaining sum. 
The final result for $J_e$ is given in equation (\ref{eq:Je}).

\subsubsection{Part f}
This part is finite, so it can be done in coordinate space. We
use again the integral representation (\refI{eq:greenint}) 
of the Green function to obtain 
\begin{eqnarray}
I_{f}[a\epsilon]&=&
\frac{m^{2}}{8\pi^{2}}\frac{(u-\frac{1}{2})(-4mE
)^{1-u}}{\cos[\pi (u-1)]}\int_{0}^{\infty}dt_{1}
\int_{0}^{\infty}dt_{2} \,(1+t_{1}+t_{2})^{-2u}
\nonumber \\ 
\nonumber &&\hspace*{-1cm}
\times\,\left[\left(\left(\frac{1+t_{1}}{t_{1}}
\right)^{\!\lambda}-1-\lambda\ln\left(\frac{1+t_{1}}{t_{1}}\right)\right)
\left(\left(\frac{1+t_{2}}{t_{2}}\right)^{\!\lambda}-1-
\lambda\ln\left(\frac{1+t_{2}}{t_{2}}\right)\right)\right]_{\Big| u=a\epsilon}
\\ \nonumber 
&=&-\,\frac{m^3E}{4\pi^2}\Bigg\{\bigg(1-\lambda\PG(-\lambda)\bigg)^2 + a
\epsilon\Bigg[2\lambda\bigg(1-\lambda\PG(-\lambda)\bigg)
\bigg(\PG(-\lambda)^2-\Psi_1(-\lambda)+\frac{\pi^2}{2}\bigg)
\\
&& +\,\bigg(2-\ln(-4mE)\bigg)\bigg(1-\lambda\PG(-\lambda)\bigg)^{\!2}
\nonumber\\
&& +\,\sum_{k=1}^{\infty}\frac{2}{k}\bigg((\lambda-k)
\bigg(\PG(k)-\PG(k-\lambda)\bigg)+k\lambda\Psi_1(k)\bigg)^{\!2}\,
\Bigg]\Bigg\}\,.
\end{eqnarray}
To arrive at the second equation we first integrate over $t_1$. 
The resulting hypergeometric function is expressed as a sum, and 
the remaining $t_2$ integral is done. Performing as many sums as possible, 
we are left with one sum from the hypergeometric function,
which has to be evaluated numerically.

\subsection{Double insertion of the Coulomb and 
$1/r^2$ potential}
\label{app:detailsNAC}

\subsubsection{Part b} This part has a divergence in the
left one-loop vertex subgraph and a double all-order Coulomb summation to the 
right of the $1/r^2$ insertion. The divergence is extracted in the same way 
as for part c of the $1/r^2$ potential single insertion. First, the 
vertex subgraph is calculated in $d$ dimensions. Then, the result is 
expanded in $\epsilon$. The remaining (finite) part can be done in $d=4$
dimensions. Again this automatically factorizes the $1/\epsilon$ pole in
the required form. For the finite part the integral is Fourier
transformed into coordinate space. For ease of notation we show the 
expressions for $I_{a+b}[\frac{1}{2}+\epsilon,1+a\epsilon]$ 
instead of $I_{b}[\frac{1}{2}+\epsilon,1+a\epsilon]$. In the end 
$I_{a}[\frac{1}{2}+\epsilon,1+a\epsilon]$ is subtracted 
to get the correct result. Carrying out the steps described above, we 
obtain
\begin{eqnarray}
 I_{a+b}[\frac{1}{2}+\epsilon,1+a\epsilon]&=&\frac{m}{8\pi^2}
\Bigg[\frac{1}{\epsilon}+2(1-\ln2)\Bigg]\,
I[1+a\epsilon]+\frac{m^3}{128\pi^4}\int_0^{\infty} \!dr\,
e^{-r\sqrt{-mE}}\int_0^1 \frac{dx}{\sqrt{x}}
 \nonumber \\
&&\hspace{-3cm}
 \frac{\partial}{\partial u}\int_0^{\infty} \!dp \,
\frac{2pr \sin(pr)}{\pi}\frac{(\mu^2)^u}{(xp^2-mE)^u}
\sum_{k=1}^{\infty}\frac{L_{k-1}^{(1)}(2r\sqrt{-mE})H(a\epsilon,k)}
{k(k-\lambda)}\,,
\end{eqnarray}
where $I[1+a\epsilon]$ is the single Coulomb insertion function defined 
in (\ref{eq:i1u}). The derivative in $u$ is used to generate a logarithm, 
so that we can use (\ref{eq:lur}), and it is understood that the derivative 
is taken at $u=0$. For the Green function between the two 
potential insertions, we used the same representations as in the case of 
the Coulomb double insertion, which explains the recurrence of the 
function $H(u,k)$ defined in (\ref{eq:defH}). After employing 
(\ref{eq:lur}), the remaining $x$- and then the 
$r$-integration can also be done, resulting in 
\begin{eqnarray}
I_{a+b}[\frac{1}{2}+\epsilon,1+a\epsilon] &=&\frac{m}{8\pi^2}
\Bigg[\frac{1}{\epsilon}+2(1-\ln2)\Bigg] I[1+a\epsilon]
+\frac{m^3}{16\pi^2}\sum_{k=1}^{\infty}\frac{N(k)H(a\epsilon,k)}
{k(k-\lambda)}\,,
\nonumber \\
\end{eqnarray}
where 
\begin{eqnarray}
N(k)&\equiv &
\frac{\partial}{\partial u}\,\Bigg\{(\mu^2)^u
\frac{2^{\frac{5}{2}-u}\,
(-mE)^{\frac{1}{4}-\frac{u}{2}}}{8\pi^2\sqrt{\pi}\,\Gamma(u)}
\nonumber\\
&&\times\,
\int_0^{\infty} \!dr \,e^{-r\sqrt{-m E}}
\,r^{-\frac{1}{2}+u}\,K_{\frac{1}{2}-u}(r\sqrt{-mE})
L_{k-1}^{(1)}(2r\sqrt{-mE})\Bigg\}_{|u=0}
\nonumber\\
&=&\frac{k}{2\pi^2}\left(1-\PG(k+1)-\frac{1}{2}\ln\left(\frac{-4mE}{\mu^2}
\right)\right).
\end{eqnarray}
The remaining sum containing $N(k)$ and $H(a\epsilon,k)$ is divergent, 
because it includes the overall divergence from part a. However, part a 
can simply be removed from $I_{a+b}[\frac{1}{2}+\epsilon,1+a\epsilon]$ 
by subtracting the $\lambda\rightarrow 0$ limit of the 
expression.\footnote{Note that $L_\lambda 
=-\frac{1}{2}\ln(-4mE/\mu^2)$ and therefore $N(k)$ 
is independent of $\lambda$. However, there is implicit 
$\lambda$-dependence in $H(a\epsilon,k)$.} Hence, the result for part b is 
\begin{eqnarray}
I_b[\frac{1}{2}+\epsilon,1+a\epsilon]&=&\frac{m}{8\pi^2}
\Bigg[\frac{1}{\epsilon}+2(1-\ln2)\Bigg] I_b[1+a\epsilon]
 \nonumber\\
 &&\hspace*{-2cm}
+\,\frac{m^3}{32\pi^4}\sum_{k=1}^{\infty}
\big(1-\PG(k+1)+L_\lambda\big)
\Bigg[\frac{H(a\epsilon,k)}{k-\lambda} -\lim_{\lambda\rightarrow
0} \frac{H(a\epsilon,k)}{k-\lambda}\Bigg].
\end{eqnarray}
To obtain the counterterm-including double insertion 
$J_b[\frac{1}{2}+\epsilon,1+a\epsilon]$ according to its definition 
(\ref{eq:defJdouble}) the above expression needs to be expanded 
to $O(\epsilon)$.\footnote{Once again, the  $O(\epsilon)$ terms 
missed by having used the four-dimensional Coulomb Green functions
are $a$-independent and therefore irrelevant for the $J$-functions.}
Inserting the formulas for $H^{(n)}(k)$ (the derivatives (\ref{eq:Hder}) of
$H(a\epsilon,k)$) most parts of the sum can be done analytically; the
the remaining sums are calculated numerically with high accuracy. 
The final expression for $J_b[\frac{1}{2}+\epsilon,1+a\epsilon]$ 
is given in (\ref{eq:doubleNAb}).

\subsubsection{Part c} 

This part is finite and can be calculated in four dimensions. The 
procedure is the same as for the Coulomb double insertion. The only 
differences are  that one has to subtract the
Green functions without gluon exchange in the left vertex, 
see figure~\ref{fig:NACsubdiagrams}, 
and the exponent of the momentum-space potential insertion is 
$1/2+\epsilon$ instead of $1+\epsilon$. 
The result using the definition (\ref{eq:defH}) of the $H$-functions is
\begin{eqnarray}
I_c[\frac{1}{2}+\epsilon,1+a\epsilon]&=&
\frac{m^3}{(4\pi)^3}\frac{\lambda}{m C_F\alpha_s}\,\sum_{k=1}^{\infty}
\frac{H_{\rm sub}(-\frac{1}{2},k)H(a\epsilon,k)}{k(k-\lambda)}\,. 
\label{eq:NACpart3_start}
\end{eqnarray}
$H_{\rm sub}$ is a subtracted version of $H$ and given by
$H_{\rm sub}=H-(H)_{\lambda\rightarrow 0}$. For $u=-1/2$ it 
can be calculated analytically:
\begin{eqnarray}
H_{\rm sub}\!\left(-\frac{1}{2},k\right)&=&\frac{2m C_F \alpha_s}{\pi \lambda}
\Bigg(k\PG(k+1)-(k-\lambda)\PG(k-\lambda)-\lambda\PG(-\lambda)\Bigg)\,.
\end{eqnarray}
Inserting this result into (\ref{eq:NACpart3_start}) and performing 
the expansion to $O(\epsilon)$, most parts of the last summation 
over $k$ can be done and we end up with (\ref{eq:doubleNAc}).

\section{Corrections to the S-wave 
quarkonium energy levels and wave functions}
\label{sec:enpsicorr} 

As discussed in section~\ref{sec:poleresummation}, the S-wave energy levels 
and wave functions at the origin can be obtained from the Laurent 
expansion of the the perturbative PNRQCD calculation of the PNRQCD 
vector current correlation function by comparing it with the expansion 
of the single pole of the full non-perturbative correlation function around 
the leading-order bound state pole location. We expand the exact bound 
state energy and wave function squared in the strong coupling 
in the form:
\begin{eqnarray}
E_n &=& E_n^{(0)} \left(1+\frac{\alpha_s}{4\pi} e_1
+\left(\frac{\alpha_s}{4\pi}\right)^{\!2} e_2
+\left(\frac{\alpha_s}{4\pi}\right)^{\!3} e_3 + \ldots\right),
\label{eq:energyexpansion}
\\
|\psi_n(0)|^2 &=& |\psi_n^{(0)}(0)|^2 \left(1+\frac{\alpha_s}{4\pi}
f_1 +\left(\frac{\alpha_s}{4\pi}\right)^{\!2} f_2
+\left(\frac{\alpha_s}{4\pi}\right)^{\!3} f_3 + \ldots\right)
\label{eq:wfexpansion}
\end{eqnarray}
with $\alpha_s=\alpha_s(\mu)$ in the $\overline{\rm MS}$ scheme as usual, 
and 
\begin{eqnarray}
&&E_n^{(0)} =-\frac{m (\alpha_s C_F)^2}{4 n^2},
\qquad\quad
|\psi_n^{(0)}(0)|^2 =\frac{(m \alpha_s C_F)^3}{8\pi n^3}
\end{eqnarray}
the leading-order values.

In the following we summarize the results up to the third order. 
As mentioned in the main text, none of these results is new. However,  
since they are scattered over several places in the literature and 
not always easy to assemble without risk of error, we find it useful 
to put them together here.

We separate the pure Coulomb, non-Coulomb (including the double insertion 
with the NLO Coulomb potential), and
ultrasoft corrections adopting the same notation as in 
\cite{Beneke:2005hg,Beneke:2007gj,Beneke:2007pj}:\footnote{In
\cite{Beneke:2005hg} the non-Coulomb and ultrasoft corrections to
the energy levels were combined.} $e_{i}=e_i^{C}+e_i^{nC}+e_i^{us}$
and $f_{i}=f_i^{C}+f_i^{nC}+f_i^{us}$. The non-Coulomb part appears
first at second order, and the ultrasoft part at third order, so
that $e_1^{nC}=e_1^{us}=e_2^{us}=0$ and
$f_1^{nC}=f_1^{us}=f_2^{us}=0$. All results will be given for the
spin-triplet S-wave state.

\subsection{Energy levels}

The results for the pure Coulomb-potential corrections to the 
S-wave spin-triplet energy levels read:
\begin{eqnarray}
e_1^C &=&  4 \beta_0 \, L_n +  c_{E,1}^{C},
\\
e_2^C&=&      12\beta_0^2\, L_n^2
            + L_n \,\bigg(\!-8 \beta_0^2 + 4 {\beta_1}
                      + 6 \beta_0 c_{E,1}^{C}\bigg)
            + c_{E,2}^{C} ,
\\
e_3^C&=&  32 \beta_0^3 \, L_n^3
          +  L_n^2\, \bigg(\! - 56 \beta_0^3
                          + 28 \beta_0 {\beta_1}
                          + 24 \beta_0^2 c_{E,1}^{C}
                   \bigg)
              + L_n\, \bigg( 16 \beta_0^3
                        - 16 \beta_0 {\beta_1}
                        + 4 {\beta_2}
                       \nonumber \\
&& - 12 \beta_0^2 c_{E,1}^{C}
                        + 6 {\beta_1} c_{E,1}^{C}
                        + 8 \beta_0 c_{E,2}^{C}
                   \bigg)
 + c_{E,3}^{C} + 32 \pi^2 C_A^3
   \bigg(L_n+S_1
   \bigg)\,.
\end{eqnarray}
The logarithm $32 \pi^2 C_A^3 L_n$ in the last line is the logarithm 
whose scale-dependence cancels with the ultrasoft contribution below, 
while all other logarithms are related to the running of the strong 
coupling. The constants $c_{E,i}^{C}$ are given by
\begin{eqnarray}
c_{E,\,1}^{C}&=& 2 a_1 + 4 S_1 \beta_0,
\\
c_{E,\,2}^{C}&=&
  a_1^2
+ 2 a_2 + 4 S_1 \beta_1 + 4 a_1 \beta_0\bigg[\, 3 S_1-1 \,\bigg]
\nonumber \\
&& + \beta_0^2
     \Bigg[\, S_1 \bigg(12 S_1-8 - \frac{8}{ n}\bigg)
          + 16 S_2
          - 8 n S_3
          + \frac{2 \pi^2}{ 3}
          + 8 n {\xi(3)}
     \,\Bigg],
\\
c_{E,\,3}^{C} &=&
 2 a_1 a_2
+ 2 a_3 + 2 a_1^2 \beta_0\Big[\, 4 S_1-5 \,\Big] + 4 a_2 \beta_0
\Big[\, 4 S_1-1 \,\Big] + 4 a_1 \beta_1  \Big[\, 3 S_1-1 \,\Big]
\nonumber \\
&& + 4 S_1 \beta_2 + \beta_0 \beta_1
      \Bigg[\,
              S_1 \bigg(28 S_1-16 - \frac{24}{ n}\bigg)
            + 36 S_2
            - 16 n S_3
            +  \frac{7 \pi^2}{3}
            + 16 n {\xi(3)}
    \,\Bigg]
\nonumber \\
&& + a_1 \beta_0^2
      \Bigg[\, S_1 \bigg( 48 S_1-56 - \frac{32}{ n} \bigg)
         + 64 S_2
         - 32 n S_3
         + 8
         + \frac{8 \pi^2}{3}
         + 32 n {\xi(3)}
     \, \Bigg]
\nonumber \\
&& + \beta_0^3
     \Bigg[\, S_1 \Bigg(
                     S_1 \bigg( 32 S_1 -56 - \frac{32}{ n}\bigg)
                    + 96 S_2
                    - 64 n S_3
                    + 16
                    + \frac{16}{n}
                    + \frac{32 \pi^2}{ 3}
                    + 64 n {\xi(3)}
                  \Bigg)
\nonumber \\
&&
             + S_2 \Bigg( 8 n S_2
                        + 16 n^2 S_3
                        -32
                        - \frac{16}{ n}
                        - \frac{40 n \pi^2}{ 3}
                        - 16 n^2 {\xi(3)}
                   \Bigg)
             + S_3 \Bigg(96 + 16 n + 8 n^2 \pi^2\Bigg)
\nonumber \\
&&
            - 104 n S_4
            + 48 n^2 S_5
            - 144 S_{2, 1}
            + 224 n S_{3, 1}
            - 32 n^2 S_{3, 2}
            - 96 n^2 S_{4, 1}
            -\frac{4 \pi^2}{3}
            + \frac{2 n \pi^4}{45}
\nonumber \\
&&
            +{\xi(3)} \bigg(32 - 16 n - 8 n^2 \pi^2\bigg)
            + 96 n^2 {\xi(5)}
      \,\Bigg].
\end{eqnarray}
The expressions for the non-Coulomb contribution at second and third order 
are
\begin{eqnarray}
e_2^{nC}/(16\pi^2) &=& \frac{C_A C_F}{n} + \frac{{C_F}^2}{n}
      \left(\frac{2}{3}-\frac{11}{16n}
      \right),
\\
e_3^{nC}/(64 \pi^2) &=& - \frac{49 n_f T_F C_A C_F}{ 36 n} + \frac{4
{C_A}^2 C_F}{3 n} \Bigg[  \frac{197}{48}
      + \ln 2
      + 2 L_n
      - 2S_1
\Bigg]
\nonumber \\
&&\hspace*{-2cm} +
 \frac{C_A {C_F}^2}{n}
\Bigg[ \frac{563}{ 108}
          - \frac{\ln n}{ 2}
          + \frac{8\ln 2}{ 3}
          - \frac{37  S_1}{ 6}
          + \frac{\ln(C_F\alpha_s)}{ 2}
          + \frac{20L_n}{ 3}
\nonumber \\
&& \hspace*{-2cm}    + \frac{1}{n} \bigg(
           \!- \frac{97}{ 72}
          - \frac{2 L_n}{ 3}
          - \frac{2 S_1}{3}
          \bigg) \Bigg]
+ \frac{{C_F}^3}{3 n} \Bigg[
      \!-\frac{9}{2}
      + \frac{7}{2 n}
      - \ln n
      - 7 S_1
      + \ln(C_F\alpha_s)
      +8 L_n
\Bigg]
\nonumber \\
&&\hspace*{-2cm} +\frac{ {C_F}^2 n_f T_F}{9 n} \Bigg[\! -\frac{4}{3}
         + \frac{5}{ 2 n}\Bigg] + \frac{2T_F {C_F}^2}{15
         n}+a_1\Bigg[\frac{C_A C_F}{2n}+\frac{3C_F^2}{4n}\bigg(1-\frac{3}{4n}\bigg)\Bigg]
\nonumber \\
&&\hspace*{-2cm} + \beta_0 \Bigg[ C_A C_F
  \bigg( \frac{2 }{ n} L_n
        -\frac{\pi^2}{6}
        + \frac{1}{2 n}
        +  S_2
   \bigg)
+ {C_F}^2 \Bigg(
      \bigg(\!-\frac{11}{8 n^2}
            +\frac{4}{3n}
            \bigg) L_n
\nonumber \\
&&\hspace*{-2cm}
    - \frac{11 S_1 }{ 8 n^2}
    +   \frac{2}{3} S_2
    + \frac{2}{n}
    + \frac{1}{24\,n^2}
    -\frac{\pi^2}{ 9}
 \Bigg)
\Bigg].
\end{eqnarray}
The ultrasoft correction is:
\begin{eqnarray}
\label{eq:ultrasoftenergy} e_3^{us}/(64 \pi^2) &=& \frac{4 {C_A}^2
C_F}{3 n} \Bigg[ -\frac{25}{12}
      -2\ln 2
      -\ln n
      - 2 L_m
      +3 \ln (C_F \alpha_s)
      + S_1
\Bigg]
\nonumber \\
&& \hspace*{-2cm} + \frac{{C_A}^3}{6} \Bigg[-\frac{5}{6}
      - \ln 2
     - 2\ln n
     + 4\ln\left(C_F \alpha_s \right)
     - 3\ln(\mu/m)
     - 2 S_1
\Bigg]
\nonumber \\
&&\hspace*{-2cm} +
 \frac{C_A {C_F}^2}{3n}
\Bigg[- 10-20L_m+8S_1-20\ln 2 -8\ln n +32 \ln(C_F \alpha_s)
\nonumber \\
&&\hspace*{-2cm}           + \frac{1}{n} \bigg(
            \frac{5}{ 3}+2L_m +2\ln 2-4\ln(C_F \alpha_s)
\bigg) \Bigg]
\nonumber \\
&&\hspace*{-2cm} + \frac{{C_F}^3}{3 n} \Bigg[
      -2 n L_E(n)-8L_m
      -\frac{20}{3}
      - 8 \ln 2
      + 16 \ln(C_F\alpha_s)
\Bigg].
\end{eqnarray}
The scale-dependence of the logarithm of $\mu/m$ in the second line 
cancels against the corresponding one in the third-order Coulomb 
contribution mentioned above. We further 
introduced the ``Bethe logarithm'' $L_E(n)$. It is  not known
in an analytic form for arbitrary $n$. The values for the first
states are:
\begin{equation}
L_E(n) = \left(-81.5379, -37.671, -22.4818, -14.5326, -9.52642,
  -6.0222,\ldots\right).
\end{equation}

Only the non-Coulomb contribution is 
spin-dependent. The expression for the spin-singlet energy level 
can be found in \cite{Beneke:2005hg}, 
which quotes the result from~\cite{Penin:2005eu}.

\subsection{Wave function at the origin}
The results for the Coulomb corrections to the wave function read
\begin{eqnarray}
f_1^C &=&
 6 \beta_0\, L_n
+ c^C_{\psi,1},
\\
f_2^C &=&
   24 \beta_0^2 \, L_n^2
+  L_n\,\bigg(- 12 \beta_0^2
            + 6 {\beta_1}
            + 8 \beta_0 c^C_{\psi,1}  \bigg)
+ c^C_{\psi,2},
\\
f_3^C &=&
  80 \beta_0^3 \, L_n^3
+  L_n^2 \,\bigg( -108 \beta_0^3
                + 54 \beta_0 {\beta_1}
                +  40 \beta_0^2 c^C_{\psi,1}  \bigg)
\nonumber \\  
&& +\,   L_n\, \bigg( 24 \beta_0^3
             - 24 \beta_0 {\beta_1}
              + 6 {\beta_2}
              - 16 \beta_0^2 c^C_{\psi,1}
              + 8 {\beta_1} c^C_{\psi,1}
              + 10 \beta_0 c^C_{\psi,2}
                    \bigg)
\nonumber \\ 
&& + \,c^C_{\psi,3}
 + 48 \pi^2 {C_A}^3
  \bigg[L_n
         +\frac{1}{3}\left(S_1+ 2 n S_2-1-\frac{n \pi^2}{3}\right)
  \bigg]\,,
\end{eqnarray}
The logarithm $48 \pi^2 C_A^3 L_n$ in the last line is the logarithm 
whose scale-dependence cancels with the ultrasoft contribution below, 
while all other logarithms are related to the running of the strong 
coupling. The constants are defined as 
\begin{eqnarray}
c^C_{\psi,\,1}&=&
 3 a_1
+2 \beta_0 \Bigg[\,  S_1 +  2 n S_2-1 - \frac{ n \pi^2}{ 3}
\,\Bigg]\,,
\\
c^C_{\psi,\,2}&=& 3 a_1^2 + 3 a_2 + 2 a_1\beta_0 \Bigg[\, 4 S_1+ 8 n
S_2-7- \frac{4 n \pi^2}{ 3}\,\Bigg] + 2  \beta_1 \Bigg[\, S_1 + 2 n
S_2-1 -  \frac{ n \pi^2}{ 3} \,\Bigg]
\nonumber \\
&& \hspace*{-1.05cm} + \beta_0^2
  \Bigg[\,
           S_1 \Bigg( 8 S_1
                     + 16 n S_2
                     -20
                     - \frac{12}{ n}
                     - \frac{8 n\pi^2}{ 3}
                \Bigg)
         + S_2 \Bigg(  4 n^2 S_2
                     + 8
                     - 8 n
                     - \frac{4 n^2 \pi^2}{ 3}
               \Bigg)
\nonumber \\
&& \hspace*{-1.05cm}
         + 28 n S_3
         - 20 n^2 S_4
         - 24 n S_{2, 1}
         + 16 n^2 S_{3, 1}
         +  4
         + \frac{(3+4 n)\pi^2}{ 3}
         + \frac{n^2 \pi^4}{ 9}
         + 20 n {\xi(3)}
  \, \Bigg]\,.
\\
c^C_{\psi,\,3}&=&
  a_1^3
+ 6 a_1 a_2 + 3 a_3 + 10 a_1^2 \beta_0
    \Bigg[\,   S_1
           +  2 n S_2
           -  \frac{31}{10}
           -  \frac{ n \pi^2}{ 3}
  \,\Bigg]
+ 10 a_2 \beta_0
    \Bigg[\,  S_1
           +   2 n S_2
\nonumber \\
&&\hspace{-1.05cm}
           -  \frac{8}{5}
           -  \frac{n \pi^2}{ 3}
    \,\Bigg]
+ 8 a_1 \beta_1
     \Bigg[\,   S_1
           +  2 n S_2
           - \frac{7}{4}
           - \frac{ n \pi^2}{ 3}
    \, \Bigg]
+ 2 \beta_2
    \Bigg[\,  S_1
          + 2 n S_2
          - 1
          - \frac{ n \pi^2}{ 3}
  \, \Bigg]
\nonumber \\
&&\hspace{-1.05cm} + \beta_0 \beta_1 \Bigg[\,        S_1 \Bigg( 22 S_1
                         + 40 n S_2
                         - 44
                         - \frac{36}{ n}
                         - \frac{20 n \pi^2}{ 3}
                     \Bigg)
                +S_2 \Bigg(  8 n^2 S_2
                            + 14
                            - 16 n
                            - \frac{8 n^2 \pi^2}{ 3}
                     \Bigg)
\nonumber \\
&&\hspace{-1.05cm}
                + 64 n S_3
                - 40 n^2 S_4
                - 56 n S_{2, 1}
                + 32 n^2 S_{3, 1}
                + 8
                +\frac{(21 + 16 n) \pi^2}{ 6}
                + \frac{2 n^2 \pi^4}{ 9}
                + 48 n {\xi(3)}
\,\Bigg]
\nonumber \\
&&\hspace{-1.05cm}
 +  a_1 \beta_0^2
\Bigg[\,         S_1 \Bigg( 40 S_1
                           + 80 n S_2
                           - 116
                           - \frac{60}{ n}
                           - \frac{40 n \pi^2}{ 3}
                     \Bigg)
               + S_2 \Bigg(   20 n^2 S_2
                            + 40
                            - 72 n
                            - \frac{20 n^2 \pi^2}{ 3}
                     \Bigg)
\nonumber \\
&&\hspace{-1.05cm}
               + 140 n S_3
               - 100 n^2 S_4
               - 120 n S_{2, 1}
               + 80 n^2 S_{3, 1}
               + 48
               + (5 + 12 n) \pi^2
               + \frac{5 n^2 \pi^4}{ 9}
               + 100 n {\xi(3)}
\Bigg]
\nonumber \\
&&\hspace{-1.05cm} + \beta_0^3 \Bigg[\,
       S_1 \Bigg(
               4 S_1 \Big( 4 S_1
                        +  16 n S_2
                        -  19
                        - \frac{6}{ n}
                        - \frac{8 n \pi^2}{ 3}
                     \Big)
              + 8 S_2 \Big( 3 n^2 S_2
                        +  2
                        -  14 n
                        -  n^2 \pi^2
                     \Big)
\nonumber \\
&&\hspace{-1.05cm}
                + 104 n S_3
              - 120 n^2 S_4
               - 112 n S_{2, 1}
               + 96 n^2 S_{3, 1}
               + 80
               + \frac{64}{ n}
               + \frac{(58 + 56 n) \pi^2}{ 3}
               +\frac{2 n^2 \pi^4}{ 3}
\nonumber \\
&&\hspace{-1.05cm}
               + 120 n {\xi(3)}
           \Bigg)
     + S_2 \Bigg( -  4 n (17 +2 n) S_2
                  +  72 n^2 S_3
                  -  96 n^2 S_{2, 1}
                  +  64 n^3 S_{3, 1}
                  -  96
                  + 16 n
\nonumber \\
&&\hspace{-1.05cm}
                  - \frac{24}{ n}
                  - \frac{8 (5 - n) n \pi^2}{ 3}
                  - 8 n^2 {\xi(3)}
           \Bigg)
    +  S_3 \Bigg( - 16 n^3 S_3 +64 - 16 n - 20 n^2 \pi^2+ 32 n^3 {\xi(3)}
           \Bigg)
\nonumber \\
&&\hspace{-1.05cm}
     + S_4 \Bigg(68 n + 40 n^2 + \frac{64 n^3 \pi^2}{ 3}
          \Bigg)
     - 312 n^2 S_5
     + 144 n^3 S_6
     + S_{2, 1} \Bigg( 48 n
                     - 120
                     + 16 n^2 \pi^2
               \Bigg)
\nonumber \\
&&\hspace{-1.05cm}
     -32 S_{3, 1}\Bigg(\frac{15n }{2}
                     +  n^2
                     + \frac{n^3 \pi^2}{ 3}
               \Bigg)
     + 384 n^2 S_{3, 2}
     + 576 n^2 S_{4, 1}
     - 224 n^3 S_{4, 2}
     - 256 n^3 S_{5, 1}
\nonumber \\
&&\hspace{-1.05cm}
     + 256 n S_{2, 1, 1}
     + 64 n^2 S_{2, 2, 1}
     - 64 n^3 S_{2, 3, 1}
     - 448 n^2 S_{3, 1, 1}
     + 192 n^3 S_{4, 1, 1}
     - 8
     - \frac{8(2 +  n) \pi^2}{ 3}
\nonumber \\
&&\hspace{-1.05cm}
     - \frac{ (83 + 10 n)n \pi^4}{ 45}
     + \frac{4 n^3 \pi^6}{ 105}
     + {\xi(3)} \Bigg(48 - 80 n - 12 n^2 \pi^2 -16 n^3 {\xi(3)}\Bigg)
     - 40 n^2 {\xi(5)}
\, \Bigg] \,.
\nonumber\\
\end{eqnarray}
The results for the non-Coulomb part read
\begin{eqnarray}
\frac{f_2^{nC}}{16\pi^2} &=& C_F^2 \Bigg[ \frac{2}{3}
L_n-\frac{15}{8 n^2} +\frac{4}{3 n} + \frac{22}{9} -\frac{2}{3} S_1
\Bigg] + C_F C_A \Bigg[
 L_n+\frac{2}{n}+\frac{5}{4}-S_1
\Bigg],
\\
\frac{f_{3}^{nC}}{64\pi^2} &=& \Bigg[\,  \frac{7}{6} C_F^{3}
        + \frac{37}{12} C_AC_F^{2}
        + \frac{4}{3}C_A^{2}C_F
        + \beta _0
          \bigg(\,\frac{4}{3}C_F^{2}
                  + 2C_AC_F\bigg)
\Bigg]\, L_n^{2}
\nonumber \\
&&\hspace*{-1cm} + \, \Bigg[
 \, C_F^{3}\,
 \bigg(-\frac{3}{2}
         +\frac{14}{3n}
         -\frac{7S_1}{3}
 \bigg)
+  C_A C_F^{2}\,
 \bigg(\, \frac{226}{27}
         +\frac{8\ln{2}}{3}
         +\frac{37}{3n}
         -\frac{5}{3n^2}
         -\frac{37S_1}{6}
 \bigg)
\nonumber \\
&& \hspace*{-0.3cm}+\,C_A^{2}C_F\,
 \bigg(\, \frac{145}{18}
         +\frac{4\ln{2}}{3}
         +\frac{16}{3n}
         -\frac{8S_1}{3}\,
 \bigg)
+ C_F^{2}\,T_F
  \bigg(\frac{2}{15}
         -\frac{59}{27}\,n_f
  \bigg)
\nonumber
\\
&& \hspace*{-0.3cm} -\, \frac{109}{36}C_AC_FT_F\,n_f + \beta_0\,
  \Bigg\{C_F^{2} \,
            \bigg(
                  \frac{16}{3}
                 +\frac{10}{3n}
                 -\frac{75}{16n^2}
                 -\frac{\pi^2n}{9}
                 -\frac{4S_1}{3}
                 +\frac{2nS_2}{3}
            \bigg)
\nonumber
\\
&& \hspace*{-0.3cm} +\, C_A C_F\,
            \bigg(
                  \frac{15}{8}
                 +\frac{5}{n}
                 -\frac{\pi^2 n}{6}
                 -2S_1
                 + nS_2
            \bigg)
   \Bigg\}\,
\Bigg] \, L_n
+ \Bigg[\,\frac{1}{3}C_F^{3}
       +\frac{1}{2}C_AC_F^{2}\,\Bigg]\,
L_{m} L_n \nonumber
\\
&&\hspace*{-1cm} +\, \Bigg[\frac{1}{12}C_F^{3}
       +\frac{1}{8} C_AC_F^{2}\,\Bigg] \,
L_m^{2}
+ \Bigg[
   C_F^{3}\,
   \bigg(
          \frac{1}{12}
         +\frac{2}{3n}
         -\frac{S_1}{3}
   \bigg)
+ C_AC_F^{2}\,
   \bigg(
         -\frac{5}{9}
         +\frac{1}{n}
         -\frac{S_1}{2}
  \bigg)
\nonumber
\\
&& \hspace*{-0.3cm} +\, \frac{1}{15}C_F^{2}T_F \Bigg]\, L_m + \frac{
c_{\psi,3}^{nC}}{64\pi^2}\,.
\end{eqnarray}
The constant term of $f_{3}^{nC}$ is:
\begin{eqnarray}
\frac{c_{\psi, 3}^{nC}}{64\pi^2}
&=&
\Bigg[-\frac{44}{9}
      -\frac{109\pi^2}{864}
      -\frac{25}{6 n}
      +\frac{35}{12 n^2}
      + S_1 \,\bigg(\frac{3}{2}
                   -\frac{14}{3 n}
                   +\frac{7 S_1}{6}
              \bigg)
      -\frac{7 S_2}{6}
\Bigg] 
C_F^{ 3}
\nonumber
\\
&& \hspace{-1cm} +\,
\Bigg[\,
      \frac{64265}{3888}
     -\frac{2437\pi^2}{5184}
     +\frac{1475}{108n}
     +\frac{\pi^2}{9n}
     -\frac{321}{32n^2}
     + \ln{2}\,
       \bigg(\frac{16}{3}
            +\frac{16}{3n}
            -\frac{4\ln{2}}{3}
       \bigg)
\nonumber
\\
&&  -\,S_1\,
      \bigg(\,\frac{226}{27}
             +\frac{8\ln{2}}{3}
             +\frac{37}{3n}
             +\frac{1}{n^2}
             -\frac{37S_1}{12}
      \bigg)
   -S_2\, 
      \bigg(\frac{37}{12}
             +\frac{2}{3n}
      \bigg)
\Bigg]
C_AC_F^{2}
\nonumber
\\
&& \hspace{-1cm}
+\,\Bigg[
       \frac{1459}{162}
      +\frac{41\pi^2}{432}
      +\frac{133}{9n}
      +\ln{2}\,
       \bigg(
             \frac{8}{3}
            +\frac{8}{3n}
            -\frac{2\ln{2}}{3}
       \bigg)
      -\frac{4S_2}{3}
\nonumber
\\
&&       - \,S_1 
       \bigg(\frac{145}{18}
+\frac{4\ln{2}}{3}
              +\frac{16}{3n}
              -\frac{4S_1}{3}
       \bigg)
\Bigg] 
C_A^{2}C_F
+
\Bigg[
       \frac{1}{15}
      +\frac{4}{15 n}
      -\frac{2S_1}{15}
\Bigg]
C_F^{2}T_F
\nonumber
\\
&& \hspace{-1cm} 
+\,\Bigg[
      -\frac{599}{162}
      -\frac{5\pi^2}{108}
      -\frac{109}{18n}
      + \frac{109S_1}{36}
\Bigg] 
C_AC_FT_F\,n_f
\nonumber
\\
&& \hspace{-1cm} 
+\,\Bigg[
      -\frac{3473}{486}
      +\frac{5\pi^2}{81}
      -\frac{118}{27n}
      +\frac{125}{24n^2}
      + \frac{59S_1}{27}
\Bigg] 
C_F^{2}T_F\,n_f
\nonumber
\\
&& \hspace{-1cm} 
+\,\beta_0\,\Bigg[\,
\Bigg\{
    \frac{1027}{648}
   +\frac{19}{6 n}
   +\frac{25}{24n^2}
   -\frac{35\pi^2}{108}
   -\frac{11\pi^2n}{27}
   +\frac{5\pi^2}{16n}
   +\frac{4nS_3}{3}
   -\frac{2n S_{2,1}}{3}
\nonumber
\\
&& \hspace*{0.5cm}-\,S_1
   \bigg(\frac{10}{9}
         +\frac{1}{3n}
         +\frac{45}{16n^2}
         -\frac{\pi^2 n}{9}
         +\frac{2nS_2}{3}
   \bigg)
   + S_2
    \bigg(1
         +\frac{22n}{9}
         -\frac{15}{8n}
   \bigg) 
\Bigg\}\,
C_F^{ 2} 
\nonumber
\\
&&+\,\Bigg\{
    \frac{7}{24}
   -\frac{91\pi^2}{144}
   -\frac{1}{4n}
   -\frac{5\pi^2n}{24}
   - S_1 
   \bigg(\frac{3}{8}
        +\frac{1}{2n}
        -\frac{\pi^2n}{6}
        + nS_2
   \bigg)
  + S_2
   \bigg(
      \frac{3}{2}
     +\frac{5n}{4}
   \bigg)
\nonumber
\\
&& \hspace*{0.5cm} -\,n S_{2,1}
  + 2 n S_3
\Bigg\}\,
C_A C_F
\Bigg].
\end{eqnarray}
These results were given previously in \cite{Beneke:2007gj}, but now we
inserted the $O(\epsilon)$ terms of the one-loop coefficients of the 
$1/m^2$ potentials, see (\refI{eq:vp1epsterms})-- (\refI{eq:vm1epsterms}), 
as well as the previously unknown  $O(\epsilon)$ term $b_{2}^{(\epsilon)}$ 
of the two-loop $1/r^2$ potential, see (\refI{eq:b2epsnew}). 
Finally, the ultrasoft correction is 
\begin{eqnarray}
f_3^{us}/({64\pi^2})&=&
\Bigg[-2C_A^2C_F-{\frac{16}{3}}C_AC_F^2-{\frac{8}{3}}C_F^3\Bigg]\ln^2{\alpha_s}
\nonumber\\
&& \hspace*{-1.8cm}
+\Bigg[-\frac{5}{6}C_A^2C_F-{\frac{11}{6}}C_AC_F^2-{\frac{1}{3}}C_F^3\Bigg]L_m^2
+\Bigg[{\frac{8}{3}}C_A^2C_F+{\frac{20}{3}}C_AC_F^2+{\frac{8}{3}}C_F^3\Bigg]L_m \ln{\alpha_s}
\nonumber\\
&& \hspace*{-1.8cm}
+\Bigg[C_A^3+\left({\frac{52}{9}}-{\frac{8}{3}}\ln2
-4H_{n}\right)C_A^2C_F +\left(6 -\frac{10}{3n^2}
-{\frac{4}{3}}\ln2 -\frac{32}{3}H_{n} \right)C_AC_F^2
\nonumber\\
&& \hspace*{-1.3cm}+\left(-{\frac{52}{9}} - \frac{4}{3n^2} +8\ln2
-{\frac{16}{3}}H_{n} \right)C_F^3\Bigg]\ln\alpha_s
\nonumber\\
&& \hspace*{-1.8cm}
 +\Bigg[-{\frac{3}{4}}C_A^3+\left(-{\frac{11}{3}}+{\frac{5}{3}}\ln2
+{\frac{8}{3}}H_{n}\right)C_A^2C_F +\left(-{\frac{3}{2}}
+\frac{5}{3 n^2} +{\frac{1}{3}}\ln2 +{\frac{20}{3}}H_{n}
\right)C_AC_F^2
\nonumber\\
&& \hspace*{-1.3cm} +\left(5 +\frac{2}{3 n^2} -6\ln2
+{\frac{8}{3}}H_{n} \right)C_F^3\Bigg]L_m+\delta^{us}_n,
\label{eq:wf-ultrasoft}
\end{eqnarray}
where $H_n=\ln\frac{C_F}{n}-\frac{2}{n}+S_1(n)$. 
Numerical values for $\delta^{us}_n$ are given in 
table~\ref{tab:delta-us}.
The scale-dependence of the term $-\frac{3}{4} C_A^3 L_m$ in the last 
two lines cancels against the corresponding one in the third-order Coulomb 
potential contribution $f_3^C$. 

\begin{table}[t]
  \begin{center}
    \begin{tabular}{|c|c|c|c|c|c|c|}
 \hline
 {$n$}& {1}& {2}& {3}& {4}& {5} & 6 \\
 \hline
 $\delta^{us}_n$ & 353.06 & 256.62 & 224.26 & 206.88 & 195.48 & 187.16\\
 \hline
    \end{tabular}
   \caption{\label{tab:delta-us} \small Numerical result
   for the non-logarithmic part of the ultrasoft contribution,
   as defined in (\ref{eq:wf-ultrasoft}). $n$ denotes the principal quantum
   number.}
  \end{center}
\end{table}

\addcontentsline{toc}{section}{\numberline{}References}
\bibliographystyle{JHEP-2}
\bibliography{paper}

\end{document}